# Automatic Text Document Summarization using Semantic-based Analysis

*Thesis submitted to the Jawaharlal Nehru University*

*for the award of the degree of*

## DOCTOR OF PHILOSOPHY

**By**

**CHANDRA SHEKHAR YADAV**

**SCHOOL OF COMPUTER AND SYSTEMS SCIENCES**
**JAWAHARLAL NEHRU UNIVERSITY**
**NEW DELHI – 110067**
**INDIA**
**JULY 2018**

*Dedicated to My*

*Parents*

जवाहरलाल नेहरु विश्वविद्यालय

**SCHOOL OF COMPUTER AND SYSTEMS SCIENCES**

**JAWAHARLAL NEHRU UNIVERSITY**

# Declaration

This is to certify that the thesis entitled “Automatic Text Document summarization using semantic based Analysis” is being submitted to the School of Computer and Systems Sciences, Jawaharlal Nehru University, New Delhi, in partial fulfillment of the requirements for the award of the degree of **Doctor of Philosophy**, is a record of bonafide work carried out by me under the supervision of **Dr. Aditi Sharan**. This thesis contains less than 100000 words in length, exclusive Tables, Figures and bibliographies.

The matter embodied in the thesis has not been submitted in part or full to any University or Institution for the award of any other degree or diploma.

**Chandra Shekhar Yadav**
Enrollment No.
School of Computer and Systems Sciences
Jawaharlal Nehru University
New Delhi-110067, India

जवाहरलाल नेहरु विश्वविद्यालय

**SCHOOL OF COMPUTER AND SYSTEMS SCIENCES**

**JAWAHARLAL NEHRU UNIVERSITY**

# Certificate

This is to certify that the thesis entitled "Automatic Text Document summarization using semantic based analysis" submitted by **Mr. Chandra Shekhar Yadav**, to the School of Computer and Systems Sciences, Jawaharlal Nehru University, New Delhi, for the award of degree of **Doctor of Philosophy**, is a research work carried out by him under the supervision of **Dr. Aditi Sharan**.

**Supervisor**
Dr. Aditi Sharan
School of Computer and Systems Sciences
Jawaharlal Nehru University
New Delhi-110067, India

**Dean**
Prof. D. K Lobiyal
School of Computer and Systems Sciences
Jawaharlal Nehru University
New Delhi-110067, India

# Acknowledgments

No goal is achievable without guidance. For this, I am very glad to express my sincere gratitude and utmost regards to my supervisor **Dr. Aditi Sharan** for his guidance with many helpful discussions and his intellectual inputs to make thesis work worthy. His extensive research experiences were very helpful in my thesis. The most important thing is his helping nature and friendly behavior that contributed an important share in the fulfillment of this work. The methodology, philosophy, and problem solving methods suggested by him have been a great help in this work and would be afterwards too. I am grateful to my all the teachers of SC & SS, JNU especially **Prof. K.K. Bharadwaj, Prof. C. P. Katti, Prof. Karmeshu, Prof. D.P. Vidyarthi, Prof. S. N. Minz, Prof. R. K. Agrawal, Prof. D K Lobiyal, Prof. T. V. Vijay Kumar** and **Dr. Buddha Singh** for their Guidance, support and motivation whenever I feel depressed due to rejection or hard comments of my research papers.

I would like to express my thanks to Dean, SC & SS JNU for his support to pursue my work in the school. Also, I extend my thanks to the school administration, especially Chandra Sir, Meena Ma'am and Ashok Sir, librarian of SC & SS library and the Dr. B. R. Ambedkar central library of JNU for supporting me.

The work was carried out in Lab 001 which has a very healthy and friendly work culture. The regular lab discussions, exchange of ideas and opinions are part of it. I am thankful to Dr. Hazra Imran (UBC, canada), Ms. Manju Lata Joshi, Dr. Nidhi Malik, Dr. Jagendra Singh, Dr. Mayank Saini, Dr. Sifatullah Siddiqi, Dr. Sonia, Mr. Vikrant Vaish, Mr. Ashish, Mrs. Sheeba Siqqique, Ms. Bhawana Gupta all the support during my stay in Lab 001, SC & SS, JNU. Special thanks to Mr. Rakesh Kumar and Miss. Payal Biswas for encouraging and helping me in writing the thesis. It was wonderful to stay in the Lab and lifetime memorable for me.

The students in SC&SS are very much helpful, supportive as well as courageous to each other. I am thankful to all my seniors, juniors, fellow students and friends. I would like to thank Dr. Kapil Gupta (NIT-K), Dr. Neetesh Kumar (IIITM Gwalior), Dr. Vipin (MGCU), Dr. Navjot (MNNIT), Dr. Yogendra Meena (DU), Dr. Raza Abbas Haidri (DU), Dr. Gaurav Baranwal (BHU), Mr. Prem Yadav (DTE-UP), Mr. Harendra

Pratap (DU), Mr. Sumit Kumar, Mr. Krishan Veer (DU), Dr. Dinesh Kumar (CUH), Mr. Utkarsh (NIT-D), Mr. Dhirendra (DTU), and those people who always encouraged and motivate me whenever I feel depressed during my work.

I am thankful to Director SLIET, Dr. Shailendra Jain and all my colleagues in computer science department, Dr. Major Singh (H.O.D, CSE), Dr. Damanpreet Singh, Dr. Birmohan Singh, Dr. Manoj Sachan, Dr. Sanjeev Singh, Mrs. Gurjinder Kaur, Dr. Vinod Verma, Mr. Jaspal Sigh, Mr. Manminder Singh, Mr. Rahul Gautam, Mrs. Preetpal Kaur, Dr. Sanjeev Singh (electrical, SLIET), Mr. Pankaj Das (electronics, SLIET), Dr. Amit Rai (chemical, SLIET) and Mr. Jonny Singla (mechanical, SLIET).

Thanks to my wonderful family for bearing with me as I am. I was blessed with the love, compassion, supplications, and support of my family members during this period. My deepest gratitude goes to them for their selfless love, care and indulgence throughout my life, this thesis was simply impossible without them. My heart goes out in reverence to my respected father **Mr. Padam Singh Yadav** and dear mother **Mrs. Suvita Devi** for their blessings, sacrifices, tremendous affection and patience. I am also indebted to my sisters **Mrs. Rinkesh Yadav,** and my younger brother **Mr. Kesari Singh Yadav** for their unconditional love and blind faith in me and also for their continuous motivation in my tough time.

Finally, I would like to express thanks to each person who is directly or indirectly related to my work. Also, I would like to thank the University Grant Commission (UGC), and Council of Scientific & Industrial Research (CSIR), India for its research fellowship during the period.

**Chandra Shekhar Yadav**
Enrollment No.
School of Computer and Systems Sciences
Jawaharlal Nehru University
New Delhi-110067, India

# Abstract

Since the advent of the web, the amount of data on wen has been increased several million folds. In recent years web data generated is more than data stored for years. One important data format is text. To answer user queries over the internet, and to overcome the problem of information overload one possible solution is text document summarization. This not only reduces query access time, but also optimize the document results according to specific user's requirements.

Summarization of text document can be categorized as abstractive and extractive. Most of the work has been done in the direction of Extractive summarization. Extractive summarized result is a subset of original documents with the objective of more content coverage and lea redundancy. Our work is based on Extractive approaches.

In the first approach, we are using some statistical features and semantic-based features. To include sentiment as a feature is an idea cached from a view that emotion plays an important role. It effectively conveys a message. So, it may play a vital role in text document summarization.

The second work in extractive summarization dimensions based on Latent Semantic Analysis. In this document are represented in the form of a matrix, where rows represent concepts that cover different dimensions and columns represents documents. LSA has the ability to be mapped to the same concept space. LSA can hold synonyms relations, since the mapping of the same concepts. Based on SVD decomposition and concepts of entropy we find most informative concepts and sentences.

Since LSA cannot hold polysemy relations so to extend this work we have used wordnet relations to handle all relationships among words/sentences. In third work, a lexical network has been created and most informative sentences extracted based on that.

To extend lexical chain based work, we have designed an optimization function to optimize content coverage and redundancy along with length constraints. Since the nature of function is linear, constraints are also linear, so we have applied Integer Linear Programming to find a solution.

# Contents

# List of Figures

# List of Tables

# Chapter 1: Introduction to Automatic Text Document Summarization

## 1.1 Introduction

The Internet is defined as the worldwide interconnection of individual networks that is being operated by government, industry, academia, and private parties. Originally the Internet served to interconnect laboratories engaged in government research, and since 1994 it has been expanded to serve millions of users and a multitude of purposes in all parts of the world. In December 1995, only 16 million internet users were present that was 0.4 % of the world population, and in June 2017 this became 3,885 million people that are 51.7 % of world population. According to an IBM Marketing Cloud study, 90% of the data on the internet has been created since 2016. People, businesses, and devices have all become data factories that are pumping out incredible amounts of information to the web each day. The size of digital data can be understood by the given statistics.

I. Since 2013, the number of tweets each minute has increased 58% to more than 455,000 tweets per minute in 2017.
II. Every minute on Facebook, 510,000 comments are posted, 293,000 statuses are updated, and 136,000 photos are uploaded.
III. 3,607,080 Google searches are conducted worldwide each minute of every day.
IV. Worldwide, 15,220,700 texts are sent every minute!
V. 1,209,600 new data are producing social media users each day.
VI. 656 million tweets per day!

According to Internet and Mobile Association of India (IMAI) and LiveMint's article published on 2 March 2017, the number of Internet users in India were expected to reach 450-465 million by June 2017, up 4-8% from 432 million in December 2016. Due to all this, information is increasing day by day, that leads to information overload. This is termed as information glut and data smog. Information overload occurs when the amount of input to a system exceeds its processing capacity. Decision makers have fairly limited cognitive processing capacity. Consequently, when information overload occurs, it is likely that a reduction in decision quality will occur. Information overload can be dealt with up to a certain level by the representation of concise information. As most of the information on the web is textual, so efficient text summarizer can concisely represent textual information on the web.

(Radev, Hovy, & McKeown, 2002) have outlined a summary as "a text that is produced from one or more texts, which conveys important information in the original text(s), and that is no longer than half of the original text(s) and usually significantly less than that." This simple definition captures three important aspects.

1. The summaries may be produced from a single document or multiple documents.
2. The summaries should preserve important information.
3. The summaries should be short.

As explained by (Alguliev, Aliguliyev, & Mehdiyev, 2011) automatic text document summarization is a task of interdisciplinary research area from computer science including artificial intelligence, statistics, data mining, linguistic, and psychology. (Torres-Moreno, 2014) has defined an automatic summary as “a text generated by software that is coherent and contains a significant amount of relevant information from the source text”. (Sakai & Sparck-Jones, 2001) defines a summary as “a reductive transformation of source text into a summary text by extraction or generation”. (Mani & Maybury, 2001) states that “a summary is a document containing several text units (words, terms, sentences or paragraphs) that are not present in the source document.”

Text summarization has various applications in accounting, research, and efficient utilization of results. A text document summarization based real life system is “Ultimate Research Assistant” was developed by (Hoskinson, 2005). Their system performs text mining on Internet search. In other work (Takale, Kulkarni, & Shah, 2016) have highlighted the applications of text document summarization in search engine as Google. Another system Newsblaster, proposed by (McKeown et al., 2003) that automatically collects, cluster, categorize and summarize news from different websites like CNN, Reuters, etc. This provides a facility for users to browse the results. (Hovy & Lin, 1999) introduced SUMMARIST system to create a robust text summarization system, a system that works on three phases which can describe in the form of an equation-like "Summarization = Topic Identification + Interpretation + Generation." An application of summarization is News summarization by “inshorts App” that is shown in Figure 1.1, and google snippet generation in Figure-1.2.

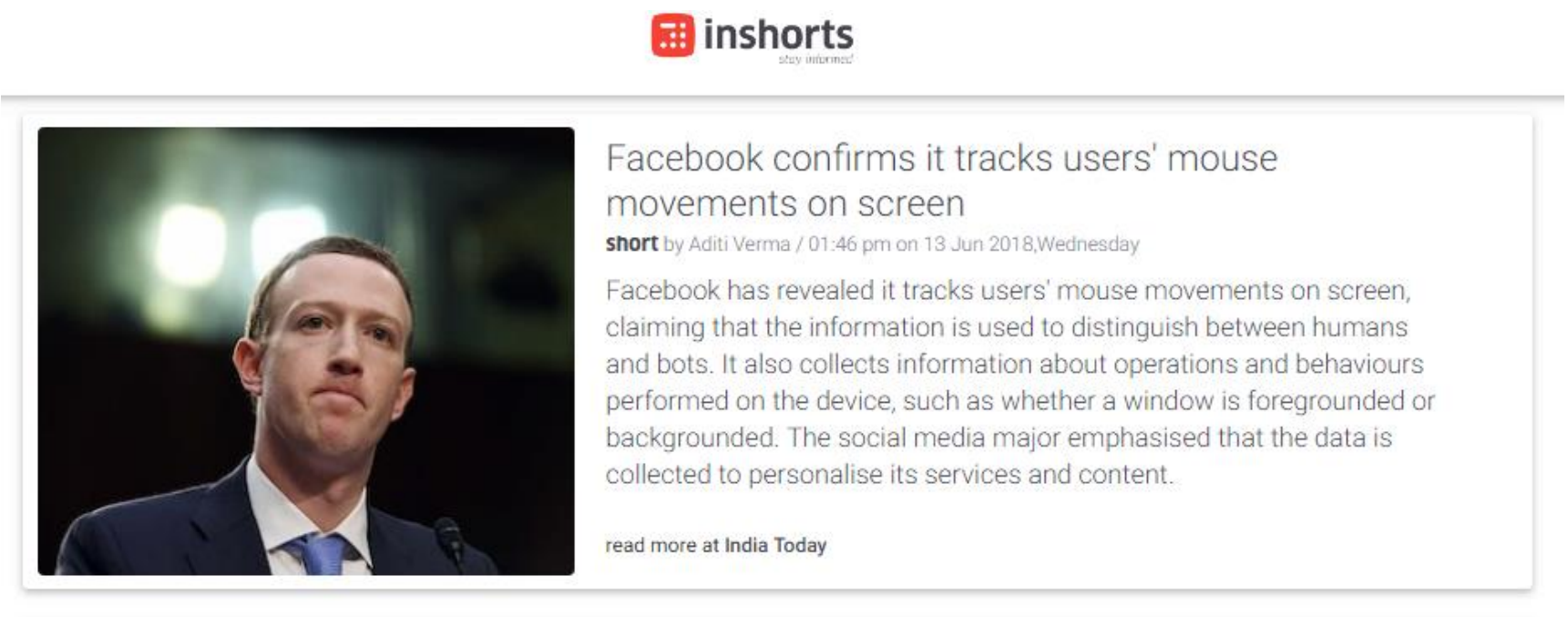

**Figure-1.1:** Showing inshorts App's news

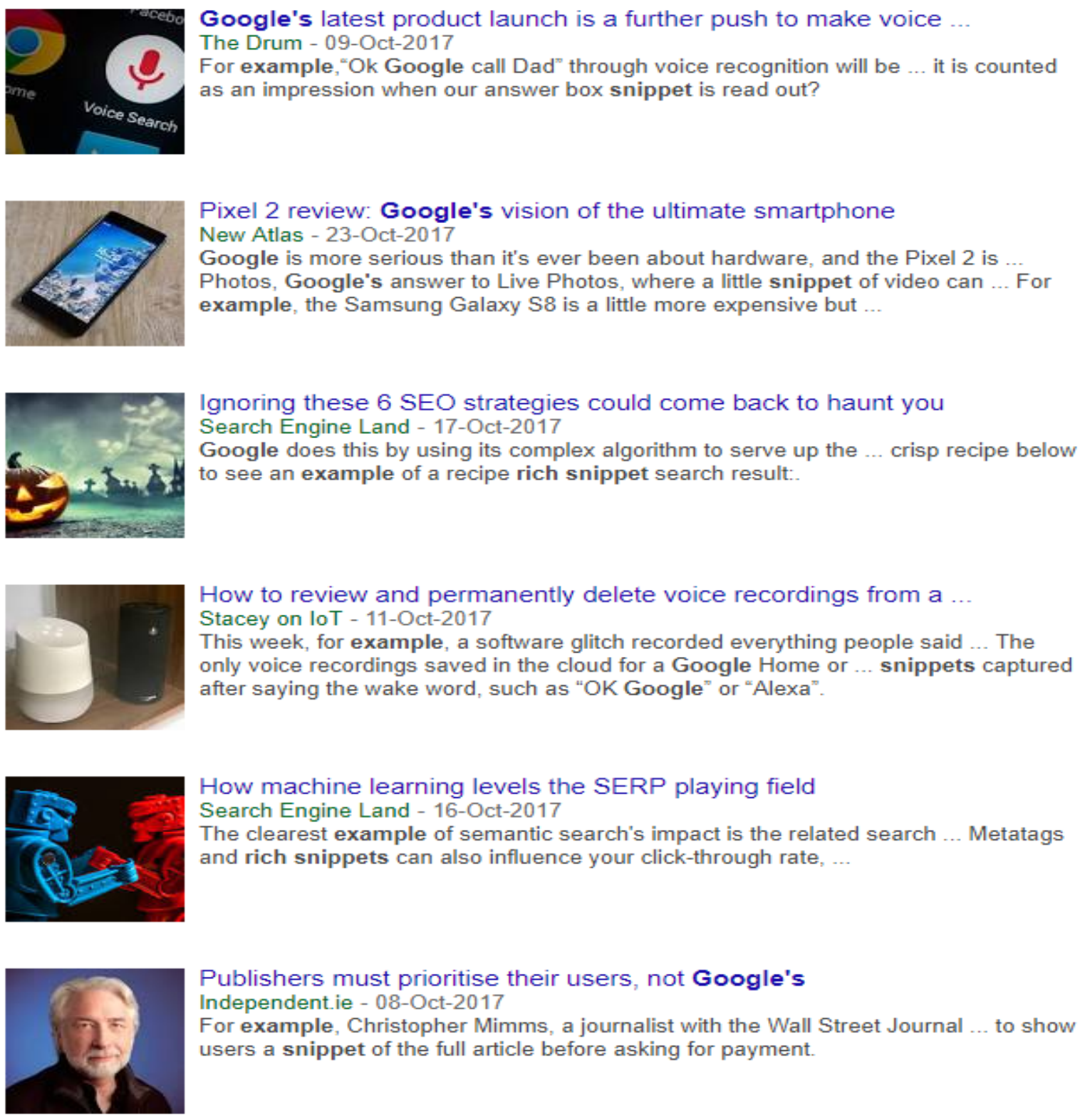

**Figure-1.2**: Google Snippet Example

Broadly summarization task can be categorized into two type extractive summarization and abstractive summarization. Abstractive summarization focus on the human-like summary. Extractive summarization is based on extractive entities, entities may be a sentence, subpart of a sentence, phrase or a word. Till now this extractive based summarization, relies on standard features like sentence position, sentence length, frequency of words, combination of Local_Weight and Global_Weight as TF-IDF score, selection of candidate word as Nouns, Verbs, cue words,

Digit/ numbers present in the text, Uppercase, bold letter words, Sentiment of words/sentences, aggregate similarity, centrality, etc. The goal of feature-based summarization either alone or a mix of different strategy is to find salient sentences which can include in the summary. This process can be shown in Figure 1.3.

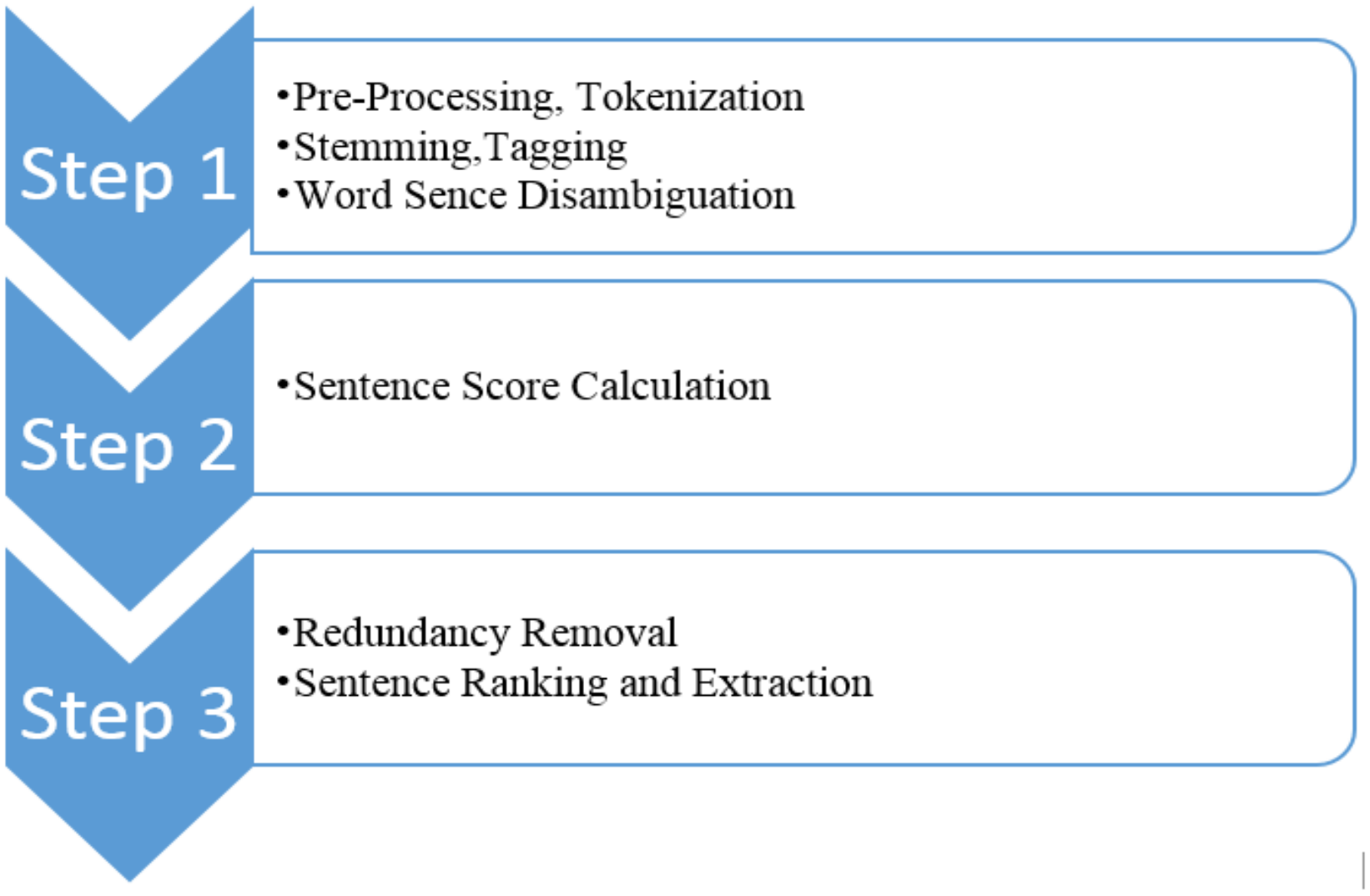

**Figure-1.3**: Three step process for Text Document Summarization

## 1.2 Flavors of summarization

Due to overlapping of summarization techniques, these cannot categorize fairly. In this ***section,*** we are trying to present a reasonable overview of existing techniques.

### 1.2.1 Extractive and Abstractive Summarization

Broadly summarization task can be categorized into two type, abstractive summarization, and extractive summarization. Abstractive summarization is a more human-like summary, which is the actual goal of text document summarization. As defined by (Mani & Maybury, 1999), and (Wan, 2008) abstractive summarization needs three things as Information Fusion, Sentences Compression, and Reformation. The actual challenge in Abstractive summarization is generation of new sentences, new phrases, along with produced summary must retain the same meaning as the same source document has. According to (Balaji, Geetha, & Parthasarathi, 2016) abstractive summarization requires semantic representation of data, inference rules, and natural language generation. They have proposed a semi-supervised bootstrapping approach to identify relevant components for an abstractive summary generation. A study was done by (Goldstein, Mittal, Carbonell, & Callan, 2000) state that human-generated summary also varies from person to person, the reason of this is maybe the setup of the human mind, domain knowledge, and interest in the particular domain, etc.

Extractive summarization is based on extractive entities, entities may be a sentence, subpart of sentence, phrase or a word. Till now most work is done on extractive summarization because, extraction is easy because this is based on some scoring criteria of words, sentences, phrases, and evaluation of extractive summary is easy because it is just based on word counts or word sequences. Our work is focused on extractive based technique.

### 1.2.2 Single and Multi-Document Summarization

According to (Ou, Khoo, & Goh, 2009) single document summarization can be defined as a "process of representing the main content of one document", and Multi-document Summarization also a process "of representing the main content of a set of related documents on a topic, instead of only one document". There are two possible approaches for multi-document summarization in the first approach, combine all documents in the single document then apply single document summary. The second possible approach generates a summary for each document, then combine all summary into one document later perform single document summarization on the combined summary to get a multi-document summary. Whereas, according to (Sood, 2013), It is important to note that concatenation of individual single document summaries may not necessarily produce a multi-document summary. Since, the issue with the later approach (first generate single summary and then again combine to generate a summary) is that in this process relative sentences position changes, and coherence lost, and this research gap opens new dimensions in research.

### 1.2.3 Query-Focused and Generic Summarization

A text document may contain several topics, like social or economic development, political views, common people's views, environment, or entertainment. Someone may be interested only in one angle, so there is a need for specific text from all given text. To full fill, this requirement user may give a query "Q" on document "D" after certain similarity the system will return desired documents. This is called query-focused summarization. First (Tombros & Sanderson, 1998) have proposed query-focused summarization for text document to develop an information retrieval system. While (White, Jose, & Ruthven, 2003) have extended it for use in web document summarization by combining it with other features including text-formatting of the page along with query dependent features. The rationale behind their approach, with which we concur, is that the words in the query should be included in the generated summary. Another type of summarizer system is generic summarizer that is regardless of user need.

### 1.2.4 Personalized summarization

According to (Dolbear et al., 2008), personalization can be defined as, this is the technology that enables a summarizer system to harmonize between differently available contents, its applications as well as user interaction modalities to a user's stated and system's learned preferences. The main objective of personalization is to enable the system for content offerings to be closely targeted user's desires. This can be achieved via different methods like content filtering that extract contents appropriate to a user's preferences from a set of available content, and give recommendations that provides content to a user based on various criteria which may include the user's previous acceptance of related content or on the consumption of related content by a peer group.

### 1.2.5 Guided Summarization

This is an extension of query-focused summarization, but instead of the single question, there are set of question. Guided summarization may be seen as template-based summarization. The template is a set of question that fired on a text document, and the system returns a summary in the form of question answers. Sometimes this method works well if the developer has good domain knowledge and accurate predictor of the question shortly. If we consider any disaster example, then set of question or theme will be based on, the cause of the accident, how many killed, how many are in critical and normal condition, relief measure, is any political visit held during this, and compensation paid to effective people. It leads to the production of the much-focused summaries concerning the questions raised.

### 1.2.6 Indicative, Informative, and critical summary

(Hahn & Mani, 2000), has defined several kinds of summary as, Indicative summaries follow the classical information retrieval approach: They provide enough content to alert users to relevant sources, which users can then read in more depth. Informative summaries act as substitutes for the source, mainly by assembling relevant or novel factual information in a concise structure. Critical summaries (or reviews), besides containing an informative gist, incorporate opinion statements on content. They add value by bringing expertise to bear that is not available from the source alone. A critical summary of the Gettysburg Address might be: The Gettsyburg Address, though short, is one of the greatest of all-American speeches, with its ending words being especially powerful "that government of the people, by the people, for the people, shall not perish from the earth."

## 1.3 State of art approach in summarization

According to the state of approaches, summarization procedure can be classified into following part; this is not limited to that.

**Linguistic Structure:** Cohesion has introduced by (Halliday & Hasan, 1976), it captures the intuition. This is a technique for "sticking together" different textual unit of the text. Cohesion can achieve through the use of semantically related terms, like coreference, conjunctions, and ellipsis. Among the different cohesion building devices 'lexical cohesion' is the most easily identifiable and most frequent type, and it can be a very important source for the 'flow' of informative content.

**Centroid and Cluster:** In this approach, documents are divided into several units, units may be document itself, paragraphs, and sentences. Based on some criteria some clusters can be created. Some famous criteria's are like Cosine, NGT, Vector-based similarity. After clustering, summarizer system picks one unit from each cluster that is considered representative of that cluster, and later that added to summary. This approach can be applied for single and multi-documents.

**Machine Learning:** Generally, we talk about extractive summarization. It is based on either statically based feature or linguistic features or its hybridization. Machine learning based summarization is more effective because it learns features weights from given data, and later learned weight can be used on test data. Only required for this is labeled data.

**Multi-Objective:** The main concern about the summary is that it should be more informative and length constraints. Informative constraints can be designed by reducing redundancy and increasing coverage. So, in this approach mostly all authors designed a function in such a way to reduce redundancy, increase coverage, and length constraints. Later that function can be optimized using different techniques.

## 1.4 Corpus Description

In Chapter 2 we used a dataset that was created by us, details about this dataset are mentioned in ***section 1.4.1***, this experiment also repeated on standard dataset DUC 2002. Rest of the works relies on only DUC-2002 dataset. Details about DUC dataset is described in section 1.4.2.

### 1.4.1 Corpus Description for Hybrid Approach

"On 16 June 2013, was a multi-day cloudburst centered on the North Indian state of Uttarakhand caused devastating floods along with landslides and became the country's worst Natural Disaster. Though some parts of Western Nepal, Tibet, Himachal Pradesh, Haryana, Delhi and Uttar-Pradesh in India experienced the flood, over 95% of the casualties occurred only in Uttarakhand. As of 16 July 2013, according to figures provided by the Uttarakhand Government, more than 5,700 people were presumed dead" Uttrakhand flood (2015). Corpus is self-designed, taken from various newspapers ex. "The Hindu," "Times of India." This Dataset is also published in paper (C.S. Yadav, Sharan, & Joshi, 2014), (Chandra Shekhar Yadav & Sharan, 2015). Here we are showing some statistically, and linguistic statics about our DataSet used.

**Statistical statistics**

Total No. of Sentences in document 56,

Length of the document after stop word removed: 1007,

Total number of distinct words: 506,

Minimum sentence Length 6 words,

Maximum sentence Length 57 words,

Average sentence length is 1454/56 = 25.96.

In our experiment, we used a SQL stopword list, which is available at http://dev.mysql.com/doc/refman/5.5/en/fulltext-stopwords.html. By seeing the Figure-1.5, we can interpret that 45 sentences are between length 10 and 40, and 36 sentences are between length 15 and 35.

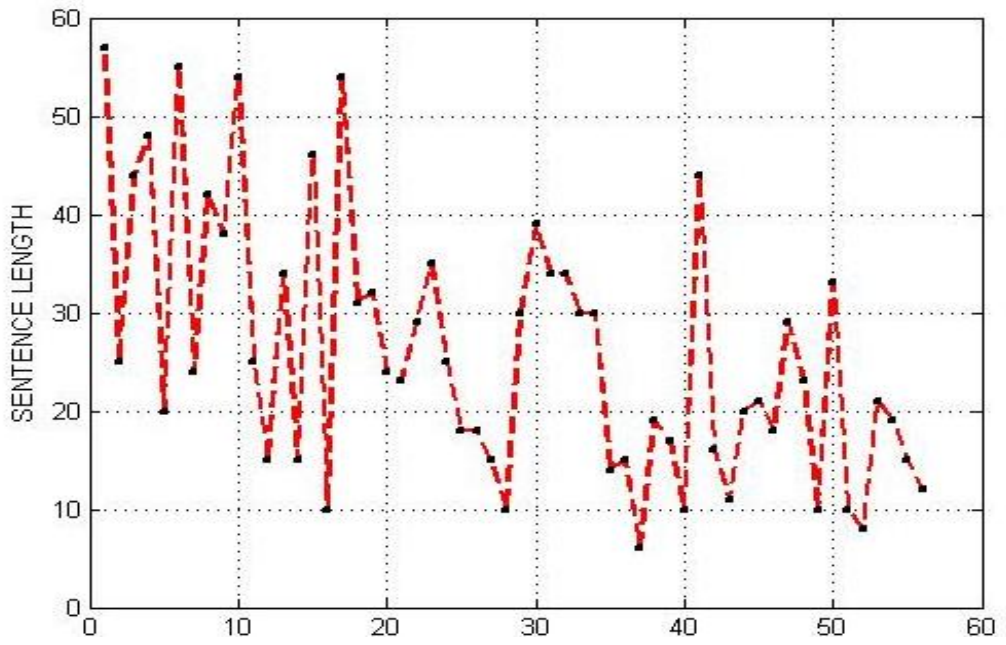

**Figure-1.5**: Sentence Length (Y-axis) Vs Sentence Number (X-axis)

**Linguistic statistics**

In linguistic me are analyzing number and nature of significant entities, it is like.

(1) 'NN': 208, 'NNP': 196, 'NNS': 131 ;

(2) 'DT': 150 ; (3) 'JJ': 70, 'JJR': 7 'JJS': 2 ;

(4) 'VB': 37, 'VBN': 64, 'VBD': 48 'VBZ': 38, ',': 38, 'VBG': 37, , 'VBP': 22

where different abbreviation stands for [ "NN-Noun, singular/mass, NNS-Nounplural, NNP-Proper noun singular, NNPS-Proper noun plural, VB-Verb, VBD-verb past tense, VBG-verb gerund, VBN-verb past participle, VBP-verb non-3rd person singular, VBZ-verb 3rd person singular, JJ-Adjective, JJR-Adjective comparative, JJS-Adjective superlative, DT-Determinant"]. [**NOTE 1**: 'X':10 means, X is entity type and 10 is its count].

### 1.4.2 DUC 2002 Dataset

The document sets are produced using data from the Text Retrieval Conference (TREC) disks used in the question-answering track in TREC-9. This dataset includes data from, Wall Street Journal (1987-1992), AP newswire (1989-1990), San Jose Mercury News (1991), Financial Times (1991-1994), LA Times from disk 5 and FBIS from disk 5. Each set average has ten documents, with at least ten words, no maximum length is defined. There is single text document abstract for each document with around a hundred words long. The multi-document abstract is divided into four parts according to two hundred, one hundred, fifty, and ten words long. Each document is divided into four sets, and set categories are following.

1. Single natural disaster event and created within at most a seven-day window.
2. Single event in any domain and created within at most a seven-day window.
3. Multiple distinct events of a single type (no limit on the time window).
4. Documents that contain biographical information mostly about a single individual.

## 1.5 Summary Evaluation

To find a good summary lot of work done, but to decide the quality of the summary still a challenging task due to several dimensions, like length constraints, different writing styles, and lexical usage, i.e., the context in which used. Research is done by (Goldstein et al., 2000) they conclude that (1) "even human judgment of the quality of a summary varies from person to person", (2) only little overlap among the sentences picked by people, (3) "human judgment usually doesn't find concurrence on the quality of a given summary". Hence, it is sometimes confusing, i.e., tedious to measure the quality of the text summary. Summary evaluation can be done by content-based, and task-based both are explained in the sub sections.

### 1.5.1 Content-Based

Content-based measure evaluate summary on the presence of textual units, i.e. n-gram in peer summary and standard summary. Example of content-based measures is ROUGE, BLUE, Pyramid.

#### *1.5.1.1 ROUGE*

For evaluation, most of the researchers are using the "Recall-Oriented Understudy for Gisting Evaluation" (ROUGE) introduced by (C.-Y. Lin, 2004), and DUC has officially adopted this for summarization evaluation model. ROUGE compares system generated summary with different

$$ROUGE-N=\frac{\sum_{S\in\{\text{ReferencesSummaries}\}}\sum_{\text{gram}_n\in S}Count_{match}(n-gram)}{\sum_{S\in\{\text{ReferencesSummaries}\}}\sum_{\text{gram}_n\in S}Count(n-gram)} \quad (1.1)$$

model summaries. It has been considered that ROUGE is an effective approach to measure document summarizes so widely accepted. ROUGE measure overlaps words between the system summary and standard summary (gold summary/human summary). Overlapping words are measured based on N-gram co-occurrence statistics, where N-gram can be defined as the continuous sequence of N words. Multiple ROUGE metrics have been defined for the different value of N and different models (like LCS, weighted ROUGE, S*, SU*, with and without lower/upper case matching, stemming, etc.). Standard ROUGE-N is defined by Equation-1.1,

Here N stands for the length of the N-gram, Count($gram_n$) is the number of N-grams present in the reference summaries, and the maximum number of N-grams co-occurring in the system summary, the set of reference summaries is $Count_{match}(gram_n)$ ROUGE measures generally gives three basic score Precision, Recall, and F-Score. Since the ROUGE-1 score is not sufficient indicator for summarizer performance, so another variation of ROUGE is; ROUGE-N, ROUGE-L, ROUGE-W, ROUGE S*, ROUGE SU*. In our evaluation, we are using six ROUGE measure (N= 1to 2, L, W, S*, and SU*), W=1.2 taken. Since DUC- 2002 task about to generate single document summary about 100 words, so we are evaluating Summary of first 100 words. As mentioned earlier (in the abstract) we are using DUC-2002 Dataset, category two which is about a single event in any domain and created within almost a seven-day window (as per DUC-2002 guidelines). Recall, and Precision, defined by following Equations-1.2, and 1.3, since we are using simple F-Score so, in our evaluation, we put β=1 In our results we are showing only F-Score.

$$\Pr ecision = \frac{Count_{match}(Sentence)}{Count_{candidate}(Sentence)} \tag{1.2}$$

$$\operatorname{Re} call = \frac{Count_{match}(Sentence)}{Count_{bestsentence}(Sentence)} \tag{1.3}$$

F-Score is given by the Harmonic mean of Precision and Recall, in Equation-1.4 we are representing fuzz F-score.

$$F - score = \frac{(1+\beta^2) \times \operatorname{Re} call \times \Pr ecision}{\operatorname{Re} call + \beta^2 \times \Pr ecision} \tag{1.4}$$

ROUGE-N measures n-grams, uni-gram, bi-gram, tri-gram and higher order n-gram overlap, ROUGE-L measure LCS (Largest common subsequence), the advantage ROUGE-L over ROUGE-N is that it doesn't require consecutive matches, and this doesn't define n-gram length in prior. If X is reference summary and Y is candidate summary, and its length is m and n respectively, then lcs based precision, recall, F-score can be defined by equation 1.5-1.7. In DUC (document understanding conference) $\beta$ is set to large quantity as 0.8.

$$R_{lcs} = \frac{LCS(X,Y)}{m} \tag{1.5}$$

$$P_{lcs} = \frac{LCS(X,Y)}{n} \tag{1.6}$$

$$F_{lcs} = \frac{(1+\beta^2)R_{lcs}P_{lcs}}{R_{lcs} + \beta^2 P_{lcs}} \tag{1.7}$$

Another variant is ROUGE-S where S stands for skipping bigram. Skip bigram allows maximum two words gap between lexical units. This can be understood by an example for the phrase "cat in the hat" then the skip-bigrams are following "cat in, cat the cat hat, in the, in hat, the hat." SKIP2(X, Y) is the number of skip bigram matches between X and Y, C is combination function, $\beta$ is to control relative importance of $P_{skip2}$ and $R_{skip2}$. Skip bigram based Precision, Recall and F-score are given by equation 1.8-1.10.

$$R_{skip2} = \frac{SKIP2(X,Y)}{C(m,2)} \tag{1.8}$$

$$P_{skip2} = \frac{SKIP2(X,Y)}{C(n,2)} \tag{1.9}$$

$$F_{skip2} = \frac{(1+\beta^2)R_{skip2}P_{skip2}}{R_{skip2} + \beta^2 P_{skip2}} \tag{1.10}$$

Another measure is ROUGE-SU*, it measures skip-bigram count between peer summary and standard summary to find out the similarity between these two summaries. This measure is quite sensitive to word order without considering consecutive matches. Uni-gram matches are also included in this measure, to give credit to a candidate sentence if the sentence does not have word pair co-occurring with its reference. Recall, Precision and F-measure are calculated in the following manner.

*1.5.1.2 BLEU*

The BLEU method was proposed for automatic evaluation of machine translation system. The primary programming task for a BLEU implementer is to compare *n*-grams of the candidate with the *n*-grams of the reference translation and count the number of matches. These matches are position independent. The more the matches, the better the candidate translation is.

*1.5.1.3 Pyramid*

Two kinds of the summary are generated one is system generated, i.e., peer summary and another human generated the summary, i.e., reference summary. Using reference summary content units (SCU) is find out, and using a set of the same words pyramid is constructed. In this evaluation method, peer summary contributor's, i.e., each lexical unit in a Summary or SCU are matched against SCU in the pyramid. The advantage of the pyramid method is that it evaluates summary along with it tells the idea of how the summary is chosen. Best result in this method is obtained with unigram overlap similarity and single link clustering. In this whole process to evaluate summary user required many reference summaries.

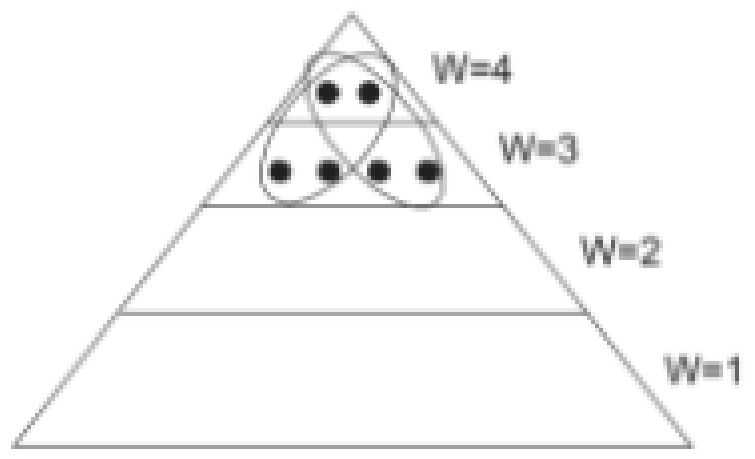

**Figure-1.4**: Two of six optimal summaries with 4 SCUs

In Figure 1.4, higher weight SCU is placed on the top of the pyramid, and less weighted SCUs of weight is placed at the bottom. This is reflecting the fact that fewer SCUs are more probable in all the summaries, compare to two, three and so on.

### 1.5.2 Task-Based

They try to measure the prospect of using summaries for a certain task. We mention the three most important tasks – document categorization, information retrieval, and question answering. For a given text document first, we have to develop a summarizer system and get a concise summary from it. Let summary is "S", now, according to task-based evaluation, we have to fire queries on "S". For example in question answering task, for a given set of question, we will measure precision and recall of queries response. It will decide the quality of summary and summarizer system.

### 1.5.3 Readability

In Text analysis conference (TAC 2009) and TAC-2010 "Automatically Evaluating Summaries of Peers" (AESOP) task, the focus was on developing automatic metrics that can measure summary content on the system level. In TAC 2011, a new task is introduced to evaluate for participant's ability to measure summary readability, both on the level of summarizers as well as on individual summaries. To measure the readability of the summaries, it accessed based on five linguistic-based criteria such as Grammaticality correctness, Nonredundancy of lexical units, Referential clarity, focus of summary text, and structure and coherence. Humans evaluated peer summaries based on these five linguistic questions and assigned a different score on a five-point scale one to five, where one represents worst and five for the best summary.

## 1.6 Thesis objective

This Thesis is divided into six chapters. This work is about extractive summarization techniques, with a focus on how semantic features can be used for summarization. First chapter about introduction of text summarization. In second chapter, we are proposing a hybrid model for a single text document summarization. This model is an extraction-based approach, which is a combination of statistical and semantic technique. The hybrid model depends on the linear combination of statistical measures, sentence position, TF-IDF, Aggregate similarity, centroid, and semantic measure. In this work we will show the impact of sentiment feature in summary generation and will find an optimal feature weight for better results. For comparison, we will generate different system summaries using proposed work, MEAD system, Microsoft system, OPINOSIS system, and Human generated summary. Evaluation of the summary will be done by content-based measure ROUGE.

In the third chapter, we are proposing three models (based on two approaches) for sentence selection that relies on LSA. In the first proposed model, two sentences are extracted from the right singular matrix to maintain diversity in the summary. Second and third proposed model is based on Shannon entropy, in which the score of a Latent/concept (in the second approach) and Sentence (in the third approach) is extracted based on the highest entropy. In this work we will propose a new measure to measure the redundancy in the text.

In the fourth chapter, we will present Lexical network based a new method for ATS. This work is divided into three different objectives. In the first objective, we will construct a Lexical Network. In the second objective after constructing the Lexical Network, we will use different centrality measures to decide the importance of sentences. Since WSD is an intermediate task in text analysis, so in third objective, we will do an analysis how the performance of centrality measure is changing over the change of WSD technique in an intermediate step and cosine similarity threshold in a post-processing step.

In the fifth chapter, we will present an optimization-based criteria for Automatic Text document Summarization. This is based on three steps; First preprocessing of sentences and output goes to the Second stage that concern about Lexical Network creation. The Output of module-2 is Lexical-Network and Importance of sentences given by Betweenness Centrality score. In the Final Module, we will decide some optimization criteria using a combination of centrality and Lexical Network. To solve this objective criterion, we will use ILP (Integer Linear Programming) to find a solution, i.e., which sentences to extract in summary.

Aligned with the means stated above, the objectives of this thesis are as follows:

- Proposing a statistical and semantic feature-based hybrid model for text document summarization,
- Proposing a LSA based model, which also captures the linguistic feature of the text,
- Proposing a new summary evaluate measure based on information contains,
- To maintain the syntactic and semantic property of the text, create a lexical network to find out lexical unit for summarization,
- Based on the previous objective, create a new objective function to optimize and expect a better summary.

# Chapter 2: Hybrid Approach for Single Text Document Summarization using Statistical and Sentiment Features

## 2.1 Introduction

In this chapter, we are proposing a hybrid method for single text document summarization. That is, a linear combination of statistical features proposed in the past and a new kind of semantic feature that is sentiment analysis. The idea is to include sentiment analysis as a feature in summary generation is derived from the concept that, emotions play an important role in communication to effectively convey any message. Hence, it can play a vital role in text document summarization. For comparison, we are using different system summaries as MEAD system, Microsoft system, OPINOSIS system, and human-generated summary. Evaluation is done using content-based measure ROUGE.

## 2.2 Literature Work

Till now, most of the research has done in the direction of extractive summarization based approaches. In extractive summarization the important the task is to find informative sentences, a subpart of sentence or phrase and include these extractive elements into the summary. Here we are presenting work done in two categories, early category work and recent work done.

The early work in document summarization has started on single text document, by (Luhn, 1958). He has proposed a frequency-based model, in which frequency of words plays a crucial role, to decide the importance of any sentence in the given document. Another work, of (Baxendale, 1958), was introduced a position based statistical model. In his research, he has found that, starting and ending sentences are more informative in summary generation. Position based measure works well for newspapers summarization, but is not better for scientific research paper documents. In continuation of position based work, (Edmundson, 1969) suggests that sentences in the first and last paragraphs and the first and last sentences of each paragraph should be assigned higher weights than other sentences in a document. While (Kupiec, Pedersen, & Chen, 1995) have assigned relatively a higher weight to the first ten paragraphs and last five paragraphs in a document. But, (Radev, Jing, Sty, & Tam, 2004), have followed different positional value position, i.e., $P_i$ of an $i^{th}$ sentence is calculated using the Equation 2.1,

$$P_i = \frac{(n-i-1)\times C_{\max}}{n} \tag{2.1}$$

here n is representing the number of sentences in the document, i represents the $i^{th}$ sentence/position of the sentence inside the text, and $C_{max}$ is the score of the sentence that has the maximum centroid value.

(Radev, Blair-Goldensohn, & Zhang, 2001) have proposed MEAD system for single and multi-document summarization. Their sentence score depends on three features centroid, TF*IDF, and position. For each sentence, these three features find out and importance of a sentence is decided by the sum of all the features. The position score, which proposed by them is linear and monotonic decreasing function.

(Ganapathiraju, Carbonell, & Yang, 2002) have considered keyword-occurrence as a feature, because as per their understanding keywords of the document represent the theme of the document, title-keywords are also indicative of the theme. They have assigned higher score to first and the last location. Uppercase word feature containing acronyms/ proper names are included for summary generation. Indicative phrases like "this report …", short length sentences, a sentence with the pronoun "she, they, it" are used to reduce the score of the sentence and generally not included in the summary.

(Rambow, Shrestha, Chen, & Lauridsen, 2004) have proposed a method for e-mail summarization that is based on some conventional feature which is common and used by other authors and some new features. Conventional features are an absolute position, centroid based on TF*IDF, IDF, length of sentence.

(Jagadeesh, Pingali, & Varma, 2005) have divided features into two type sentence level, and word level. Sentences level features include the position of sentences in the given document, the presence of the verbs in the sentences, referring pronouns in sentences, and length of the sentence in terms of a number of words. Word level features include term frequency (TF), word length, parts of speech tag, and familiarity of the word. As per their analysis, smaller words has higher frequency occur more frequently than the larger words, so to negate this effect they considered the word length as a feature for summarization feature. The familiarity of the word is derived from the WordNet relations. Familiarity can indicate the ambiguity of the word. As per author words which have less familiarity were given higher weight. The sigmoid function is used to calculate the importance of the word given by Equation-2.2.

$$\frac{1}{1+e^{-8(\frac{1}{fam}-0.5)}} \tag{2.2}$$

Some other features used by them are Named entity tag, Occurrence in headings or subheadings, and Font style. The Score of the sentence is given by combining all the features. In most of the work, it is widely considered that leading sentences are more important compared to preceding sentences. But, according to (Ouyang, Li, Lu, & Zhang, 2010), this is always not true for actual data. Importance of sentences varies according to the user and user writing style. Instead of sentences position, they focus on word position and claims that word position features are superior to traditional sentence position features. They have defined different word position features: direct proportion, inverse proportion, geometric sequence, and binary function. Finally, the score to sentence is given by Equation-2.3, where pos($w_i$) is found using one of the features.

$$Score(s) = \sum_i \frac{\log freq(w_i) \cdot pos(w_i)}{|s|} \tag{2.3}$$

(Karanikolas & Galiotou, 2012) have defined features into three category term weighting, position, and keyword based. Term weighting is done for sentence weighting and comprises different ways as local and global weighting, TF*IDF, TF*ISF, TF*RIDF. For term weighting, he has proposed three different ways shown in following Equation-2.4, 2.5 and 2.6,

$$t_{ij} = \frac{F_{ij}}{\max F_i} \tag{2.4}$$

$$t_{ij} = 0.5 + 0.5 * \frac{F_{ij}}{\max F_i} \tag{2.5}$$

$$t_{ij} = \frac{F_{ij}}{\sum F_i} \tag{2.6}$$

here $t_{ij}$ is representing the weight of the $j^{th}$ term in the document $D_i$, $F_{ij}$ is the frequency of the $j^{th}$ term in the document $D_i$, max $F_i$ is the frequency of the most frequent term in document $D_i$, and $F_i$ is the sum of frequencies of the index terms existing in document $D_i$. For TF*IDF, and TF*ISF standard approach followed. They have introduced new feature RIDF that is residual IDF. Residual IDF of a $j^{th}$ term in each document $D^i$ is defined as the difference between the observed IDF & expected IDF under the assumption that the terms follow a Poisson distribution. To give sentences position, they have followed a new kind of model that is proposed in The News Articles algorithm (Hariharan, 2010). Their score method considers both, paragraph location and sentence location in the paragraph.

$$Position_score = \left(\frac{(SP - P + 1)}{SP}\right) * \left(\frac{(SIP - SSIP + 1)}{SIP}\right) \qquad (2.7)$$

Where SP is the number of paragraphs in the document, P is the position of the paragraph, SIP is the number of sentences in the paragraph, and SSIP is the sentence position inside the paragraph. The third kind of feature was title words / keywords. The final score is given by combining all the features scores.

Feature used by (Luo, Zhuang, He, & Shi, 2010) were, position of the sentence, the length of the sentence, likelihood of the sentence, the number of thematic words, the number of low frequency words, the LSA-Score of the sentence, the number of two gram keywords, number of words appearing in other sentences, the entropy of the sentence, the relevance of the sentence. They have defined relevance measures as intra-sentence relationships between sentences, and entropy based feature denotes the quantity of information implied by the sentence. They have mentioned that long sentences are likely to cover a number of aspects in the document compared to short sentences. Therefore, the long sentence has comparative more entropy than a short length sentence. Hence, a large entropy of sentence possibly implies a large converge.

(Shimada, Tadano, & Endo, 2011) have proposed a method for multi-aspects review summarization based on evaluative sentence extraction. They proposed three features, ratings of aspects, TF-IDF value, and the number of mentions with a similar topic. Ratings of aspects were divided into different levels from Low to high, and the rating has given between one to five.

(Zhang, Li, Gao, & Ouyang, 2013) have considered following words for summarization features like cue words and phrases, abbreviations and acronyms, non-cue words, opinion words, vulgar words, emoticons such as O:),-:), twitter-specific symbols, #, @, and RT.

(Tofighy, Raj, & Javad, 2013) have used six features like word frequency, keywords in the sentence, headline word, cue word, no of cue word in sentence/ no of cue word in a paragraph, sentence location, and sentence length. To give the sentence position scores they have used the following method which gives equal importance to first and last, second and second last. Position score is represented by the following Equation-2.8,

$$Max(\frac{1}{i},\frac{1}{n-i+1}) \tag{2.8}$$

(PadmaLahari, Kumar, & Prasad, 2014) have proposed the first feature is keyword based. In their work keyword are nouns and determined using TF*IDF. The keyword has found using morphological analysis, noun phrase extraction, and clustering and scoring. Second, feature are position based, and the third feature is term frequency, that is calculated using both the unigram and bigram frequency. Only nouns are considered for computing of bigram frequencies. The fourth feature is the length of the word, the fifth feature is Parts of Speech Tag, in which tags are ranked and assigned weights that are based on the information contribution of the sentence. Other linguistic features are a proper noun and pronouns.

(Rautray, Balabantaray, & Bhardwaj, 2015) have proposed eight features to score the sentence. The first feature is title feature that is based on similarity (overlapping) between the sentence & the document title, divide by, a total number of words in sentences and title. The second feature is sentence length, longer sentences given more weight compared to small length sentences. The third feature is frequency based, the fourth feature is position based that depends on both positions in a paragraph, and paragraph's position. The fifth feature is an aggregate similarity, the sixth feature is based on counting on proper nouns, seventh is thematic word score based on word frequency, eight is the Numerical data-based score.

(Roul, Sahoo, & Goel, 2017) have used length of the sentence, weight of the sentence that is given by TF * IDF, sentence density, presence of named entities in the sentence (belong to the number of categories like names of people/ organization/locations, quantities, etc.). Presence of cue-phrases in the sentence as “in summary”, “our investigation”, “in conclusion”, “the paper describes”, “important”, “the best”, “hardly”, significantly, “in particular” etc.), Relative offset of the sentence (Sentences that are located at the beginning or towards, the end of a document tends to be more imperative as they carry relevant information like definitions and conclusions. Such type of sentences receive a score  either 1, or 0), Presence of title words in the sentence (score of the sentence is given by common textual units between the document title and total number of words in the title), Presence of specially emphasized text i.e. quoted text in the sentence (generally situated within " " (double quotation) marks then it receives the score  either score 1, else 0), presence of upper case letters in the sentence (these uppercase words or phrases are usually to refer to the important acronyms, like names, and places, etc. Such sentences have also received either score 1, else 0), sentence density represented by, ration of total count of keywords in a sentence and the total count of words which, including all stop words of the sentence.

## 2.3 Background

In this section, we are describing a different model used for weight learning. In this section, we have divided our dataset into training (90%), and testing (10%).

### 2.3.1 Random Forest

Random forest used predictive modeling and machine learning technique. It is an ensemble classifier made using many decision models. Ensemble models combine the results from different models. It is a versatile algorithm capable of performing both Regression and classification. This performs an implicit feature selection. According to (Breiman, 2001) "Random forests are a combination of tree predictors such that each tree depends on the values of a random vector sampled independently and with the same distribution for all trees in the forest". The algorithm of random forest is presented by Figure 2.1, (Boulesteix, Janitza, Kruppa, & König, 2012). One characteristic of this algorithm is that, due to a large number of trees generated in this technique, therefore, no issue of overfitting and it is always convergent.

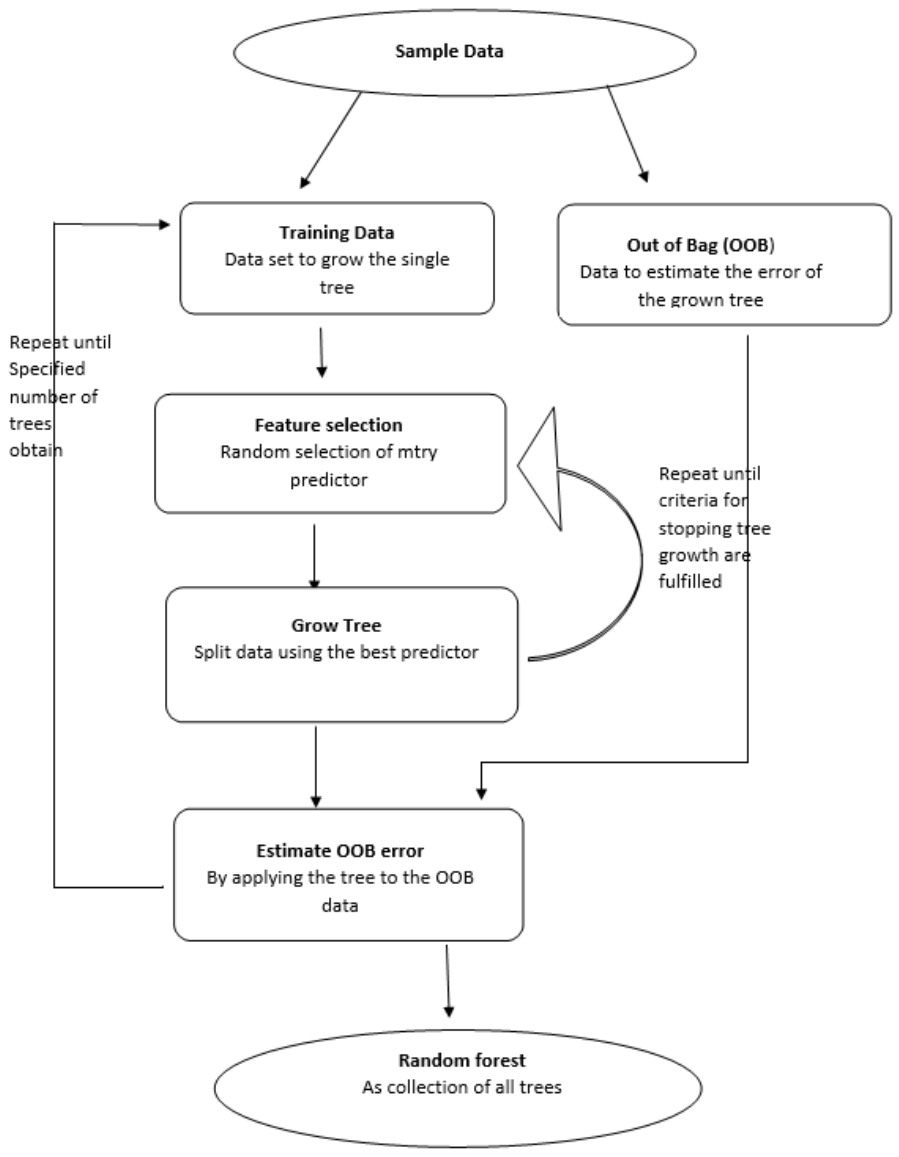

Figure 2.1: Random forest algorithm

### 2.3.2 Binary Logistic Regression

If, Y be a binary response variable or dependent feature. $Y_i$ is True or 1 if condition satisfies otherwise $Y_i$ is False or 0, and X= ($X_1$, $X_2$, $X_3$ … $X_k$) is set of independent features, $X_i$ may be discrete, continuous, or a combination. $x_i$ is the observed value of independent $i^{th}$ observation. $\beta_0$ *and* $\beta_1$ are to learn. The model is represented by Equation 2.9, 2.10. The equation-2.11 shows how parameter estimation done by logistic regression.

$$\pi_i = \Pr(Y_i = 1 \mid X_i = x_i) = \frac{\exp(\beta_0 + \beta_1 x_i)}{1 + \exp(\beta_0 + \beta_1 x_i)} \quad 2.9$$

*or*

$$\log it(\pi_i) = \log\left(\frac{\pi_i}{1-\pi_i}\right) = \beta_0 + \beta_1 x_i \quad 2.10$$

Parameter estimation is done by maximizing equation 2.11,

$$L(\beta_0, \beta_1) = \prod_{i=1}^{N} \pi_i^{yi}(1-\pi_i)^{n_i - y_i} = \prod_{i=1}^{N} \frac{\exp\{yi(\beta_0 + \beta_1 x_i)\}}{1 + \exp(\beta_0 + \beta_1 x_i)} \quad 2.11$$

## 2.4 Features used in text document summarization

We are proposing a hybrid model for salient sentence extraction for "single text document summarization". This is based on two types of features, statistical features, i.e., location, frequency (TF-IDF), aggregate similarity, centroid and semantic feature that is sentiment feature. In this ***section***, we are presenting detailed features, description used in our sentence selection approach.

### 2.4.1 The location Feature (Score1)

(Baxendale, 1958) has introduced a position feature. Although his work was almost manual. Later, this measure used widely in sentence scoring. The Author has concluded that leading sentences of an article are important. The model given by them is explained by Equation-2.12, where N is a total number of sentences. The used model is: (Where: 1< i < N, and $Score(S_i)$ =(0,1])

$$Score(S_i) = 1 - \frac{i-1}{N} \qquad (2.12)$$

### 2.4.2 The aggregation similarity Feature (score2)

(Kim, Kim, & Hwang, 2000) have defined aggregate similarity as, "the score of a sentence is as the sum of similarities with other all sentence vectors in document vector space model." It is given by Equation-2.13 and 2.14.

$$Sim(S_i, S_j) = \sum_{k=1}^{n} W_{ik} \bullet W_{jk} \qquad (2.13)$$

$$Score(S_j) = \sum_{j=1, i \neq j}^{n} Sim(S_i, S_j) \qquad (2.14)$$

Where $W_{ik}$ is defined as the binary weight ok $k^{th}$ word in an $i^{th}$ sentence. Similarity measure plays an important role in text document summarization. In literature, it is proposed that, different similarity measure affects the outcome. In our implementation, we are using cosine similarity based criteria. Let we have two sentences vector, $S_i$= [$W_{i1}$, $W_{i2}$,.$W_{im}$] and $S_j$=[ $W_{j1}$, $W_{j2}$,….$W_{jm}$]. Standard Cosine similarity between $S_i$ and $S_j$, given by Equation-2.15. Value of i and j vary from 1 to m.

$$Sim(S_i, S_j) = \frac{\sum_{k=1}^{m} W_{ik} \bullet W_{jk}}{\sqrt{\sum_{k=1}^{m} W_{ik}^{2} \sum_{k=1}^{m} W_{jk}^{2}}} \qquad (2.15)$$

### 2.4.3 Frequency Feature (score3)

The early work in document summarization started on "Single Document Summarization," by (Luhn, 1958) at IBM in the 1950s. The Author has proposed a frequency-based model. The frequency of word plays a crucial role, to decide the importance of any word or sentence in a given document. In our method, we are using the traditional method of "TF-IDF" measure defined by Equation-2.16, i.e., TF stands for term frequency, IDF for inverse document frequency.

$$W_i = TF_i \times IDF_i = Tf_i \times \log \frac{ND}{df_i} \qquad (2.16)$$

Where $TF_i$ is the term frequency of the $i^{th}$ word in the document, ND represents a total number of documents, and $IDF_i$ is the document frequency of the $i^{th}$ word in the whole data set. In our implementation to calculate the importance of word $W_i$, for TF we are considering the sentence as a document and for IDF entire document as a Dataset.

### 2.4.4 Centroid Feature (score4)

(Radev et al., 2004) have defined centroid as "a centroid is a set of words that are statistically important to a cluster of documents." As such, centroids can be used both to identify salient sentences in a cluster and classify relevant documents. The centroid score $C_i$ for sentence $S_i$ is computed, as the sum of the centroid scores $C_{w,i}$ of all words appeared in the particular sentence. That is presented in Equation-2.17.

$$C_i(S_i) = \sum_{w} C_{w,i} \qquad (2.17)$$

### 2.4.5 Sentiment Feature (score5)

In previous ***sections***, we mentioned statistical measures used by us, and in this part we are elaborating semantic based feature. We are calling this feature as a semantic feature because in this a set of things are related to one other. As defined by (Mani & Maybury, 1999) semantic summary generation may be done using shallow level analysis and deep level analysis. In the shallow approach to the most analysis done on the sentence level is syntactic, but important to note that, word level analysis may be semantic level. In deep analysis, at least a sentential semantic level of representation is done. So, our approach i.e. sentiment feature is semantic and low-level analysis (because at the entity level).

For finding sentiment score for a sentence, first we have found out all the entities present in the sentence, then, find sentiment scores of each entity and then do the sum of all entity's sentiment score (i.e., sentiment strength). If the sentiment of entity is neutral then we are scorning it as zero, if entity's sentiment is positive, then considering as same and adding to find the total score of a sentence, but if the sentiment score is negative we are multiplying it by -1 (minus one) to covert in positive score then adding this score to find the total score given in Equation-2.18. Reason for considering negative score to positive score is that we are interested only in sentiment strength which may be positive or negative, i.e., if sentiment score of an entity if "-0.523" it means sentiment of entity is negative, and strength is "0.523". Detail procedure is explained in ***section*** 2.5.2's Fifth feature. Here | A | representing mode (A) i.e. |-A | = | A | = A.

$$Score5 = \sum_{i=1}^{n} |Sentiment(Entity_i)| \tag{2.18}$$

## 2.5 Summarization Procedure

Our summarization approach is based on salient sentence selection and extraction. The importance of any sentence is decided by the combined score given by the sum of statistical measures and semantic measure. In the next section 2.5.1, we are explaining our approach (algorithm) used in this work and section 2.5.2 with detail explanation. Basically, our work of summarization has been divided into three Passes (1) sentence scoring, (2) sentence extraction and (3) evaluation.

### 2.5.1 Algorithm

This Algorithm is divided into three passes,

**PASS-1:** Sentence scoring according to linear combinations of different measures.

**PASS-2:** Salient sentence Extraction (Summary Generation).

**PASS-3:** Evaluation of Summary.

***PASS 1: Sentence Scoring***

**Input:** Documents

**Output:** Scored sentences

Step 1: Score the sentence given with five different measures. The outcome is M × N Matrix (M no. of sentences, N no. of measures) (1) Aggregate cosine similarity, (2) Position, (3) Sentiment of sentence, (4) Centroid score of the sentence, (5) TF×IDF.

Step 2: Normalized columns of the matrix

Step 3: Add all the features for every sentence, we are calling this sum as a score of the sentence.

Step 4: Sort according to score, the highest score representing most significant sentence.

***PASS 2: Algorithm for Redundancy***

**Input:** Number of sentences descending according to total score

**Output:** Extracted sentences

Parameter Initialization: (1) Summary= "" // Empty summary, (2) Similarity Threshold "θ", (3) L // required length summary.

Step 1: Summary = (Topmost scored sentence)

Step 2: For i=1 to (number of sentences)

if (Similarity(Summary, ith sentence) <θ ) AND (Length (summary) < L) Then

Summary= Summary + ith sentence

Step 3: Rearrange Summary sentences, as given in Source Document to maintaining cohesiveness.

***PASS 3: Evaluation of Summary***

***Input:*** Different summaries as "Standard Summaries" and "Peer summaries."

***Output:*** Precision, Recall, and F-score

*Step 1:* Generate different summary, different length using MEAD, MICROSOFT, OPINOSIS, HUMAN (5 humans) and Our-Proposed algorithm.

*Step2:* For experiment 1

Model summary: MEAD, MICROSOFT, OPINOSIS

Peer summary: Our-Proposed algorithm

For Experiment 2:

Model summary: Human generated summary

Peer summary: MEAD, MICROSOFT, OPINOSIS, Our-Proposed method

*Step 3:* Used Rouge-N (N=1 to 10), ROUGE-L, ROUGE-W (we set W=1.5), ROUGE S*, ROUGE SU* measure to find Precision, Recall, and F-Measure.

Figure 2.2 Algorithm for summarization

### 2.5.2 Detailed Approach Description

Here we are describing detail approach used as described in section 2.5.1.

***PASS 1. Sentence scoring and Extraction:***

In algorithm (defined in ***section 2.5.1)*** most things are covered and gives the main idea of Algorithm. Still, some micro points are needs to specify. Pass-1 is the sum of the linear combination of five different measures, four are statistically dependent (i.e., Aggregate Similarity, Position, TF ×IDF, Centroid) and the fifth measure is semantically dependent (i.e., Sentiment). Demonstration of working of this model is shown in Table-2.1.

(1) The first feature is the position of the sentences. Position is an important indicator for informative sentence. It has been analyzed that first or leading sentences, mostly contain important information. In our implementation we are using equation-2.12. Results are shown in **Table**-2.1's second column.

(2) The second feature is, TF × IDF. We are using the standard formula as defined in the previous ***section***. Normalized TF-IDF score is given in **Table**-2.1's third columns.

(3) The third feature is an Aggregate similarity (cosine) score of a sentence vector can be calculated as the "sum of similarities with other all sentence vectors in document Vector Space Model." The significance of this is to find sentences which are highly similar to all other sentences. After representing all sentences in a vector space, and then find vector cosine similarity with all other sentences as a defined standard formula by equation-2.15. Normalized Aggregate cosine similarity in **Table**-2.1 column four.

Since other scores (Centroid, Position, Sentiment) are between [0, 1], so we need to normalized score. Normalization of values means "adjusting to values measured on different scales to one notionally common scale" that removes the chance to be biased w.r.t. some values. In our implementation, we are just using column normalization instead of matrix normalization. Normalization of a column vector $X=[X_1, X_2 \ldots\ldots X_n]$ is done using Equation-2.19. Where $X_i$ is the $i^{th}$ element in the column, and n is the size of the column.

$$X_i = X_i \times \frac{1}{\sqrt{\sum_{i=1}^{n} X_i^{\,2}}} \tag{2.19}$$

$$A = \begin{bmatrix} 1 & 1 & .4 \\ 2 & 4 & .4 \\ 3 & 5 & .3 \end{bmatrix} \qquad B = \begin{bmatrix} 1 \times \frac{1}{\sqrt{1^2+2^2+3^2}} & 1 \times \frac{1}{\sqrt{1^2+4^2+4^2}} & .4 \\ 2 \times \frac{1}{\sqrt{1^2+2^2+3^2}} & 4 \times \frac{1}{\sqrt{1^2+4^2+5^2}} & .4 \\ 5 \times \frac{1}{\sqrt{1^2+2^2+3^2}} & 5 \times \frac{1}{\sqrt{1^2+4+5^2}} & .3 \end{bmatrix}$$

Let “A” is a given matrix, which size is 3×3 and column one and two has doesn't have values between [0,1], then we are doing normalization of only column one, and two but not third column and B is the give normalized matrix in our case.

(4) The fourth feature is centroid based, (Radev et al., 2004) defined as "Centroid as a set of words that arestatisticall important to a cluster of documents." In our approach using MEAD centroid score output as our input. The centroid value of a sentence is given by the summation of each word's centroid score present in the sentence.

(5) The fifth feature is "Sentiment score,". This is a novelty in our work, to find this feature we are dependent on Alchemy API which is available at http://www.alchemyapi.com/. We have considered that it is finding sentiment score is a semantic approach and fall under shallow level approach as defined in ***section*** 2.4.5. For any sentence or words, we can define three kinds of sentiment (a) Neutral, (b) Negative, and (c) Positive. Neutral sentiment value means that words or that sentence sentiment score are zero, most important to note that it is easy to find sentiment score based on cue word like good, bad, pleasant, etc., but still due to so much complexity in text, words, limitation of NLP, etc. it is not possible to find the correct sentiment score. Sometimes even it is also not possible to detect sentiment due to hidden sentiments. The overall working of the fifth feature can be understood clearly by following documents.

**Document 2.1:** "NAAC Accredited JNU with the CGPA of 3.91 on a four-point scale of A grade (highest grade by NAAC)", and Sentiment of this is "NEUTRAL."

**Document 2.2:** "JNU ranked in top 100 in Times Higher Education Asia and BRICS Top Ranking”, and Sentiment of this document is Positive, and the score is 0.499033. [NOTE-2 NAAC stands for National Assessment and Accreditation Council (NAAC), and BRICS stands for five nations Brazil, Russia, India, China, and South Africa]

Here Document-2.1 and Document-2.2 both are representing positive news about JNU. But the sentiment of document-2.1 is neutral, and sentiment of document-2.2 is Positive with .499033 score. Still, we have to discover an approach which can find correct sentiment (hidden sentiment). Some results are displayed in **Table**-2.1 with sentiment results.

In our implementation to find the sentiment score of a sentence, we are using alchemy API, first finding all entities present in the sentence and their sentiment score, then we add all entity's sentiment score. For example, consider document number 49 "Meanwhile, BJP spokesperson Prakash Javadekar has said that party president...etc " has five entities as follow (1) Prakash Javadekar : Person Name: -0.212091 (2) Rajnath Singh : Person Name: -0.212091 (3) Uttarakhand: State/County: -0.212091 (4) BJP: Company : -0.212091, (5) president: JobTitle: -0.212091 [NOTE-3 Triplet X: Y: Z representing, X is Entity Name, Y is Entity Type, Z Sentiment Score.], and to give sentiment score of sentence 49 we add all, Score5= $\sum_{i=1}^{n} Sentiment(Entity_i)$ so sentiment score (score5)= -1.060455 for sentence 49. But, if we see the **Table**-2.1 result in row 49 and column 3 (sentiment score) the value is 1.060455. Here we are considering only positive sentiment scores, if the entity sentiment score is negative, then by multiplying -1, to convert it into positive score. The obvious goal of the procedure is to give equal importance if magnitude is same. Let consider one the childhood story about “The fox and the grapes", in Figure-2.3.

> One afternoon a fox was walking through the forest and spotted a bunch of grapes hanging from over a lofty branch. Just these sweat and juicy grapes to quench my thirst, he thought. Taking a few steps back, the fox jumped and just missed the hanging grapes. Again the fox took a few paces back and tried to reach them but still failed. Finally, giving up, the fox turned up his nose and said, they're probably sour anyway, and proceeded to walk away".

**Figure-2.3 The Fox and The Grapes story.**

If here we consider two sentences (given below, document 2.3 and 2.4) as a document, both are important in the story and about the same things grapes.

**Document 2.3:** Just these sweat and juicy grapes to quench my thirst, and
**Document 2.4:** They’re probably sour anyway.

With Alchemy system if we find a sentiment of both these sentences, then the sentiment of document-2.3 is positive, and the score is 0.707112 where, the sentiment of document-2.4 is negative with -0.598391. For us the only magnitude is important, reasons to consider as + value are (1) we are interested to find sentiment strength, it may be negative or positive, and both are important for us, and (2) if we will add a negative score to find the total score then the value will reduce.

In the next step, we are finding the total score of a sentence, by adding all scores. The total score can be represented by the Equation-2.20, given below. In our implementation $w_k=1$ (k=1 to n). Detail result of the individual score is given in Table-2.1, and the last column is the total score of all scores.

$$TotalScore(S_i) = \sum_{k=1}^{5} W_k \times score_k \qquad (2.20)$$

**Table-2.1:** Different Features scores and Total Score for sentences

| **Sent No.** | **Position Score** | **TF * IDF** | **Aggregate Cosine sim.** | **Centroid Score** | **sentiment Score** | **SUM OF ALL** |
|---|---|---|---|---|---|---|
| 0 | 1 | 0.175146269 | 0.154823564 | 0.750856924 | 0.661311 | 2.742137758 |
| 1 | 0.98245614 | 0.127104001 | 0.156975892 | 0.307405503 | 0 | 1.573941537 |
| 2 | 0.964912281 | 0.170528144 | 0.185519412 | 0.50702023 | 0.394824 | 2.222804067 |
| 3 | 0.947368421 | 0.182701719 | 0.174777773 | 0.663083509 | 0.217356 | 2.185287422 |
| 4 | 0.929824561 | 0.119835913 | 0.144151017 | 0.440031566 | 0 | 1.633843058 |
| 5 | 0.912280702 | 0.184106205 | 0.176538454 | 1 | 0.389024 | 2.661949361 |
| 6 | 0.894736842 | 0.128152365 | 0.144782619 | 0.353263066 | 0 | 1.520934892 |
| 7 | 0.877192982 | 0.172677585 | 0.14243688 | 0.423917828 | 0.889318 | 2.505543275 |
| 27 | 0.526315789 | 0.098431452 | 0.034336443 | 0.120511453 | 0 | 0.779595137 |
| 49 | 0.140350877 | 0.148292229 | 0.122318812 | 0.423393814 | 1.060455 | 1.894810731 |
| 50 | 0.122807018 | 0.091023788 | 0.054947802 | 0.048475042 | 0 | 0.31725365 |
| 51 | 0.105263158 | 0.076719171 | 0.044899933 | 0.102930791 | 0 | 0.329813053 |
| 52 | 0.087719298 | 0.126084523 | 0.149793298 | 0.206132535 | 0.327554 | 0.897283655 |
| 53 | 0.070175439 | 0.127388217 | 0.037996007 | 0.137435141 | 0.363308 | 0.736302803 |
| 54 | 0.052631579 | 0.113474586 | 0.064967719 | 0.114433381 | 0 | 0.345507265 |
| 55 | 0.035087719 | 0.100535505 | 0.066218503 | 0.247865211 | 0 | 0.449706939 |

***PASS 2. Redundancy***

To remove redundancy we have used the same model as proposed by (Sarkar, 2010) which, the topmost sentence (according to total score defined in the Equation 2.20) is add in summary. We added next sentence in the summary if the similarity is less than threshold θ. The algorithm is described in ***section*** 2.5.1's pass-2, input in this pass is a number of sentences which are sorted according to descending total score. We need to initialize some parameter to get the desired summary, parameter like summary initially empty, given similarity threshold θ and L for the desired length summary. Even in our system, L stands for the maximum length of the desired summary, but due to the limitation (here the length of sentences), we can guarantee minimum (maximum (length (summary))). We will add the next sentence, in summary, if still summary length is less than L and similarity (new-sentence, summary) <θ. The Output of this step is a summary with minimal redundancy and length nearly equal to L, but the position of the sentence is zigzag that lost the sequence and cohesiveness. To maintain the sequence, we need to rearrange the sentences according to given in the initial index.

In **Table**-2.2 we are representing the summary generated by our system, in which the similarity threshold θ is .1, and the desired summary length is 15 %. We can define arbitrary L in a number of words or percentage of summary required. Here we have chosen θ small. If we put θ large like 0.4 or 0.5 then the sentences, which are, in summary, will depend only on the total score (as in **Table**-2.1). In other words, the summary only depends on total scores as shown in **Table**-2.1, but our objective is also to minimize redundancy. *[NOTE 4: Before calculating sim(new sentence, summary), we are eliminating stopwords, stopwords play a big role to increase the similarity between two sentences. With different stopwords list we will get different similarity score].* MEAD, Microsoft, and Our-Model generated summary with different length are shown below in **Table**-2.2, 2.3, 2.4. The truth is we don't have the model behind Microsoft generated summary. This summarizer is inbuilt inside Microsoft office package. When we observed that, Microsoft summarizer is not reducing redundancy sentence 4 and 18 are almost similar sentences, it is shown in **Table**-**2.**3.

***PASS 3. Evaluation:***

(Goldstein et al., 2000) have concluded two things (1) "even human judgment of the quality of a summary varies from person to person", (2) "human judgment usually doesn't find concurrence on the quality of a given summary", hence it is difficult to judge the quality of a summary. For evaluation of any summary, we need two summaries first one is a system generated the summary, and another summary is user-generated (Model summary or Standard Summary).

To generate different model summary we used three approaches (1) we give our text data set to 5 people and tell them to write a summary in about to 20% to 40% words, (2) We generate summary by MEAD tool, in this approach we taking linear combination of position and centroid score, score(mead)=($w_1$×centroid)+($w_2$×position) with $w_1$=$w_2$=1, and (3) third model summary is generated by OPINOSIS by (Ganesan, Zhai, & Han, 2010) (summary given in Figure 2.5) (4) Microsoft System.

To evaluate summary, we are using ROUGE evaluation package. DUC adopts ROUGE for official evaluation metric for both single text document summarization and multi-document summarization. ROUGE finds Recall, Precision, and F-Score for evaluation results. Based on N-gram co-occurrence statistic, it (ROUGE) measures how much the system generated summary (machine summary) overlaps with the standard summary (human summaries/model summary). Where an N-gram is a contiguous sequence of N words. In our evaluation, we are adopting different measures of ROUGE, as ROUGE-N (N=1 to 10), ROUGE-W, ROUGE-L, ROUGE-S*, and ROUGE-SU*.

## 2.6 Experiment and Results

In this ***section,*** we are presenting four experiments done by us.

In the first experiment, we took our own summary (generated by algorithm discussed in ***section*** 2.5.1) as system generate summary and another summary as a Model summary. In the second experiment we are comparing different system generated summary w.r.t. Human summary. In the third Experiment, we are showing the significance of Sentiment feature. IN fourth experiment we are finding best feature weight combination using regression and Random Forest.

### 2.6.1 Experiment 2.1

As explained in *section* 2.4.1's third pass we have created four types of the model summary (1) Human summary (via we gave data set to 5 persons to summarize it, based on their experience with instruction to summarize it within 20% to 40% words length. Due to limitations and user experiences, the generated summary varies from 24 % to 52% words length), (2) MEAD, (3) Microsoft, and (4) OPINOSIS system. Different system generated summaries are given in Table-2.2, 2.3, 2.4. Since OPINOSIS summarizer is the abstractive type, in Figure 2.5 we are giving summarization result length 10% summary generated by OPINOSIS System. Table-2.2, 2.3, 2.4 presenting different summaries generated by different systems.

Table-**2.2**: Our-System generated summary (using proposed Algorithm)

| Sent No | Extracted sentences (Our proposed Model) 15% summary length |
|---|---|
| 0 | "Uttarakhand-Flood Missing untraced till July 15 will be presumed dead: Bahuguna acing giants time, the Uttarakhand government on Thursday decided that those missing in the flood-ravaged state will be presumed dead if they remain untraced till July 15 and asked officials to remain vigilant in the wake of warning of heavy rains over the next two days". |
| 5 | "A team of seven mountaineers is also engaged in a combing operation in areas adjoining the shrine in search of bodies while over 50 members of a team |
| 27 | Fifty-five helicopters have been pressed into service for rescue work". |
| 35 | "There has been large scale destruction of property and loss of lives in this disaster". |
| 37 | "The CRPF rank and file joins the countrymen in conveying its deepest concern for the victims of the tragedy". |
| 42 | "We are keeping them in the morgue and documenting their details". |
| 44 | "Central Army Commander Lt Gen Anil Chait said on Friday that about 8,000 to 9,000 people are still stranded in Badrinath". |
| 52 | "He also praised the efforts of the armed forces and the Indo-Tibetan Border Police, saying they were doing a laudable job". |

Table-**2.3**: Microsoft system generated summary

| SENT NO | Extracted sentences (microsoft summary) 10 % |
|---|---|
| 4 | "Meanwhile, the Indian Air Force flew 70 civil administration personnel to the Kedarnath temple premises to clean the surroundings there |
| 18 | "The Indian Air Force (IAF) today flew 70 civil administration personnel to the Kedarnath temple premises to clean the surroundings after |
| 26 | "The Indian Air Force (IAF) has deployed 13 more aircraft for relief and rescue operations". |
| 29 | "New Delhi: Indian Air Force has airlifted over 18,000 persons and dropped more than 3 lakh kg of relief material in flood-hit Uttarakhand |
| 32 | "The Central Reserve Police Force (CRPF) on Saturday announced it will contribute one day's salary of its personnel to the Prime Minist |
| 41 | "Rescue teams and police personnel have recovered 48 dead bodies from the River Ganga in Haridwar". |

Table-**2.4**: MEAD system generated summary

| SENT NO | Extracted sentences (mead system summary) 15 % |
|---|---|
| 0 | Uttarakhand-Flood Missing untraced till July 15 will be presumed dead: Bahuguna acing gainst time, the Uttarakhand government on Thursday decided that those missing in th |
| 1 | "Chief Minister Vijay Bahuguna said the exact number of people missing after the tragedy is 3,064 and the deadline for finding them is July 15". |
| 2 | "Considering the magnitude of the crisis, the state Cabinet has decided that if the missing persons are not found by July 15, we will presume that they are dead and the rocess of paying compensation to their next of kin will begin," he said. |
| 3 | "With the MeT department issuing a warning of heavy rains at places in Kumaon region over the next two days, Bahuguna said that for the next 50 hours the administration needs to be highly vigilant, adding 250 National Disaster Response Force personnel have been deployed in these areas". |
| 4 | "Meanwhile, the Indian Air Force flew 70 civil administration personnel to the Kedarnath temple premises to clean the surroundings there". |
| 5 | "A team of seven mountaineers is also engaged in a combing operation in areas adjoining the shrine in search of bodies while over 50 members of a team of experts and volunteers is stationed in Kedarnath to clean the temple premises of tonnes of debris under which more bodies may be lying, an official said". |
| 8 | "Mass cremation of bodies in Kedarghati held up for the past few days on Thursday started with 23 more consigned to flames, taking the number of bodies disposed of so far to 59 even as a team of experts worked on removal of debris and extricating bodies from under them at the Himalayan shrine". |
| 14 | "Despite continuing bad weather in affected areas amid a MeT department prediction of heavy rains in the next 48 hours at places, efforts were on to airdrop relief material in affected villages totally cut off after the calamity in the worst-hit Rudraprayag, Chamoli and Uttarkashi districts". |
| 29 | "New Delhi: Indian Air Force has airlifted over 18,000 persons and dropped more than 3 lakh kg of relief material in flood-hit Uttarakhand since June 17 in its biggest ever helico |

*Indian air force has deployed 13 more aircraft for relief and rescue operations. a total of 93 sorties and dropped about 12,000 kgs of relief and equipment , said .Indian air force flew 70 civil administration personnel to the temple premises to the surroundings there . for the pastfew days on Thursday started with 23 more consigned to flames .people have so far been evacuated from the flood and landslide-hit areas of Uttarakhand so far and so far .gen chait said on Friday that about 8,000 to 9,000 people are still stranded in Badrinath .efforts to help those in distress in different inaccessible parts of the state and the hill state .rs18 crore to support the victims of the massive.*

**Figure-2.4:** OPINOSIS generated summary

In the first experiment, we took our own summary (generated by algorithm discussed in ***section*** 2.5.1) as system generate summary and another summary as a Model summary. In the next step, we find different Rouge scores (N=1 to 10, ROUGE-L, ROUGE-W where W=1.5, ROUGE S* and ROUGE -SU*) as defined by (C. Lin & Rey, 2004) and followed by others like (Sankarasubramaniam, Ramanathan, & Ghosh, 2014). ROUGE scores are given in Chapter-1 with Equation-1.1. It measures three things Recall, Precision and F-Score for any System generated summary and Model Summary (or Reference summary).

We are comparing our system generated summary, with other's (as model summary) same length summary. The result of this Experiment-1 is given in **Table**-2.5, **Table**-2.6, and **Table**-**2.**7, for 10%, 20%, 30% length respectively (due to the limitation of space we are providing only three **Table**s). Figure-2.8 showing F Measure with different model summaries of the length of nearly 30% and our summary length is nearly 27%. In simple term we can define "High precision means that an algorithm retrieved substantially more relevant than irrelevant" and, "High recall means that an algorithm returns most of the relevant result" ("Precision and recall," wiki.). From Figure-2.5, 2.6 and 2.7 (for 30% summary length) we are getting high Precision, F-Score w.r.t. MEAD reference summary and high Recall w.r.t Microsoft generated a summary.

**Table-2.5**: Summary generated by our algorithm considered as system summary another summary as a model summary

| **MEASURE** | **MEAD - 10%** | | | **Microsoft 10%** | | | **OPIOSIS 10%** | | |
|---|---|---|---|---|---|---|---|---|---|
| **10 % summary** | **R** | **P** | **F** | **R** | **P** | **F** | **R** | **P** | **F** |
| **ROUGE-1** | 0.46 | 0.71 | 0.56 | 0.345 | 0.275 | 0.306 | 0.508 | 0.464 | 0.485 |
| **ROUGE-2** | 0.35 | 0.54 | 0.42 | 0.032 | 0.025 | 0.025 | 0.258 | 0.235 | 0.246 |
| **ROUGE-3** | 0.33 | 0.51 | 0.4 | 0 | 0 | 0 | 0.209 | 0.19 | 0.199 |
| **ROUGE-4** | .32 | .47 | .39 | 0 | 0 | 0 | .187 | .17 | .178 |
| **ROUGE-5** | 0.32 | 0.49 | 0.38 | 0 | 0 | 0 | 0.171 | 0.156 | 0.163 |
| **ROUGE-6** | 0.31 | 0.48 | 0.38 | 0 | 0 | 0 | 0.155 | 0.141 | 0.147 |
| **ROUGE-7** | 0.31 | 0.47 | 0.37 | 0 | 0 | 0 | 0.138 | 0.126 | 0.132 |
| **ROUGE-8** | 0.3 | 0.47 | 0.36 | 0 | 0 | 0 | 0.122 | 0.111 | 0.116 |
| **ROUGE-9** | 0.29 | 0.46 | 0.35 | 0 | 0 | 0 | 0.105 | 0.095 | 0.1 |
| **ROUGE-10** | 0.29 | 0.45 | 0.35 | 0 | 0 | 0 | 0.088 | 0.08 | 0.084 |
| **ROUGE-L** | 0.45 | 0.69 | 0.54 | 0.307 | 0.244 | 0.272 | 0.474 | 0.433 | 0.453 |
| **ROUGE-W** | 0.03 | 0.35 | 0.06 | 0.017 | 0.076 | 0.029 | 0.041 | 0.154 | 0.065 |
| **ROUGE-S*** | 0.16 | 0.37 | 0.22 | 0.097 | 0.061 | 0.075 | 0.22 | 0.183 | 0.2 |
| **ROUGE-SU*** | 0.16 | 0.37 | 0.22 | 0.1 | 0.063 | 0.077 | 0.223 | 0.186 | 0.203 |

**Table-2.6**: Summary generated by our algorithm as system summary, another summary of a model summary

| **MEASURE** | **MEAD 20%** | | | **OPIOSIS 20%** | | |
|---|---|---|---|---|---|---|
| **21 % summary** | **R** | **P** | **F** | **R** | **P** | **F** |
| **ROUGE-1** | 0.399 | 0.619 | 0.485 | 0.508 | 0.521 | 0.515 |
| **ROUGE-2** | 0.251 | 0.39 | 0.306 | 0.267 | 0.273 | 0.27 |
| **ROUGE-3** | 0.222 | 0.345 | 0.27 | 0.207 | 0.212 | 0.21 |
| **ROUGE-4** | 0.214 | 0.334 | 0.261 | 0.181 | 0.185 | 0.183 |
| **ROUGE-5** | 0.211 | 0.329 | 0.257 | 0.166 | 0.17 | 0.168 |
| **ROUGE-6** | 0.207 | 0.323 | 0.253 | 0.151 | 0.155 | 0.153 |
| **ROUGE-7** | 0.204 | 0.318 | 0.248 | 0.137 | 0.14 | 0.138 |
| **ROUGE-8** | 0.2 | 0.313 | 0.244 | 0.122 | 0.125 | 0.123 |
| **ROUGE-9** | 0.196 | 0.308 | 0.24 | 0.107 | 0.11 | 0.108 |
| **ROUGE-10** | 0.193 | 0.302 | 0.235 | 0.092 | 0.094 | 0.093 |
| **ROUGE-L** | 0.391 | 0.607 | 0.475 | 0.485 | 0.496 | 0.49 |
| **ROUGE-W** | 0.027 | 0.287 | 0.05 | 0.04 | 0.167 | 0.065 |
| **ROUGE-S*** | 0.151 | 0.365 | 0.214 | 0.201 | 0.211 | 0.206 |
| **ROUGE-SU*** | 0.152 | 0.367 | 0.215 | 0.203 | 0.203 | 0.208 |

**Table-2.7**: Summary generated by our algorithm as system summary, another summary of a model summary

| **Measure** | **MEAD 25 %** | | | **MEAD 30%** | | | **Microsoft 25%** | | | **Microsoft 30%** | | | **OPIOSIS 30%** | | |
|---|---|---|---|---|---|---|---|---|---|---|---|---|---|---|---|
| **27 % summary** | **R** | **P** | **F** | **R** | **P** | **F** | **R** | **P** | **F** | **R** | **P** | **F** | **R** | **P** | **F** |
| **ROUGE-1** | 0.49 | 0.69 | 0.57 | 0.42 | 0.70 | 0.53 | 0.64 | 0.57 | 0.60 | 0.55 | 0.61 | 0.58 | 0.50 | 0.54 | 0.52 |
| **ROUGE-2** | 0.35 | 0.49 | 0.41 | 0.30 | 0.50 | 0.37 | 0.45 | 0.40 | 0.42 | 0.36 | 0.41 | 0.38 | 0.24 | 0.27 | 0.25 |
| **ROUGE-3** | 0.33 | 0.46 | 0.38 | 0.28 | 0.46 | 0.35 | 0.39 | 0.35 | 0.37 | 0.32 | 0.35 | 0.34 | 0.18 | 0.20 | 0.19 |
| **ROUGE-4** | 0.32 | 0.45 | 0.37 | 0.27 | 0.45 | 0.34 | 0.37 | 0.34 | 0.35 | 0.30 | 0.34 | 0.32 | 0.15 | 0.17 | 0.16 |
| **ROUGE-5** | 0.31 | 0.44 | 0.36 | 0.26 | 0.44 | 0.33 | 0.36 | 0.33 | 0.34 | 0.29 | 0.33 | 0.31 | 0.14 | 0.15 | 0.15 |
| **ROUGE-6** | 0.30 | 0.43 | 0.36 | 0.26 | 0.43 | 0.32 | 0.35 | 0.32 | 0.34 | 0.28 | 0.32 | 0.30 | 0.13 | 0.14 | 0.13 |
| **ROUGE-7** | 0.30 | 0.42 | 0.35 | 0.25 | 0.42 | 0.32 | 0.34 | 0.31 | 0.33 | 0.28 | 0.31 | 0.29 | 0.11 | 0.12 | 0.12 |
| **ROUGE-8** | 0.29 | 0.41 | 0.34 | 0.25 | 0.41 | 0.31 | 0.34 | 0.30 | 0.32 | 0.27 | 0.30 | 0.29 | 0.10 | 0.11 | 0.10 |
| **ROUGE-9** | 0.28 | 0.41 | 0.33 | 0.24 | 0.41 | 0.30 | 0.33 | 0.30 | 0.31 | 0.26 | 0.30 | 0.28 | 0.09 | 0.10 | 0.09 |
| **ROUGE-10** | 0.28 | 0.40 | 0.33 | 0.23 | 0.40 | 0.29 | 0.32 | 0.29 | 0.30 | 0.26 | 0.29 | 0.27 | 0.07 | 0.08 | 0.08 |
| **ROUGE-L** | 0.47 | 0.67 | 0.56 | 0.41 | 0.69 | 0.51 | 0.60 | 0.54 | 0.57 | 0.51 | 0.57 | 0.54 | 0.47 | 0.51 | 0.49 |
| **ROUGE-W** | 0.03 | 0.30 | 0.06 | 0.03 | 0.29 | 0.05 | 0.05 | 0.26 | 0.08 | 0.04 | 0.26 | 0.07 | 0.04 | 0.15 | 0.06 |
| **ROUGE-S*** | 0.22 | 0.44 | 0.29 | 0.17 | 0.47 | 0.25 | 0.39 | 0.32 | 0.35 | 0.29 | 0.36 | 0.32 | 0.19 | 0.23 | 0.21 |

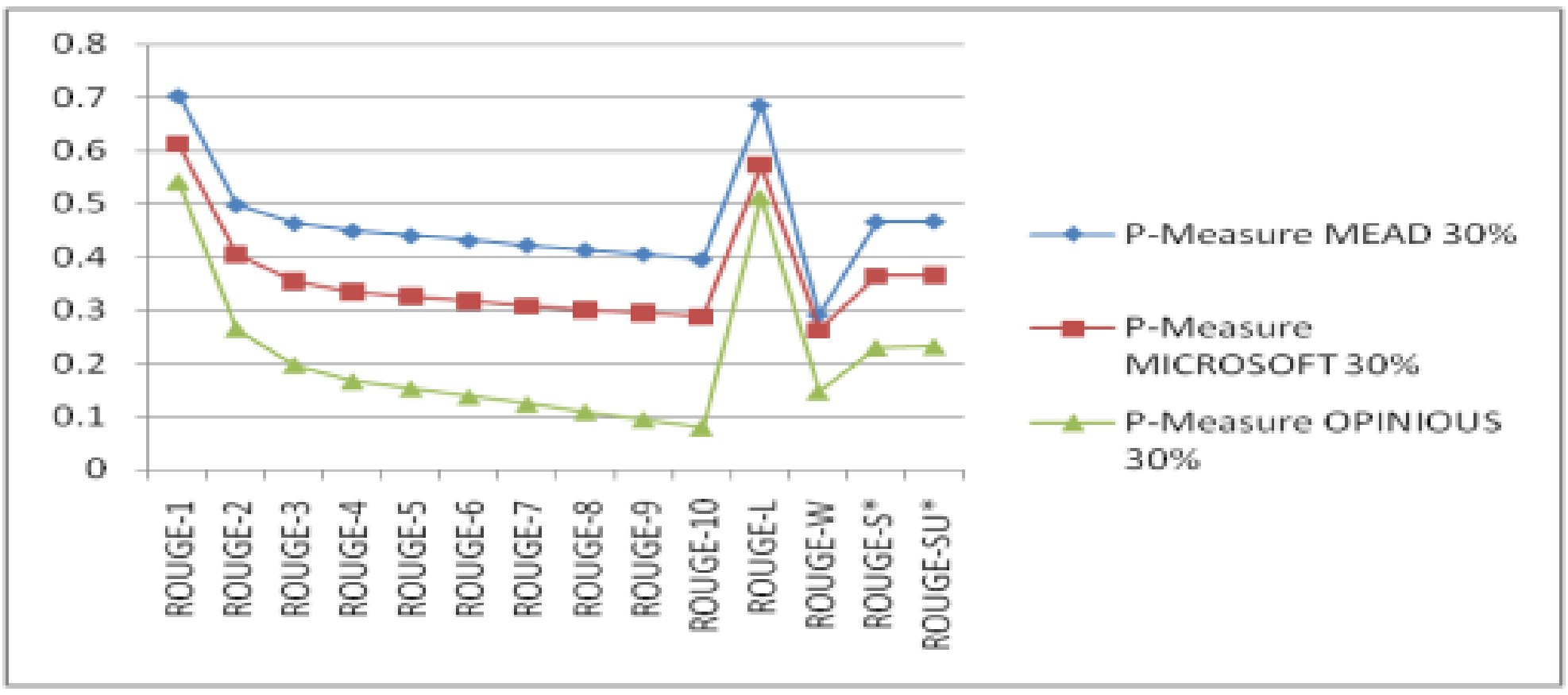

**Figure-2.5**: Precision curve

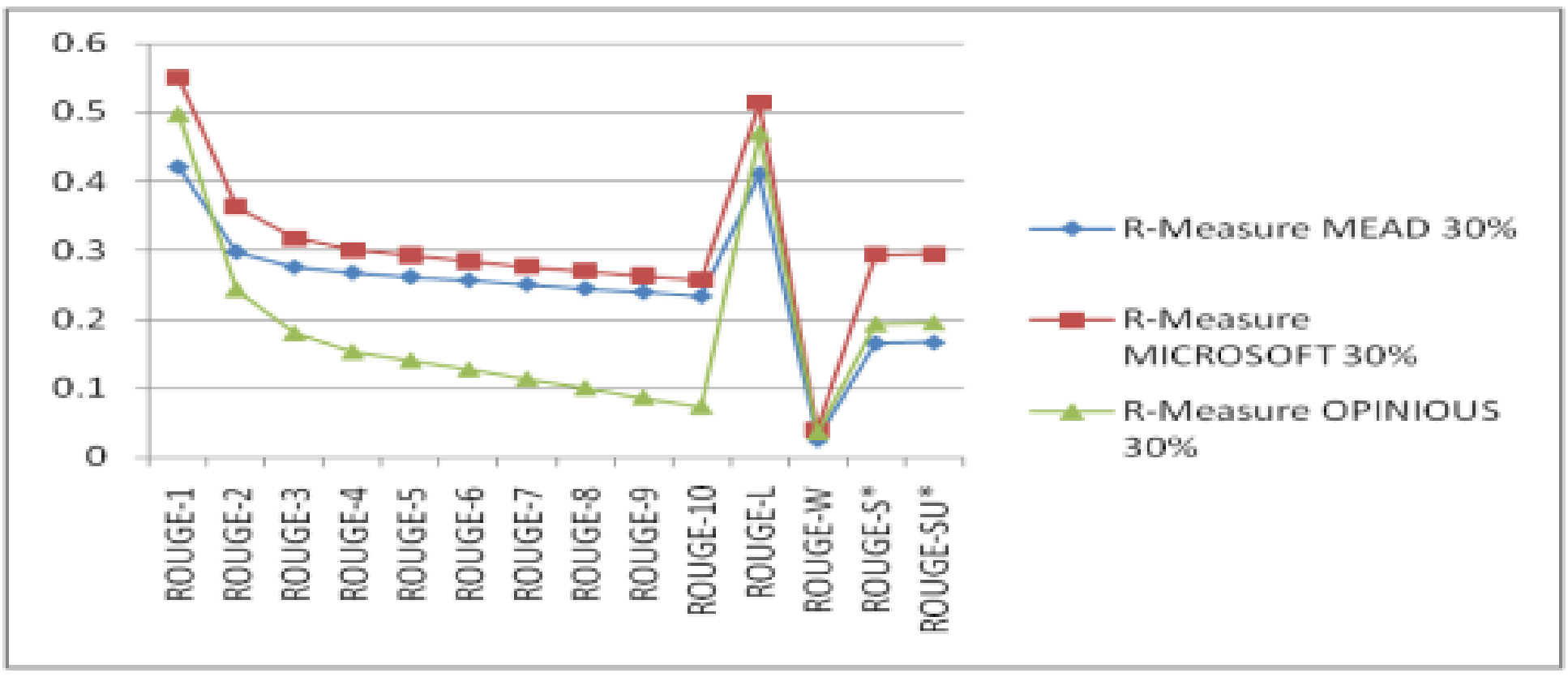

Figure-2.6: Recall curve

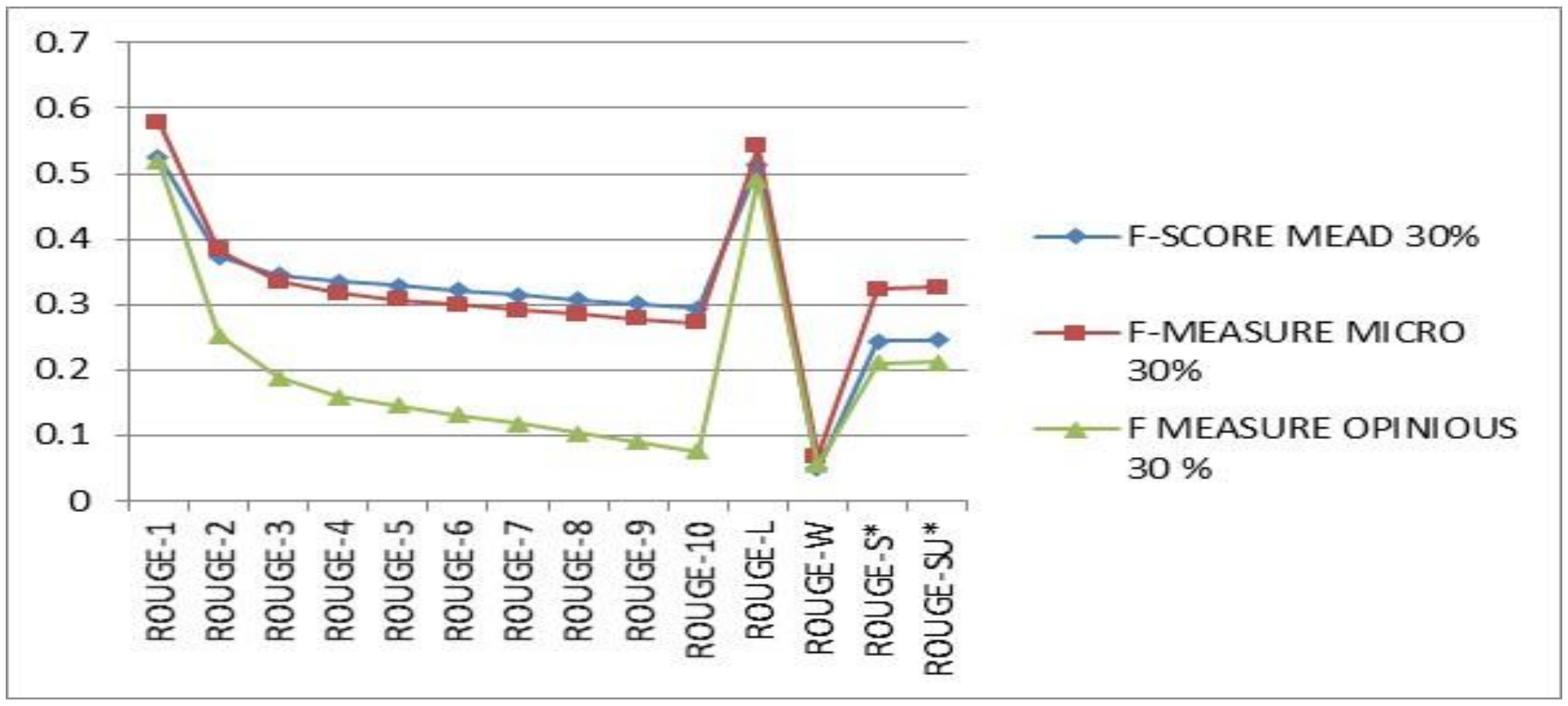

Figure-2.7: Showing F-Score

### 2.6.2 Experiment 2.2

In the second experiment we are comparing different system generated summary w.r.t. Human summary (or in other words, here Model (Gold/ Reference) summary is a Human generated summary) and other summaries are system generated summary (i.e., MEAD, MICROSOFT, OPINOSIS, OUR-ALGO are system generated summary). The result is shown in Table-2.8 and Table-2.9. Table-2.8 is representing different ROUGE scores for the summary length of 24%.

**Table-2.8**: Summary generated by different as system summary, human-generated summary as a model summary

| Measure | **My-Method 24** | | | **MEAD - 24** | | | **MICRO-24%** | | | **OPINIOUS -24%** | | |
|---|---|---|---|---|---|---|---|---|---|---|---|---|
| **USER Summary- 24 %** | **R** | **P** | **F** | **R** | **P** | **F** | **R** | **P** | **F** | **R** | **P** | **F** |
| **ROUGE-1** | 0.45 | 0.49 | 0.47 | 0.53 | 0.38 | 0.44 | 0.07 | 0.59 | 0.12 | 0.47 | 0.50 | 0.48 |
| **ROUGE-2** | 0.18 | 0.20 | 0.19 | 0.21 | 0.14 | 0.17 | 0.01 | 0.13 | 0.03 | 0.17 | 0.18 | 0.17 |
| **ROUGE-3** | 0.11 | 0.12 | 0.11 | 0.10 | 0.07 | 0.08 | 0.00 | 0.03 | 0.00 | 0.09 | 0.09 | 0.09 |
| **ROUGE-4** | 0.07 | 0.08 | 0.08 | 0.06 | 0.04 | 0.05 | 0.00 | 0.00 | 0.00 | 0.06 | 0.06 | 0.06 |
| **ROUGE-5** | 0.05 | 0.06 | 0.06 | 0.04 | 0.03 | 0.03 | 0.00 | 0.00 | 0.00 | 0.05 | 0.05 | 0.05 |
| **ROUGE-6** | 0.04 | 0.04 | 0.04 | 0.02 | 0.02 | 0.02 | 0.00 | 0.00 | 0.00 | 0.03 | 0.03 | 0.03 |
| **ROUGE-7** | 0.03 | 0.03 | 0.03 | 0.01 | 0.01 | 0.01 | 0.00 | 0.00 | 0.00 | 0.02 | 0.02 | 0.02 |
| **ROUGE-8** | 0.02 | 0.02 | 0.02 | 0.01 | 0.01 | 0.01 | 0.00 | 0.00 | 0.00 | 0.01 | 0.02 | 0.01 |
| **ROUGE-9** | 0.01 | 0.01 | 0.01 | 0.00 | 0.00 | 0.00 | 0.00 | 0.00 | 0.00 | 0.01 | 0.01 | 0.01 |
| **ROUGE-10** | 0.01 | 0.01 | 0.01 | 0.00 | 0.00 | 0.00 | 0.00 | 0.00 | 0.00 | 0.01 | 0.01 | 0.01 |
| **ROUGE-L** | 0.41 | 0.45 | 0.43 | 0.49 | 0.35 | 0.41 | 0.06 | 0.54 | 0.11 | 0.44 | 0.46 | 0.45 |
| **ROUGE-W** | 0.03 | 0.13 | 0.04 | 0.03 | 0.11 | 0.05 | 0.01 | 0.20 | 0.01 | 0.03 | 0.13 | 0.05 |
| **ROUGE-S*** | 0.20 | 0.24 | 0.22 | 0.28 | 0.14 | 0.18 | 0.00 | 0.29 | 0.01 | 0.16 | 0.18 | 0.17 |
| **ROUGE-SU*** | 0.20 | 0.24 | 0.22 | 0.28 | 0.14 | 0.19 | 0.00 | 0.30 | 0.01 | 0.16 | 0.19 | 0.17 |

**Table-2.9**: Summary generated by different system considered as system summary, human-generated summary as a model summary

| **MEASURE** | **My-Method-40%** | | | **MEAD - 40%** | | | **MICRO-40%** | | | **OPINIOUS-40%** | | |
|---|---|---|---|---|---|---|---|---|---|---|---|---|
| **User summary 40 %** | **R** | **P** | **F** | **R** | **P** | **F** | **R** | **P** | **F** | **R** | **P** | **F** |
| **ROUGE-1** | 0.60 | 0.71 | 0.65 | 0.82 | 0.58 | 0.68 | 0.70 | 0.72 | 0.71 | 0.48 | 0.61 | 0.54 |
| **ROUGE-2** | 0.45 | 0.53 | 0.48 | 0.69 | 0.48 | 0.57 | 0.53 | 0.54 | 0.54 | 0.23 | 0.29 | 0.26 |
| **ROUGE-3** | 0.40 | 0.48 | 0.44 | 0.64 | 0.45 | 0.53 | 0.48 | 0.49 | 0.48 | 0.15 | 0.19 | 0.17 |
| **ROUGE-4** | 0.38 | 0.45 | 0.41 | 0.62 | 0.43 | 0.51 | 0.46 | 0.47 | 0.46 | 0.12 | 0.15 | 0.13 |
| **ROUGE-5** | 0.36 | 0.42 | 0.39 | 0.60 | 0.42 | 0.50 | 0.44 | 0.45 | 0.45 | 0.11 | 0.13 | 0.12 |
| **ROUGE-6** | 0.34 | 0.40 | 0.37 | 0.59 | 0.41 | 0.48 | 0.42 | 0.44 | 0.43 | 0.09 | 0.12 | 0.10 |
| **ROUGE-7** | 0.32 | 0.38 | 0.35 | 0.57 | 0.40 | 0.47 | 0.41 | 0.42 | 0.41 | 0.08 | 0.10 | 0.09 |
| **ROUGE-8** | 0.30 | 0.36 | 0.33 | 0.56 | 0.39 | 0.46 | 0.39 | 0.41 | 0.40 | 0.07 | 0.08 | 0.07 |
| **ROUGE-9** | 0.28 | 0.34 | 0.31 | 0.54 | 0.38 | 0.45 | 0.38 | 0.39 | 0.38 | 0.06 | 0.07 | 0.06 |
| **ROUGE-10** | 0.27 | 0.32 | 0.29 | 0.53 | 0.37 | 0.44 | 0.36 | 0.38 | 0.37 | 0.05 | 0.06 | 0.05 |
| **ROUGE-L** | 0.58 | 0.69 | 0.63 | 0.82 | 0.57 | 0.67 | 0.69 | 0.72 | 0.70 | 0.47 | 0.59 | 0.52 |
| **ROUGE-W** | 0.03 | 0.23 | 0.06 | 0.06 | 0.22 | 0.09 | 0.04 | 0.25 | 0.07 | 0.02 | 0.16 | 0.04 |
| **ROUGE-S*** | 0.344 | 0.48 | 0.4 | 0.68 | 0.337 | 0.5 | 0.49 | 0.523 | 0.51 | 0.184 | 0.294 | 0.227 |
| **ROUGE-SU*** | 0.345 | 0.48 | 0.4 | 0.69 | 0.337 | 0.5 | 0.49 | 0.524 | 0.51 | 0.185 | 0.295 | 0.228 |

From **Table**-2.8 we can say that,

- We are getting the high F-Score comparison to MEAD, MICROSOFT system, and OPINOSIS system. Except for ROUGE-W in MEAD's ROUGE-1 and OPINOSIS's ROUGE-W. F-score is represented in Figure-2.8.
- We are getting high PRECISION to compare to MEAD and OPINOSIS but, Microsoft system, leading in ROUGE-1, ROUGE-L, ROUGE-W, ROUGE-S*, ROUGE-SU* only.
- We are getting High RECALL comparison to MEAD in ROUGE-3 to ROUGE-10 and higher Recall comparison to OPINOSIS and MICROSOFT in all measures except OPINOSIS getting ROUGE-1 higher than OUR-System.
- In Figure-2.8 we are representing a comparison of different system generated summary (24 % length) using F-Measure, and Figure-2.9 (comparison of 40% length summary) and we representing here only F-Score. From Table-2.9 we can say that,
- MEAD system and MICROSOFT perform better in term of RECALL, but our system is performing better compared to OPINOSIS.
- OUR Method gets Higher PRECISION to compare to OPINSIS's all ROUGE score (P) and higher PRECISION achieved compared to MEAD except ROUGE-6 to ROUGE-10.
- We are getting low F-Score compares to MEAD and MICROSOFT system but higher w.r.t OPINOSIS.

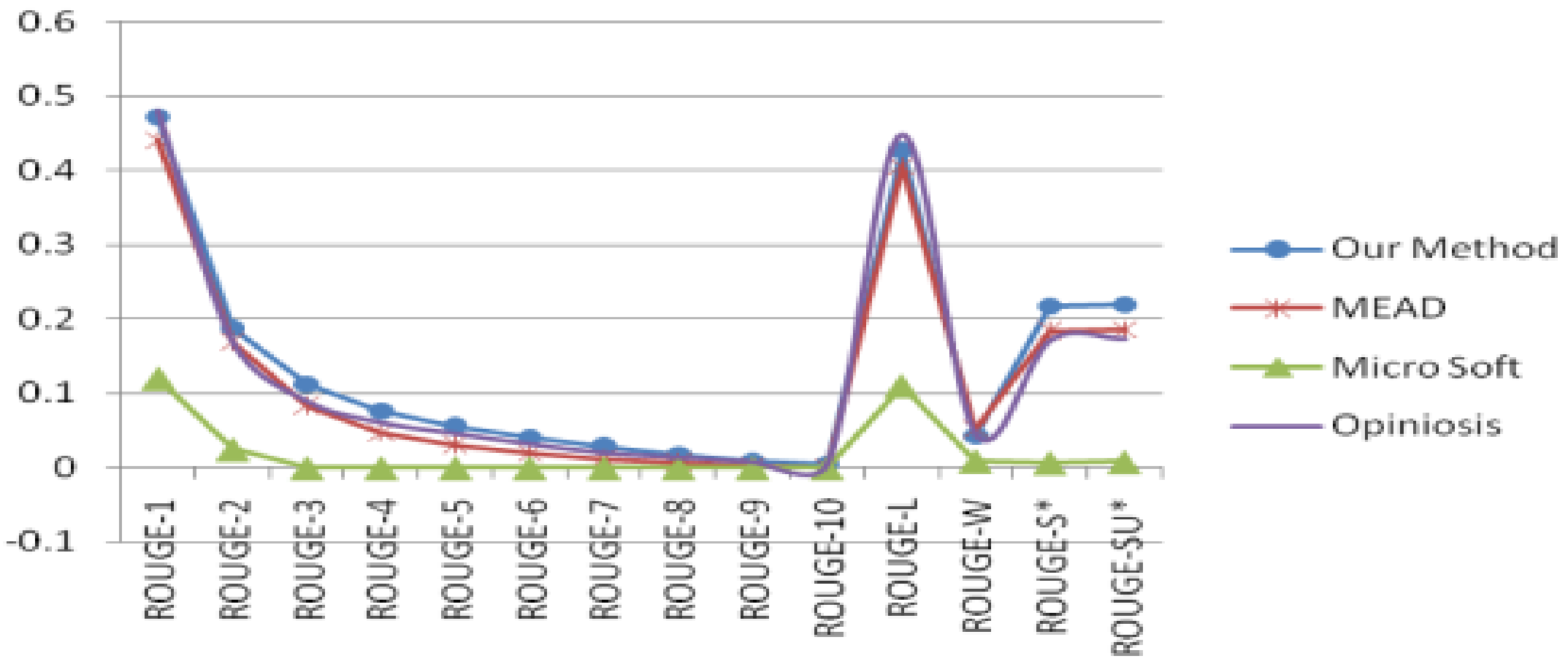

**Figure-2.8**: F-score 24 % summary

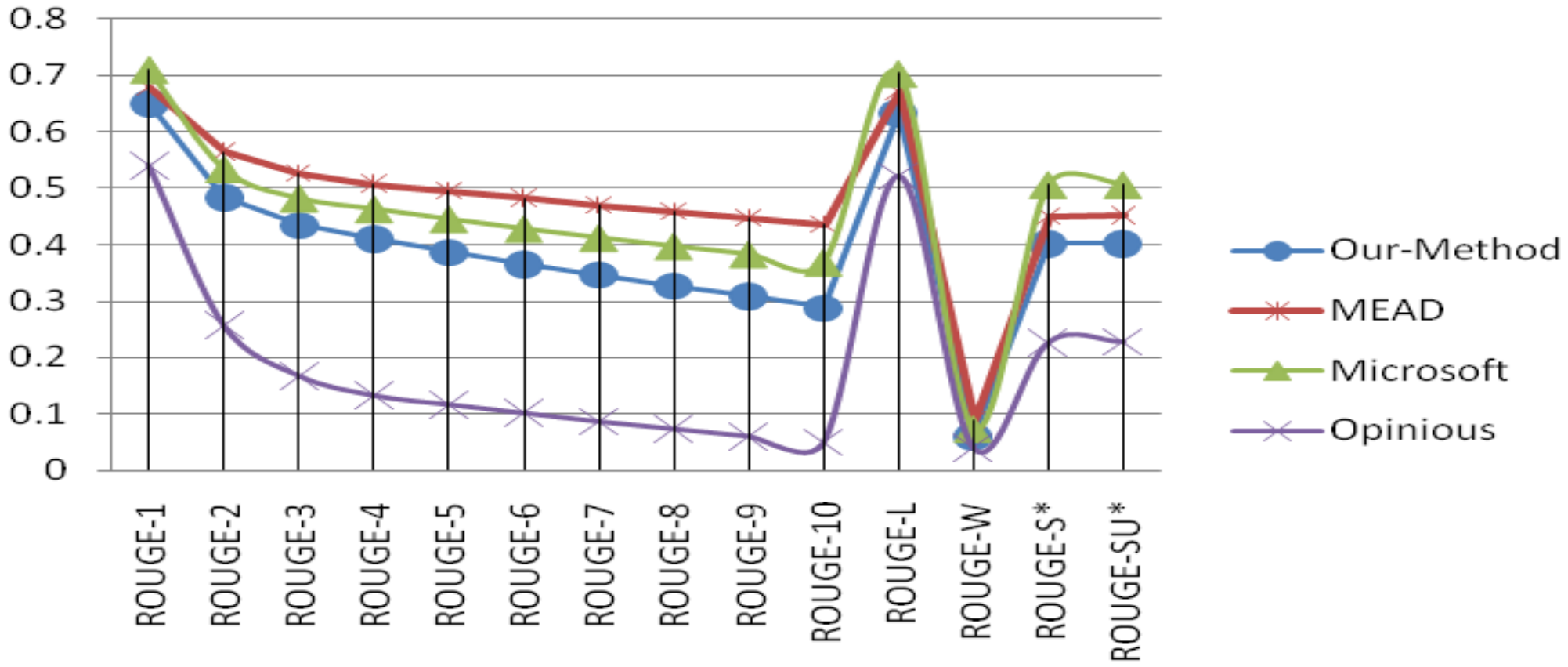

**Figure-2.9**: F-Score 40% summary

### 2.6.3 Experiment 2.3

In the third Experiment, we are showing the significance of Sentiment feature. The purpose of this experiment to show, is really sentiment score performs a significant role in salient sentence extraction. To generate a good quality summary of limited words like hundred words is a tedious task. In our experiment, we have considered five different features. We have tried all combinations of all five features and using this combination we are trying to prove this feature is playing a significant role in summarization. If a number of features are n then, with at least one feature, the total number of combinations possible=$2^n$-1, so in our experiment, we are trying all 31 combinations (calling 31 approaches). Here we are generating a summary using single stand-alone feature-based summary and summary in which sentiment score is playing a role. Here first we generate approximate 100 word summary, to evaluate this summary, we have taken three human-generated summaries as a Gold / Reference summary. Motivated by DUC -2002 task, we are evaluating only the first 100 words of the summary. Let: 1- stands for TF-IDF feature; 2-stands for Aggregate similarity score; 3-stands for Position based feature; 4- stands for the Centroid-based feature; 5-stands for Sentiment based score; 1+2+3 feature showing a collective score of TF-IDF, Aggregate similarity, Position based features.

Table- **2.10:** Different ROUGE score for summary generated using different Approaches.

| **Approach** | **1** | **2** | **3** | **4** | **5** | **9** | **12** | **14** | **15** | **6** | **18** | **10** | **20** |
|---|---|---|---|---|---|---|---|---|---|---|---|---|---|
| **Measures** | **1** | **2** | **3** | **4** | **5** | **1+5** | **2+5** | **3+5** | **4+5** | **1+2** | **1+2+5** | **2+3** | **2+3+5** |
| ROUGE-1 | 0.563 | 0.279 | 0.622 | 0.265 | 0.566 | 0.576 | 0.569 | 0.585 | 0.563 | 0.56 | 0.582 | 0.625 | 0.582 |
| ROUGE-2 | 0.459 | 0.077 | 0.5 | 0.056 | 0.427 | 0.433 | 0.429 | 0.433 | 0.447 | 0.457 | 0.433 | 0.504 | 0.433 |
| ROUGE-L | 0.54 | 0.242 | 0.592 | 0.216 | 0.537 | 0.533 | 0.54 | 0.54 | 0.53 | 0.537 | 0.533 | 0.595 | 0.536 |
| ROUGE-W | 0.319 | 0.105 | 0.342 | 0.092 | 0.315 | 0.314 | 0.316 | 0.316 | 0.314 | 0.318 | 0.313 | 0.343 | 0.315 |
| ROUGE-S* | 0.324 | 0.065 | 0.403 | 0.059 | 0.34 | 0.348 | 0.343 | 0.361 | 0.329 | 0.321 | 0.357 | 0.409 | 0.358 |
| ROUGE-SU* | 0.329 | 0.069 | 0.407 | 0.062 | 0.344 | 0.352 | 0.348 | 0.365 | 0.334 | 0.326 | 0.362 | 0.413 | 0.362 |

| **Approach** | **13** | **21** | **19** | **23** | **17** | **24** | **22** | **26** | **16** | **27** |
|---|---|---|---|---|---|---|---|---|---|---|
| **Measures** | **3+4** | **3+4+5** | **2+3+4** | **2+3+4+5** | **1+2+4** | **1+2+4+5** | **1+2+3+4** | **1+2+3+4+5** | **1+2+3** | **1+2+3+5** |
| ROUGE-1 | 0.504 | 0.563 | 0.508 | 0.563 | 0.284 | 0.566 | 0.504 | 0.563 | 0.553 | 0.595 |
| ROUGE-2 | 0.35 | 0.446 | 0.355 | 0.447 | 0.062 | 0.449 | 0.35 | 0.446 | 0.43 | 0.44 |
| ROUGE-L | 0.454 | 0.523 | 0.459 | 0.524 | 0.232 | 0.527 | 0.454 | 0.523 | 0.527 | 0.545 |
| ROUGE-W | 0.254 | 0.312 | 0.258 | 0.312 | 0.098 | 0.313 | 0.254 | 0.312 | 0.313 | 0.318 |
| ROUGE-S* | 0.201 | 0.329 | 0.204 | 0.328 | 0.065 | 0.331 | 0.201 | 0.329 | 0.33 | 0.367 |
| ROUGE-SU* | 0.207 | 0.334 | 0.21 | 0.333 | 0.069 | 0.336 | 0.207 | 0.334 | 0.334 | 0.372 |

Here we are presenting 23 different features combination to find a summary. By seeing Table-2.10, we can conclude that Position based feature (approach 3- highlighted in green) is performing best among all. This is due to, that in all three human reference summary (out of five summary three are extractive type summary- extractive type summary is available at address 36 ), which are used for evaluation contains top sentences 1, 3, and 4 which is almost 100 words, and the model which we are using for score position giving higher preference for leading sentences. It is widely considered that position based score can't perform well in all cases for example in scientific article. So we need some more features. From **Table-**2.10 this is clear when we are taking Sentiment feature (id 5), along with other features we are getting an improved summary. More Rouge score means more accurate summary. The conclusion of this experiment is that,

- Out of 11 approaches (when sentiment feature added), nine times we are getting an improved summary by adding Sentiment feature. For example, if we take collective features (1+2+3+4), and by adding sentiment feature (1+2+3+4+5), we are getting improved results (highlighted in red color).
- Here Position based feature performing best among all approaches, the reason is given above and we can't depend only on Position based feature, so we need more features.
- In Approach 10 (2+3, i.e., Aggregate and Position), when we add Sentiment score (done in approach 20, i.e., 2+3+5, highlighted in Blue color), the performance is reduced, this is due to Position based score not preferred, i.e., Position based feature is not dominating here as in Approach 3.
- Results are obtained from three human summaries as Reference summary, and summary obtains from different 31 approaches consider as system summary, along with document are available at (C.S. Yadav et al., 2014). To remove biases and evaluate the first 100 world summary, we use -l 100 i.e. initial 100-word summary, and to evaluate ROUGE-W we have taken W=1.2.

We have performed this experiment on DUC 2002 dataset, and after the results are shown in Table-2.11. We have found out that after incorporating sentiment feature the performance has been improved. This experiment is performed with and without stop words.

**Table-2.11:** Different ROUGE score is shown on DUC dataset

| **ROUGE** | **Location** | **Location + Sentiment** | **Location + Stop word removed** | **Location +Sentiment Stopword removed** |
|---|---|---|---|---|
| Rouge-1 | .399 | .400 | .322 | .326 |
| Rouge-2 | .164 | .167 | .174 | .168 |
| Rouge-3 | .103 | .111 | .105 | .100 |
| Rouge-4 | .074 | .085 | .068 | .061 |
| Rouge-L | .3622 | .363 | .300 | .315 |
| Rouge-W 1.2 | .1700 | .172 | .174 | .174 |
| Rouge-S | .1543 | .149 | .112 | .098 |
| Rouge-S* | .159 | .154 | .119 | .105 |

### 2.6.4 Experiment 2.4

In this experiment, we are finding the best optimal feature weights to get more improved results. To decide the feature weights, we are using a supervised algorithm that is Random forest and logistic regression. In this, Internal estimates monitor error, strength, correlation, and it is used to evaluate the variable importance. This experiment has been performed on DUC dataset. In Figure-2.10 we are showing correlation between independent features those are used by us. By visualizing, we can say there is no correlation or less correlation between features.

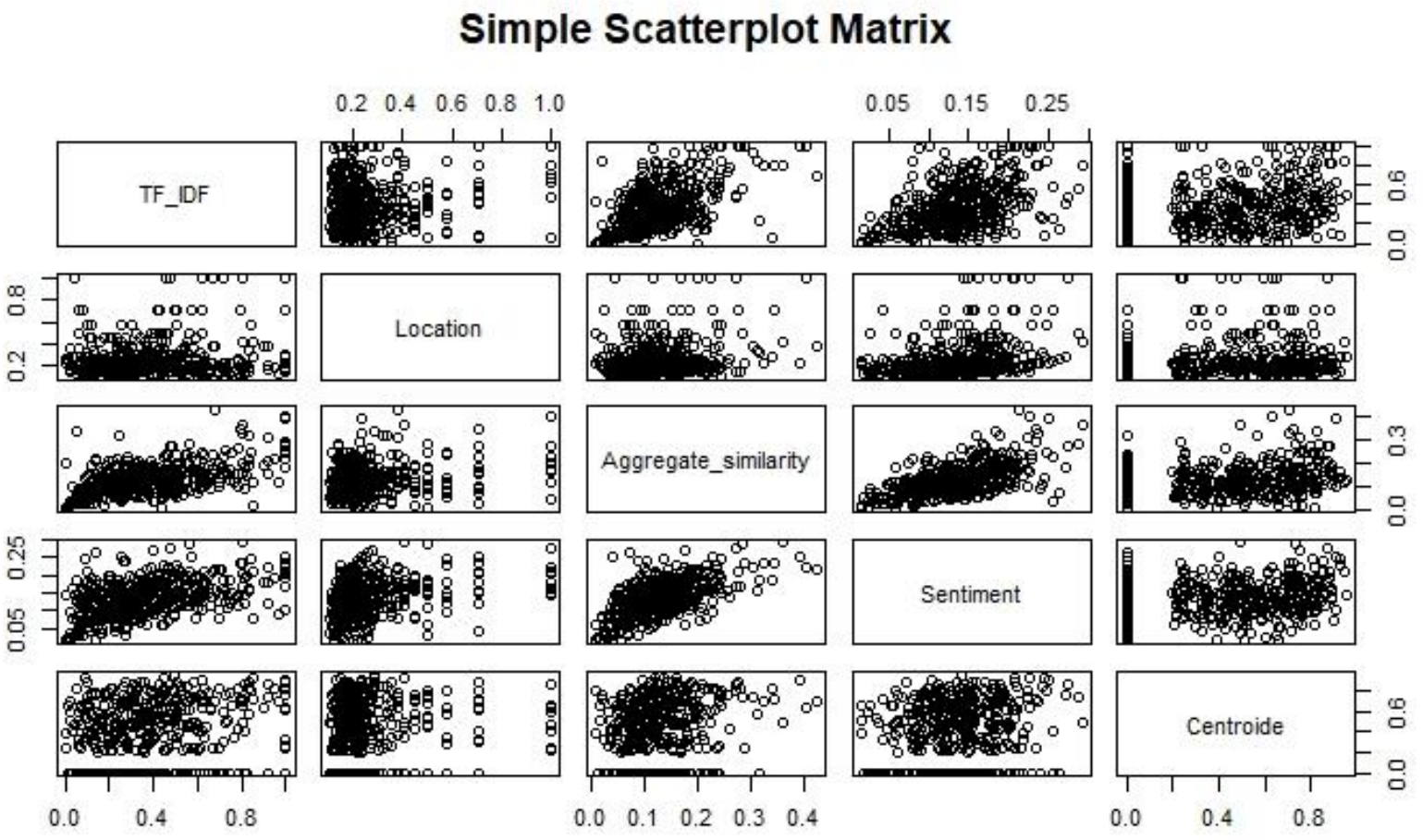

Figure-2.10: showing the correlation between feature used for summarization.

We have divided our data into training and testing and develop a model to find best feature weight combination. In this regard, we have developed two models Random forest and Logistic Regression. In Table-2.12 features weights are suggested with corresponding model currency.

Table 2.12: Showing optimal feature weights using different models

| Model | Location | TF-IDF | Centroid | Aggregate Similarity | Sentiment | Model Accuracy |
|---|---|---|---|---|---|---|
| **Random Forest** | 20.12116 | 22.53912 | 12.60026 | 22.84219 | 39.18935 | 76.34 % |
| **Logistics Regression** | 0.4469 | 0.8440 | .4545 | .1878 | .4978 | 79.4 % |

From, Table 2.12, we can say that in logistic regression model sentiment score weight is 39.18 is the highest feature weight, this signifies importance of sentiment is highest here. In Logistic Regression based model sentiment feature weight is second highest. Between, these two models Random forest and Logistic Regression, Logistic regression suggesting feature weights 0.4446 (w1), 0.8440 (w2), 0.4545 (w3), 0.1878 (w4), 0.4978 (w5) corresponding to location, TF-IDF, Centroid, Aggregate similarity, and sentiment, with model accuracy 79.4%. So, here we are testing the model-2.4.1 and 2.4.2, represented by corresponding equation 2.21 and 2.22.

$$Model\ 2.4.1 = location + TF - IDF + Centroid + Aggregate\ similarity + sentiment \quad (2.21)$$

$$Model\ 2.4.2 = (w1 \times location) + (w2 \times TF - IDF) + (w3 \times Centroid) + (w4 \times Aggregate\ similarity) + (w5 \times sentiment) \quad (2.22)$$

**Table 2.13: Showing performance of Model 2.4.1**

| | P | R | F | (95%-conf.int. P) | | (95%-conf.int. R) | | (95%-conf.int. F) | |
|---|---|---|---|---|---|---|---|---|---|
| **ROUGE-1** | 0.26441 | 0.31649 | 0.28798 | 0.21369 | 0.31704 | 0.2562 | 0.37572 | 0.23196 | 0.3444 |
| **ROUGE-2** | 0.08008 | 0.09595 | 0.08726 | 0.04648 | 0.11428 | 0.05594 | 0.13739 | 0.05075 | 0.12418 |
| **ROUGE-3** | 0.04196 | 0.05031 | 0.04574 | 0.01684 | 0.06922 | 0.02053 | 0.08278 | 0.0184 | 0.0751 |
| **ROUGE-4** | 0.02387 | 0.02886 | 0.02612 | 0.00684 | 0.04486 | 0.00845 | 0.05465 | 0.00752 | 0.04917 |
| **ROUGE-L** | 0.22948 | 0.27484 | 0.25 | 0.17986 | 0.28181 | 0.21613 | 0.33504 | 0.19677 | 0.30583 |
| **ROUGE-W 1.2** | 0.08987 | 0.20152 | 0.12422 | 0.07084 | 0.10903 | 0.15856 | 0.24421 | 0.09782 | 0.1501 |
| **ROUGE-S*** | 0.06524 | 0.09306 | 0.07657 | 0.04118 | 0.09 | 0.05922 | 0.12735 | 0.04845 | 0.10529 |
| **ROUGE-SU*** | 0.0714 | 0.10131 | 0.08362 | 0.04651 | 0.09694 | 0.06647 | 0.13607 | 0.05464 | 0.11282 |

**Table 2.14: Showing Performance o Model 2.4.2**

| | **P** | **R** | **F** | **(95%-conf.int. P)** | | **(95%-conf.int. R)** | | **(95%-conf.int. F)** | |
|---|---|---|---|---|---|---|---|---|---|
| **ROUGE-1** | 0.27434 | 0.32673 | 0.29803 | 0.2238 | 0.32937 | 0.26956 | 0.38534 | 0.24459 | 0.35532 |
| **ROUGE-2** | 0.07775 | 0.09222 | 0.08431 | 0.04633 | 0.11253 | 0.05553 | 0.13197 | 0.05049 | 0.12092 |
| **ROUGE-3** | 0.03799 | 0.04498 | 0.04116 | 0.01443 | 0.06371 | 0.01799 | 0.07584 | 0.01602 | 0.06891 |
| **ROUGE-4** | 0.02178 | 0.02605 | 0.02371 | 0.00584 | 0.04246 | 0.00719 | 0.05129 | 0.00641 | 0.04638 |
| **ROUGE-L** | 0.23945 | 0.2854 | 0.26022 | 0.18912 | 0.29345 | 0.22891 | 0.34442 | 0.20707 | 0.31598 |
| **ROUGE-W 1.2** | 0.09045 | 0.20129 | 0.12472 | 0.07247 | 0.10917 | 0.16333 | 0.24096 | 0.10022 | 0.14953 |
| **ROUGE-S*** | 0.06795 | 0.09564 | 0.07922 | 0.04367 | 0.09407 | 0.06444 | 0.12807 | 0.05162 | 0.10764 |
| **ROUGE-SU*** | 0.07349 | 0.10303 | 0.08555 | 0.0486 | 0.10038 | 0.07095 | 0.13615 | 0.05746 | 0.11449 |

Model 2.4.1 is simple linear model, which is proposed and tested in previous sections on self-made data set. In this section we have implemented that model on DUC dataset. In this experiment 2.4, we have implemented the previous model (as 2.4.1), and new model (2.4.1) after feature weight learning using regression analysis. Results are attached in Table 2.13 and 2.14. Performance of model is decided based on Precision (P), Recall (R), and F-Score (F), and 95% confidence interval shows the probability of lying values of P, R, F-score between given range. Figure 2.11 shows comparative performance analysis based on content based ROUGE measure. These tables 2.13 and 2.14, and figure 2.11 give a proof of significant change in results in model 2.4.2 over 2.4.1.

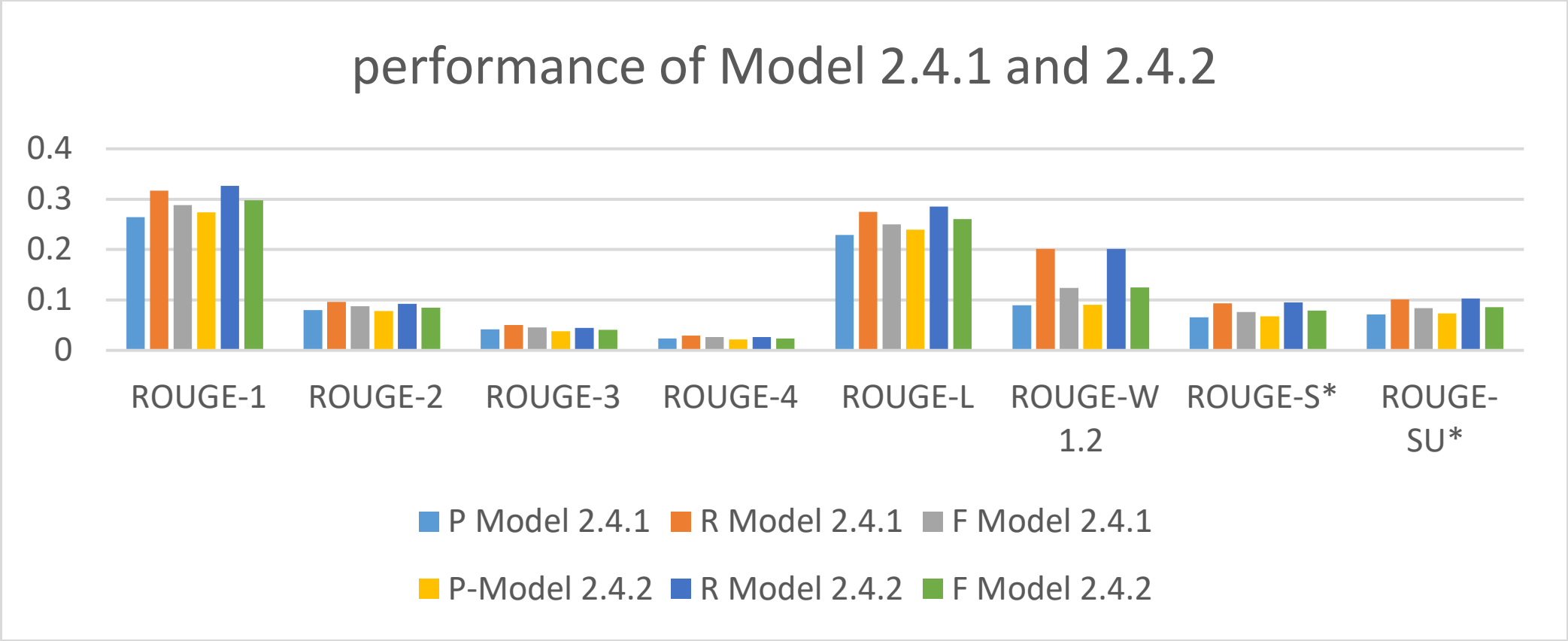

Figure: 2.11 Comparative performance of Model 2.4.1 and Model 2.4.2

## 2.7 Concluding Remark

In this work, we have presented a hybrid model for text document summarization which is based on a linear combination of different statistical measures and semantic measures. In our hybrid approach, we have considered statistical measures like sentence position, centroid, TF-IDF as well as a semantic approach (doing sentiment analysis) that is based on word-level analysis. Sentiment score of a sentence is given as the sum of sentiment score of every entity present in a sentence. We are getting three polarities for any entity as Neutral, Negative and Positive. If any entity sentiment is negative, then we are multiplying every score by -1 to treat it as positive score. The reason for doing this we want to select a sentence in which strong sentiment is present it may be either negative or positive, and both have the same importance for us. To calculate the score or importance for a sentence, we have just added all the scores for every sentence and pick up the highest scoring sentence. In the next step, if the similarity between summary and sentences is lower than the threshold to maintain diversity, then, we have added it in the summary. Our stopping criteria is summary length constraints.

To generate several summaries of different length, we have used methods like MEAD, Microsoft, OPINOSIS and HUMAN generated summary. Evaluation is done by ROUGE measure. In this chapter, we have done four experiments. In the first approach, we took our proposed algorithm based generated summary as system summary and all other as a model summary. In this experiment, it has shown that we are getting high precision almost every time. This signifies, that we covered most relevant results. In the second experiment, we have compared different system generated a summary (MEAD, Microsoft, OPINOSIS, and our algorithm) to the Model summary (human generated). In this we find that our explained algorithm performed well for 24% generated summary for almost every time but, in 40% MEAD system generates a summary leading in some way but here also we are getting higher RECALL comparatively to MEAD. In the third experiment, ***section*** 2.6.3 we have shown that when we are adding sentiment score as a feature, we are getting improved results to compare to without a sentiment score. The third experiment is showing that sentiment score has a contribution in the extraction of more appropriate sentences. In the fourth experiment we have divided our data into training and testing and proposed a feature weighted approach using Random forest and Regression. The logistic regression model is giving better accuracy. Since, Logistic Regression is producing better model, so feature weights are assigned as per regression model. A new experiment again performed to show the model improvement using feature weight analysis.

# Chapter 3: A new Latent Semantic Analysis and Entropy-based Approach for Automatic Text Document Summarization

## 3.1 Introduction

In this chapter, we are proposing Latent Semantic Analysis (LSA) based model for text summarization. In this work, we have proposed three models those are based on two approaches. In the first proposed model, two sentences are extracted from the right singular matrix to maintain diversity in the summary. Second and third proposed model is based on Shannon entropy, in which the score of a Latent/concept (in the second approach) and Sentences (in the third approach) are extracted based on the highest entropy.

The advantage of these models is that these are not Length dominating model, giving better results, and low redundancy. Along with these three new models, entropy-based summary evaluation criteria are proposed and tested. We are also showing that our entropy based proposed model is statistically closer to DUC-2002's standard/gold summary. In this work, we are using dataset taken from Document Understanding Conference-2002.

## 3.2 Background

In this ***section***, we are presenting an introduction to LSA, Advantages, and Limitation of LSA, how LSA can be used for summarization, and how to find Information content by using the principle of Information theory.

### 3.2.1 Introduction to Latent Semantic Analysis

In text mining direction, early application of LSA has started by (Deerwester, Dumais, Furnas, Landauer, & Harshman, 1990), their objective was indexing of text document, now it has been used for various applications in automatic text document summarization. First, the objective is to convert a given document "D" into matrix representation "A." Matrix "A" is Term-Document matrix in which elements $A_{ij}$ represents the weighted term frequency of $i^{th}$ *term*, in $i^{th}$ in the *document*. This can be designed by combining local and global weight as Equation-3.1.

$$A_{ij} = Local_weight_{ij} \times Global_weight_{ij} \quad (3.1)$$

LSA is a vector space approach that involves the projection of the given matrix. Matrix $A_{M\times N}$, usually M>>N, is represented in reduced dimension r denoted by "$A_r$" such that r < M. Using LSA, input matrix "A" is decomposed as Equation-3.2, and Figure-3.1.

$$A = U \times \sum \times V^T \quad (3.2)$$

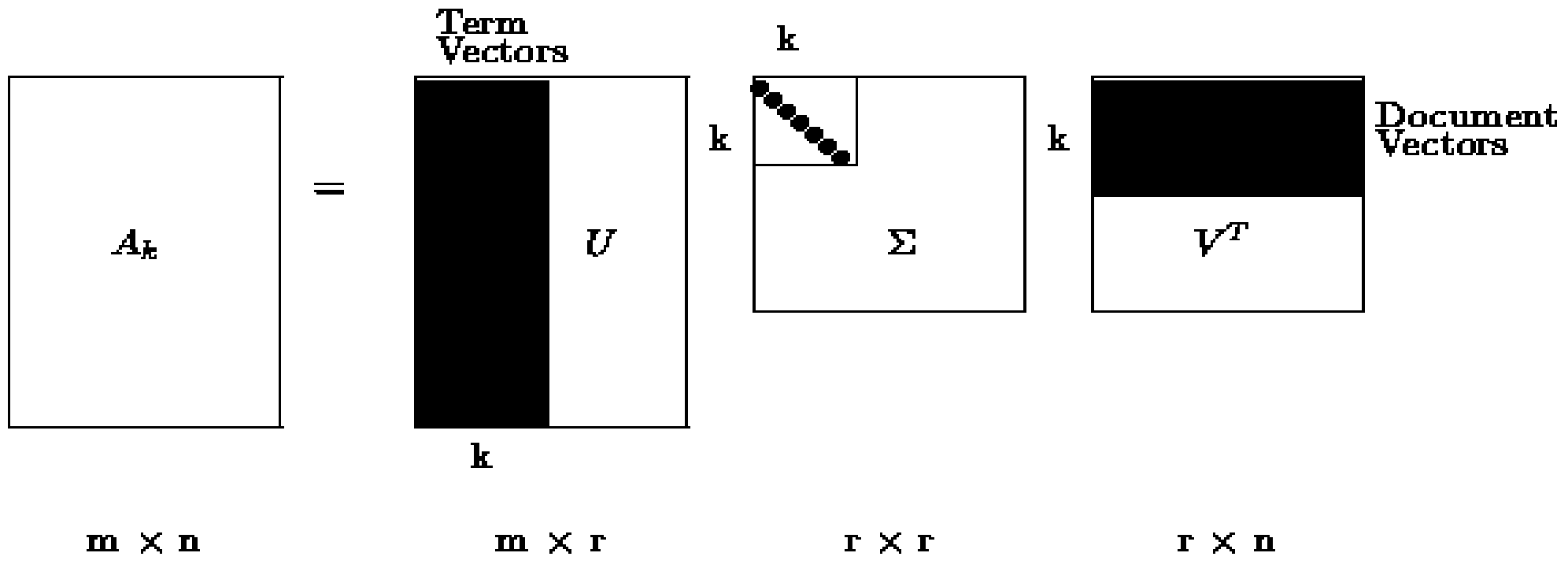

**Figure-3.1**: Showing Mechanism of LSA, Matrix "A" decomposed into U, $\sum$, and $V^T$ Matrix

Where U is an M×N column orthonormal matrix, which columns are called left singular vector, $\sum_{N(N}$ is a N×N diagonal covariance matrix, whose diagonal elements are non-negative singular values, which are sorted in descending order as in Equation-3.3. $V^T$ right singular matrix also an orthonormal matrix with size N×N, which columns are called right singular vector. Let, Rank(A) = r, then matrix $\sum$ matrix's properties can be express as Equation-3.3, where $\sigma_i$ ,1 < i< r and r ≤ N is representing diagonal elements,

$$\sigma_1 \geq \sigma_2 \geq \sigma_3 .... \geq \sigma_r \text{ , and}$$
$$\sigma_{r+1} = ....\sigma_{N-1} = \sigma_N = 0 \quad (3.3)$$

Several methods decomposition of factorization is available, like ULV low rank orthogonal, Semi discrete decomposition, and SVD. We are using LSA based on SVD decomposition, the reason of this is,

1. SVD decompose matrix "A" into orthogonal factors that represent both types and documents. Vector representation of both types and documents are achieved simultaneously.

2. Second, the SVD sufficiently capture for adjusting the representation of types and documents in the vector space by choosing the number of dimensions.
3. Computation of the SVD in manageable for the large data set.

The interpretation of applying the SVD to the term by sentence matrix “A” can be made from two viewpoints,

1. Transformation point of view, the SVD derives a mapping between the M-dimensional space spanned by the weighted Term Frequency vectors and the r dimensional singular vector space.
2. The semantic point of view, the SVD derives the latent semantic structure from the document represented by matrix “A”.

There are some obvious reasons for using LSA, but sometimes this has some limitations in implementations like this,

**Advantage**:

- First, all the textual units, including documents and word have the ability to be mapped to the same concept space. In this concept space, we can cluster words, documents, and easy to find out how these clusters coincide so we can retrieve documents based on words and vice versa.
- Second, the concept of space has immensely fewer K dimensions compared to the designed original matrix M, i.e., M>>K.
- Third, LSA is an inherently global algorithm, but, LSA can also be usefully combined with a more local algorithm to be more useful.

**Limitations**:

- LSA considers a Frobenius norm and Gaussian distribution which is not fit for all problems. In our problem, our data follow a Poisson distribution, i.e. depends on word counts as studied by (S. M. Katz, 1996).
- LSA can't handle Polysemy (multiple meanings).
- LSA depend on singular value decomposition (SVD) which is computationally expensive and hard to update if the case of new documents required to add.

### 3.2.2 LSA and Summarization

LSA is an algebraic, statistical technique. SVD over given document can be understood regarding independent concepts. SVD is a method, which models relationships between words and sentences. The beauty of LSA is that it can find out semantically similar words and sentences, it has the capability of noise reduction, which leads to improved accuracy. Our LSA based Summarization is presented in Figure-3.2 which comprises three major steps (1) Input Matrix creation, (2) SVD, and (3) Sentence Extraction/Selection, which is briefly explained in this ***section***.

*Note: * Document and Sentence are interchangeable, for us sentences are documents.*

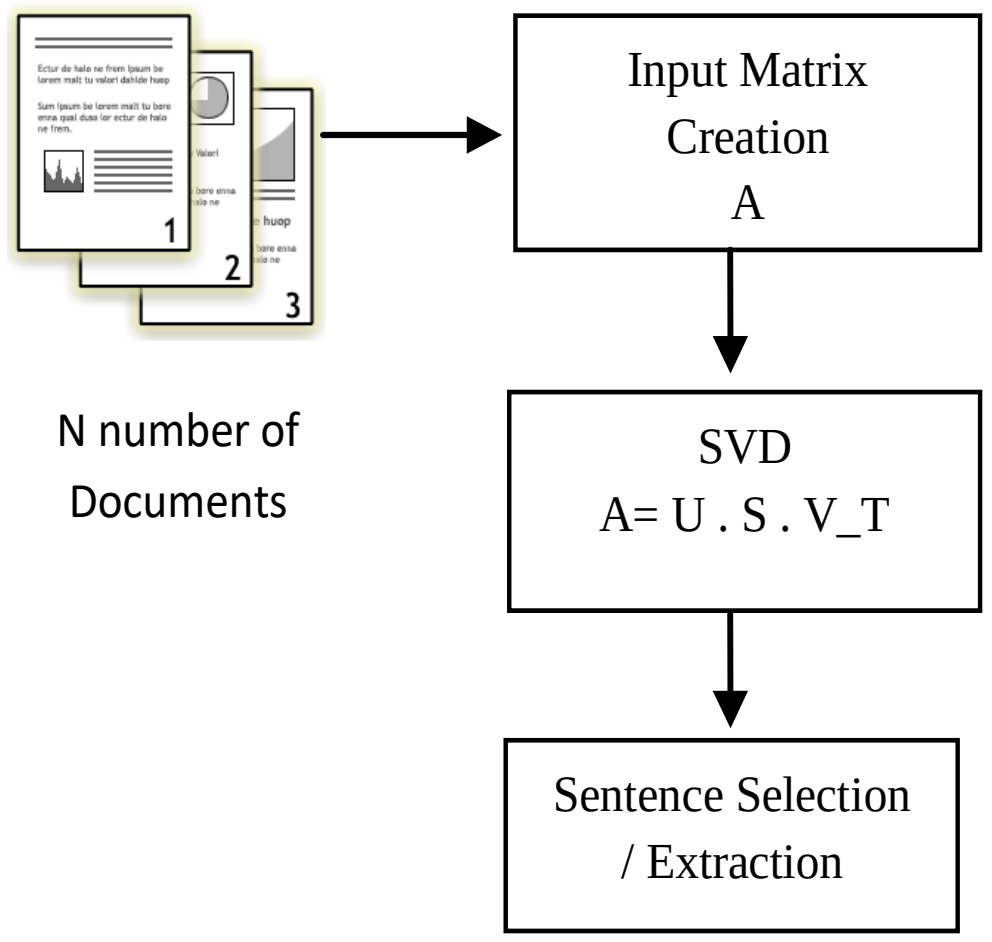

**Figure-3.2**: LSA based Summarization procedure

**Step-1: Input Matrix Creation,**

In most of the papers, Input matrix creation is based on a bag of word approach or word count approach, because it is easy to create, as well as widely accepted. Input matrix may be considered as a combination of Local weight ($L_{ij}$) and Global weight($G_i$). Local weight criteria define weight in the range of document it may be a sentence, paragraph or set of paragraphs, and global weight is decided among all documents. Let we want to use a bag of word approach, and then some common local weighting methods are mentioned in Table-3.1,

| S.NO. | Local Weight Approach | Global Weight Approach |
|---|---|---|
| **1.** | Binary : $L_{ij}$= 1 if term exists in the document else $L_{ij}$=0, | Binary: $G_i$= 1, if term in document i |
| **2.** | Term-Frequency: $L_{ij}=TF_{ij}$, the number of occurrence of term i in document j, | $Normal\ G_i = \frac{1}{\sqrt{\sum_j TF_{ij}^2}}$ |
| **3.** | Log : $L_{ij}$= log($TF_{ij}$ +1) , $TF_{ij}$ is the number of occurrence of term i in document j, | GF-IDF: , where $gf_i$ is the total number of times term i occur in the whole collection, and $df_i$ is $i^{th}$ the number of documents in which term i occurs. |
| **4.** | $Augnorm\ L_{ij} = \left(\frac{TF_i}{Max(TFi_j)}\right) + \frac{1}{2}$ | $IDF - Gi = \log_2 \frac{n}{1+df_i}$ |
| **5.** | Semantic: In matrix A with size $M \times N$, location $A_{ij}$ can connect to location $A_{fg}$ , if there is any semantic connection available between words, m < i, j, f, g,< n. Semantic connection examples are Synonyms, Anatomy, Polysemy, Hypernymy, Hyponym, etc. | $Entropy\ G_i = 1 + \sum_j \frac{p_{ij} \log p_{ij}}{\log n}, where\ p_{ij} = \frac{tf_{ij}}{gf_i}$ <br><br> where $gf_i$ is the total number of times term i occur in the whole collection, $TF_{ij}$ is the number of occurrence of term i in document j, n total documents, $p_{ij}$ probability of occurrence of $i^{th}$ term in $j^{th}$ document. |

Table-3.1: Local and Global Weight model

**Step2: Decomposition**

LSA uses SVD for matrix decomposition. If we have an input matrix "A" (M, N, M- number of words a,d N- number of sentences). Then SVD decomposes "A" matrix decomposed into three matrixes such as left singular "U" described as Words $\times$ Concepts with $M \times N$ size, $\sum$ is a matrix with $N \times N$ size represents scaling values and "V" matrix $N \times N$ size Sentences $\times$ Concepts.

**Step3: Sentence selection**

Right singular matrix gives information about sentences and concepts if we want information about sentence we can use "V." Detailed about selection method is presented in related work. Some authors like (Gong & Liu, 2001), used "S" and "V" both for sentence selection.

### 3.2.3 Introduction to Entropy and Information

Entropy was originally defined by (Shannon, 1948) to study the amount of information contained in the transmitted message. In information theory, entropy may be considered as the measure of the amount of the information that is missing from transmitted information and received information and is sometimes called as Shannon entropy.

The definition of the information entropy can express regarding a discrete set of probabilities $P(X_i)$. For example, in the case of a transmitted message, these probabilities $P(X_i)$ were the probabilities that a particular message $X_i$ actually transmitted, and the entropy of the message system was a measure of how much information is present in the message. Let in a message M; there are n symbols as $m^1$, $m^2$,...., $m^n$ are present. This message M transmitted from source "S" to destination "D." Suppose further that the source S transmits the symbols $m^i$, $1<i<n$ with probabilities $P(m^1)$, $P(m^2)$, ..., $P(m^n)$, respectively. If symbol $m^i$ is repeated T times, then the probability of occurrence of $m^i$ will be given by $T \bullet P_i \approx T \times P_i$. Thus, in T independent observations, the total information "I" given by Equation-3.4. (It is assumed that the symbols are emitted independently)

$$I = \sum_{i=1}^{n"} (T \bullet P_i) \log \frac{1}{P_i} \tag{3.4}$$

The average information of symbols I=T will be calculated by Equation-3.5, 3.6, and 3.7. In our case, we are also using the same entropy function.

$$\frac{1}{T} = \frac{1}{T} \sum_{i=1}^{n} (T \bullet P(m_i)) \log \frac{1}{P(m_i)} \tag{3.5}$$

$$I = \sum_{i=1}^{n} P(m_i) \log \frac{1}{P(m_i)} \tag{3.6}$$

$$I = -\sum_{i=1}^{n} P(m_i) \log P(m_i) \tag{3.7}$$

## 3.3. Related Work

In this ***section***, we are mentioning five previous different models, which have been used for comparison purpose. All these five models have based on LSA based decomposition. Each model is presented along with its pros & cons.

### 3.3.1 Model-1 / Gongliu model

This model has been proposed by (Gong & Liu, 2001). Authors have used $V^T$ matrix for sentence selection. According to them each row of $V^T$ representing one topic/concept and corresponding to that topic selects only one sentence, which has the highest corresponding value on that topic. They have added extracted sentences in the summary, and the process is repeated until the required length summary is achieved. We have observed the following advantages and drawbacks in the model-1.

**Salient:**

- New and unique approach.

**Drawback**:

- In this approach, the summary is depended only on $V^T$, as per author convenient, they can select some rows, i.e., from reduced dimensions of $V^T$. The maximum selected summary sentences will be equal to the reduced dimension r. If a case if { ||summary sentences|| > reduced dimension-r}, then, no way is suggested to select more sentences.
- Top concepts represent more information compared to bottom concepts. Sometimes in this approach extraction of sentences that may belong to less important concepts.
- Let system choose an $i^{th}$ concept, for sentence selection. From the selected concept if two sentences have high values like .80 and .79, then Gong's approach will choose only one sentence. As we know both are highly related, but still the second sentence is not respected in this approach.
- All selected concepts give equal importance, but some may not so much importance in the $V^T$.
- This is well known that concepts are independent of each other, so this is expected that those sentences are extracted from concepts are also independent. Ideally, it is not possible in text data especially due to linking of name entities, stopwords and pronoun words present in the text. In simple words, if we want diversity in summary, then our matrix “A” should be noiseless data i.e after removing stop words. So the better output from this approach depends only on the input matrix “A”.

### 3.3.2 Model-2 / Murray Model

This model proposed by, (Murray, Renals, & Carletta, 2005), have used both $V^T$ and S-matrix for sentences selection. Instead of selecting only one sentence from $V^T$ which depends on the highest index value, selects the number of sentences based on matrix S. The number of sentence selection from one concept depends on getting the percentage of the related singular values over the sum of all singular values (or in reduced space r). We have observed the following advantage and drawbacks in the model-2.

**Salient**:

- Overcomes the problem of (Gong & Liu, 2001) approach, which selects only one sentence from each concept.

**Drawback**:

- Some starting values from ∑ matrix will play a significant role and will dominate.

### 3.3.3 Model-3 / SJ Model

(Steinberger & Ježek, 2004) have proposed another model for document summarization. They are also using both ∑ and V matrix (transpose of $V^T$) for sentence selection. To add a sentence, in summary, the author finds the length of a sentence using "S" and "V" in reduced space r that is given by Equation-3.8. "S" matrix is multiplied by "V" matrix to give more emphasis on topics, i.e., by property "S" is sorted in decreasing order, and so the first topic is multiplied with higher value compare with later occur topics.

$$Length\ of\ \ Sentence\ i = \ S_{\ i} = \sqrt{\sum_{j=1}^{n} V_{ij} * \Sigma_{jj}} \tag{3.8}$$

We have observed the following advantage and drawbacks in the model-3.

**Salient**:

- Sentence selection is based on new reduced space, so consideration about only preferred concepts and highest length sentence.

**Drawbacks**:

- Long length sentences may dominate by calculating $S_i$.
- No explicit criteria to increase diversity in the summary.
- Even all sentences are somehow related to all the concepts, in preprocessing authors have considered some sentence as unrelated sentences.

### 3.3.4 Model-4 / Omg-1 Model

Another model cross method was proposed by (Ozsoy, Alpaslan, & Cicekli, 2011), in which $V^T$ matrix is preprocessed that represents only core sentences as per Equation-3.9 (remove sentences that have index value less than mean). Then $V^T$ matrix is multiplied with $\sum$ matrix to give importance for topics shown in Equation-3.is calculatedces length are calculated by adding columns of $V^T$ matrix i.e. Equation-3.11. Highest length sentence is added to the summary, this process is repeated until the required length summary.

$$V^{'T} = \Pr eprocessed(V^T) \quad (3.9)$$

$$Wr = \sum_r \bullet V_r^T \quad \text{// Reduces space r} \quad (3.10)$$

$$Sentence\ i's\ length = S_i = \Sigma_{j=1,i=1}^{j=r,j=n} W_{ji} \quad (3.11)$$

We have observed the following advantages and drawbacks in the model-4.

**Drawback**:

- Diversity, in summary is not considered, i.e., no explicit criteria.
- Length dominating summary, i.e., long length sentences extracted and added to the summary.

### 3.3.5 Model-5 / Omg-2 Model

Another Topic-based approach has proposed by (Ozsoy et al., 2011), in that preprocessing is same as done in cross approach w.r.t. $V^{T,}$ i.e., $V^T$ is representing only core sentences for each topic.

Instead of sentence length approach as explained as Model-3/SJ Model, they have found out important concept. Based on those concepts sentence selection is made. The important concept is found by Concept $\times$ Concept matrix. This matrix is formed by summing up the cell values that are common to these concepts. We have observed the following advantage and drawbacks in the model-5.

**Salient**:

- Finding the main topic based on Concept $\times$ Concept Matrix is giving better results.
- Higher concept score is showing that a concept is much more related to other concepts.

**Drawback**:

- Still following (Gong & Liu, 2001) approach from sentence selection, i.e., select only one sentence from one concept.
- If some chosen a concept-k for sentence selection, in this concept if two sentences have high values, Gong's approach choosing only one sentence. With one example if there is a case that, concept-k is related to 0.80 with sentence $S_i$ and, 0.78 with sentence $S_{i+1}$. As we know, these are highly related then also this is not respected.
- Even some sentences are somehow related to concepts, in preprocessing author put this to zero is same as showing that respective sentence and concept are unrelated.
- This is said that con-0 > con-1 > con-2.....> con-n. (Sign A >B showing A is preferred over B), property not considered.
- Extracted Sentences are assumed to be implicitly diverse.

## 3.4. Proposed Model

Sentence selection is done from either $V^T$ matrix or with a combination of diagonal matrix $\Sigma$ and $V^T$ matrix. As we know the concepts in $V^T$ is independent of each other. So, this is inherently assumed that among extracted sentence have minimum similarity with each other. Matrix $\Sigma$ singular values are always in sorted order, and this is assumed by (Gong & Liu, 2001) $\text{concept}_i$ is preferred over $\text{concept}_{i+1}$, but approach followed by (Steinberger & Ježek, 2004), cross approach by (Ozsoy et al., 2011) , and (Ozsoy, Cicekli, & Alpaslan, 2010) are not considering this phenomenon and sentence selection is based on the longest length sentence given by Equation-3.8 and 3.11 respectively. In both of these approach topics/concept with the highest strength is chosen.

We are proposing two different approaches first one is based on vector concept/relatedness and termed as proposed _model-1, and the second approach is based on entropy that is further explored and represented with proposed_model-2, and proposed_model-3. Before explaining these model or approaches first we are taking an example in *section* 3.4.1, representing a document as D ≈ {$S_1$, $S_2$ …$S_9$}. Creating a matrix "A" of **Words × Sentences.** Over that Matrix "A" we are applying SVD decomposition, and this is represented by $A^{decomposed}$, that can further use in different ways to generate a summary.

In *section* 3.4.1 first we are taking a document D, in example-3.1, creating matrix word frequency based matrix "A," and computing SVD(A)=U.∑.$V^T$, in *section* 3.4.2 proposed model-1, in *section* 3.4.3 proposed model-2, and *section* 3.4.4 contains proposed model-3.

### 3.4.1 Working of LSA with an Example

In this *Section*, first we are representing an LSA-based ***Example-3.1,*** i.e., sentences are numbered from 1 to 9, and corresponding (Word × Sentences) Matrix "A" is represented in **Table** **3.**1. This example is taken from *"http://tiny.cc/w1uczx"*, the reason for doing this is, it Widely available, and this example expresses the property of LSA very well in reduced dimensions 3 (r=3), i.e., representation of given document Term-Document matrix, is same as LSA in 3-dimensional space only. This has been shown at the same link via graph representation in Figure-3.5. Let consider an example with nine given sentences,

***Example-3.1:***

1. "The Neatest Little Guide to Stock Market Investing."
2. "Investing For Dummies, 4th Edition".
3. "The Little Book of Common Sense Investing: The Only Way to Guarantee Your Fair Share of Stock Market Returns."
4. "The Little Book of Value Investing."
5. "Value Investing: From Graham to Buffett and Beyond."
6. "Rich Dad's Guide to Investing: What the Rich Invest in, That the Poor and the Middle Class, Do Not!".
7. "Investing in Real Estate, 5th Edition".
8. "Stock Investing For Dummies."
9. "Rich Dad's Advisors: The ABC's of Real Estate Investing: The Secrets of Finding Hidden Profits Most Investors Miss."

Here, we considered a document D that is equal to D= {$S_1$ U $S_2$......U $S_9$} where $S_i$ representing $i^{th}$ sentence, and considering only underlined words *(Book, Dads, Dummies, Estate, Guide, Investing, Market, Real, Rich, Stock, Value)* as a set of words. Finally, based on presence/not presence words in sentences $S_1$, $S_2$… $S_9$ input Matrix “A” (Word × Sentence matrix) is constructed that is given in Table-3.2. Rows of “A” representing the presence of words in a different sentence with frequency, and columns representing the presence of words. In this example keyword selection is arbitrary, but for keywords selection, we can follow many approaches as proposed by (Beliga, Meštrović, & Martinčić-Ipšić, 2016), (Sharan, Siddiqi, & Singh, 2015).

| **Index Words** | **Titles** | | | | | | | | |
|---|---|---|---|---|---|---|---|---|---|
| | **S 1** | **S 2** | **S 3** | **S 4** | **S 5** | **S 6** | **S 7** | **S 8** | **S 9** |
| **Book** | - | - | 1 | 1 | - | - | - | - | - |
| **Dads** | - | - | - | - | - | 1 | - | - | 1 |
| **Dummies** | - | 1 | - | - | - | - | - | 1 | - |
| **Estate** | - | - | - | - | - | - | 1 | - | 1 |
| **Guide** | 1 | - | - | - | - | 1 | - | - | - |
| **Investing** | 1 | 1 | 1 | 1 | 1 | 1 | 1 | 1 | 1 |
| **Market** | 1 | - | 1 | - | - | - | - | - | - |
| **Real** | - | - | - | - | - | - | 1 | - | 1 |
| **Rich** | - | - | - | - | - | 2 | - | - | 1 |
| **Stock** | 1 | - | 1 | - | - | - | - | 1 | - |
| **Value** | - | - | - | 1 | 1 | - | - | - | - |

**Table-3.2**: Frequency based Words × Documents matrix “A.”

Now $A^{decomposed}$= SVD (A) =U .∑ .$V^T$, in reduced space r =3, “A” is given by composition of this matrix in Figure-3.3, where "U" denotes is Left decomposed Matrix, and "V" for the right decomposed matrix, ∑≈ "S" Matrix Non-Negative Diagonal Matrix.

| | | | |
|---|---|---|---|
| **Book** | 0.15 | -0.27 | 0.04 |
| **Dads** | 0.24 | 0.38 | -0.09 |
| **Dummies** | 0.13 | -0.17 | 0.07 |
| **Estate** | 0.18 | 0.19 | 0.45 |
| **Guide** | 0.22 | 0.09 | -0.46 |
| **Investing** | 0.74 | -0.21 | 0.21 |
| **Market** | 0.18 | -0.30 | -0.28 |
| **Real** | 0.18 | 0.19 | 0.45 |
| **Rich** | 0.36 | 0.59 | -0.34 |
| **Stock** | 0.25 | -0.42 | -0.28 |
| **Value** | 0.12 | -0.14 | 0.23 |

| | | |
|---|---|---|
| 3.91 | 0 | 0 |
| 0 | 2.61 | 0 |
| 0 | 0 | 2.00 |

| D1 | D2 | D3 | D4 | D5 | D6 | D7 | D8 | D9 |
|---|---|---|---|---|---|---|---|---|
| 0.35 | 0.22 | 0.34 | 0.26 | 0.22 | 0.49 | 0.28 | 0.29 | 0.44 |
| -0.32 | -0.15 | -0.46 | -0.24 | -0.14 | 0.55 | 0.07 | -0.31 | 0.44 |
| -0.41 | 0.14 | -0.16 | 0.25 | 0.22 | -0.51 | 0.55 | 0.00 | 0.34 |

Concepts

**Figure-3.3:** LSA decomposition on "A" matrix, in reduced dimension 3.

### 3.4.2 Proposed_Model-1 / MinCorrelation Model

This model is based on proposed_approach-1, and that is using $V^T$ right decomposed matrix. As explained in (Gong & Liu, 2001) approach author used $V^T$ matrix, and only one sentence is selected from each concept. We can fix k number to decide k-sentences to extract from each concept. In our proposed_Approach-1/proposed_model-1, first, we are selecting a concept as done by (Gong & Liu, 2001). Then, instead of selecting just one sentence that is highly related to that concept, select two sentences in such a way that one most related and second least related to that concept i.e. more related means higher value corresponding to entry (concept, sentence), less related means less value corresponding to entry (concept, sentence). This approach is to cover two different topics/sentences from the same concept. Graphically, this procedure can be understood by, Figure-3.4 in a two-dimensional space. Figure-3.4 has four vectors {V1, V2, V3, V4}, and each vector is maintaining some angular distance from X-axis. If corresponding to X-axis, we want to select two vectors, then our choice will be {V1, V4}, because V1 is closed to X-axis and V4 is least related to V1. Corresponding to X-axis, we want to select two vectors, then our choice is {V1, V4}, if we are interested in selecting three vectors, then the selection set will be {V1, V4, V2}. In our application V1, V2, etc. are a reference to documents.

If the total number of the vector is N' then objective may give by Equation-3.12. Here co-relation may be some similarity measure or other measure, but, in this work, co-relation is measured from co-occurrence matrix find by $V^T$, by LSA,

$$Total_co-relation_score = Minimun\left( \sum_{j=1}^{N'} \sum_{i=1, i \neq j}^{N'} Co-relation(Vi,\ Vj) \right)^{*} \qquad (3.12)$$

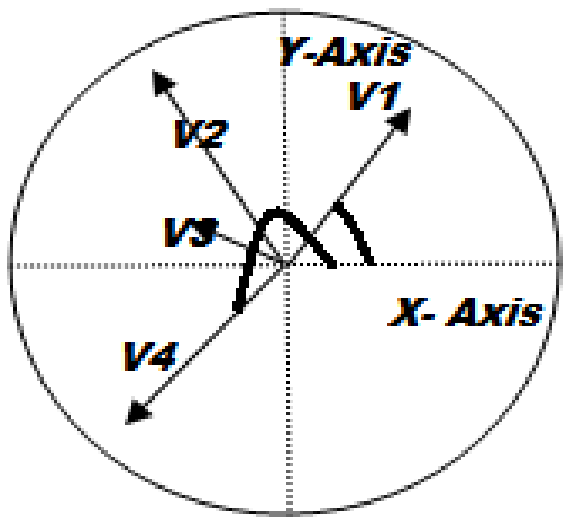

**Figure-3.4**: Vector representation in 2-D space**

*Objective Function denoted by Equation-3.12 return set of vector (sentences) those are minimal co-related.
**Above given Example can be understood by Equation-3.12, and Figure-3.4. In Figure 3.4 co-relation can be considered as angular distance between vectors.

### 3.4.3 Proposed_Model-2/ LSACS Model

Proposed_Model-2/ LSA based concept selection (LSACS) is based on proposed_approch-2, which is based on entropy technique. In this model, we have to find out concepts that contain more information. How much information is contained in a concept is decided based on entropy. Detail about information and Entropy are presented in ***section*** 3.2.5. Concepts/hidden latent with high entropy is showing more degree of freedom that means, the particular concept is related to the number of dimensions/concepts/sentences. Since concepts are hidden in SVD so we cannot show these latent, but just for consideration if we consider dimensions X, Y and Z as a latent/concept, then we are interested to find which dimension is more representative (anyone), and we are finding this using Shandon's entropy proposed in (Shannon, 1948). In Figure-3.5 sentences are represented using triangle sign Δ (blue color, from $S_1$ to $S_9$, in example-3.1 $D_1$ to $D_9$), words are represented by star sign * (red color, underlined words in example 3.1). The X, Y, and Z axis (which are assumed as like hidden latent, and not possible to show in the physical diagram) is represented by the arrow sign in red color. Below we are presenting the proposed algorithm. The objective of this "to find the one latent/axis among X, Y, or Z which represents more information."

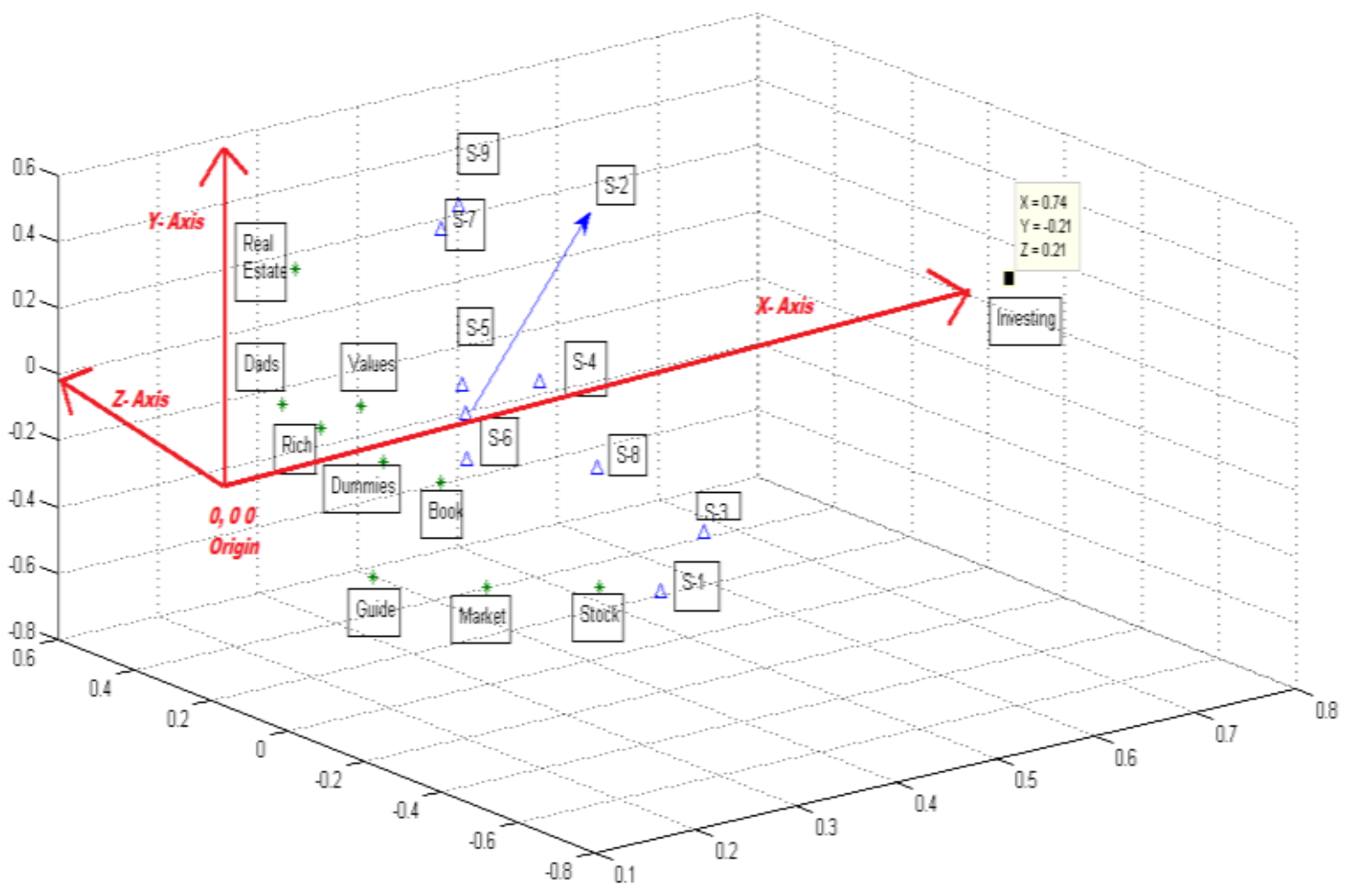

**Figure-3.5**: U and $V^T$ Matrix in 3-D space, representing Words and Sentences

**Algorithm-3.1:**

**Input:**

Document D, i.e., the content of words, sentences, paragraphs;

L- Predefined length of summary; initially L=0; $L^{max}$= 100

$\theta$ -cosine similarity threshold = 0.4,

**Output:** Summary

1. Decompose the given document D into sentences {$S_1$, $S_2$,.. $S_n$}, extract keywords and use these keywords to form the master sentence "S."
2. For given document D, construct a matrix "A," i.e., Term (Sentence matrix. "A" can design by various ways we are using TF-IDF approach.
3. Perform the SVD on A as

$$A = U \times \sum \times V^T$$

4. For every sentence in reduced space r, by using right singular matrix "$V^T$" and $\sum$, compute "W" matrix. Columns of W are representing sentences and rows are representing concepts/latent.
   I. Case-1: $W = V_r^{'T}$ similar to Gong Approach or Model-1
   II. Case -2: $W = S_r \times V_r^{'T}$ similar to S&J approach or Model-3*
5. Preprocess on W by selecting only core sentences corresponding to concepts, by strikeout least related sentences. // (Table-3.4)
6. For each row in W, compute P($X_i$), where $P(X_i) = X_i \Big/ \sum_i^n X_i$ , this P($X_i$) is representing the probability of sentence j appear in concept i.

7. Find the information contains in each concept using Shannon's entropy method, as follow ***section*** 3.2.5

$$-\sum P(X_i).\log(P(X_i))$$

8. Select most informative concept $C_p$, $1 \le P \le r$ , r reduce space,
9. Using W matrix, and selected concept $C_p$, select a sentence $S_i$ that is most related to $C_p$:

    if: (" $L<L^{max}$ " && "similarity ($S_i$, $S_j$) < $\theta$ "), i.e. $S_j \in$ summary sentences,

        add $S_j$ to the summary and L=L+length($S_j$),

        goto 9

    else: choose next sentence

In Algorithm-3.1, step-1 is representing preprocessing on given document D. In step two creations of matrix "A" is done by "lacal_weight (global_weight", here TF*IDF approach is followed by us. In step three SVD decomposition is performed on "A" matrix. SVD (A) decomposed the matrix into three matrices, in which right singular matrix represents information about words, left singular matrix informs about sentences, and (is covariance matrix that can be used to decide the importance of parameters, i.e., the importance of concepts/sentences/words.

Step-4 have two cases of how to design W (based on two sub-approaches given by (Gong & Liu, 2001), i.e., Model-1 and (Steinberger & Ježek, 2004) i.e., Model 2). In W, each row is representing concepts and column for sentences.

In **Table-3.3**, we are demonstrating case-2 of W formation, that using covariance matrix $\sum$ and $V^T$. In this example for demonstration purpose, we are using k=3, i.e., reduced dimension three. **Table-**3.3 is representing the relatedness of Sentences S-1 to S-9, corresponding to Concepts-1 to Concept-3 (reduced dimension r = 3). In step-5, based on the negative sign we will eliminate these entries, i.e., "neglect those sentences which are least informative." This is represented in **Table-**3.4. In step 6 we are finding the probability of relatedness between sentences and concepts over W. The value in W[i][j] is representing $P(X_{ij})$, i.e., the probability of $j^{th}$ sentence in $i^{th}$ concept. In each concept, all sentences are represented in terms of probability such that, where $X_i$ denotes $i^{th}$ concept. This is shown in **Table-**3.5.

$$\sum_{i=1}^{m} P(X_i) = 1 \tag{3.13}$$

In step-7 finding the information contained in by each concept that is given by the Equation-3.7, and in **Table-**3.5. **Table-**3.5 is representing freedom of Concept or information in the Concept (Step-7). In Step-8, we will choose the highest informative concept, and from a selected concept in Step-9 sentences are added to the summary based on certain cosine similarity criteria matched that is given by Equation-3.15. This step-9 is iteratively repeated until length constraints satisfied.

| **Sentences** | **S-1** | **S-2** | **S-3** | **S-4** | **S-5** | **S-6** | **S-7** | **S-8** | **S-9** |
|---|---|---|---|---|---|---|---|---|---|
| **Concept-1** | 1.368 | 0.860 | 1.329 | 1.016 | 0.860 | 1.915 | 1.094 | 1.133 | 1.170 |
| **Concept-2** | - 0.835 | - 0.391 | - 1.200 | - 0.626 | - 0.365 | 1.435 | - 0.182 | - 0.809 | 1.148 |
| **Concept-3** | - 0.82 | 0.28 | - 0.32 | 0.5 | 0.44 | - 1.02 | 1.1 | 0 | 0.68 |

**Table-3.3**: W in reduced space r=3

| **Sentences** | **S-1** | **S-2** | **S-3** | **S-4** | **S-5** | **S-6** | **S-7** | **S-8** | **S-9** | **Row_Sum** |
|---|---|---|---|---|---|---|---|---|---|---|
| **Concept-1** | 1.368 | 0.860 | 1.329 | 1.016 | 0.860 | 1.915 | 1.094 | 1.133 | 1.170 | 10.745 |
| **Concept-2** | ~~-0.835~~ | ~~-0.391~~ | ~~-1.200~~ | ~~-0.626~~ | ~~-0.365~~ | 1.435 | ~~-0.182~~ | ~~-0.809~~ | 1.148 | 2.583 |
| **Concept-3** | ~~-0.82~~ | 0.28 | ~~- 0.32~~ | 0.5 | 0.44 | ~~-1.02~~ | 1.1 | 0 | 0.68 | 3 |

**Table-3.4:** W is processed by retaining only positive related sentences, and rowsum is calculated to find the probability

| **Sentences** | **S-1** | **S-2** | **S-3** | **S-4** | **S-5** | **S-6** | **S-7** | **S-8** | **S-9** | **Total Entropy** $-\sum P(X_i).\log(P(X_i))$ |
|---|---|---|---|---|---|---|---|---|---|---|
| **P(Concept-1 w.r.t. sentences)** | **.1273** | **.080** | **.1236** | **.0945** | **.08** | **.1782** | **.1018** | **.1054** | **.1088** | **2.795223** |
| **P(Concept-2 w.r.t. sentences)** | - | - | - | - | - | **.555** | - | - | **.444** | **0.99152** |
| **P(Concept-3 w.r.t. sentences)** | - | **.0933** | - | **.166** | **.1466** | - | **.366** | **0** | **.2266** | **2.171480** |

**Table-3.5**: In reduced space-r information contained by each concept, and corresponding sentence

In this proposed work, we are interested to find a concept that is more closely related to all sentences, i.e. relatedness is considered in term of probability. Value$_{ij}$ indexed at location (i$^{th}$ concept, j$^{th}$ Sentence) showing that concept$_i$ is related to Sentence$_j$ with strength value$_{ij}$. A higher value is representing that; a concept is more related to a particular sentence. According to our proposed approach, the concept is more informative if that is related to more number of sentences with high score. For example let a $Concept_i$, related to $Sentence_j$ with probability $P(X_{ij})$, then the information contain in $Concept_i$ is given by Equation-3.14, here N' representing refined sentences

$$I' = \sum_{j=1}^{j=N'} P(X_{ij}) * \log\left( 1 / P(X_{ij}) \right) \tag{3.14}$$

Since entropy is given by $-\sum P(X_i).\log(P(X_i))$, the concept that has more negative value contains more information. So we will choose Concept-1 for sentence selection, and sentences are selected from this in an iterative way, till required length summary achieved. For given an example, the order of sentences to include in the summary will be {S6, S1, S3, S9, S8, S7, S4, S2, S5}.To resolve a tie between two sentences/Concepts we are using First Come First Serve (FCFS) approach.

To maintain the diversity, we have set a similarity threshold $\theta$ to add the next sentence in the summary. The similarity may be user-defined, we are using cosine similarity which is given by Equation-3.15. By experimenting we have found, if $\theta$ is too small then summary contains too much diversity resulting low performance, so we have to set it carefully. In the experiments, we chose $\theta = 0.4$. We are interested to find cosine similarity between summary sentences and newly selected sentences (next to add in summary). Here k (1 to m) representing the length of master sentence, i, j for $i^{th}$ and $j^{th}$ sentence (if total n sentences).

$$Cosine_similarity(S_i, S_j) = \frac{\sum_{k=1}^{m} W_{ik} \bullet W_{jk}}{\sqrt{\sum_{k=1}^{m} W_{ik}^2 \sum_{k=1}^{m} W_{jk}^2}} \quad i, j = 1 \text{ to n} \tag{3.15}$$

### 3.4.4 Proposed_Model-3 / LSASS Model

Proposed_Model-3/ LSA based Sentence Selection (LSASS) is also based on Entropy-based proposed_approch-2. In this proposed model, we have to find sentences that contain the highest information. A sentence with high entropy is showing more degree of freedom that means, the particular sentence is related to the number of dimensions/concepts/sentences. In Figure-3.5 sentences are represented using ^ sign corresponding label in rectangle diagrams, words are represented by * sign with the corresponding label in rectangle diagram. Below we are presenting the proposed algorithm.

**Algorithm-3.2:**

**Input:** Document D, i.e. the content of words, sentences, paragraphs;

L- Predefined length of summary; initially L=0; $L^{max}$= 100

$\theta$ -cosine similarity threshold = 0.4,

**Output:** Summary

1. Decompose the given document D into sentences {$S_1$, $S_2$, .. $S_n$, extract keywords and use these keywords to form the master sentence S.
2. For given document D, construct matrix "A" i.e. Term × Sentence matrix.
3. Perform the SVD on "A" as follow,

$$A = U \times \sum \times V^T$$

4. For every sentence in reduced space- k, by using preprocessed $V^T$ and S, compute matrix multiplication W = $V^T \times$ S. Column of W is representing sentences and row representing concepts.
5. Preprocess W, select only core sentence, by strikeout negative related sentences. //(**Table**-3.7)
6. For each column in W, compute $P(Y_i)$ where $P(Y_i)$ representing the probability of a sentence to appear in latent i 1<i<r, (r-reduced space)

$$P(Y_i) = Y_i \Big/ \sum_i^k Y_i$$

7. Find the information contains in each sentence using **Shannon's entropy method,** as follow

$$-\sum P(Y_i).\log(P(Y_i))$$

8. Select next most informative Sentence $S_j$ 1<j<m,

   if ("L< $L^{max}$" && "similarity (Si, Sj) < $\theta$ ")

   //Si $\in$ summary sentences and, $S_j$ $\in$ sentence to be added in the summary

   add $S_j$ to the summary, and L=L+len($S_j$),

   goto step-8

   else: goto step-8

In Algorithm-3.2, Step-1 to Step-3 are concerned about preprocessing and SVD decomposition. In the next step, a new matrix W is obtained by combining the Right decomposed Matrix "V" and covariance matrix "∑" (W = $\sum \times V^T$) and further reducing the dimension to k. The rows of Matrix W signify concepts in the document, and the column signifies the sentences of the document. In **Table**-3.6 we are representing the score of Sentences S-1 to S-9, corresponding to Concepts-1 to Concept-3 (here k=3). In step 5, the matrix is further reduced by eliminating the rows containing negative values. Here the negative values signify less informative sentences (shown in Table-3.7). In Step-6 we are interested to find the probability of concepts to appear in sentences. We are calculating the probability value of W[i][j], which denotes by $P(X_{i,j})$. $P(X_{ij})$ is representing the probability of a concept i to appear in sentence j (shown in Table-3.8). For each sentence, the sum of the probability of concepts is 1 as Equation-3.16, here $X_i$ denotes $i^{th}$ concept.

$$\sum_{i=1}^{m} P(X_i) = 1 \qquad (3.16)$$

In Step-7, we are willing to find out information contained in a sentence corresponding to all concepts. It is computed by summing the columns corresponding to the sentence. The resultant value signifies the information contained in the sentences (Table-3.8). In step-8, the sentence is extracted based on information content and added in the summary. This step will continue until similarity threshold criteria hold and required summary length constraint is satisfied. If a $Sentence_j$, related to $concept_i$ with probability $P(Y_i)$, then the information contained will be given by Equation-3.17,

$$I' = \sum_{i=1}^{i=r} P(Y_i) * \log(1/P(Y_i)) \qquad (3.17)$$

| Sentences | S-1 | S-2 | S-3 | S-4 | S-5 | S-6 | S-7 | S-8 | S-9 |
|---|---|---|---|---|---|---|---|---|---|
| **Concept-1** | 1.368 | 0.860 | 1.329 | 1.016 | 0.860 | 1.915 | 1.094 | 1.133 | 1.7204 |
| **Concept-2** | - 0.835 | - 0.391 | - 1.200 | - 0.626 | - 0.365 | - 1.435 | - 0.182 | - 0.809 | 1.148 |
| **Concept-3** | - 0.82 | 0.28 | - 0.32 | 0.5 | 0.44 | - 1.02 | 1.1 | 0 | 0.68 |

**Table-3.6**: W in reduced space r=3

| Sentences | S-1 | S-2 | S-3 | S-4 | S-5 | S-6 | S-7 | S-8 | S-9 |
|---|---|---|---|---|---|---|---|---|---|
| **Concept-1** | 1.368 | 0.860 | 1.329 | 1.016 | 0.860 | 1.915 | 1.094 | 1.133 | 1.7204 |
| **Concept-2** | ~~-0.835~~ | ~~-0.391~~ | ~~-1.200~~ | ~~-0.626~~ | ~~-0.365~~ | ~~-1.435~~ | ~~-0.182~~ | ~~-0.809~~ | 1.148 |
| **Concept-3** | ~~-0.82~~ | 0.28 | ~~-0.32~~ | 0.5 | 0.44 | ~~-1.02~~ | 1.1 | 0 | 0.68 |
| **Coloumn_Sum** | 1.368 | 1.14 | 1.329 | 1.516 | 1.3 | 1.915 | 2.194 | 1.133 | 3.5484 |

**Table-3.7**: W is processed by keeping only positive related sentences, and column sum is find to measure the probability

| Sentences | S-1 | S-2 | S-3 | S-4 | S-5 | S-6 | S-7 | S-8 | S-9 |
|---|---|---|---|---|---|---|---|---|---|
| **P(Sentence w.r.t. Concept-1)** | 1 | .7543 | 1 | .670 | 0.6615 | 1 | .4986 | 1 | .4848 |
| **P(Sentence w.r.t Concept-2)** | - | - | - | - | - | - | - | - | .3235 |
| **P(Sentence w.r.t Concept-3)** | - | 0.2456 | - | 0.3298 | 0.3384 | - | .501 | 0 | .1916 |
| **Entropy** $-\sum P(Y_i)*\log(P(Y_i))$ | 0 | 0.804333 | 0 | 0.914894 | 0.92336 | 0 | 1.0001 | 0 | 1.48984 |

**Table-3.8**: Information contained by each sentence in reduced space-r

We can re-write Equation-3.17 as, so entropy can also give by, $-\Sigma P(Y_i)\log(P(Y_i))$. The sentence that contains more information/highest entropy extracted, and added in summary after measuring the similarity between summary sentences and selected sentences. So we will include Sentence-9 in summary, and then Sentence-5, up to the required length summary is achieved. For given an example, the order of sentences to include in the summary will be {S9, S7, S5, S4, S2, S1, S3, S6, S8}. To resolve a tie between two sentences, we are using First Come First Serve (FCFS) approach. To maintain diversity, we can set a similarity threshold to add the next sentence in summary, if the similarity between previous sentences and the next sentence is less than "$\theta$." Similarity may be user-defined we are using cosine similarity which is given by this formula in Equation-3.15.

### 3.4.5 Proposed model to measure redundancy in summary

The text is a collection of units of words; the summary has important characteristics like (1) Coverage, (2) Redundancy, and (3) Summary length. Coverage and Redundancy are reciprocal to each other when we increase other, another reduces. A lot of research like (Alguliev, Aliguliyev, & Isazade, 2013b) (Alguliev, Aliguliyev, & Isazade, 2013a) (Alguliev et al., 2013b) happen in this direction to obtain an optimal solution using multi-objective optimization.

In our approach, we are generating just 100 word summary because 100 words reference summary is available for each file. In this ***section***, we are calculating information based on n-gram content, entropy information, from 1-gram to 3-gram. R($I_{text}$) denotes redundant information in text files, and it measures using the Equation-3.18,

$$R(I_{text}) = \sum_{n>1}^{n=3} count(n-gram) \times \log(\frac{1}{count(n-gram)}) \tag{3.18}$$

By experiments we find that to reach some conclusion, we have to consider only if n>1, if we consider for n=1 then count(1-gram) will dominate in a text file because of trivial presence. The experiment is done on DUC-2002 files so to find average information, i.e., Redundancy in the dataset we are using Equation-3.19, where S is a total number of documents. Count(n-gram)$^{s}$ representing a count of n-grams for s$^{th}$ sentence,

$$R(I_{text}) = \frac{1}{S} \times \left( \sum_{sen=1}^{S} \sum_{n>1}^{n=3} count(n-gram)^{s} \times \log(\frac{1}{count(n-gram)^{s}}) \right) \tag{3.19}$$

## 3.5. EXPERIMENT AND RESULTS

In this ***section*** we are presenting a wide number of experiments done by us, following.

### 3.5.1 Experiment-3.1

In the ***first experiment***, we have implemented all previously proposed models (Model-1 Model-2, Model-3, Model-4, and Model-5, that are presented in ***section*** 3.3) and compare their performance in this experiment. Its comparative performance in shown in **Table**-3.9. Among these five pre-proposed models highest efficiency (w.r.t. ROUGE score) is given by Model-4, that is proposed by (Ozsoy et al., 2011) and lowest ranked model is Model-1 by (Gong & Liu, 2001). Since all five models are based on LSA, and it is assumed that concepts are independent of each other so extracted sentences also best representative of the document. Even two concepts are independent there may be a number of units/words are common to each other. If we are interested in more coverage (or less redundancy), we need to update our sentence selection approach. This is done in experiment second using proposed_model-1, proposed_model-2, and proposed_model-3 in ***Section*** 3.5.2, 3.5.3 and 3.5.4.

| Original | Model-1 | Model-2 | Model-3 | Model-4 | Model-5 |
|---|---|---|---|---|---|
| **R-1** | .266 | .367 | .343 | ***.396*** | .290 |
| **R-2** | .063 | .122 | .105 | ***.161*** | .075 |
| **R-L** | .224 | .312 | .290 | ***.345*** | .244 |
| **R-W 1.2** | .106 | .149 | .138 | ***.169*** | .116 |
| **R-S*** | .063 | .114 | .100 | ***.135*** | .074 |
| **R-SU*** | .067 | .119 | .104 | ***.140*** | .078 |

**Table-3.9**: ROUGE score of previously proposed models

### 3.5.2 Experiment-3.2

In the ***second Experiment***, to get more diversity, i.e., or less redundancy, we have used proposed Approach-1 (proposed in ***section*** 3.5.2) to select k sentences from the same concepts (we set k=2). First, we will select a concept, then, we select only two sentences from each concept one is most related, and one is the most unrelated w. r. t. chosen concept/latent. This approach is an extension of Model-1 and Model-4 (follow ***section*** 3.3), so we presented our approach as *"Model-1+Proposed_Approach-1"* and *"Model-4+Proposed_Approach-1"*. Using **Table-**3.9 and **Table-**3.10, we can say that, our proposed Approach is not only performing better to respective model *('model-1+proposed_approach-1' w.r.t model-1, and 'model-4+proposed_Approach-1' w.r.t. model-4)* but also *"Model-4+Proposed_Approach-1"* is performing better compared to all previous Models. Results are shown in **Table-**3.10 and Figure-3.6. Figure-3.6 is showing how much gain in performance by using the proposed approach.

| ROUGE | Model-1 + | Model-4 + |
|---|---|---|

| Score | Proposed_Approach-1 | Proposed_Approach-1 |
|---|---|---|
| **R-1** | .353 | .409 |
| **R-2** | .116 | .164 |
| **R-L** | .299 | .354 |
| **R-W 1.2** | .142 | .170 |
| **R-S*** | .106 | .146 |
| **R-SU*** | .111 | .152 |

Table-**3.10**: Performance of proposed Models with proposed Approach-1 or proposed model-1

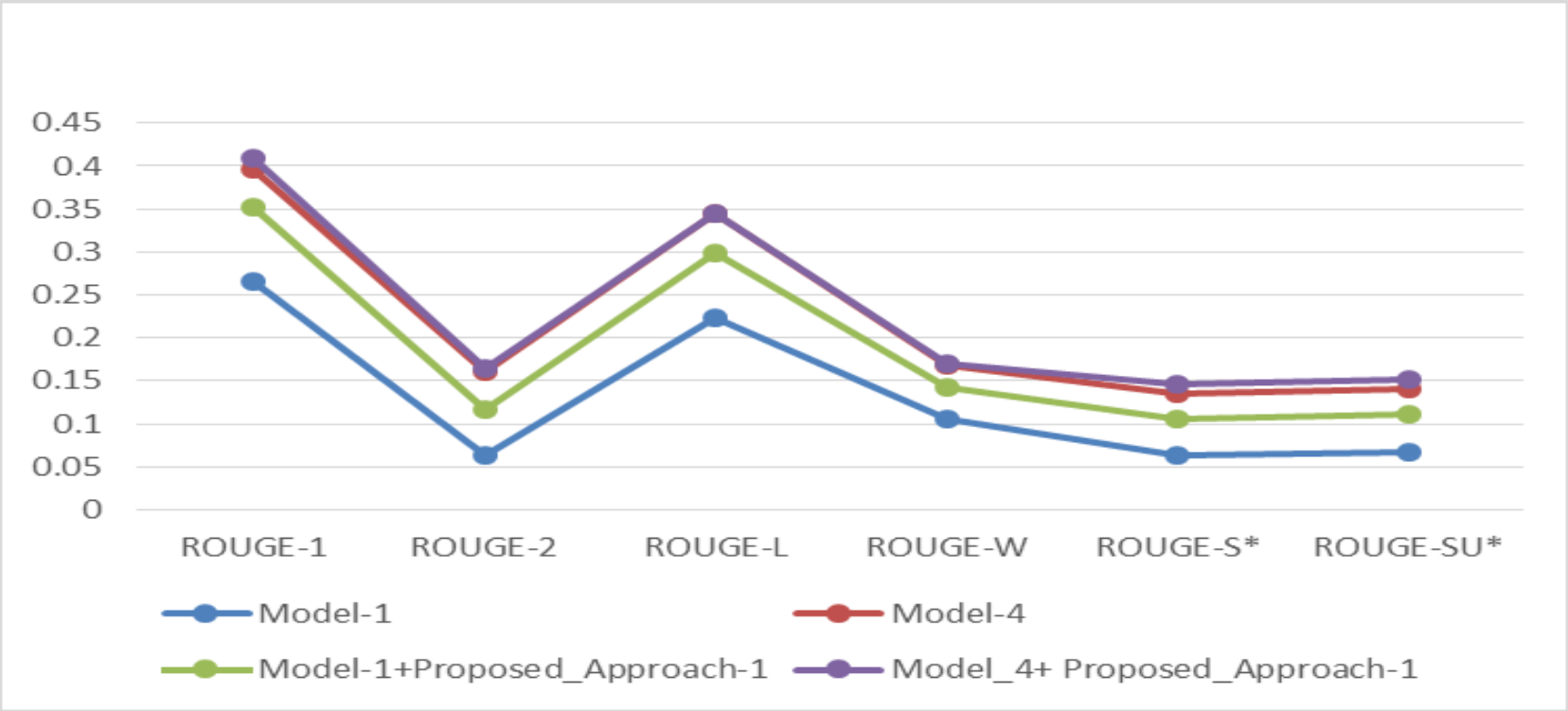

Figure-**3.6**: Showing improved performance using our proposed approach-1or proposed model-1

### 3.5.3 Experiment-3.3

Next ***third experiment***, which is based on Entropy-based approach (second approach) as defined in ***Section*** 3.4.3 (Proposed_Model-2), and ***Section*** 3.4.4 (Proposed_Model-3). Proposed_Model-2 have two cases as, (1) case-1: using $V^T$ which is termed as "Model-1+Entropy", (2) case-2: using $S \times V^T$, which is termed as "Model-3+Entropy". The sole objective of this model is to find a concept that is more related to all the sentences, and summary sentences are extracted from this concept/latent.

In continuation with the third experiment, the next model is *"Proposed_Model-3"* that is termed as "Model-4+Entropy". The objective of this model is to find sentences which are strongly related to all the concepts, and that is decided based on Entropy. All entropy-based model's performance is shown in **Table-**3.11, and we are getting improved results w.r.t. previously proposed models (**Table-**3.9, exception case Model-4), and proposed_model-1 (**Table-**3.10, exception case "Model-4+Proposed_model-1").

We are generating, and evaluating only hundred word length summary. Since long sentences are representative of more information, and Model-4 extract long sentences so this is length dominating model, hence performing better. The entropy-based model proposed by us (***Section*** 3.4.3 and ***Section*** 3.4.4) is not length, dominating so, the performance of Model-4 is sometimes better to compare to our proposed approaches. Model-4 playing better role compare to Entropy-based model, but not always when the user is interested in 30%, 40% length summary. This is true that long sentence contains more information, but when repetition (selection) of long sentences starts there is more chance to have more redundancy (and less coverage), therefore performance will negatively impact. This is shown in the next experiment. So if we want evergreen model, from these existing choices, better will be a selection of an Entropy-based model for summarization problem. The performance of Previous approaches Model-1, Model-3, Model-4, and by incorporating entropy in these models is shown by the graph in Figure-3.7.

| ROUGE Measure | Model-1+Entropy | Model-3+Entropy | Model-4+Entropy |
|---|---|---|---|
| **R-1** | .370 | .364 | .397 |
| **R-2** | .118 | .122 | .135 |
| **R-L** | .314 | .307 | .336 |
| **R-W 1.2** | .145 | .146 | .159 |
| **R-S*** | .120 | .114 | .131 |
| **R-SU*** | .125 | .118 | .136 |

**Table-3.11**: Performance of Model with Entropy-based Model

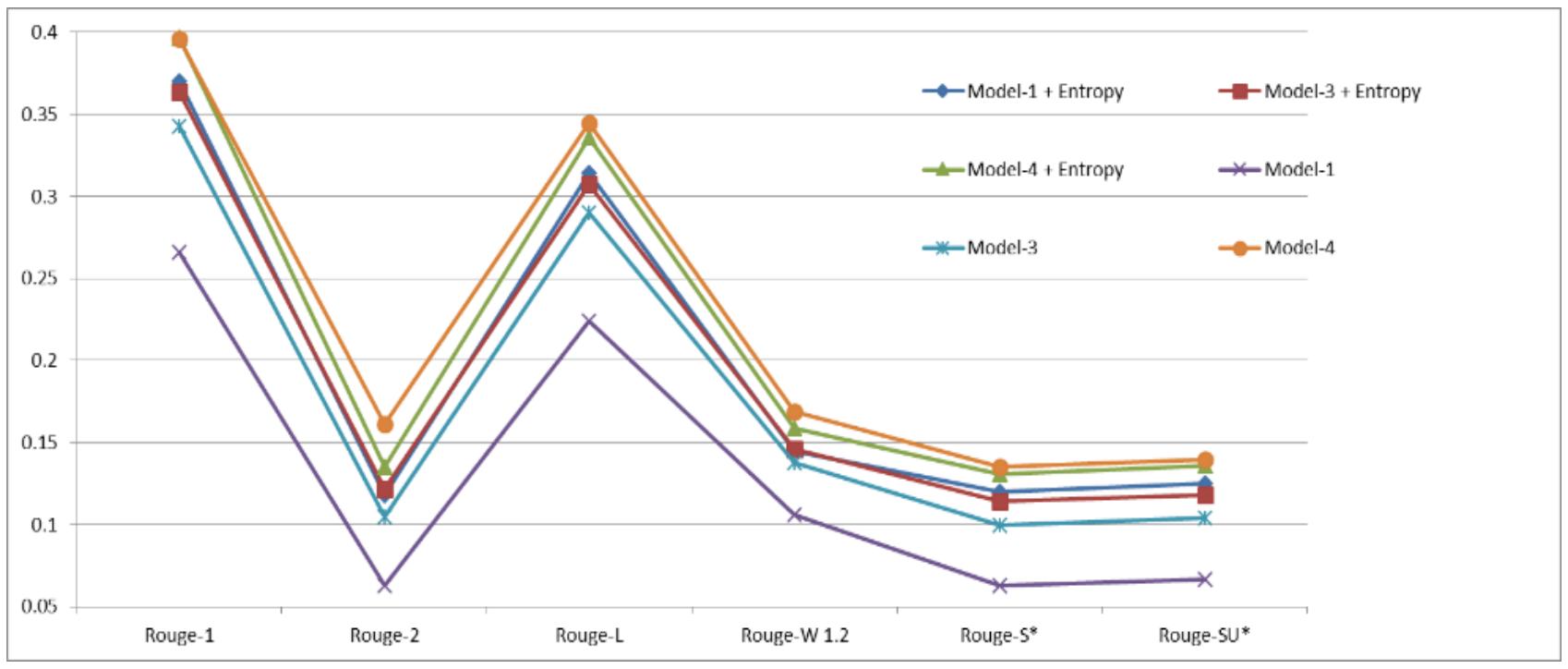

**Figure-3.7**: Showing improved performance using our Entropy-based proposed approach-2

### 3.5.4 Experiment-3.4

***Experiment four*** is showing redundancy and information contained in different summaries generated by different previous proposed models and our proposed approaches. Since Information is represented by words and in short Equation-3.20 is followed,

$$Infromation \ \alpha \ \mathrm{Count}(units) \ \alpha \ \mathrm{Co}verage \ \alpha \ \frac{1}{\mathrm{Re}\,dundancy} \qquad (3.20)$$

Figure-3.8 and **Table-**3.12 we are showing that as we are increasing the number of words in a text, ROUGE score also increasing. Sometimes it is reduced because of count($gram_n$) is increasing, and $Count_{match}$($gram_n$) reducing simultaneously (follow ROUGE score formula given by Equation-11), shown by the pair in the bold red letter. We represented information in terms of entropy given in Equation-3.6, and Equation-3.7. In **Table-**3.13 the information score (represented by Equation-3.19 and Equation-3.20) is proportional to the ROUGE score given in **Table-**3.9, **Table-**3.10, and **Table-**3.11, which is obvious (trivial case) and we are trying to show that redundancy is also increasing in Text along this, which is not our target of summarizer system.

| **WORDS** | **5** | **10** | **15** | **20** | **25** | **30** | **35** | **40** | **45** | **50** | **55** | **60** | **65** | **70** | **75** | **80** | **85** | **90** | **95** | **100** | **105** | **110** | **115** |
|---|---|---|---|---|---|---|---|---|---|---|---|---|---|---|---|---|---|---|---|---|---|---|---|
| **ROUGE-1** | .175 | .248 | .293 | .329 | .344 | **.352** | **.348** | .352 | .364 | .370 | .379 | **.387** | **.386** | .390 | .393 | .397 | .4901 | .406 | .407 | .409 | .414 | .419 | .423 |
| **ROUGE-L** | .169 | .235 | .262 | .292 | .296 | **.299** | **.297** | .302 | .311 | .315 | .322 | **.330** | **.335** | .339 | .340 | .344 | .346 | **.352** | **.350** | .354 | .356 | .360 | .361 |

**Table-3.12**: Generally by increasing, summary length ROUGE score is also increasing, and sometimes reducing

In this experiment, we are trying to measure the redundancy of information which is given by Equation-3.20. This Equation-is based on n-gram count entropy function explained in ***Section*** 3.2.5, more score means more redundancy. We are counting n-grams only if count(n-gram) > 1, the reason of this by experiment we find that if we take all count(n-gram) > 0, then hard to reach any conclusion. We are following a simple approach, giving equal weight to all n-grams.

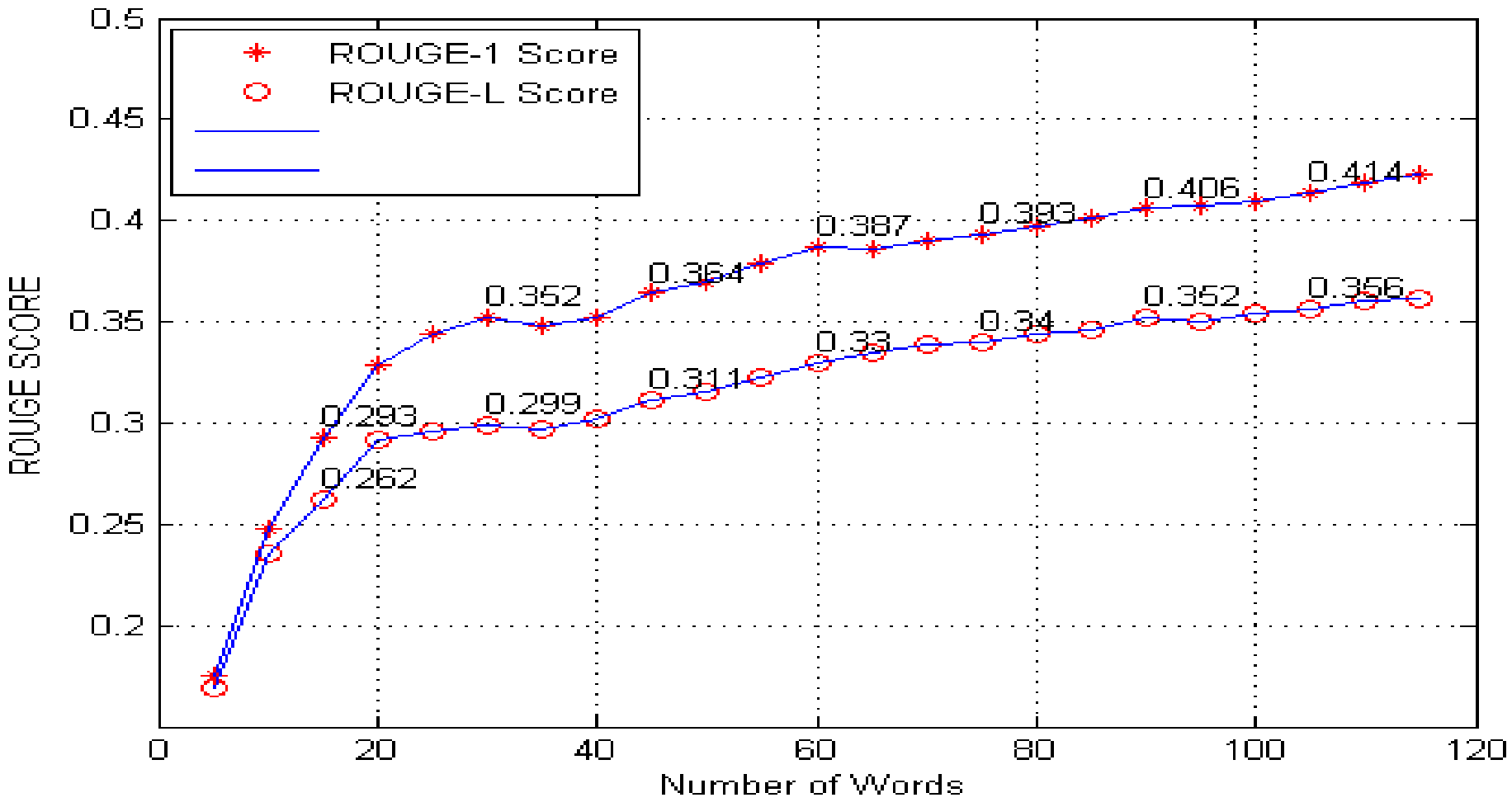

**Figure-3.8**: ROUGE Score for the Entropy-based system, showing as the number of words increasing ROUGE score also increasing/ decreasing.

But statistical as “n” go large, like 2,3,4,5..n, then it will contain more information and if the respective count increase than redundancy will increase more, i.e., coverage will be less. From this experiment, we find out that even Model-4 is performing better (follow **Table-**3.9) but, this contains more redundancy so low coverage, shown in **Table-**3.13, and **Table-**3.14.

| Model | Previously proposed Models | Proposed_Model-1 | Entropy-based Proposed_Approach-2 | Gold |
|---|---|---|---|---|
| **Model-1** | 1.3010137 | 1.39947 | 1.49261 | 1.4658 |
| **Model-2** | 1.745250 | -- | -- | |
| **Model-3** | 1.6795 | -- | 1.45620 | |
| **Model-4** | 3.10085 | 1.5391 | 1.4490 | |
| **Model-5** | 1.36435 | -- | -- | |

**Table-3.13**: Showing average information contains in summary generated by different systems

| **Model** | **Previous Models** | | | **Proposed Approach-1** | | | **Entropy-based Proposed Approach-2** | | | **Gold/Peer Summaries** | | |
|---|---|---|---|---|---|---|---|---|---|---|---|---|
| **Average (Count Frequency) > 1** | **1-gram** | **2-gram** | **3-gram** | **1-gram** | **2-gram** | **3-gram** | **1-gram** | **2-gram** | **3-gram** | **1-gram** | **2-gram** | **3-gram** |
| **Model-1** | 37.12 | 3.53 | 1.53 | 38.69 | 5.20 | 1.56 | 41.67 | 8.30 | 2.46 | | | |
| **Model-2** | 41.76 | 9.25 | 4.69 | -- | -- | -- | -- | -- | -- | | | |
| **Model-3** | 40.48 | 6.92 | 3.43 | -- | -- | -- | 39.65 | 6.02 | 1.64 | **37.14** | **4.53** | **1.32** |
| **Model-4** | ***51.10*** | ***20.46*** | ***15.17*** | ***41.17*** | ***5.35*** | ***1.66*** | 39.26 | 5.02 | 1.38 | | | |
| **Model-5** | 37.51 | 4.53 | 1.56 | -- | -- | -- | -- | -- | -- | | | |

**Table-3.14**: Average n-gram presents in different systems generated summary

Since, Rouge score directly depends on Count_matched(n-gram), if redundancy increases ROUGE also score increases. From **Table**-3.14, previously proposed models (Model-4 2$^{nd}$ best performer) has more redundancy i.e. more number of n-grams, hence this is performing better, which is too much greater than peer/gold summary n-grams counts (37.14, 4.53, 1.32 **Vs** 51.10, 20.46, 15.17), instead of this our entropy based proposed method is close to peer/ standard summary. The closeness of two texts can find using Equation-3.21, where golde$_{n\text{-}gram}$ refers n-gram present in gold summary, and system$_{\text{-}gram}$ refers to summary generated from different systems

$$\sqrt[2]{\sum_{n=1}^{3}(gold_{n-gram} - system_{n-gram})^2} \qquad (3.21)$$

Results are shown in **Table**-3.10, 3.11, 3.12, 3.13, and 3.14 can be summarized in **Table**-3.15. In this (**Table**-3.15), we are representing only ROUGE-L score, Information_entropy calculated for different described systems/models, and an average count of 1-gram, 2-gram, 3-gram with frequency(count)>1. From **Table**-3.15 we can interpret different results like,

- Our proposed system "Model-4+ Entropy" is performing better compare to other in term of ROUGE Score.
- The Model-4 system proposed by (Ozsoy et al., 2011), is performing at second place, but redundancy is high compared to allotherrsystemsm.
- From Table-3.14 we can conclude that Statistically, our second, entropy-based proposed approach is closer to the Gold / Peer summary, i.e., in term of information_enropy and count(n-gram).
- Model-1 and Model-5 are performing low, and Information_entropy/Redundancy also low.

| Rank | ROUGE-L | | Information_entropy | | COUNT(1-gram) | | Count(2-gram) | | Count(3-gram) | |
|---|---|---|---|---|---|---|---|---|---|---|
| | Model | Score | Model | Score | Model | Score | Model | Score | Model | Score |
| 1 | Proposed_approach-1 + M4 | .354 | Model-4 | 3.10 | Model-4 | 5.10 | Model-4 | 20.46 | Model-4 | 15.17 |
| 2 | Model-4 | .345 | Model-2 | 1.74 | Model-3 | 48.48 | Modl-2 | 9.25 | Model-2 | 4.69 |
| 3 | Entropy+M4 | .336 | Model-3 | 1.67 | Model-2 | 41.76 | Entropy+M1 | 8.30 | Model-3 | 3.49 |
| 4 | Entropy+M1 | .314 | Proposed_approach-1 + M4 | 1.53 | Entropy+M1 | 41.67 | Entropy+M3 | 6.02 | Entropy+M1 | 2.46 |
| 5 | Model-2 | .312 | Entropy+M1 | 1.49 | Proposed_approach-1 + M4 | 41.17 | Model-3 | 6.92 | Proposed_approach-1 + M4 | 1.66 |
| 6 | Entropy+M3 | .307 | Entropy+M3 | 1.45 | Entropy+M3 | 39.65 | Proposed_approach-1 + M4 | 5.35 | Entropy+M3 | 1.64 |
| 7 | Proposed_approach-1 + M1 | .299 | Entropy+M4 | 1.44 | Entropy+M4 | 39.26 | Proposed_approach-1 + M1 | 5.20 | Model-5 | 1.56 |
| 8 | Model-3 | .290 | Proposed_approach-1 + M1 | 1.39 | Proposed_approach-1 + M1 | 38.69 | Entropy+M4 | 5.02 | Proposed_approach-1 + M1 | 1.56 |
| 9 | Model-5 | .244 | Model-5 | 1.36 | Model-5 | 37.51 | Model-5 | 4.53 | Model-1 | 1.53 |
| 10 | Model-1 | .224 | Model-1 | 1.30 | Model-1 | 37.12 | Model-1 | 3.53 | Entropy+M4 | 1.38 |

**Table-3.15:** Summarization of Table-3.9, 3.10, 3.11,3.13, 3.14 (*M stands for Model)

## 3.6 Concluding Remark

In this work, we have proposed two new approaches (three new models) for automatic text document summarization and a novel Entropy based approach for summary evaluation. Both the approaches for summary generation is based on SVD based decomposition. In the first approach (proposed_model-1), we are using right singular matrix "$V^T$" for processing, and selects a concept one by one (top to bottom till required). Previous approaches are focused on selecting only one sentence of the highest information content. In our first approach, we are selecting two sentences

w.r.t each concept such that sentence-1 is highest related to concept and sentence-2 least related to the concept. This approach is based on assumption that by doing this we are covering two different topics. As a result, it leads to more coverage and diversity.

The second approach is based on Entropy, which formulate into two different models (proposed_model-2 and proposed_model-3). In proposed_model-2, first we are selecting a highest informative concept and from that concept, we are selecting summary sentences. In proposed_model-3 repeatedly we are selecting highest informative sentences, i.e. a sentence which is related to all the concepts with high score. The advantage of the Entropy-based model is that these are not length dominating models, giving a better ROUGE score, statistically closer to the standard/ gold summary.

During experiment, we have found out that ROUGE score depends only on the count of matched words, on increasing the summary length sometimes ROUGE score decreases, and on increasing redundancy ROUGE score also increases. We have pointed out that ROUGE score doesn't measure redundancy i.e. count matched sentences. We have also realized the need for new measure for summary evaluation that provide a tradeoff between redundancy & countmatch, and Entropy-based criteria has proposed. During testing of new proposed measure on different summary generated by previous models, and our proposed models we have found that our entropy based summary is closer to standard summary. From the experiment results, it is clear that our model works well for summary evaluation (especially for higher length summary), because as summary length increases redundancy also increases and in this measure we are measuring redundancy. Currently we are giving equal importance to all n-gram, but theoretically and practically we should give more weight to higher n-gram because of high redundancy of information (in case of repetition).

# Chapter 4: LexNetwork based summarization and a study of: Impact of WSD Techniques, and Similarity Threshold over LexNetwork

## 4.1 Introduction

In this chapter, we are presenting a lexical chain based method for Automatic Text Document summarization. This work is divided into three objectives. In the first objective, we are constructing a Lexical Network in which nodes are representing sentences and edges drawn based on Lexical and Semantic Relations between these sentences. In the second objective after constructing the Lexical Network, we are applying different centrality measures to decide the importance of the sentences. Sentences have extracted and added in the summary based on these measures. The third study (third objective) done in this work is based on WSD. Since WSD is an intermediate task in text analysis, so we are presenting how the performance of centrality measure is changing over the change of WSD technique in an intermediate step and cosine similarity threshold in post-processing step.

## 4.2 Related Work

(Morris & Hirst, 1991) have proposed a logical description for the implementation of Lexical Chain using Roget Thesaurus, and (Barzilay & Elhadad, 1997) had developed the first text document summarizer using lexical chain. Their Algorithm was exponential time taking. (Silber & McCoy, 2000) had followed the research of (Barzilay & Elhadad, 1997) for lexical chains creation and proposed a linear time algorithm O(n) for lexical chain creation. According to (Kulkarni & Apte, 2014) " the concept of using lexical chains helps to analyze the document semantically and the concept of correlation of sentences." Using lexical chain text summary generation has three stages, first step candidate word selection for chain building as Noun, Verb, second step is Lexical chain construction and chain scoring model to represent the original document and, the third step is chain selection and chain extraction for summary generation. In literature chain scoring strategy is mostly based on TF-IDF, the distinct number of words, position in the text. (Gurevych & Nahnsen, 2005) have proposed a Lexical chains construction in which candidate words are selected based on POS tagging of Nouns, and WordNet1.7 used. (Gonzàlez & Fort, 2009) have used WordNet or EuroWordNet lexical databases proposed an algorithm for Lexical Chain construction, which is based on a global function optimization through relaxation labeling. Authors have categorized relations into an Extra Strong, Strong and Medium Strong relations. Chains are distinguished/extracted based on Strong, Medium and Lightweight.

(Pourvali & Abadeh, 2012) have proposed an algorithm for single document summarization based on two different knowledge source WordNet and Wikipedia (for words which are not present in the WordNet). (Erekhinskaya & Moldovan, 2013) have used different knowledge source WordNet (WN), eXtended WordNet (XWN), and eXtended WordNet Knowledge Base (XWN KB). (Gonzàlez & Fort, 2009) have used WordNet or EuroWordNet lexical databases proposed an algorithm for Lexical Chain construction, which is based on a global function optimization through relaxation labeling. Authors have categorized relations into an Extra Strong, Strong and Medium Strong relations. Chains are distinguished/extracted based on Strong, Medium and Lightweight. (Y. Chen, Wang, & Guan, 2005) have used Chinese WordNet HowNeT. (Kulkarni & Apte, 2014) have used WordNet relations. (Y. Chen, Liu, & Wang, 2007) have used Chinese language resources like HowNet and TongYiCiCiLin. (Stokes, 2004) have used lexical cohesion to generate very short summaries for given number of news articles. They have used different relations, and distinct weights are assigned like for Extra strong relation (repetition) assigned 1.0, Strong relation (synonym) assigned 0.9, Strong relation (hypernym, hyponym, meronym, holonym, antonym) assigned 0.7, medium strength relation assigned 0.4, statistical relation assigned 0.4. After that author identifies the highest scoring noun and proper noun chains using the above relations. Along with these works, another closely related work has done by (Vechtomova, Karamuftuoglu, & Robertson, 2006), and (Ercan & Cicekli, 2007). The Author has considered keywords and sort version of document summary and proposed Lexical chain for Keyword Extraction. (Steinberger, Poesio, Kabadjov, & Jez, 2007) have used anaphoric information (another linguistic task) in Latent Semantic Analysis and show anaphoric task giving better results for summarization purpose. (Chandra Shekhar Yadav & Sharan, 2015), (C.S. Yadav et al., 2016) have used position, TF-IDF, Centrality, Positive sentiment and Negative sentiment based Semantic feature, centrality. (J. Yeh, 2005) (J.-Y. Yeh, Ke, Yang, & Meng, 2005) have used static information like position, positive and negative keyword, centrality, and the resemblance to the title to generate an extractive summary. Along with this LSA and GA also integrated into their work.

(Doran, Stokes, Carthy, & Dunnion, 2004) also, have proposed lexical chain based summarization in which the chain's score is calculated based on the frequency and relationship present between words/chain member (using WordNet). The score of word pairs depends on the *"sum of the frequencies of the two words, and multiplied by the relationship score between them."* Synonym relations have assigned value 0.90, specialization or generalization and part-whole or whole-part 0.70, Proper nouns chain scores strength depends on the type of match, 1.0 assigned for an exact match, 0.80 for a partial match, and 0.70 for a fuzzy match. Sentences have ranked according to the sum of the scores of the words in each sentence and later extracted.

(Medelyan, 2007) has presented a graph-based approach for computing lexical chains, where nodes are document's terms and edges reflecting some "semantic relations" between these nodes. Based on graph diameter (given by the "longest shortest distance" between any two different nodes in the graph,) concept strong cohesive, weakly cohesive and moderately cohesive chains are computed. (Plaza, Stevenson, & Díaz, 2012) have proposed a summarization system for the biomedical domain that represents documents like a graph designed from concepts and relations present in the UMLS Metathesaurus (Unified Medical Language System). JDI algorithm and Personalized PageRank algorithm used for WSD. (Y.-N. Chen, Huang, Yeh, & Lee, 2011) have represented the document in graph-like structure, in which node are sentences and edges are drawn by topical similarity, and the Random Walk applied for summary generation. In the same way, (Xiong & Ji, 2016)also have proposed a novel hypergraph-based Vertex-reinforced random walk. (Gonzàlez & Fort, 2009) have used WordNet or EuroWordNet lexical databases proposed an algorithm for Lexical Chain construction, which is based on a global function optimization through relaxation labeling. Authors have categorized relations into Extra Strong, Strong and Medium Strong relations. Chains are distinguished/extracted based on Strong, Medium and Lightweight.

LSA/ PLSA/LDA approaches have been proposed by (Steinberger et al., 2007), (Chiru, Rebedea, & Ciotec, 2014) study summarization result based on LSA, LDA, and Lexical chain. Two most important results from them are (1) LSA and LDA have shown the strongest correlation, and (2) the results of the lexical chain are not much correlated neither with LSA nor LDA. Therefore, according to them during performing semantic analysis for different NLP applications, lexical chains may be used as complementary to LSA or LDA. (J. Yeh, 2005) have used static information like position, positive keyword and negative keyword, centrality, and the resemblance to the title to generate an extractive summary. Along with this, LSA and GA are also integrated into their work. (Steinberger et al., 2007) have used anaphoric information (another linguistic task) in Latent Semantic Analysis and showed anaphoric task giving better results for summarization purpose.

## 4.3 Background

Lexical Network construction is the first target of this work, and during the construction of LexNetwork WSD is an intermediate task. After construction of Network, we have applied different centrality measures to score the sentences and later extract. In this ***section***, we are presenting WSD, Centrality techniques, and Lexalytics Algorithm used in this work.

### 4.3.1 Word Sense Disambiguation (WSD)

In nature, various human language exists, and one common problem for all languages is word ambiguity, i.e. multiple sense of a word according to the context in which words occurs. WSD (Word sense disambiguation) is an Intermediate task and technique in NLP applications to computationally decide which sense of a particular word is active by its use in a particular sense. (Erekhinskaya & Moldovan, 2013) have studied the ambiguity of word originates from a variety of factors like the approach of representation of the word sense, and dependency of knowledge source used like WordNet, Roget thesaurus, Extended WordNet knowledge source. Without knowledge either internal or external, it would be impossible for both humans and machines to find the correct sense. WordNet lists five senses for the word *pen*:

Pen — "a writing implement with a point from which ink flows".

Pen — "an enclosure for confining livestock".

Playpen, pen — "a portable enclosure in which babies may be left to play".

Penitentiary, pen — "a correctional institution for those convicted of major crimes".

Pen — "female swan".

In this work, we are using a variant of the Lesk algorithm for WSD proposed by (Lesk, 1986). The Lesk algorithm is based on the presumption that, the given words (to disambiguate the sense), and neighborhood of this word will tend to share a common topic. Since words are ambiguous and their sense relies on knowledge source so, in this work we are using three different algorithms for WSD. These are Lesk, Adapted Lesk, and Cosine Lesk.

#### 4.3.1.1 Simple Lesk Algorithm

A simplified version of the Lesk algorithm is to compare the ordinary dictionary definition (called gloss found in a traditional dictionary such as "Oxford advance learner's") of an ambiguous word with the terms contained in its neighborhood.

1. Initialization, (1) Put words into Word Vector for which disambiguate sense need, (2) select context window, i.e., a neighbor of the word.
2. For every possible sense of the word to disambiguate one should count some words that are in both neighborhoods of that word and in the dictionary definition of that sense.
3. The sense that is considered best possible sense which has the highest number of overlapping counts.

The advantage of this technique is that this is non-syntactic, and not dependent on global information, the disadvantage of this doesn't use previous sense used, i.e. for next word it will compute again so that it is time-consuming, and performance varies according to the selection of neighboring words.

#### 4.3.1.2 Adapted Lesk

Adapted Lesk algorithm has been proposed by (Banerjee & Pedersen, 2002) Simple Lesk algorithm uses knowledge sources as a standard dictionary for gloss definition, wherein Adapted Lesk Algorithm in place of standard dictionary author using electronic database WordNet. WordNet provides a rich hierarchy of semantic relations.

#### 4.3.1.3 Cosine Lesk

The cosine Lesk has proposed by (Tan, 2013) is a vector-space / distributional space version to calculate Lesk overlaps (also known as signatures in the original Lesk paper). For example, let us take sentence-3, and here the ambiguous word is "deposit" with two different contexts as below in Meaning-1 and Meaning-2,

Sentence 3: “I got to the bank to deposit my money.”

Meaning-1: “financial institute and withdraw and transact money.”

Meaning-2: “The strip by river where soil deposit resident.”

Instead of a global vector/distributional space built from a corpus, we can do it locally for the given ambiguous word. So given the meaning word "deposit" and the context sentence, we have the vocabulary:

Vocabulary= [financial, institute, deposit, withdraw, transact, money, strip, river, soil, resides, I go]

And the following vector will be assigned to,

Sentence-3: [0,0,1,0,0,1,0,0,0,0,1,1]

Meaning 1: [1,1,1,1,1,1,0,0,0,0,0,0]

Meaning 2: [0,0,1,0,0,0,1,1,1,1,0,0]

And then Cosine_Similarity(sentence-3, meaning1), and Cosine_Similarity(sentence-3, meaning2) is computed. The highest similarity would give us the closest meaning of the word "deposit" given the context sentence. This is cheaper to compute, and there's no need for a corpus, and also it can be done on the fly with WordNet without storing the vector space in memory before disambiguating our text. But the lack of memory usage would also mean that meaning vectors are computed and recomputed as we disambiguate and that might be wasteful regarding computing resources and time.

### 4.3.2 Lexical Chain

A lexical chain is a sequence of related words in writing, spanning short or long distances generally limited to next few lines. A lexical chain is independent of the grammatical structure of the text and in effect, it is a list of words that captures a portion of the cohesive structure of the text. Applications of lexical chains are keyphrase extraction, keyword extraction, event detection, document clustering, text summarization etc. This can be explained by the following example-4.1 and 4.2. The Figure-4.1 showing a lexical chain created for a document available at (Elhadad, 2012).

Example 4.1: Rome → capital → city → inhabitant

Example 4.2: Wikipedia → resource → web

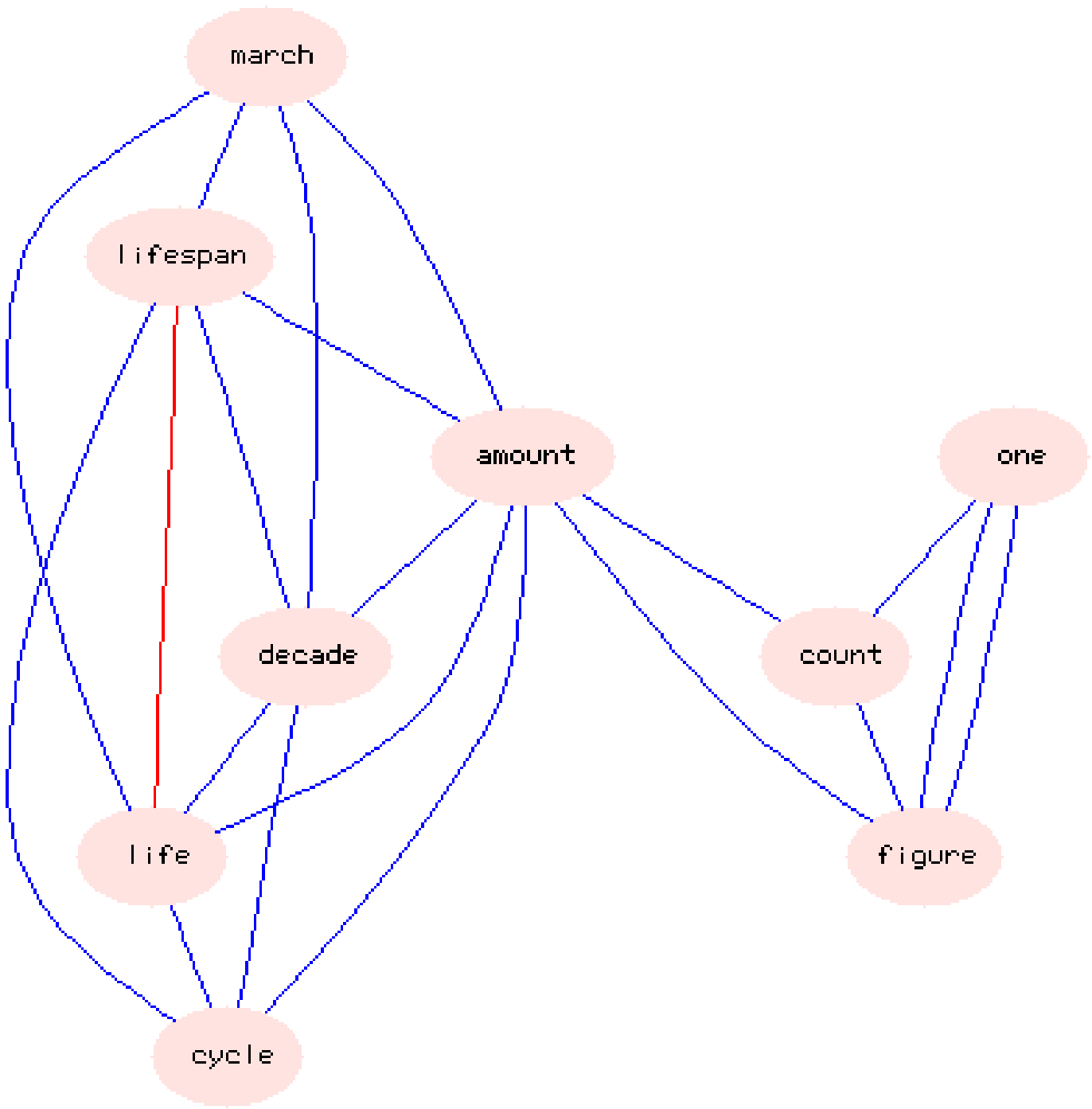

Figure-4.1: Example of lexical chain creation dataset available (Elhadad, 2012).

Mechanism of how to create has been given in (Barzilay & Elhadad, 1997). Chains are constructed between words, so it lost the meaning of the text, and grammatical structure. There is no way to find out that, particular word belongs to how many sentences. To sort out these issues we have proposed a LexicalNetwork that overcome all these problems.

### 4.3.3 Lexalytics Algorithms

Lexalytics offers text mining software for companies and enterprises around the globe, which is available at https://www.lexalytics.com/. This System can access through JAVA or Python API, and Excel Add-ons. Text Analytics tasks like Named Entity Extraction, finding the meaning of text via Classification and Tagging, Sentiment Analysis, Context finding, and Summary generation. Lexical chaining is an intermediate step in mentioned applications. Various Lexalytics algorithms heavily rely on it, like Summarization. In Lexical Chaining, nouns are selected for lexical chain construction. If the nouns are related to each other, we can find that the conceptual (lexical) chain in the given content, even when many other unrelated sentences separate those sentences. The score of all lexical chain is based on the length of the chain and the relationships (Lexical or Semantic) between nouns. In this system, Location-based features also considered deciding sentence's Priority, because of an initial sentence always more informative. Relation used for chain construction purpose are same-word, Antonym, Synonym, Meronym, Hyper/Holonyms, etc.). In this system, even if "those sentences are not adjacent to each other in the text, they are lexically related to each other and can thus be associated with each other." During Summary generation it extracts best sentences from the chain and shifts out nonessential sentences from the summary.

### 4.3.4 Centrality

In graph theory or social network analysis, to decide the relative importance of a node, one existing approach is Centrality-based measures. Some used approached are mentioned by us.

#### A) Degree Centrality

The first and simplest type of centrality is degree centrality (W. N. Venables & Team, 2018). If we define a graph G(V, E) where |V| is some nodes, and |E| stands for the number of edges, then on graph G degree centrality $C_D(V)$ of a node "V" can be defined as the number of edges incident towards a node. Graph "G" may be directed and undirected, if G is di-graph then, two type of degree is defined as In-degree, and Out-degree. For a given node v, v €V, in-degree of a node v is some links approaching towards the v, and out-degree of v is the counts of links that the node v directs towards other nodes (V-v). The degree centrality of a vertex V, for a given graph G, is defined by Equation-(4.1) (general degree centrality), and Equation-(4.2) (normalized centrality), N is the total number of vertices, in Figure-4.4 two graphs are shown with normalized degree centrality scores.

$$C_D(V) = \deg(V) \tag{4.1}$$

$$C_D(V) = \frac{\deg(V)}{N-1} \tag{4.2}$$

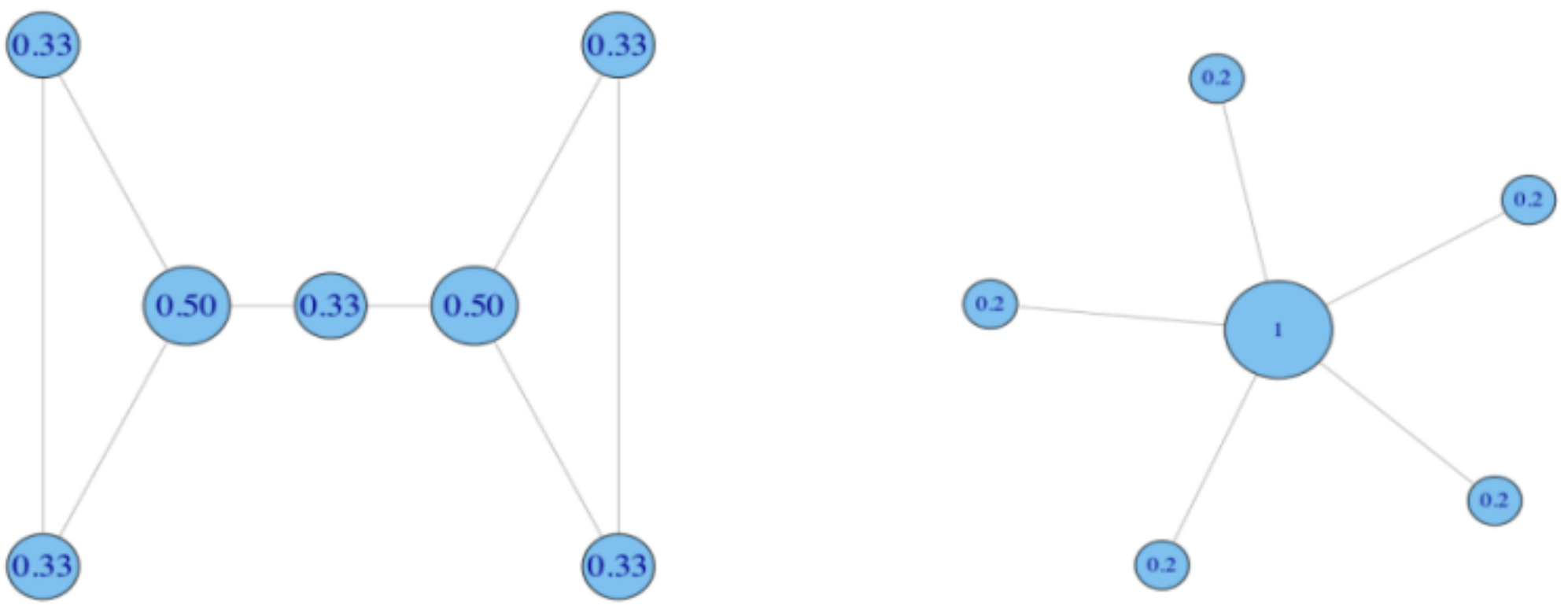

**Figure-4.4**: Showing Normalized Degree Centrality for an undirected graph

**B) Eigen Value Centrality**

In linear algebra, an Eigenvector of a square matrix (let "A") can define as "a vector that does not change its direction under the associated linear transformation." Alternatively, it can be defined as if V is a vector; then it is an Eigenvector of a square matrix "A", if AV is a scalar multiple of V and V≠0. This condition can be written as the following Equation-4.3, where $\lambda$ is the scalar known as the Eigenvalue associated with the Eigenvector V.

$$AV = \lambda V \tag{4.3}$$

A more advanced version of the degree centrality is Eigen-Vector centrality. Degree centrality of a node is just based on a simple count of the number of connections. Where, Eigenvector centrality acknowledges that not all connections are equal or in other words "The Eigenvector centrality defined in this way accords each vertex a centrality that depends both on the number of connection, i.e. degree of a node, and the quality of its connections." In social life, this is followed connections to people who are themselves influential will lend a person more influence compare to in connections with less influential persons. If we denote the centrality of $i^{th}$ vertex by $X_i$, then "we can allow for this effect by making $X_i$ proportional to the average of the centralities of $i^{th}$ network neighbors" (Newman, 2008) represented in Equation-4.4.

$$X_i = \frac{i}{\lambda}\sum_{j=1}^{n} A_{ij} X_j \tag{4.4}$$

**C) Closeness Centrality**

In Graph, theory closeness is a one of the centrality measures defined for a vertex in the graph G (V, E). Vertices which are shallow (short geodesic distances) to other vertices have given higher closeness value. Closeness can be defined in a various way, as defined by (Freeman, 1977), "the closeness centrality of a vertex is defined by the inverse of the average length of the shortest paths to/from all the other vertices in the graph," given below by Equation-4.5,

$$C_C = \frac{1}{\sum_{v \neq i} d_{vi}} \tag{4.5}$$

**D) Alpha Centrality**

If we denote an adjacency matrix "A" in which $A_{ij}$ equivalent to $a_{ij}$ means i contributes in j's status, and if x is a vector of centrality scores, then this can be written by Equation-4.6,

$$x_i = a_{1i}x_1 + a_{2i}x_2 + \ldots\ldots a_{ni}x_n \tag{4.6}$$

In Alpha centrality, status/score of an individual can be seen as a function of the status of those who choose him, given by upper Equation-4.6 which can rewrite as Equation-4.7 ($A^T$ is denoting the transpose of matrix A).

$$A^T x = x \tag{4.7}$$

In short alpha centrality's status is a linear function of another node to which node is connected. This could be understood with two examples First, in a community power study an actor's status/value is increased much if he/she nominated from those who themselves are receiving more nominations compare to other. Same as an upper example in a school also, a student's popularity is increased more by receiving voted from other students who are themselves more popular w.r.t other students. One solution to this problem is Solution 1: Eigenvector centrality, but this doesn't produce justifiable results for the networks given in Graph-1, and Graph-2 with Figure-4.5.

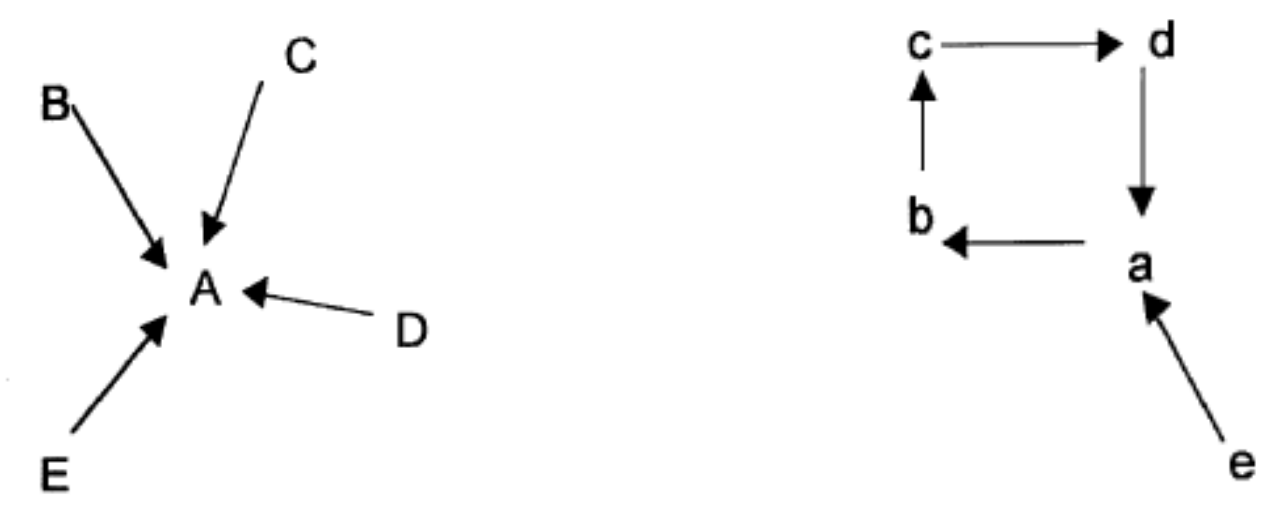

Graph -1 Graph-2

**Figure-4.5**: showing a graph with some nodes and edges

A possible solution of this kind problem is possible if we allow every individual (here nodes/ students) with some status/ score that doesn't depend on its connection to others. Like, in a class or school each student's popularity that depends on its external status characteristics. Let "e" be a vector of these exogenous sources of status or information; then we can replace the Equation-4.7 with a new Equation-4.8, in which the parameter α reflects the relative importance of endogenous versus exogenous factors in the determination of centrality.

$$x = \alpha A^T x + e \tag{4.8}$$

From Equation-4.8 this has the matrix solution Equation-4.9,

$$x = (1 - \alpha A^T)^{-1} e \tag{4.9}$$

In Figure-4.5's Graph-1, for any value of $\alpha > 0$ position of labeled vertex 'A' is the most central. In Graph-2, the order is $x_a > x_b > x_c > x_d > x_e$ for any $\alpha > 0$. The Alpha centrality based measure is almost identical to the measure has proposed by (L. Katz, 1953). Katz has suggested that "influence could measure by a weighted sum of all the powers of the adjacency matrix A," here Powers of "A" gives indirect paths connecting points.

**E) Betweenness centrality**

In a Graph G(V, E) Betweenness centrality given by (Freeman, 1977), a measure of a vertex is based on the shortest path metric. Vertices $v \in V$ those occur in some shortest paths between other vertices have higher Betweenness centrality compare to those don't. For a graph G with |V|=n vertices, the Betweenness for a vertex is computed by Equation-4.10,

$$C_B(v) = \sum_{s \neq v \neq t \in V} \frac{\sigma_{st}(v)}{\sigma_{st}} \tag{4.10}$$

In Equation-4.10 $\sigma_{st}$ denotes the counting of shortest paths from node "s" to node "t," and $\sigma_{st}(v)$ is the total number ofpassedst paths from node "s" to node "t" that is passing via a vertex v.

#### F) Bonacich's power centrality / BonPow

Let given matrix "A" is an Adjacency matrix, then Bonacich's power centrality has proposed by (Bonacich, 1987) is defined by Equation-4.11, where β is an attenuation parameter

$$CBP(\alpha,\beta) = \alpha(I-\beta A)^{-1} A1 \tag{4.11}$$

$$\beta \to \frac{1}{\lambda_{A1}} \tag{4.12}$$

From Equation-4.12, β is the reciprocal of the $\lambda_{A1}$ which is the largest Eigenvalue of Adjacency matrix "A", this is to within a constant multiple of the familiar Eigenvector centrality score; for other values of $\beta$ (else than $\lambda_{A1}$), the behavior of the measure is quite different. In particular, $\beta$ gives positive and negative weight to even and odd walks, respectively, as can be seen from the series expansion in Equation-4.13.

$$CBP(\alpha,\beta) = \alpha \sum_{k=0}^{\infty} \beta^k A^{k+1} \tag{4.13}$$

Which converges so long as Equation-4.14 holds,

$$|\beta| < \frac{1}{\lambda_{A1}} \tag{4.14}$$

From (W. N. Venables & Team, 2018), "the magnitude of $\beta$ controls the influence of distant actors on ego's centrality score, with larger magnitudes indicating slower rates of decay. (High rates, hence, imply a greater sensitivity to edge effects.) Interpretively, the Bonacich power measures correspond to the notion that the power of a vertex is recursively defined by the sum of the power of its alters. The power exponent controls the nature of the recursion involved: positive values imply that vertices become more powerful as their alters become more powerful (as occurs in cooperative relations), while negative values imply that vertices become more powerful only as their alters become weaker (as occurs in competitive or antagonistic relations). The magnitude of the exponent indicates the tendency of the effect to decay across long walks, i.e. higher magnitudes imply slower decay. One interesting feature of this measure is its relative instability to changes in exponent magnitude (particularly in the negative case)".

#### G) HITS/ HUB-Authority

Hyperlink-Induced Topic Search (HITS- known as Hubs & Authorities) developed by (Kleinberg, 1999) is a type of link analysis algorithm that rates web pages. A set of web pages can consider as a connected graph. The algorithm assigns two different scores (hub and authority score) for all pages. The Authority score estimates the value of the content of the particular page (on a node in the graph), and its Hub score estimates the value of its links (edge in the graph) to all other pages. In other words, this can interpret as that, "a good hub represents a web page that points too many other pages and a good authority represent a page that was linked by many different hubs."

The HITS algorithm relies on an iterative method and converges to a stationary solution. Each node i in the graph is assigned two non-negative scores, an authority score $x_i$ (let) and a hub score $y_i$ (let). The $x_i$ and $y_i$ are initialized with any arbitrary nonzero value, and scores will update according to iterative ways present in Equation-4.15, and Equation-4.16,

$$x_i^{(k)} = \sum_{j:(j,i)\in E} y_j^{(k-1)}, \quad and \tag{4.15}$$

$$y_i^{(k)} = \sum_{j:(i,j)\in E} x_j^{(k-1)}, \quad \text{k=1,23....} \tag{4.16}$$

Later, the weights are normalized as Equation-4.17,

$$\sum_j (x_j^{(k)})^2 = 1, \ \sum_j (y_j^{(k)})^2 = 1 \tag{4.17}$$

#### H) Subgraph Centrality

Let G (V, E) is a graph, then any subgraph G'(V', E') will hold this V' $\subseteq$ V and E' $\subseteq$ E. (Estrada & Rodriguez-Velazquez, 2005)proposed a centrality measure which based on the participation of each node in all subgraphs in the network. In their work smaller sub-graph gives more weight compared to larger ones, which makes this measure appropriate for characterizing the network motifs (define as "designates those patterns that occur in the network far more often than in random networks with the same degree sequence" by (Freeman, 1977, 1978).

Degree Centrality can be considered like as direct influence, but this is unable to cover long-term relations (or in another word indirect influence) in the network (L. Katz, 1953). There is another centrality based measures like Betweenness centrality, closeness centrality, but these measures are justifiable in a connected network because the path distance between unconnected nodes is not defined. Eigenvector Centrality (EC) also don't depend on to paths. EC measure "simulates a mechanism in which each node affects all of its neighbors simultaneously," (Bonacich, 1987), But the Eigenvalue based Centrality (EC) measure cannot consider as a measure of centrality in which nodes are ranked according to their participation in different network subgraphs.

Let in Figure-4.6, which is 3-regular graphs, with eight vertices. Vertices set {1, 2, 8} is forming part of a triangle, {4, 6} part of three squares, and {3, 5, 7} form part of only two and the rest do not form part of any. These groups are distinguishable according to their participation in the different subgraphs, although EC cannot distinguish them.

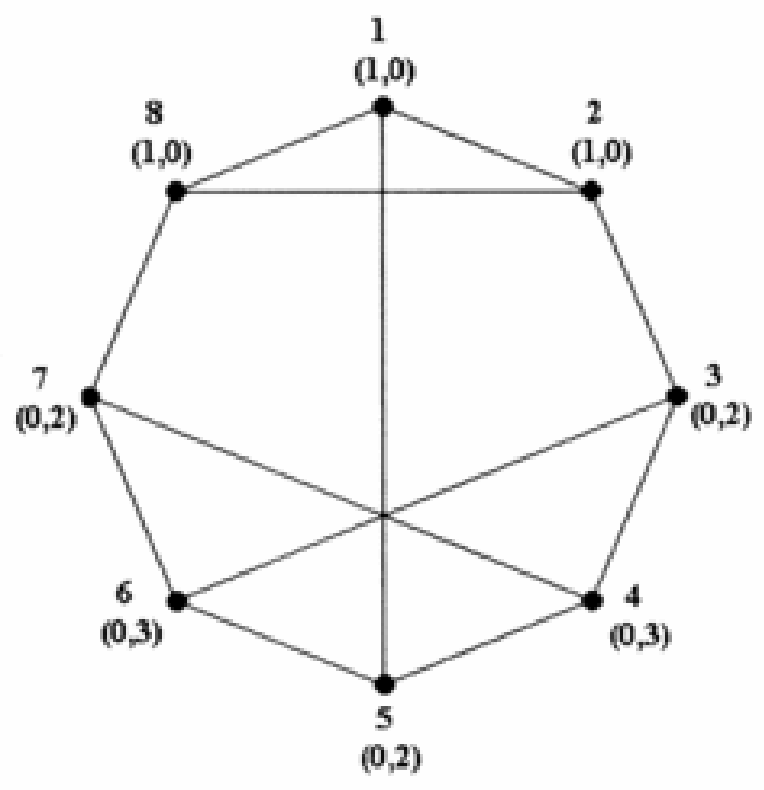

**Figure-4.6**: Showing eight node, with 3-degree regular graph

Subgraph based centrality is based on the number of closed walks starting and ending at the node. Closed walks are weighted such that "their influence on the centrality decreases as the order of the walk increases." In this technique, all closed walk is associated with a connected subgraph, which points out that this measure counts the times that a node takes part in the differently connected subgraphs of the network. Here smaller subgraphs have higher importance corresponding to larger. The subgraph centrality is represented via graph angles, as noticed by (Estrada & Rodriguez-Velazquez, 2005), given in Equation-4.18,

$$C_S(u) = \sum_{k\geq 0} \frac{(A^k)_{uu}}{k!} = \sum_{k\geq 0} \frac{1}{k!} \sum \alpha^2{}_{u,i} \mu_i^k = \sum_{i=1}^m \alpha^2{}_{u,i} \sum_{k\geq 0} \frac{\mu_i^k}{k!} = \sum_{i=1}^m \alpha^2{}_{u,i} e^{\mu_i} \qquad (4.18)$$

According to Equation-4.18 if the graph is walk-regular, then every node has an identical subgraph centrality.

$$\alpha_{u,i} = \sqrt{\sum_{j=1}^{k_i} (x_{i,j})^2{}_u} \qquad (4.19)$$

Denote the nodes of the graph G by 1... n. Let $\mu_1, \mu_2, \ldots \mu_m$ , where m be the distinct Eigenvalues of given adjacency matrix A, with multiplicities, respectively, and $x_{i,1}, x_{i,2} \ldots$ be the basis of the Eigenspace, corresponding to the Eigenvalue, i = 1,....m. Above Equation-4.19 defines the angle alpha corresponding to the node u of G and the Eigenspace.

**I) PageRank**

According to (Gleich, 2015), due to several features of Rage Rank algorithm simplicity, guaranteed existence, generality, uniqueness, and fast computation make this technique applicable far beyond its origins in Google's web-search proposed by (Page, Brin, Motwani, & Winograd, 1999). PageRank is used as a network centrality measure proposed by (Koschützki et al., 2005), and our work also motivated from this work. If we assume that a given page "A1", and other pages $T_1, \ldots T_n$, which pointing towards page ("A1") (called citations). The parameter d is a damping factor which can be set between 0 to 1 (usually set to 0.85). $C(T_i)$ is defined as the number of links going out of page A1. The PageRank score of page "A1" is given by Equation-4.20,

$$PR(A1) = (1-d) + d\left[\frac{PR(T_1)}{C(T_1)} + \ldots + \frac{PR(T_n)}{C(T_n)}\right] \qquad (4.20)$$

Our used variation is given by Equation-4.21,

$$PR(A1) = \frac{(1-d)}{N} + d\left[\frac{PR(T_1)}{C(T_1)} + \ldots + \frac{PR(T_n)}{C(T_n)}\right] \qquad (4.21)$$

PageRank technique forms a probability distribution over web pages (or in the graph), such that the sum of all web pages' PageRank score will be one.

## 4.4 Proposed Work

In this ***section***, we are presenting just a layout of our algorithm which is Four Step process, including Preprocessing, Lexical Network creation, Sentence Raking, i.e. computing score/ importance of each sentence, and Summary generation based on required length.

The First Step is Preprocessing, this step is required for proper sentence read because we cannot set an arbitrary delimiter for a regular expression like just "." or "./n" ( a dot followed by the new line, new tab). To get the better accuracy, we transform some special form like U. S. to the U.S., deleted just bank space in the form of " .". The second Step takes the input from the output of the first Step; it reads sentences one by one and creates a Lexical network among all sentences. Detail about this is presented in Algorithm 1 in next ***Section*** 4.3.3.1.

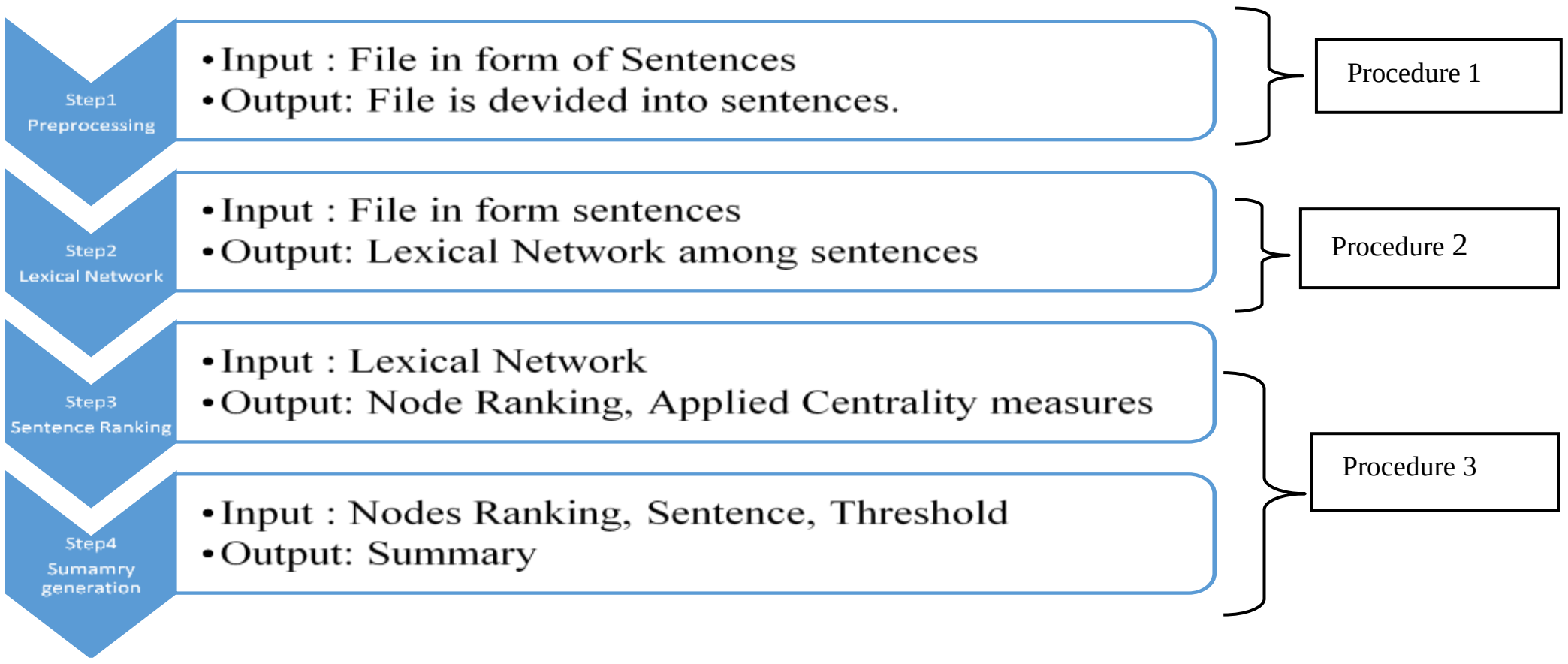

**Figure-4.2**: Steps in Summarization

In the third step, we are applying different graph-based centrality measures for finding informative sentences, and in the last Step, we set some constraints like threshold, summary length for summary generation according to a user request.

*4.4.1 Algorithm*

In this **section**, we are presenting Algorithm for a summary generation in three main Steps, basically four as per previous Figure 4.5, but in this **section** Step 3 and Step 4 are merged in Step 3 of this algorithm.

**Algorithm 4.1: Lexical Network construction**

In this **section**, we are presenting the Summarization algorithm proposed by us. Our work comprises three steps, in the first Step is preprocessing done which is a most important step in text mining that can improve, harm the efficiency of the algorithm, in Step second we are creating Lexical Network. Meanwhile, WSD is done with the help of the Lesk algorithm, and in Step 3 sentence scoring and sentence Extraction was done.

```
Input: Collection of Sentences
Output: Extracted Sentences

Procedure 1: Preprocessing
        Step1: Proper fragmentation of sentences set proper delimiter.
        Step2: Tagging of Sentences using POS tagger.
```

**Procedure 1:**

In the document, we are given some sentences if we apply any regular expression for proper bifurcation of sentences is not proper possible because of some words like Mr., Dr., U.S. So for proper sentence extraction, we are doing some preprocessing, and then we are tagging these sentences using NLTK (Natural Language Tool Kit) parser. The result of this step is stored in list format in Sentence variable which can be randomly accessed, i.e. like Array.

```
Procedure 2: Lex Network creation
        Step 1: Initialize matrix Lex_Network[|Sentences|*|Sentences|]
        Step 2:
         For ith sentence0<i< |Sentences|      // for each Actual Sentence
        {
        Extract Nouns , Verb ; called modified sentences (contains only significant units )
                {
                For ith modified sentences;        // locating corresponding same sentence
                  {
                        takeIth unit (I'=1 to |modified sentence|-1)
                        Word=Ith unit
                        Correct Sense=Lesk(Ith word ; ithActual sentence)
                        For (i+1)th modified sentences;
                        {
                        takeJth unit     0 < J < |modified sentence|
                        If relation (correct sense; Jth unit)
                                then
                                Lex_Network[ith][i+1th]=Lex_Network[ith][i+1th]+1
        } } }  }
```

```
Procedure 3:Sentence Scoring and Extraction
        Step1: Centrality_vector=centraliry_measure(Lex_Network)
        Step2: Paired (Sentence; Centrality Score)
        Step3: Sort paired Sentences based on Centrality score
        Step4: Set similarity measure Θ threshold, and summary=[] //empty summary
        Step5: Extract ith sentence from pool after Step 3 // Diversity Maintained
                        Pick jth sentence from summary pool
                                If(similarity(jth,ith)< Θ)
                                        {
                                        Add ith sentence to summary
}
```

**Procedure 2:**

In this Procedure, we are constructing a Lexical Network of N×N matrix. Where N is some sentences, i.e. N= |Sentences|. We are picking each sentence on by one and extracting only Nouns (like a proper noun), Verbs from this sentence because only these provide information, and this is added to the modified sentence

Example: From DUC 2002, HealthCare Data Set, File number WSJ910528-0127.txt

***ActualSentence*** = Sentences= "An administration task force has been studying possible health care solutions, but has yet to offer a comprehensive proposal."

***TaggedSentence*** = [('administration', 'NN'), ('task', 'NN'), ('force', 'NN'), ('studying', 'VBG'), ('possible', 'JJ'), ('health', 'NN'), ('care', 'NN'), ('solutions', 'NNS'), ('offer', 'VB'), ('comprehensive', 'JJ'), ('proposal', 'NN')]

***ModifiedSentence*** = "administration task force studying health care solutions offer proposal".

The actual Lexical Network is construed for ModifiedSentences here |ModifiedSentences|=N=|Sentences|. We are finding is any relation either Lexical, or Sematic is present between two ModifiedSentence. To find the relation present between sentences we need to disambiguate the sense of the word this is done with the help of Lesk Algorithm. Here Lesk Algorithm takes two arguments, one word to disambiguate sense and other is $i^{th}$ Actual sentence. Based on that sense this algorithm finds Hyper, Hypo, Synonyms, Meronym, and Antonym relation present in other remaining sentences. At a time only one relation will be available, and according to us, every relation is given the same importance/ weight. So we will assign an edge between sentences, and each time increase by one if more relation is present between sentences.

In **Table**-4.2, we are showing Lexical Network connection between 22 sentences (WSJ910924-0116.txt, from DUC 2002 Dataset). Last sentence ($22^{nd}$) is empty, so there is neither lexical nor semantic similarity available, so corresponding to A[22], to A[i] entries are zero (1<i<|sentences| ) vice versa. Corresponding to **Table-4.2**, Figure-4.3 showing a graphical view of this lexical network in which node are sentences and edges are a corresponding total similarity. (Note: "When we will split any text with delimiter "dl", if the total number of delimiter are n then, after split operation n+1 number of segmentation will appear, so here 22),

**Table-4.1**: Showing all sentences of WSJ910924-0116.txt, from DUC 2002 Dataset, for the creation of Lexical-Network

| | |
|---|---|
| 1 | "WASHINGTON Health and Human Services Secretary Louis Sullivan called for a summit with the chief executives of major insurance companies to discuss ways of paring the administrative costs of health care". |
| **2.** | "But in a speech here , Dr . Sullivan indicated that the Bush administration isn't likely to put forth a broad legislative proposal to overhaul the country's health care system" . |
| **3.** | "Rather , he advocated focusing on ways to improve the current system" . |
| **4.** | "Administration health officials said the meeting with insurers , tentatively scheduled for Nov . 5 , is likely to commence a series of discussions with players in the nation's $670 billion health care system on problems of cost and access" . |
| **5.** | "Administrative costs , like excessive paper work , have burdened the health care system with billions of dollars in unnecessary costs , many observers believe" . |
| **6.** | "Some studies have put the price at more than $100 billion a year , but HHS officials believe it is more like $15 billion to $25 billion" . |
| **7.** | "While numerous health care reform proposals have been introduced in Congress this year mostly by Democrats the comments of Dr . Sullivan and other HHS officials yesterday suggest the Bush administration isn't interested in striving for systematic revision and will push for limited fixes and incremental changes" . |
| **8.** | " I will not propose another grand , sweeping , speculative scheme , " Dr . Sullivan said . " Rather , I want a public debate to focus on some immediate , practical options that address our most urgent healthcare concerns" . |
| **9.** | "Some of the options , he said , are: Making health insurance more affordable to small businesses" . |
| **10.** | Easing barriers to " high quality , cost effective managed and coordinated care . " |
| **11.** | "Researching the effectiveness of various medical procedures to encourage the use of the most cost effective ones" . |
| **12.** | Altering the tax code to " increase consumer awareness of the true cost of health care and distribute current tax subsidies more equally . " |
| **13.** | "Among other things , administration officials have been looking at the possibility of limiting the tax exemption for employer paid health insurance premiums" . |
| **14.** | "Increasing the availability of primary care to the neediest people" . |
| **15.** | "Reducing the administrative costs of health care" . |
| **16.** | "The U. S . spends more per capita on health care than any country , yet more than 30 million Americans lack health insurance" . |
| **17.** | "In January 1989 , President Bush ordered an administration task force to study problems of health and access , but it has yet to propose solutions" . |
| **18.** | "Dr . Sullivan repeated his dislike for the two most widely discussed health care revision proposals: establishing a nationwide federally sponsored health care system , similar to Canada's or mandating employers to either provide basic health benefits to workers or pay a tax to finance a public health plan" . |
| **19.** | Such approaches , he said , would be inflationary , " smother competition , " and lead to " rationing and waiting lists . " |
| **20.** | "He said experimentation in health care reform should be left to the states" . |
| **21.** | " Local solutions for local problems should be our working philosophy , as should learning from local mistakes in order to avoid harm to the nation as a whole , " he said . |
| **22.** | "" |

**Table-4.2**: Showing Lexical and Semantic relations present between sentences

| SENT | 1 | 2 | 3 | 4 | 5 | 6 | 7 | 8 | 9 | 10 | 11 | 12 | 13 | 14 | 15 | 16 | 17 | 18 | 19 | 20 | 21 | 22 |
|---|---|---|---|---|---|---|---|---|---|---|---|---|---|---|---|---|---|---|---|---|---|---|
| 1 | 0 | 6 | 0 | 7 | 5 | 1 | 4 | 0 | 3 | 3 | 1 | 5 | 3 | 2 | 5 | 7 | 3 | 12 | 0 | 4 | 0 | 0 |
| 2 | 6 | 0 | 1 | 10 | 5 | 2 | 9 | 0 | 1 | 2 | 0 | 3 | 5 | 2 | 3 | 4 | 5 | 9 | 0 | 3 | 0 | 0 |
| 3 | 0 | 1 | 0 | 1 | 1 | 0 | 1 | 2 | 0 | 0 | 0 | 1 | 0 | 0 | 0 | 0 | 0 | 1 | 0 | 1 | 0 | 0 |
| 4 | 7 | 10 | 1 | 0 | 6 | 4 | 6 | 2 | 3 | 3 | 1 | 7 | 5 | 2 | 4 | 8 | 4 | 14 | 1 | 4 | 0 | 0 |
| 5 | 5 | 5 | 1 | 6 | 0 | 7 | 4 | 1 | 1 | 5 | 2 | 6 | 1 | 3 | 5 | 6 | 1 | 11 | 0 | 4 | 0 | 0 |
| 6 | 1 | 2 | 0 | 4 | 7 | 0 | 4 | 0 | 0 | 4 | 4 | 4 | 0 | 0 | 0 | 3 | 1 | 0 | 0 | 0 | 0 | 0 |
| 7 | 4 | 9 | 1 | 6 | 4 | 4 | 0 | 0 | 1 | 2 | 0 | 4 | 5 | 2 | 3 | 4 | 5 | 9 | 0 | 5 | 0 | 0 |
| 8 | 0 | 0 | 2 | 2 | 1 | 0 | 0 | 0 | 1 | 3 | 0 | 4 | 1 | 3 | 4 | 6 | 2 | 14 | 0 | 4 | 0 | 0 |
| 9 | 3 | 1 | 0 | 3 | 1 | 0 | 1 | 1 | 0 | 0 | 0 | 1 | 3 | 0 | 1 | 3 | 1 | 4 | 0 | 1 | 0 | 0 |
| 10 | 3 | 2 | 0 | 3 | 5 | 4 | 2 | 3 | 0 | 0 | 3 | 7 | 0 | 5 | 5 | 5 | 0 | 10 | 0 | 5 | 0 | 0 |
| 11 | 1 | 0 | 0 | 1 | 2 | 4 | 0 | 0 | 0 | 3 | 0 | 1 | 0 | 0 | 0 | 0 | 0 | 1 | 1 | 0 | 0 | 0 |
| 12 | 5 | 3 | 1 | 7 | 6 | 4 | 4 | 4 | 1 | 7 | 1 | 0 | 1 | 1 | 2 | 3 | 1 | 6 | 0 | 2 | 0 | 0 |
| 13 | 3 | 5 | 0 | 5 | 1 | 0 | 5 | 1 | 3 | 0 | 0 | 1 | 0 | 0 | 1 | 3 | 2 | 4 | 0 | 1 | 0 | 0 |
| 14 | 2 | 2 | 0 | 2 | 3 | 0 | 2 | 3 | 0 | 5 | 0 | 1 | 0 | 0 | 1 | 1 | 0 | 2 | 0 | 1 | 0 | 0 |
| 15 | 5 | 3 | 0 | 4 | 5 | 0 | 3 | 4 | 1 | 5 | 0 | 2 | 1 | 1 | 0 | 3 | 1 | 6 | 0 | 2 | 0 | 0 |
| 16 | 7 | 4 | 0 | 8 | 6 | 3 | 4 | 6 | 3 | 5 | 0 | 3 | 3 | 1 | 3 | 0 | 2 | 12 | 0 | 4 | 2 | 0 |
| 17 | 3 | 5 | 0 | 4 | 1 | 1 | 5 | 2 | 1 | 0 | 0 | 1 | 2 | 0 | 1 | 2 | 0 | 5 | 0 | 1 | 1 | 0 |
| 18 | 12 | 9 | 1 | 14 | 11 | 0 | 9 | 14 | 4 | 10 | 1 | 6 | 4 | 2 | 6 | 12 | 5 | 0 | 1 | 8 | 1 | 0 |
| 19 | 0 | 0 | 0 | 1 | 0 | 0 | 0 | 0 | 0 | 0 | 1 | 0 | 0 | 0 | 0 | 0 | 0 | 1 | 0 | 0 | 0 | 0 |
| 20 | 4 | 3 | 1 | 4 | 4 | 0 | 5 | 4 | 1 | 5 | 0 | 2 | 1 | 1 | 2 | 4 | 1 | 8 | 0 | 0 | 0 | 0 |
| 21 | 0 | 0 | 0 | 0 | 0 | 0 | 0 | 0 | 0 | 0 | 0 | 0 | 0 | 0 | 0 | 2 | 1 | 1 | 0 | 0 | 0 | 0 |
| 22 | 0 | 0 | 0 | 0 | 0 | 0 | 0 | 0 | 0 | 0 | 0 | 0 | 0 | 0 | 0 | 0 | 0 | 0 | 0 | 0 | 0 | 0 |

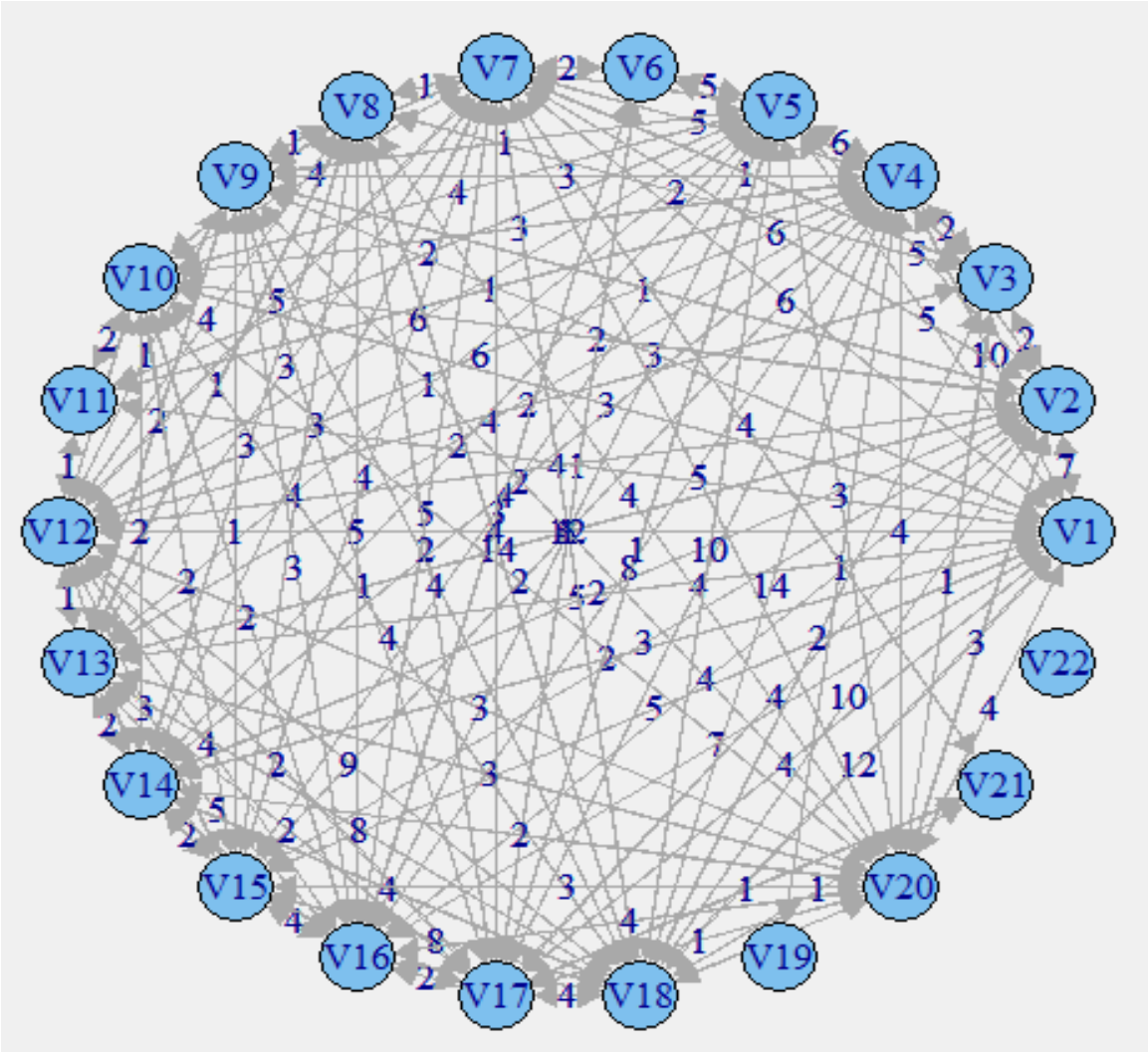

**Figure-4.3**: Table-4.2 in Graphical form, Directed Graph. V1 refers to Sentence 1 from Table-4.1, and from Table-4.2 corresponding weighted edges between two sentences are shown here.

**Procedure 3:**

After getting a Lexical Network, we are applying Centrality-based measures for sentence scoring. Sentence with high centrality score is decided more important compared to low centrality score. To extract sentence for summary generation, we decide Cosine similarity threshold (ϴ), we add the next sentence in summary, if ∀ X, X ∈ SummarySentence: Cosine-Similarity (X,

Next_Sentence) < ϴ. This Work is Extended with the impact of different WSD techniques and different similarity thresholds in the post-processing step.

To find the similarity between two sentences we are using cosine similarity. If we have two vectors A and B, each of length n then cosine similarity is given by Equation-4.22. AI, and $B_i$ are a component of vector A, and B, this similarity measure also used in (C.S. Yadav et al., 2016; Chandra Shekhar Yadav & Sharan, 2015).

$$\text{Similarity}=\cos(\Theta)=\frac{A\bullet B}{\|A\|\ \|B\|}=\frac{\sum_{i=1}^{n}A_iB_i}{\sqrt{\sum_{i=1}^{n}A_i^2}\sqrt{\sum_{i=1}^{n}B_i^2}} \qquad (4.22)$$

In graph theory or Network Analysis is to decide the relative importance of a node (or to find central Figures), one existing approach is Centrality-based measures. For example, in the social network to decide the importance of a person can do with the help of Centrality. Some Centrality-based measures are (1) Degree Centrality, (2) Eigenvector Centrality, (3) Katz Centrality, (4) PageRank, (5) Betweenness Centrality, (6) Closeness Centrality, (7) Alpha Centrality, (8) BonPow Centrality, (9) HUB/Authority (10) Subgraph Centrality etc. detailed description of different centrality measure used by us are,

## 4.5 Experiments and Results

In this section, we are presenting different experiments performed by us along with result analysis. First, we are presenting Performance of different Centrality measure over Selected Lesk Algorithm with different Threshold, then Impact of WSD Technique and Threshold for Different Centrality Measures.

### 4.5.1 Experiment-4.1: Performance of different Centrality measure over Selected Lesk Algorithm with different Threshold

In this ***section***, we are showing the performance of Different Centrality-based measures over fixed WSD algorithm like Adapted Lesk, Cosine Lesk, and Simple Lesk respectively and different cosine similarity threshold in post-processing. We have selected a cosine similarity threshold range (0%, 5%, 10%, …30%, 35%, 40%), due to the limitation of space, and sufficient need we are interested to show result only for 5 % and 10% threshold.

#### 4.5.1.1 Adapted Lesk (Th 10%)

In Table-4.3 we are using Adapted Lesk Algorithm for WSD for lexical network creation, and in the third step, we set a cosine similarity threshold of 10%. After lexical Network construction (Procedure 2 in Algorithm 2), we have applied different centrality measure like Alpha Centrality (coded as α) for different values 0≤ α ≤1.

From the corresponding graph of Table-4.3, in Figure-4.7 we can interpret that, in Alpha Centrality when 0.6≤ α ≤0.9, performance is continuously increasing, but when α =1, then the performance of summarizer system is strangely reduced. Eigen Value and Hub/Authority are performing equal with the highest performance, and Second, highest performance is done by Subgraph based Centrality.

**Table-4.3:** Performance of Different Centrality measures using Adapted Lesk as WSD, and 10% similarity threshold

| Score | α=0.1 | α=0.2 | α=0.3 | α=0.4 | α=0.5 | α=0.6 | α=0.7 | α=0.8 | α=0.9 | α=1 | Betweenness | BonPow | Closeness | Eigen Value | Hub/ Authority | Page Rank | Sub graph |
|---|---|---|---|---|---|---|---|---|---|---|---|---|---|---|---|---|---|
| ROUGE-1 | 0.287 | 0.292 | 0.272 | 0.272 | 0.266 | 0.305 | 0.317 | 0.329 | 0.334 | 0.296 | 0.301 | 0.297 | 0.296 | 0.349 | 0.349 | 0.335 | 0.345 |
| ROUGE-L | 0.264 | 0.262 | 0.249 | 0.249 | 0.245 | 0.273 | 0.288 | 0.303 | 0.306 | 0.277 | 0.271 | 0.268 | 0.267 | 0.319 | 0.319 | 0.304 | 0.309 |
| ROUGE-W | 0.117 | 0.115 | 0.109 | 0.11 | 0.107 | 0.129 | 0.138 | 0.146 | 0.148 | 0.123 | 0.115 | 0.118 | 0.116 | 0.141 | 0.141 | 0.137 | 0.141 |
| ROUGE-S* | 0.068 | 0.074 | 0.06 | 0.061 | 0.06 | 0.07 | 0.076 | 0.092 | 0.092 | 0.077 | 0.077 | 0.073 | 0.076 | 0.106 | 0.106 | 0.101 | 0.111 |
| ROUGE-SU* | 0.072 | 0.078 | 0.064 | 0.066 | 0.064 | 0.072 | 0.079 | 0.094 | 0.094 | 0.082 | 0.082 | 0.077 | 0.081 | 0.111 | 0.111 | 0.106 | 0.116 |

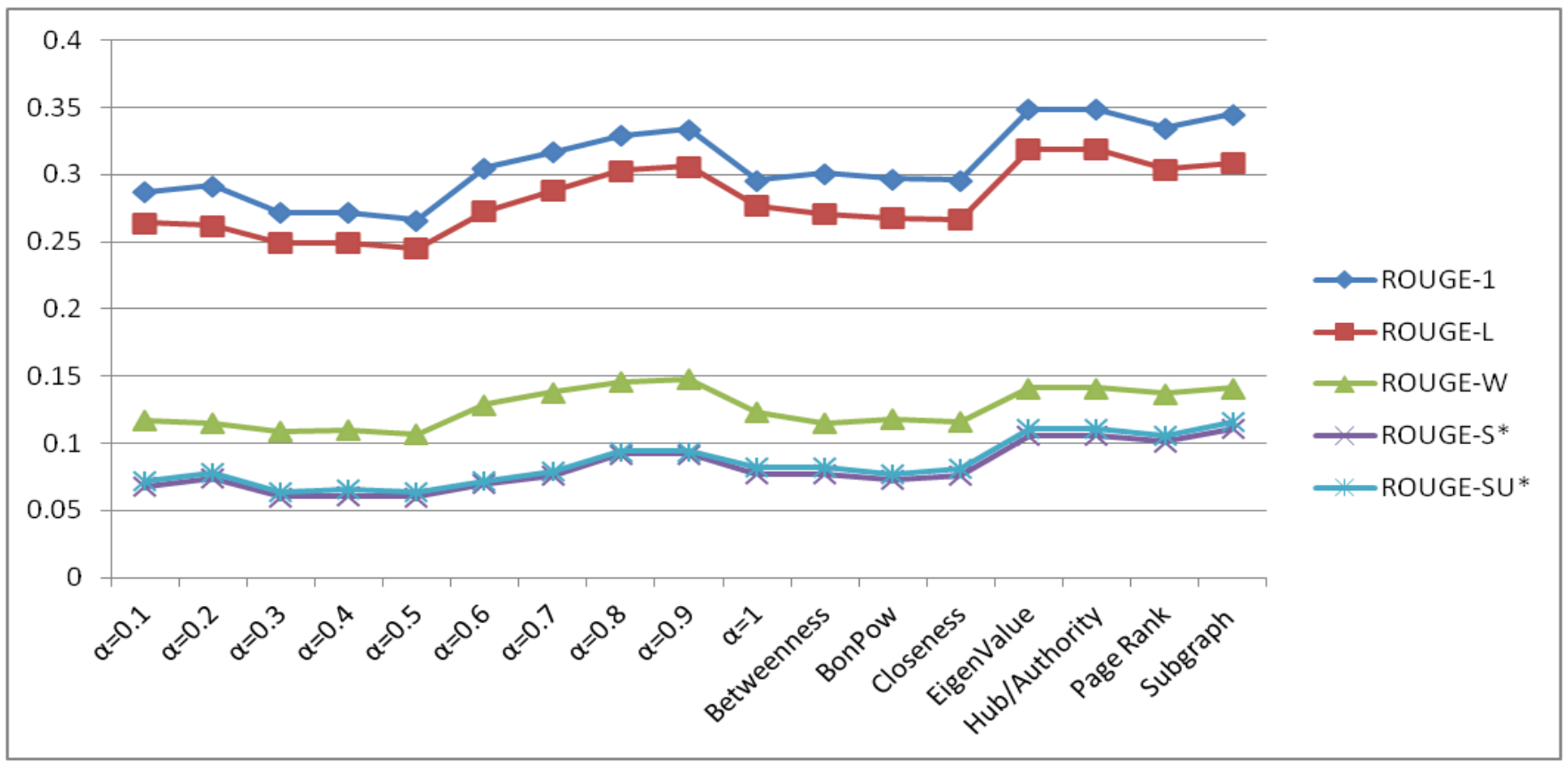

**Figure-4.7**: Performance Evaluation using Rouge Score of different centrality measures (Adapted Lesk as WSD, and 10% similarity threshold)

#### 4.5.1.2 Adapted Lesk (Th 5%)

In Table-4.4 we are using Adapted Lesk Algorithm for WSD for lexical network creation, and in the third step, we set with similarity Threshold 5%. After lexical Network construction (Procedure 2 in Algorithm, we have applied different centrality measure like Alpha Centrality (coded as α) for different values 0≤ α ≤1. From the corresponding graph of Table-4.4 in Figure-4.8, we can interpret that, in Alpha Centrality when 0.6≤ α ≤0.9, performance is continuously increasing (exception α=.7) but when α =1, then the performance of summarizer system is reduced compared to this. Eigen Value and Hub/Authority are performing equally with second highest performance, and Subgraph based Centrality achieves the highest performance.

**Table-4.4**: Performance of Different Centrality measures using Adapted Lesk as WSD, and 5% similarity threshold

| Score | α=0.1 | α=0.2 | α=0.3 | α=0.4 | α=0.5 | α=0.6 | α=0.7 | α=0.8 | α=0.9 | α=1 | Betweenness | Bon Pow | Closeness | Eigen Value | Hub/ Authority | Page Rank | Sub graph |
|---|---|---|---|---|---|---|---|---|---|---|---|---|---|---|---|---|---|
| ROUGE-1 | 0.289 | 0.278 | 0.26 | 0.275 | 0.278 | 0.31 | 0.3 | 0.312 | 0.31 | 0.291 | 0.291 | 0.289 | 0.285 | 0.322 | 0.322 | 0.314 | 0.325 |
| ROUGE-L | 0.265 | 0.245 | 0.233 | 0.246 | 0.256 | 0.28 | 0.277 | 0.286 | 0.282 | 0.269 | 0.265 | 0.26 | 0.256 | 0.299 | 0.299 | 0.284 | 0.287 |
| ROUGE-W | 0.116 | 0.107 | 0.102 | 0.109 | 0.112 | 0.122 | 0.121 | 0.127 | 0.126 | 0.12 | 0.112 | 0.11 | 0.11 | 0.136 | 0.136 | 0.129 | 0.127 |
| ROUGE-S* | 0.067 | 0.063 | 0.056 | 0.06 | 0.064 | 0.077 | 0.072 | 0.082 | 0.082 | 0.072 | 0.068 | 0.07 | 0.065 | 0.094 | 0.094 | 0.089 | 0.095 |
| ROUGE-SU* | 0.072 | 0.068 | 0.06 | 0.065 | 0.068 | 0.081 | 0.076 | 0.086 | 0.086 | 0.077 | 0.073 | 0.075 | 0.07 | 0.099 | 0.099 | 0.094 | 0.1 |

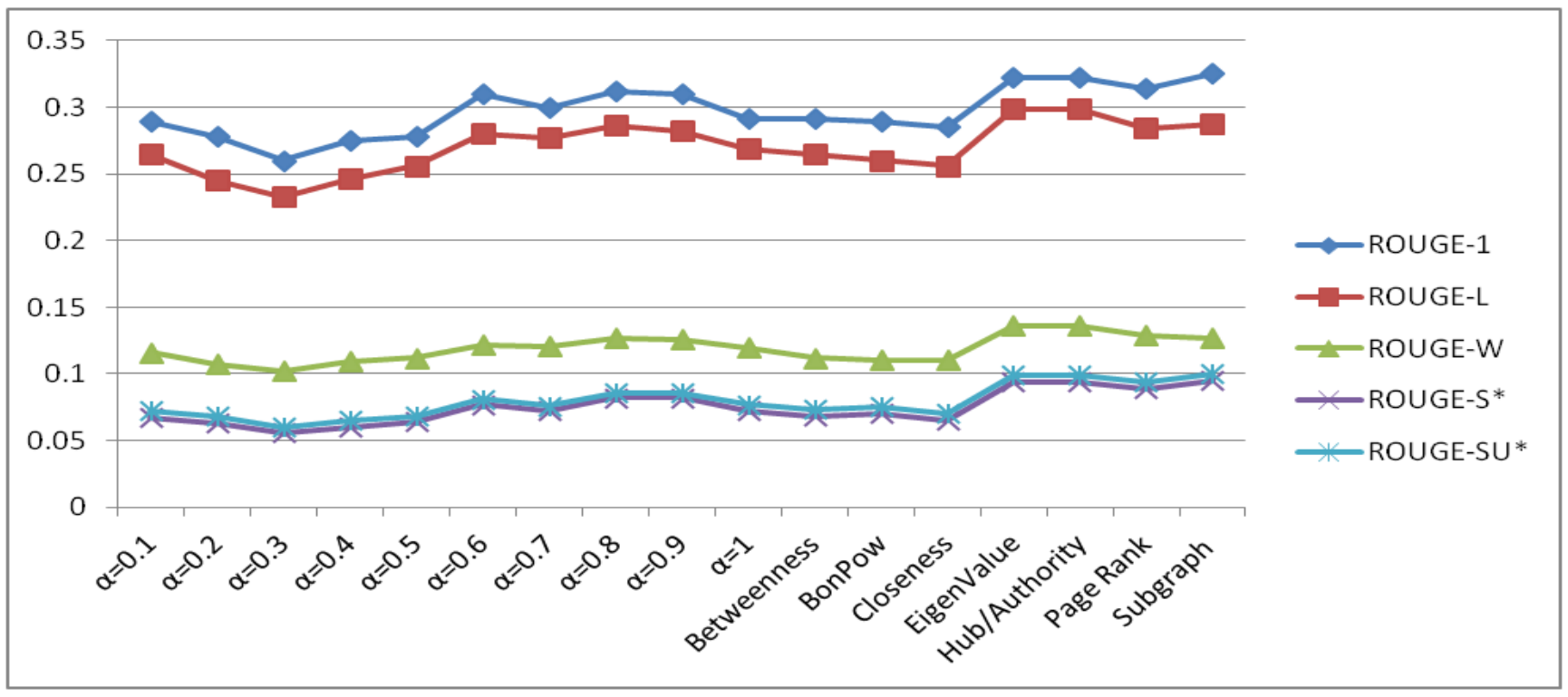

**Figure-4.8**: Performance Evaluation using Rouge Score of different centrality measures (Adapted Lesk as WSD, and 5% similarity threshold)

#### 4.5.1.3 Cosine Lesk (Th 10%)

In Table-4.5 we are using Cosine Lesk Algorithm for WSD for lexical network creation, and in the third step, we set with Cosine Similarity Threshold 10%. After lexical Network construction (Procedure 2 in Algorithm 1), we have applied different centrality measure like Alpha Centrality (coded as α) for different values 0≤ α ≤1. From the corresponding graph of Table-4.5 and Figure-4.9 we can interpret that, in Alpha Centrality when 0.6≤ α ≤0.9, performance is continuously increasing (exception α=.7) but when α =1, then the performance of summarizer system instantly reduced. Eigen Value and Hub/Authority are performing equal (third highest) with Page Rank (second highest) and Subgraph Based Centrality (top performance), but the thing here to notice that all these measures are performing almost same.

**Table-4.5**: Performance of Different Centrality measures using Cosine Lesk as WSD, and 10% similarity threshold

| Score | α=0.1 | α=0.2 | α=0.3 | α=0.4 | α=0.5 | α=0.6 | α=0.7 | α=0.8 | α=0.9 | α=1 | Betweenness | BonPow | Closeness | Eigen Value | Hub/ Authority | Page Rank | Sub graph |
|---|---|---|---|---|---|---|---|---|---|---|---|---|---|---|---|---|---|
| ROUGE-1 | 0.269 | 0.274 | 0.278 | 0.279 | 0.281 | 0.311 | 0.294 | 0.315 | 0.313 | 0.26 | 0.284 | 0.283 | 0.304 | 0.337 | 0.337 | 0.338 | 0.34 |
| ROUGE-L | 0.241 | 0.25 | 0.256 | 0.254 | 0.254 | 0.288 | 0.268 | 0.284 | 0.283 | 0.24 | 0.26 | 0.257 | 0.276 | 0.311 | 0.311 | 0.3 | 0.307 |
| ROUGE-W | 0.106 | 0.106 | 0.111 | 0.107 | 0.109 | 0.135 | 0.127 | 0.134 | 0.133 | 0.103 | 0.115 | 0.113 | 0.123 | 0.14 | 0.14 | 0.136 | 0.14 |
| ROUGE-S* | 0.058 | 0.062 | 0.061 | 0.064 | 0.065 | 0.078 | 0.068 | 0.078 | 0.076 | 0.053 | 0.065 | 0.067 | 0.077 | 0.103 | 0.103 | 0.099 | 0.103 |
| ROUGE-SU* | 0.062 | 0.067 | 0.065 | 0.069 | 0.07 | 0.081 | 0.071 | 0.081 | 0.078 | 0.058 | 0.07 | 0.071 | 0.081 | 0.108 | 0.108 | 0.104 | 0.108 |

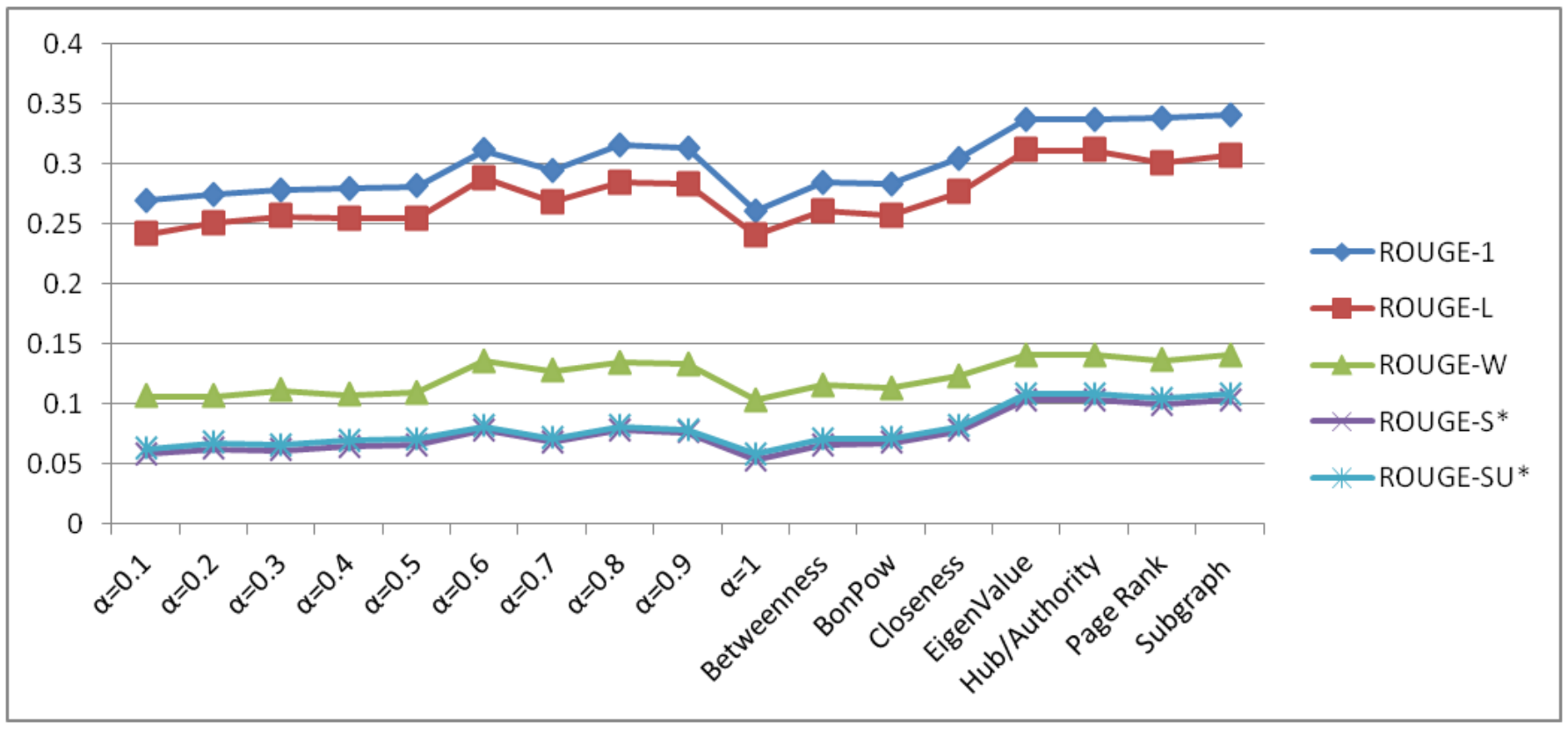

**Figure-4.9**: Performance Evaluation using Rouge Score of different centrality measures (Cosine Lesk as WSD, and 10% similarity threshold)

#### 4.5.1.4 Cosine Lesk (Th 5%)

In Table-4.6 we are using Cosine Lesk Algorithm for WSD for lexical network creation, and in the third step, we set a cosine similarity threshold 5%. After lexical Network construction (Procedure 2 in Algorithm 2), we have applied different centrality measure like Alpha Centrality (coded as α) for different values 0≤ α ≤1. From the corresponding graph of Table-4.6 and Figure-4.10 we can interpret that, in Alpha Centrality when 0.6≤ α ≤0.9, performance is continuously increasing (exception α=0.7) but when α =1, then the performance of summarizer system is strangely reduced. Eigen Value and Hub/Authority are performing equal with the highest performance, and Second, highest performance is measured by subgraph based centrality.

**Table-4.6:** Performance of Different Centrality measures using Cosine Lesk as WSD, and 5% similarity threshold

| Score | α=0.1 | α=0.2 | α=0.3 | α=0.4 | α=0.5 | α=0.6 | α=0.7 | α=0.8 | α=0.9 | α=1 | Betweenness | BonPow | Closeness | Eigen Value | Hub/ Authority | Page Rank | Sub graph |
|---|---|---|---|---|---|---|---|---|---|---|---|---|---|---|---|---|---|
| **ROUGE-1** | 0.257 | 0.28 | 0.272 | 0.251 | 0.267 | 0.313 | 0.3 | 0.311 | 0.303 | 0.249 | 0.285 | 0.259 | 0.291 | 0.327 | 0.327 | 0.309 | 0.313 |
| **ROUGE-L** | 0.23 | 0.248 | 0.246 | 0.226 | 0.244 | 0.283 | 0.272 | 0.281 | 0.274 | 0.227 | 0.261 | 0.235 | 0.262 | 0.3 | 0.3 | 0.28 | 0.278 |
| **ROUGE-W** | 0.101 | 0.105 | 0.107 | 0.097 | 0.106 | 0.133 | 0.118 | 0.122 | 0.118 | 0.097 | 0.115 | 0.103 | 0.112 | 0.136 | 0.136 | 0.126 | 0.128 |
| **ROUGE-S*** | 0.055 | 0.062 | 0.059 | 0.052 | 0.057 | 0.076 | 0.071 | 0.076 | 0.072 | 0.048 | 0.066 | 0.056 | 0.069 | 0.096 | 0.096 | 0.085 | 0.089 |
| **ROUGE-SU*** | 0.059 | 0.066 | 0.063 | 0.056 | 0.061 | 0.078 | 0.075 | 0.079 | 0.075 | 0.052 | 0.07 | 0.061 | 0.074 | 0.101 | 0.101 | 0.09 | 0.093 |

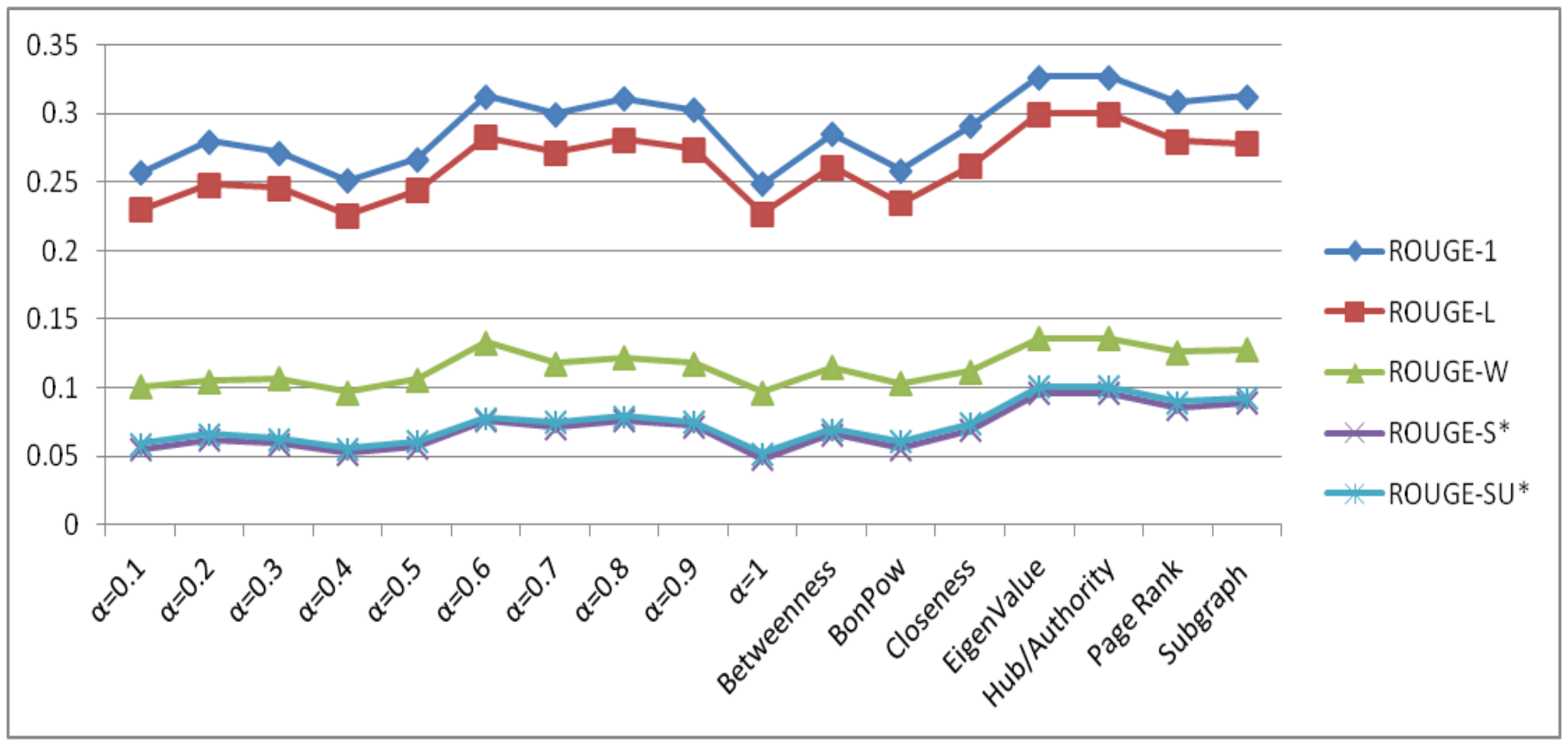

**Figure-4.10**: Performance Evaluation using Rouge Score of different centrality measures (Cosine Lesk as WSD, and 5% similarity threshold)

### 4.5.1.5 Simple Lesk ( Th 10%)

In Table-4.7 we are using Simple Lesk Algorithm for WSD for lexical network creation, and in the third step, we set a cosine similarity threshold of 10%. After lexical Network construction (Procedure 2 in Algorithm 1), we have applied different centrality measure like Alpha Centrality (coded as α) for different values 0≤ α ≤1. From the corresponding graph of Table-4.7 in Figure-4.11 we can interpret that, in Alpha Centrality when 0.6≤ α ≤0.9, performance is continuously increasing (exception α=0.7) but when α =1, then the performance of summarizer system is strangely reduced. Eigen Value and Hub/Authority are performing equal with Third highest performance, PageRank gives the second highest performance, and Subgraph based Centrality measures the highest performance.

**Table-4.7:** Performance of Different Centrality measures using Simple Lesk as WSD, and 10% similarity threshold

| Score | α=0.1 | α=0.2 | α=0.3 | α=0.4 | α=0.5 | α=0.6 | α=0.7 | α=0.8 | α=0.9 | α=1 | Betweenness | Bon Pow | Closeness | Eigen Value | Hub/ Authority | Page Rank | Sub graph |
|---|---|---|---|---|---|---|---|---|---|---|---|---|---|---|---|---|---|
| **ROUGE-1** | 0.282 | 0.269 | 0.28 | 0.269 | 0.262 | 0.315 | 0.307 | 0.317 | 0.314 | 0.269 | 0.284 | 0.278 | 0.297 | 0.336 | 0.336 | 0.34 | 0.379 |
| **ROUGE-L** | 0.255 | 0.236 | 0.251 | 0.248 | 0.237 | 0.285 | 0.279 | 0.288 | 0.287 | 0.241 | 0.254 | 0.251 | 0.269 | 0.304 | 0.304 | 0.304 | 0.341 |
| **ROUGE-W** | 0.111 | 0.102 | 0.109 | 0.105 | 0.102 | 0.135 | 0.132 | 0.137 | 0.137 | 0.105 | 0.112 | 0.11 | 0.119 | 0.137 | 0.137 | 0.136 | 0.161 |
| **ROUGE-S*** | 0.064 | 0.06 | 0.063 | 0.059 | 0.053 | 0.076 | 0.072 | 0.077 | 0.078 | 0.063 | 0.068 | 0.062 | 0.078 | 0.104 | 0.104 | 0.106 | 0.131 |
| **ROUGE-SU*** | 0.069 | 0.064 | 0.068 | 0.064 | 0.057 | 0.079 | 0.074 | 0.079 | 0.08 | 0.067 | 0.072 | 0.066 | 0.083 | 0.109 | 0.109 | 0.111 | 0.136 |

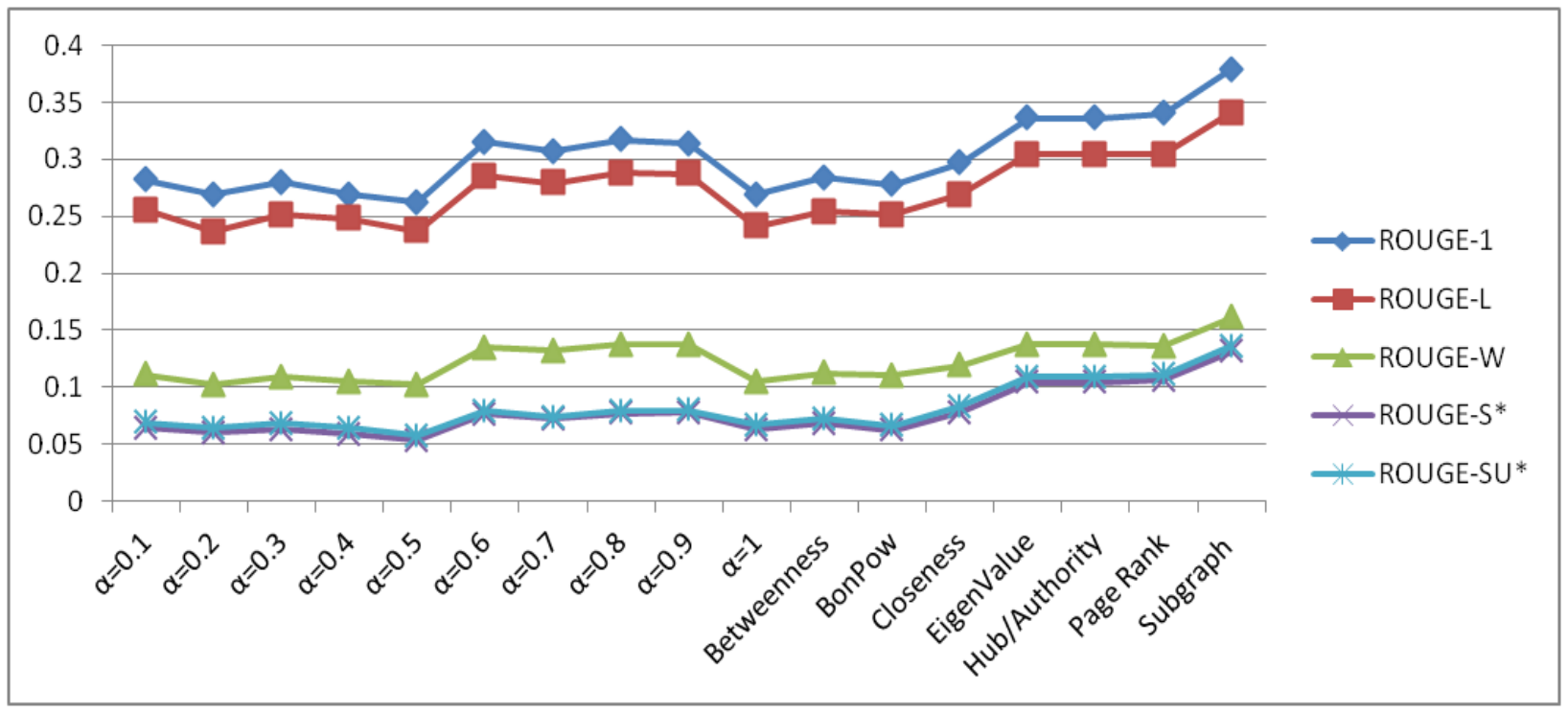

**Figure-4.11**: Performance Evaluation using Rouge Score of different centrality measures (Simple Lesk as WSD, and 10% similarity threshold)

**4.5.1.6 Simple Lesk (Th 5%)**

In Table-4.8 we are using Simple Lesk Algorithm for WSD for lexical network creation, and in the third step, we set a cosine similarity threshold 5%. After lexical Network construction (Procedure 2 in Algorithm 2), we have applied different centrality measure like Alpha Centrality (coded as α) for different values 0≤ α ≤1. From the corresponding graph of Table-4.8 in Figure-4.12 we can interpret that, in Alpha Centrality when 0.6≤ α ≤0.9, performance is continuously increasing (exception α=0.7) but when α =1, then the performance of summarizer system is strangely reduced. Eigen Value and Hub/Authority are performing equal with Third highest performance, PageRank measures second highest performance, and Subgraph based Centrality measures the highest performance.

Table-**4.8**: Performance of Different Centrality measures using Simple Lesk as WSD, and 5% similarity threshold

| Score | α=0.1 | α=0.2 | α=0.3 | α=0.4 | α=0.5 | α=0.6 | α=0.7 | α=0.8 | α=0.9 | α=1 | Betweenness | Bon Pow | Closeness | Eigen Value | Hub/ Authority | Page Rank | Sub graph |
|---|---|---|---|---|---|---|---|---|---|---|---|---|---|---|---|---|---|
| **ROUGE-1** | 0.268 | 0.264 | 0.281 | 0.256 | 0.273 | 0.315 | 0.302 | 0.313 | 0.319 | 0.258 | 0.285 | 0.268 | 0.287 | 0.312 | 0.312 | 0.322 | 0.339 |
| **ROUGE-L** | 0.249 | 0.232 | 0.266 | 0.229 | 0.243 | 0.284 | 0.273 | 0.284 | 0.289 | 0.229 | 0.259 | 0.242 | 0.262 | 0.288 | 0.288 | 0.293 | 0.314 |
| **ROUGE-W** | 0.109 | 0.1 | 0.111 | 0.098 | 0.104 | 0.126 | 0.121 | 0.125 | 0.13 | 0.098 | 0.116 | 0.107 | 0.115 | 0.136 | 0.136 | 0.131 | 0.143 |
| **ROUGE-S*** | 0.058 | 0.058 | 0.065 | 0.05 | 0.057 | 0.079 | 0.072 | 0.076 | 0.08 | 0.053 | 0.068 | 0.059 | 0.072 | 0.089 | 0.089 | 0.094 | 0.109 |
| **ROUGE-SU*** | 0.062 | 0.062 | 0.069 | 0.054 | 0.061 | 0.082 | 0.076 | 0.08 | 0.083 | 0.058 | 0.072 | 0.064 | 0.077 | 0.093 | 0.093 | 0.098 | 0.113 |

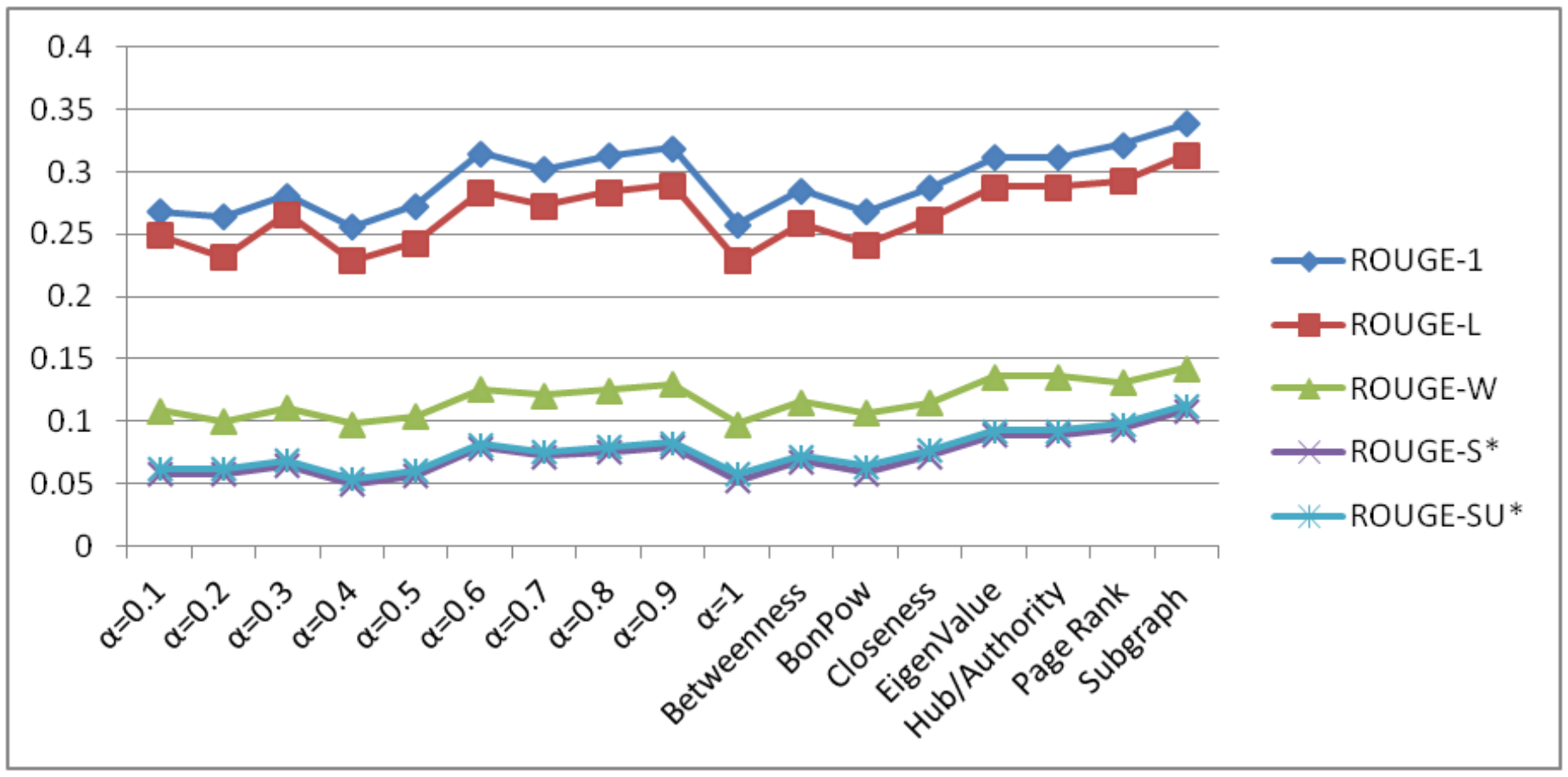

Figure-**4.12**: Performance Evaluation using Rouge Score of different centrality measures (Simple Lesk as WSD, and 5% similarity threshold)

### 4.5.2 **Experiment-4.2:** Impact of WSD Technique and Threshold for Different Centrality Measures

In this ***section***, we are presenting different WSD technique and different threshold impacts in summarization over the subgraph based centrality. Here code, T1 (technique-1) stands for Adapted Lesk, T2 (technique -2) for cosine lesk, and T3 (technique -3) for simplified lesk algorithm, where TH_10 (threshold 10%) means cosine similarity threshold 10%, and Th_5 means cosine similarity threshold is 5%.

#### *4.5.2.1 Subgraph based*

Here we are presenting the performance of Subgraph based centrality in different conditions. Subgraph based centrality put more emphasis on small subgraphs, and this covers long-term relations. Here we are presenting the subgraph based centrality measure with different WSD techniques, and thresholds. From **Table-4.**9 and Figure-4.13, we can conclude that simple WSD technique adapted lesk, and cosine lesk with threshold 10% is giving a better results, followed by simple lesk, adapted lesk, cosine lesk with threshold 5%.

**Table-4.9: Impact of WSD technique and similarity threshold over Subgraph based Centrality**

| Score | T1_TH_10 | T1_TH_5 | T2_TH_10 | T2_TH_5 | T3_TH_10 | T3_TH_5 |
|---|---|---|---|---|---|---|
| **ROUGE-1** | 0.345 | 0.325 | 0.34 | 0.313 | 0.379 | 0.339 |
| **ROUGE-L** | 0.309 | 0.287 | 0.307 | 0.278 | 0.341 | 0.314 |
| **ROUGE-W** | 0.141 | 0.127 | 0.14 | 0.128 | 0.161 | 0.143 |
| **ROUGE-S*** | 0.111 | 0.095 | 0.103 | 0.089 | 0.131 | 0.109 |
| **ROUGE-SU*** | 0.116 | 0.1 | 0.108 | 0.093 | 0.136 | 0.113 |

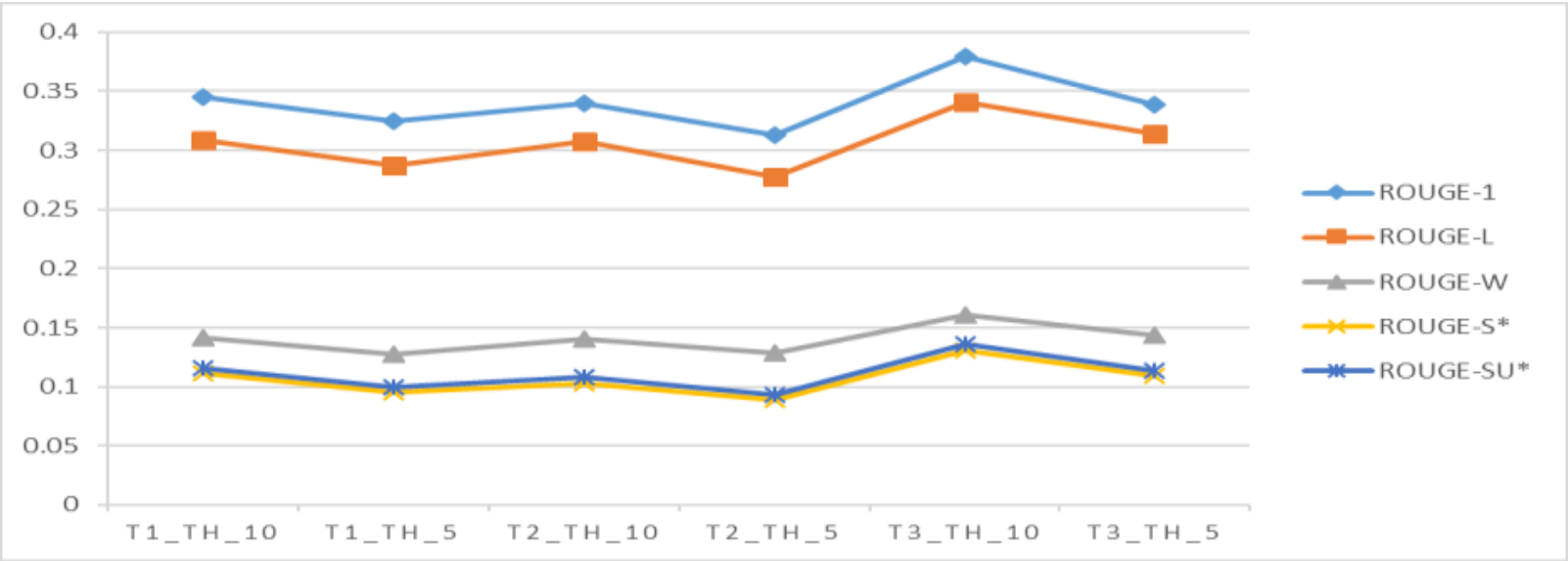

**Figure-4.13**: Impact of WSD technique and similarity threshold over Subgraph based Centrality

*4.5.2.2 Eigenvalue Centrality-based Analysis*

Eigen value centrality depends on both numbers of connections and quality of the connection. From Table-4.10 and Figure-4.14 we can conclude that Simple WSD technique, with threshold 10% cosine lesk, adapted lesk, and simple lesk, but for threshold 5% cosine, adapted, simple lesk is performing in an order higher to lower.

**Table-4.10**: Impact of WSD technique and similarity threshold over Eigen Value based Centrality

| Score | T1_TH_10 | T1_TH_5 | T2_TH_10 | T2_TH_5 | T3_TH_10 | T3_TH_5 |
|---|---|---|---|---|---|---|
| **ROUGE-1** | 0.349 | 0.322 | 0.337 | 0.327 | 0.336 | 0.312 |
| **ROUGE-L** | 0.319 | 0.299 | 0.311 | 0.3 | 0.304 | 0.288 |
| **ROUGE-W** | 0.141 | 0.136 | 0.14 | 0.136 | 0.137 | 0.136 |
| **ROUGE-S*** | 0.106 | 0.094 | 0.103 | 0.096 | 0.104 | 0.089 |
| **ROUGE-SU*** | 0.111 | 0.099 | 0.108 | 0.101 | 0.109 | 0.093 |

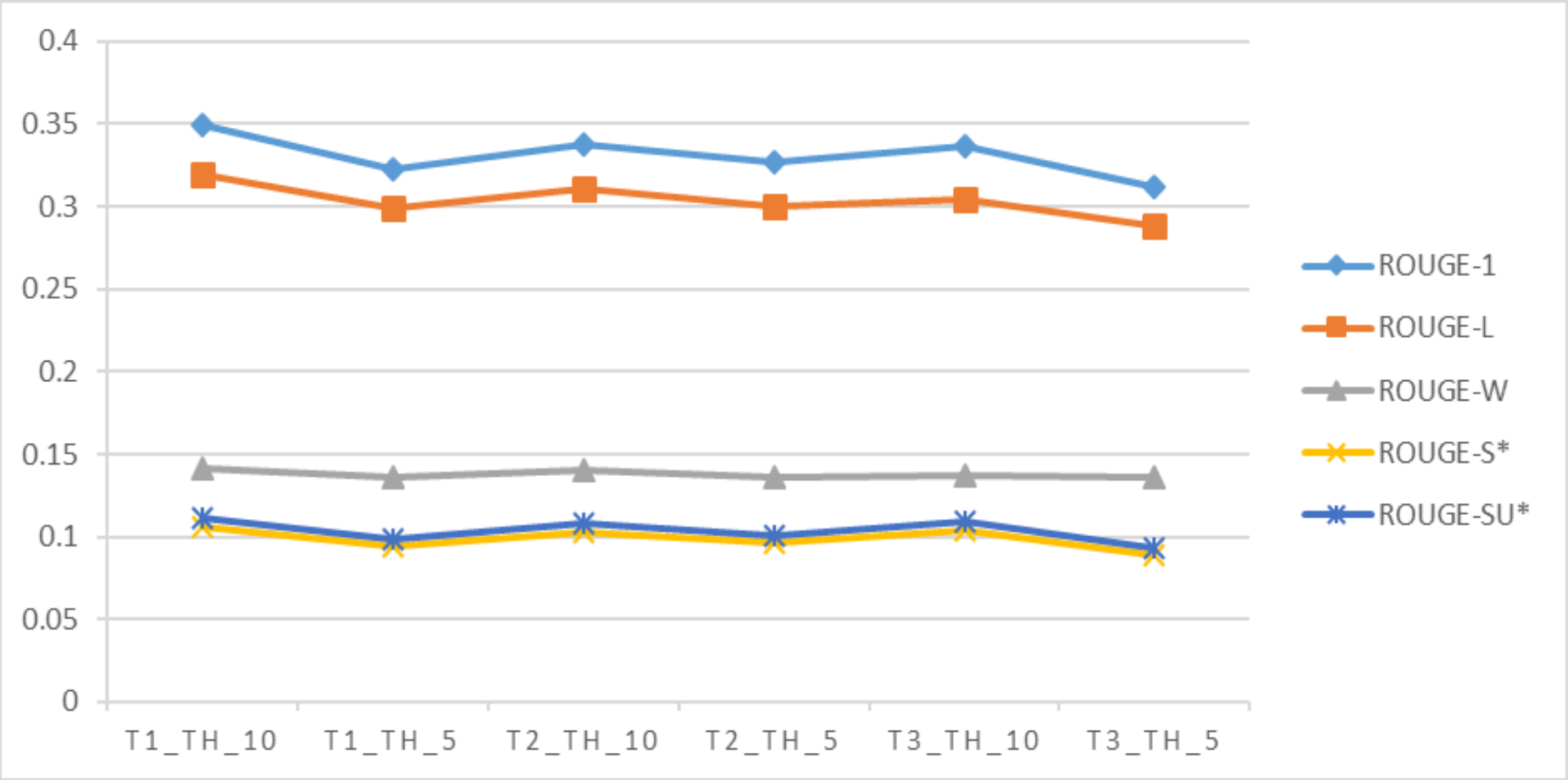

**Figure-4.14:** Impact of WSD technique and similarity threshold over Eigen Value based Centrality

*4.5.2.3 Page Rank*

Page rank is simple, ensure the guaranteed existence of a solution, and it is fast in computation. It forms a probability distribution over a network. In our experiment, we have set dumping factor 0.85. From Table-4.11 and Figure-4.15, we can conclude that Simple WSD technique, with threshold 10% Simple, Cosine Lesk, and Adapted Lesk, But for threshold, 5% Adapted Lesk, Simple Lesk, and Cosine, are performing in order highest to lowest.

**Table-4.11**: Impact of WSD technique and similarity threshold over Page Rank as a Centrality measure

| Score | T1_TH_10 | T1_TH5 | T2_TH_10 | T2_TH_5 | T3_TH_10 | T3_TH_5 |
|---|---|---|---|---|---|---|
| **ROUGE-1** | 0.335 | 0.314 | 0.338 | 0.309 | 0.34 | 0.322 |
| **ROUGE-L** | 0.304 | 0.284 | 0.3 | 0.28 | 0.304 | 0.293 |
| **ROUGE-W** | 0.137 | 0.129 | 0.136 | 0.126 | 0.136 | 0.131 |
| **ROUGE-S*** | 0.101 | 0.089 | 0.099 | 0.085 | 0.106 | 0.094 |
| **ROUGE-SU*** | 0.106 | 0.094 | 0.104 | 0.09 | 0.111 | 0.098 |

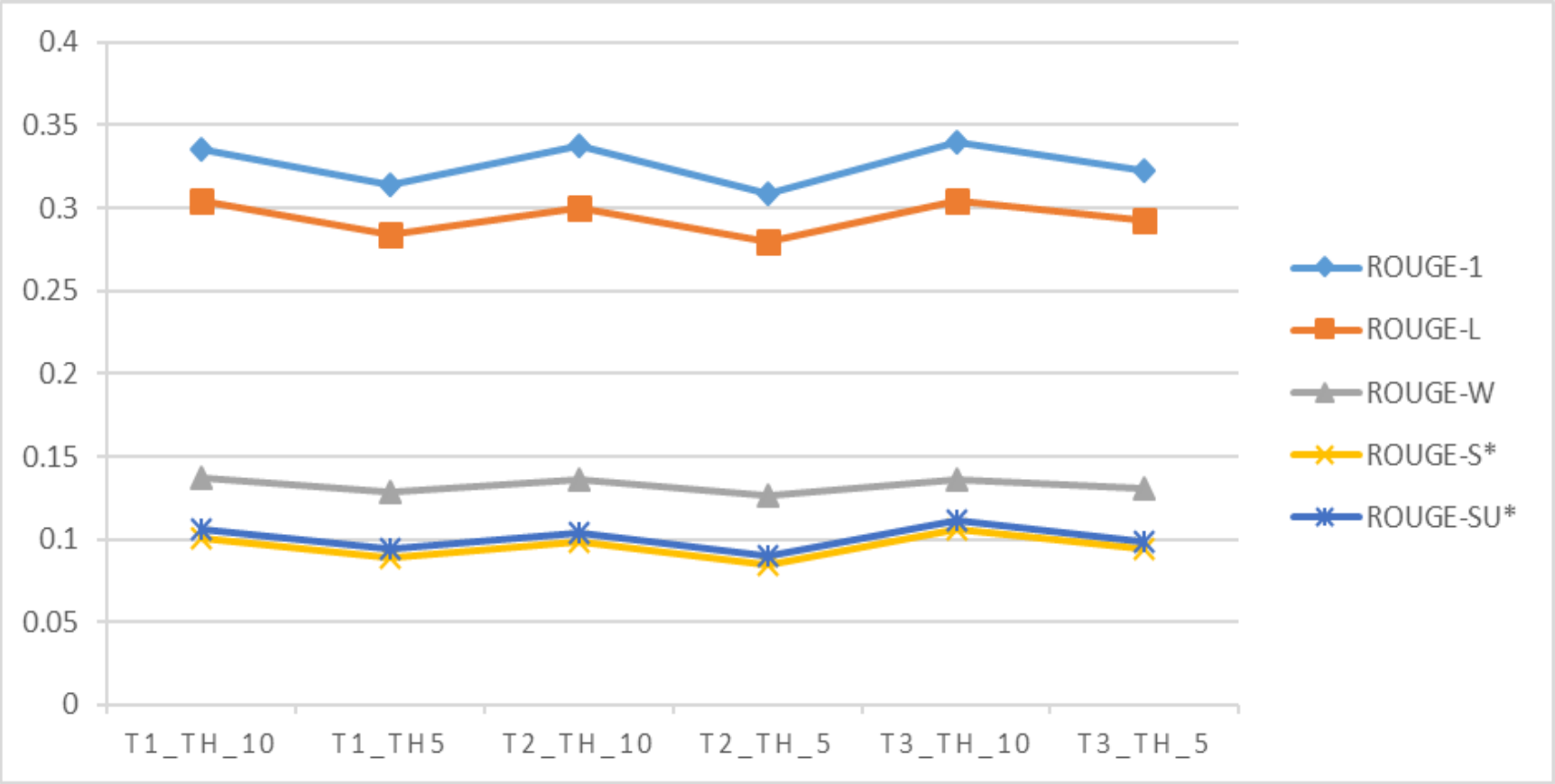

**Figure-4.15**: Impact of WSD technique and similarity threshold over Page Rank as a Centrality measure

*4.5.2.4 BONPOW*

In Bonpow centrality measure, the importance of a node is decided by alters. From Table-4.12, and Figure-4.16, we can conclude that Simple WSD technique, with threshold 10% Adapted Lesk, Cosine Lesk, and Simple Lesk, But for threshold, 5% Adapted Lesk, Simple Lesk, and Cosine, are performing in order highest to lowest.

Table-4.12: Impact of WSD technique and similarity threshold over BONPOW based Centrality

| Score | T1_TH10 | T1_TH5 | T2_TH10 | T2_TH5 | T3_TH10 | T3_TH5 |
|---|---|---|---|---|---|---|
| **ROUGE-1** | 0.297 | 0.289 | 0.283 | 0.259 | 0.278 | 0.268 |
| **ROUGE-L** | 0.268 | 0.26 | 0.257 | 0.235 | 0.251 | 0.242 |
| **ROUGE-W** | 0.118 | 0.11 | 0.113 | 0.103 | 0.11 | 0.107 |
| **ROUGE-S*** | 0.073 | 0.07 | 0.067 | 0.056 | 0.062 | 0.059 |
| **ROUGE-SU*** | 0.077 | 0.075 | 0.071 | 0.061 | 0.066 | 0.064 |

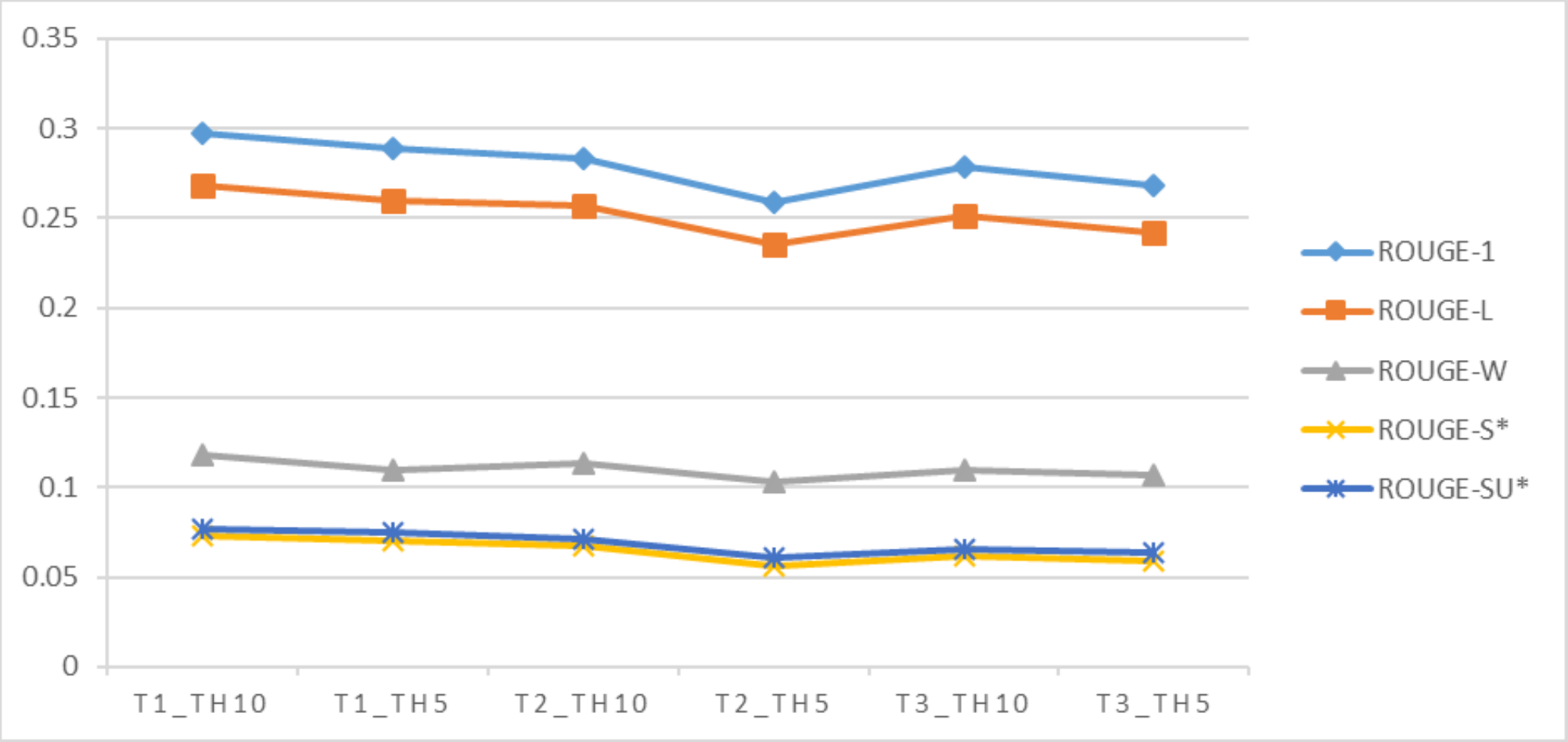

Figure-4.16: Impact of WSD technique and similarity threshold over BONPOW based Centrality

***4.5.2.5 BetweenNess***

Betweenness centrality measures the importance of each node based on a factor of participation of node between shortest paths (any nodes) in a network. From Table-4.13, and Figure-4.17 we can conclude that Adapted Lesk Algorithm with a threshold of 10% is performing better, than Adapted Lesk Algorithm with 5% threshold. Cosine Lesk 5%, Simple Lesk 5%, Cosine Lesk 10%, Simple Lesk 10% threshold.

Table-4.13: Impact of WSD technique and similarity threshold over Between-Ness Centrality

| Score | T1_TH1 | T1_TH5 | T2_TH1 | T2_TH5 | T3_TH1 | T3_TH5 |
|---|---|---|---|---|---|---|
| **ROUGE-1** | 0.301 | 0.291 | 0.284 | 0.285 | 0.284 | 0.285 |
| **ROUGE-L** | 0.271 | 0.265 | 0.26 | 0.261 | 0.254 | 0.259 |
| **ROUGE-W** | 0.115 | 0.112 | 0.115 | 0.115 | 0.112 | 0.116 |
| **ROUGE-S*** | 0.077 | 0.068 | 0.065 | 0.066 | 0.068 | 0.068 |
| **ROUGE-SU*** | 0.082 | 0.073 | 0.07 | 0.07 | 0.072 | 0.072 |

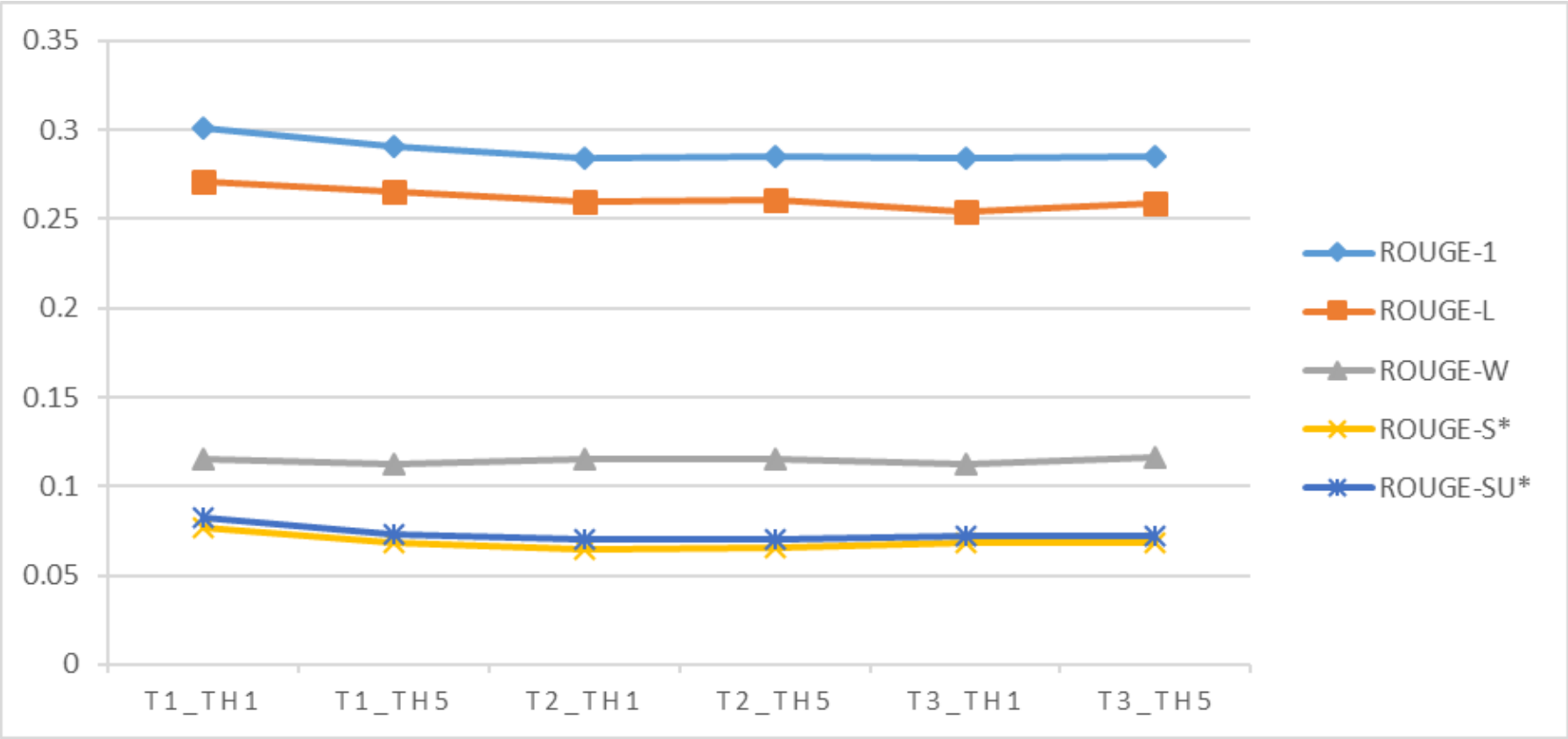

**Figure-4.17**: Impact of WSD technique and similarity threshold over Between-Ness Centrality

*4.5.2.6 Closeness*

Closeness centrality measure importance of node based on the shortest path from that node to all other nodes. From **Table-4.**14 and Figure-4.18, we can conclude that with threshold 10% Cosine Lesk, Simple Lesk, and Adapted Lesk, for threshold, 5% also Cosine Lesk, Simple Lesk, and Adapted Lesk, are performing in order highest to lowest.

**Table-4.14:** Impact of WSD technique and similarity threshold over Closeness Centrality

| Score | T1_TH1 | T1_TH5 | T2_TH1 | T2_TH5 | T3_TH1 | T3_TH5 |
|---|---|---|---|---|---|---|
| **ROUGE-1** | 0.296 | 0.285 | 0.304 | 0.291 | 0.297 | 0.287 |
| **ROUGE-L** | 0.267 | 0.256 | 0.276 | 0.262 | 0.269 | 0.262 |
| **ROUGE-W** | 0.116 | 0.11 | 0.123 | 0.112 | 0.119 | 0.115 |
| **ROUGE-S*** | 0.076 | 0.065 | 0.077 | 0.069 | 0.078 | 0.072 |
| **ROUGE-SU*** | 0.081 | 0.07 | 0.081 | 0.074 | 0.083 | 0.077 |

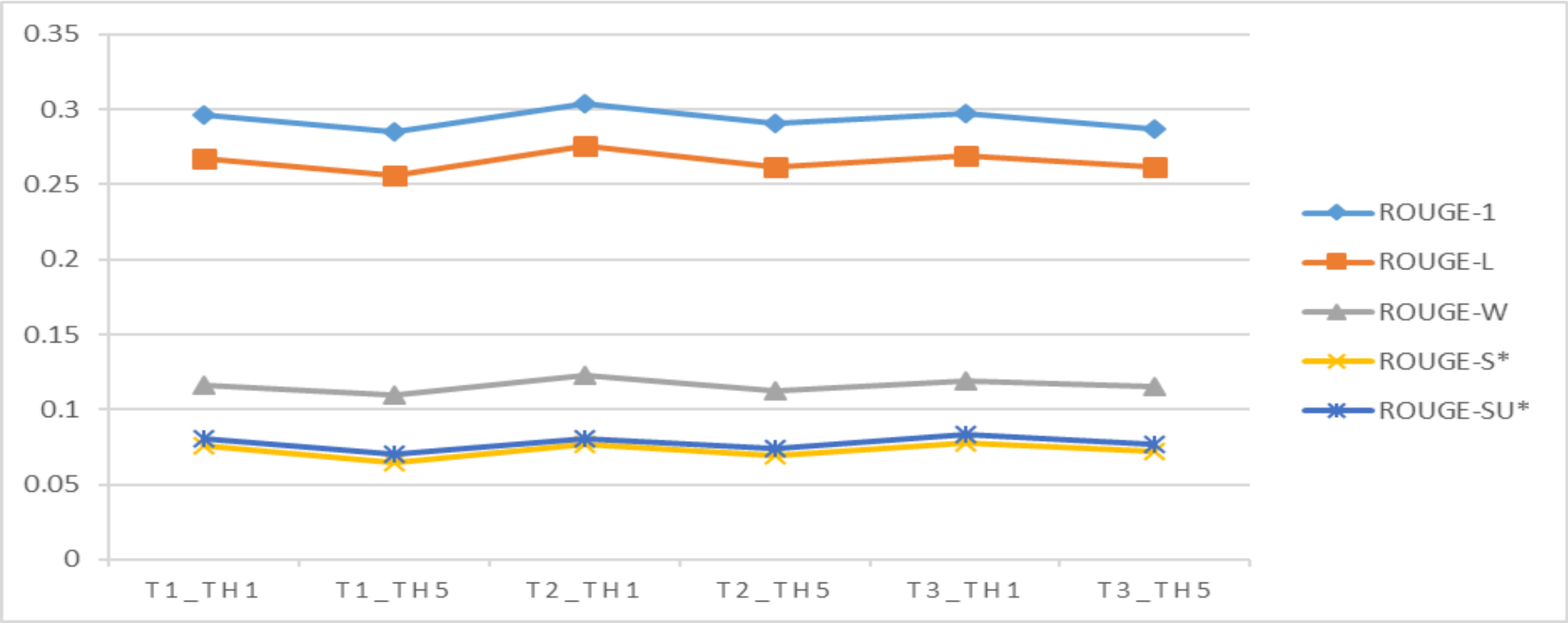

**Figure-4.18**: Impact of WSD technique and similarity threshold over Closeness Centrality

*4.5.2.7 Alpha*

The alpha centrality score depends on the importance of neighboring nodes. In Figure-4.19 we have shown Different ROUGE score using Alpha Centrality (alpha=.1 to 1) for ranking of sentences, extraction, and summarization. From Figure-4.19, this is clear that when alpha is between .1 to .5 (including), this is unpredicted **Table**-4.7, and Table-4.8 no pattern is achieved. But when alpha is 0.6 to 0.9 (including) rouge score is increasing with some exception at alpha =0.7, but at alpha=1 efficiency is again reduced (for all). *[Note: T1, T2 for Adapted Lesk, T3, T4 for Cosine Lesk, T5, T6 for Simplified Lesk Algorithm, TH_10 means Cosine Similarity threshold 10%; Th_5 means 5% threshold.]*

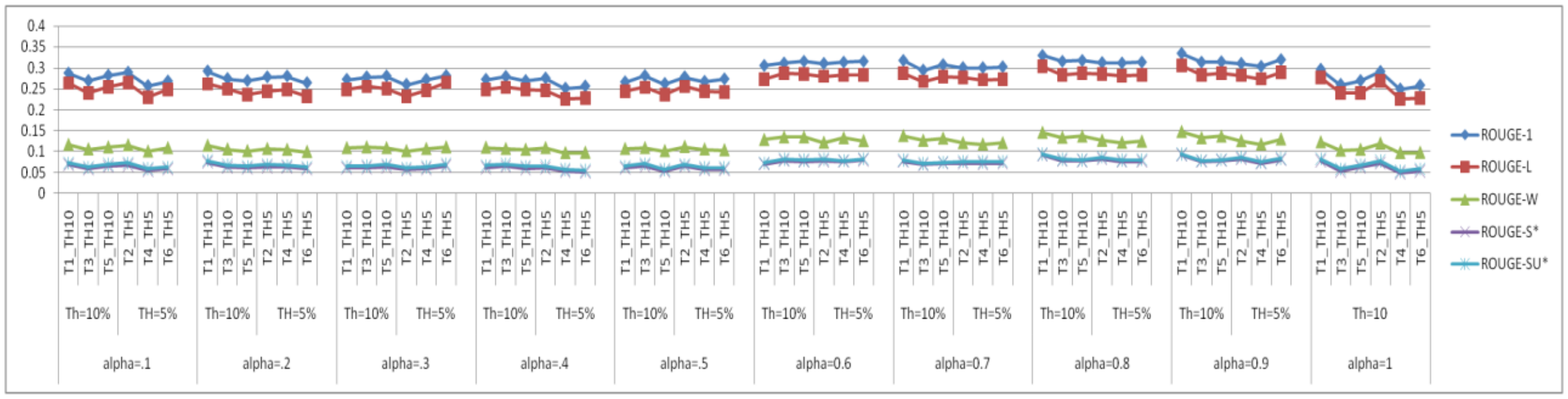

**Figure-4.19**: Impact of WSD technique and similarity threshold over Alpha Centrality (.1 ≤ alpha ≤ 1)

### 4.5.3 Experiment-4.3: Comparisons with Lexalytics algorithms

Semantrica-Lexalytics is a System (available at https://www.lexalytics.com/) that can access through JAVA and Python API for different tasks like Summarization, Sentiment, Entity, etc. The summarization task of this system depends on Lexical Chain creation. The lexical chain is created using Semantic and Lexical relations; Location also gives priority in this system. In this ***section***, we are describing that, where system proposed by us lead by Semantrica. Even this is performing better many times, and many times our proposed methods lead this is shown in **Table-4.**14, 4.15, 4.16 and Figure-4.20.

**Table-4.15**: Comparison of Semantrica-Lexalytics Algorithm with, Difference Centrality-based Measure in Which Adapted Lesk used as WSD (Showing only Improved System w.r.t Semantrica)

| Score | | Adapted Lesk,10% similarity threshold | | | | | | Adapted Lesk,5% similarity threshold | | | | | |
|---|---|---|---|---|---|---|---|---|---|---|---|---|---|
| | Sematrica-Lexalytics | α=0.8 | α=0.9 | Eigen Value | Hub/ Authority | Page Rank | Subgraph | α=0.7 | α=0.8 | α=0.9 | Hub/ Authority | Page Rank | Subgraph |
| ROUGE-1 | .3296 | 0.329 | 0.334 | 0.349 | 0.349 | 0.335 | 0.345 | 0.3 | 0.312 | 0.31 | 0.322 | 0.314 | 0.325 |
| ROUGE- | .277 | 0.303 | 0.306 | 0.319 | 0.319 | 0.304 | 0.309 | 0.277 | 0.286 | 0.282 | 0.299 | 0.284 | 0.287 |
| ROUGE- | .125 | 0.146 | 0.148 | 0.141 | 0.141 | 0.137 | 0.141 | 0.121 | 0.127 | 0.126 | 0.136 | 0.129 | 0.127 |
| ROUGE | .1094 | 0.092 | 0.092 | 0.106 | 0.106 | 0.101 | 0.111 | 0.072 | 0.082 | 0.082 | 0.094 | 0.089 | 0.095 |
| ROUGE-SU* | .1138 | 0.094 | 0.094 | 0.111 | 0.111 | 0.106 | 0.116 | 0.076 | 0.086 | 0.086 | 0.099 | 0.094 | 0.1 |

**Table-4.16**: Comparison of Semantrica-Lexalytics Algorithm with, Difference Centrality-based Measure in Which Cosine Lesk used as WSD (Showing only Improved System w.r.t Semantrica)

| Score | | Cosine Lesk as WSD, and 10% similarity threshold | | | | Cosine Lesk as WSD, and 5% similarity threshold | | | | |
|---|---|---|---|---|---|---|---|---|---|---|
| | Sematrica-Lexalytics | Eigen Value | Hub/ Authority | Page Rank | Subgraph | α=0.8 | EigenValue | Hub/ Authority | Page Rank | Subgraph |
| ROUGE-1 | .3296 | 0.337 | 0.337 | 0.338 | 0.34 | 0.311 | 0.327 | 0.327 | 0.309 | 0.313 |
| ROUGE-L | .277 | 0.311 | 0.311 | 0.3 | 0.307 | 0.281 | 0.3 | 0.3 | 0.28 | 0.278 |
| ROUGE-W 1.5 | .125 | 0.14 | 0.14 | 0.136 | 0.14 | 0.122 | 0.136 | 0.136 | 0.126 | 0.128 |
| ROUGE S* | .1094 | 0.103 | 0.103 | 0.099 | 0.103 | 0.076 | 0.096 | 0.096 | 0.085 | 0.089 |
| **ROUGE-SU*** | .1138 | 0.108 | 0.108 | 0.104 | 0.108 | 0.079 | 0.101 | 0.101 | 0.09 | 0.093 |

**Table-4.17**: Comparison of Semantrica-Lexalytics Algorithm with, Difference Centrality-based Measure in Which Simple Lesk used as WSD (Showing only Improved System w.r.t Semantrica)

| **Score** | **Sematrica - Lexalytics** | **Simple Lesk as WSD, and 10% similarity** | | | | | | | | **Simple Lesk as WSD, and 5% similarity** | | | | | |
|---|---|---|---|---|---|---|---|---|---|---|---|---|---|---|---|
| | | **α=0.6** | **α=0.7** | **α=0.8** | **α=0.9** | **Eigen Valu** | **Hub/ Authority** | **Page Rank** | **Subgraph** | **α=0.8** | **α=0.9** | **Eigen Valu** | **Hub/ Authority** | **Page Rank** | **Subgraph** |
| **ROUGE** | .3296 | 0.31 | 0.30 | 0.31 | 0.31 | 0.336 | 0.336 | 0.34 | 0.379 | 0.31 | 0.31 | 0.312 | 0.312 | 0.32 | 0.339 |
| **ROUGE** | .277 | 0.28 | 0.27 | 0.28 | 0.28 | 0.304 | 0.304 | 0.30 | 0.341 | 0.28 | 0.28 | 0.288 | 0.288 | 0.29 | 0.314 |
| **ROUGE -W 1.5** | .125 | 0.135 | 0.132 | 0.137 | 0.137 | 0.137 | 0.137 | 0.136 | 0.161 | 0.125 | 0.13 | 0.136 | 0.136 | 0.131 | 0.143 |
| **ROUGE S*** | .1094 | 0.076 | 0.072 | 0.077 | 0.078 | 0.104 | 0.104 | 0.106 | 0.131 | 0.076 | 0.08 | 0.089 | 0.089 | 0.094 | 0.109 |
| **ROUGE -SU*** | .1138 | 0.079 | 0.074 | 0.079 | 0.08 | 0.109 | 0.109 | 0.111 | 0.136 | 0.08 | 0.083 | 0.093 | 0.093 | 0.098 | 0.113 |

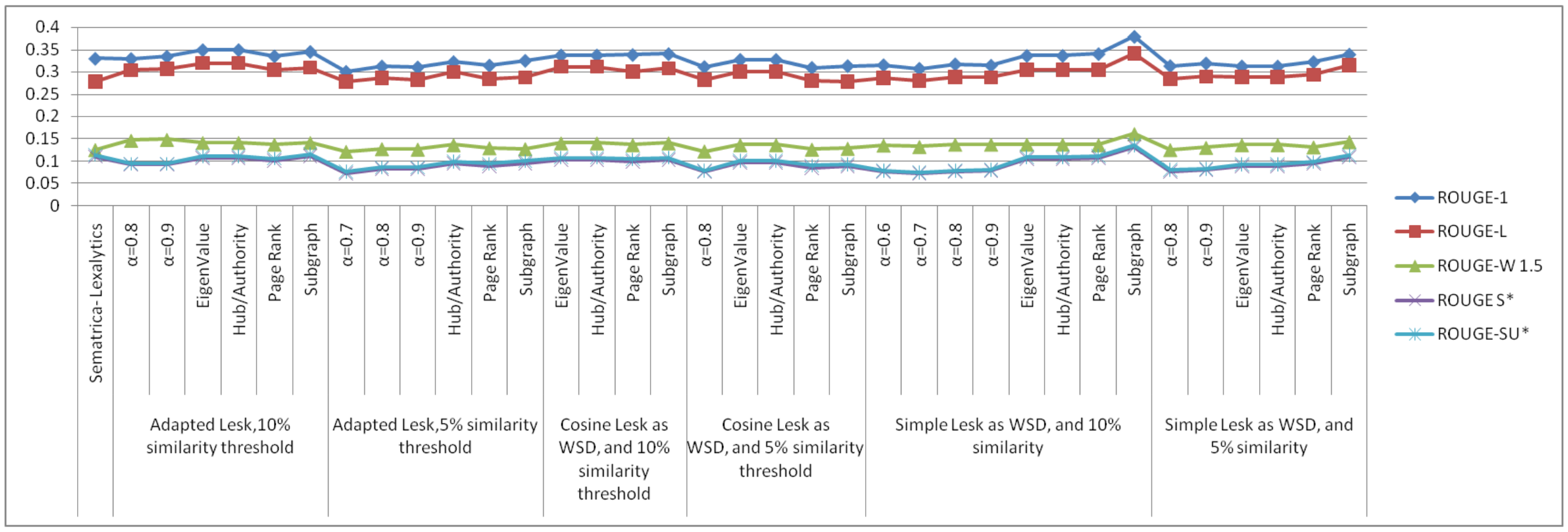

**Figure-4.20**: Semantrica-Lexalytics VS Our Proposed System's (Improved Only)

### 4.5.4 Overall Performance

From **Table**-4.3, 4.4, 4.5, 4.6, 4.7, 4.8 and Figure-4.7, 4.8, 4.9, 4.10, 4.11, 4.12 this is clear that when Alpha =1 to 0.5, we are unable to decide the performance pattern, but when .6≤ α ≤0.9, performance is increasing sometime exception is (α=.7), and again reduced at α=1. Most of the time Subgraph based centrality is giving a better result, the reason of this, subgraph measure gives importance to subgraph in the graph, the smaller subgraph is given more importance, and this is natural that in the large graph there will be some subgraphs, cycles.

In Figure-4.21 we are showing some top performances of all three WSD techniques for different threshold (here 5%, and 10%), in which Subgraph based centrality is performing better. It may look hard to interpret X-axis, for clarification on X-axis different Centrality-based measure are shown, T1,T2 stands for Adapted Lesk Algorithm., T3, T4 for Cosine Lesk, and T5, T6 for Simple Lesk algorithm, TH10 (shown only TH1), TH5 stands for cosine similarity threshold 10%, and 5% respectively. For example, very first entry on X-Axis is EigenValue, and (T1_TH1) means, we are using Adapted Lesk Algorithm for WSD (T1) and cosine similarity threshold set is 10% (TH1) using EigenValue centrality measure. Same as Highest performance is achieved by (T5_TH10, Subgraph) which can interpret as Simplified Lesk Algorithm is used (because of T5), with similarity threshold 10%, and Subgraph based centrality measured is used.

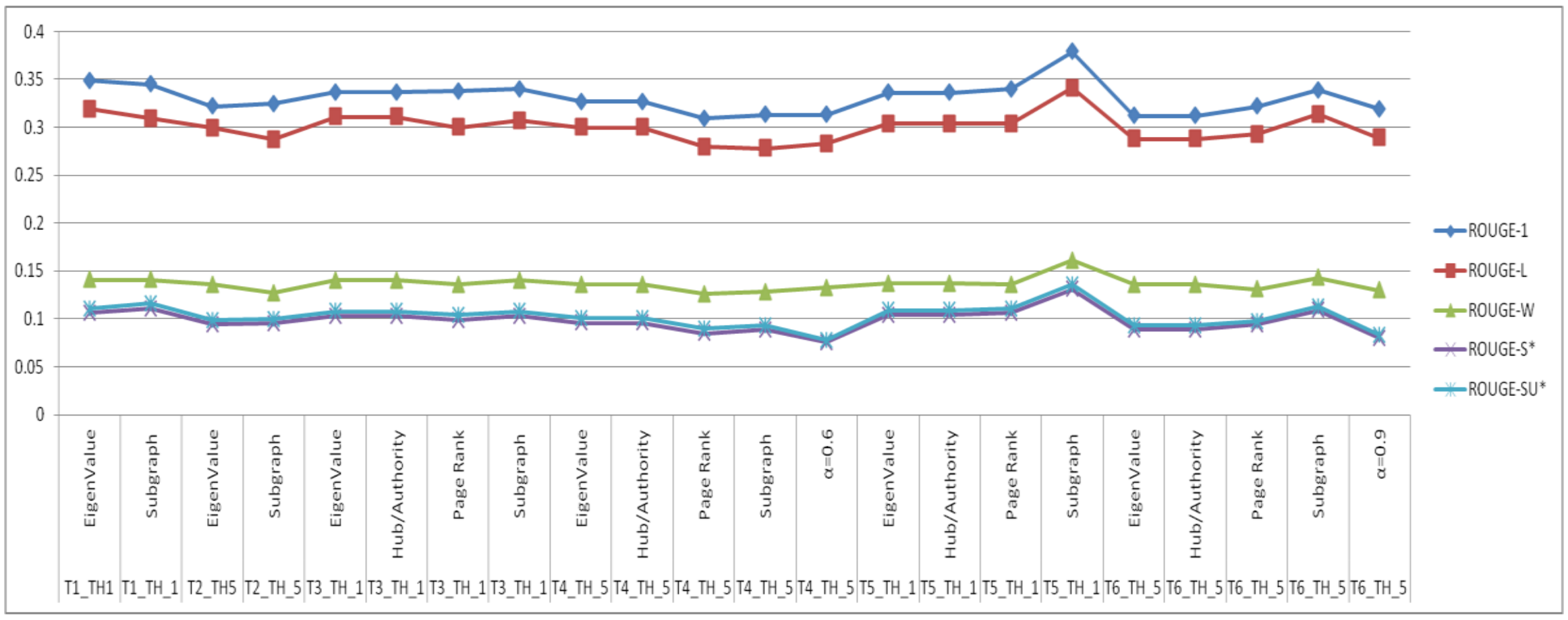

**Figure-4.21**: Showing Overall Performance of some system for the comparative study

In the next, (from **Table-4.**18) we are presenting performance ranking based on all used centrality measure like (subgraph, Eigen Value, etc.. total eight), with different WSD, and two different thresholds. Out of these eight centrality measures for 10% threshold, simple Lesk 2 times, Adapted Lesk 3 times, Cosine Lesk 1 time, and similarly for 5% threshold simple Lesk 2 times, Adapted Lesk 3 times, Cosine Lesk 2 times are performing best w.r.t. Any centrality measure. In the last row of **Table-**4.18, we have presented alpha centrality's top 21 measures out of total 48 measures (8×3×2, due to 8 different centrality measures, three different WSD technique, and two different thresholds). From this result, we can't say that only particular type WSD is better all time. But we can say that, the 10% threshold is better to compare to 5% threshold, and Subgraph based centrality is better vs. all other measures.

**Table-4.18:** Top Performance, by Centrality-based measures, WSD technique used, and similarity threshold

| Centrality Measure | Performance ranking | | Reference |
|---|---|---|---|
| | WSD Lesk Technique | Threshold | |
| Subgraph | Simple<br>Adapted<br>Cosine<br>Simple<br>Adapted<br>Cosine | 10%<br>10%<br>10%<br>5%.<br>5%.<br>5% | Table-4.9, Figure-4.14 |
| Eigenvalue and Hub/Authority | Simple<br>Cosine<br>Adapted<br>Cosine<br>Adapted<br>Simple | 10%<br>10%<br>10%<br>5%.<br>5%.<br>5%. | Table-4.10, Figure-4.15 |
| Page Rank | Simple<br>Cosine<br>Adapted<br>Adapted<br>Simple<br>Cosine | 10%<br>10%<br>10%<br>5%.<br>5%.<br>5%. | Table-4.11, Figure-4.16 |
| Bonpow | Adapted<br>Cosine<br>Simple<br>Adapted<br>Simple<br>Cosine | 10%<br>10%<br>10%<br>5%.<br>5%.<br>5%. | Table-4.12, Figure-4.17 |
| Between Ness | Adapted<br>Adapted<br>Cosine | 10%<br>5%<br>5% | Table-4.13, Figure-4.18 |
| | Simple<br>Cosine<br>Simple | 5%<br>10%<br>10% | |
| Closeness | Cosine<br>Simple<br>Adapted<br>Cosine<br>Simple<br>Adapted | 10%<br>10%<br>10%<br>5%.<br>5%.<br>5%. | Table-4.14,<br>Figure-4.19 |
| Alpha<br>(Top 20 best performances) out of 48 methods | Adapted<br>Adapted<br>simple<br>cosine<br>Adapted<br>simple<br>simple<br>Adapted<br>simple<br>simple<br>cosine<br>simple<br>cosine<br>cosine<br>Adapted<br>cosine<br>Adapted<br>simple<br>Adapted<br>Adapted<br>Cosine | alpha=0.9<br>alpha=0.8<br>alpha=0.9<br>alpha=0.6<br>alpha=0.7<br>alpha=0.8<br>alpha=0.9<br>alpha=0.8<br>alpha=0.6<br>alpha=0.6<br>alpha=0.8<br>alpha=0.8<br>alpha=0.6<br>alpha=0.9<br>alpha=0.9<br>alpha=0.8<br>alpha=0.6<br>alpha=0.7<br>alpha=0.7<br>alpha=0.9<br>alpha=0.9 | Figure-4.19<br>Th=10%<br>Th=10%<br>Th=5%<br>Th=10%<br>Th=10%<br>Th=10%<br>Th=10%<br>Th=5%<br>Th=10%<br>Th=5%<br>Th=10%<br>Th=5%<br>Th=5%<br>Th=10%<br>Th=5%<br>Th=5%<br>Th=5%<br>Th=10%<br>Th=5%<br>Th=10%<br>Th=5% |

## 4.6 Concluding Remark

In this Chapter, we have presented the Lexical Network concept for Automatic Text Document summarization, which is a little bit different from previously proposed lexical chain based techniques. In previous techniques author concentrate to create a number of lexical chains, that creates ambiguity which chain to prefer, even this problem efficiently can handle with chain scoring techniques. Still, lexical chains have problem that if one particular word *let "Donald Trump"* is coming in two or more sentences then which sentence to prefer. Another problem with this technique that they consider only nearby sentences for a chain construction, i.e. window of two or three sentences was selected, so this is unable to handle long term relationship between sentences. Our Lex-Net Handle long term relationship between sentences. Nodes are scored, and higher score sentence gave priority. So, both problems are handled in this way.

Our Lexical network is based on the number of Lexical and semantic relations. To decide the importance of sentences we have applied centrality-based measure on Lexical Network. We have found that subgraph based centrality is performing best among all. Since human language is highly ambiguous (here English) so we need to find a correct sense of a particular word in a given context. The solution to this ambiguity problem done with simplified Lesk, cosine Lesk, and adapted Lesk algorithm. In this work, we have done a study on the impact of WSD techniques and cosine similarity threshold (Θ= 0%, 5%, 10%, ….35%,40%). Due to space limitation, we are showing results only with Θ = 5% and 10%. The Less value of Θ represents more diversity compared to high valued. More diversity (if Θ is less) is not good for summary because it will maintain diversity but less relatedness between sentences which is harming good summary property. For Comparison purpose we have used Semantrica-Lexalytics Algorithm in which we have found that System proposed by us is working better many times, Especially when we are using Subgraph based Centrality.

From this work we have reached on number of conclusions like (1) for alpha centrality when alpha is 0.1 to 0.5 (inclusion) the performance of summarizer system is arbitrary up and down, but after that alpha= 0.6 to 0.9 (inclusion) for all centrality measures, for different value of threshold performance is continuously increasing (some time exception at alpha=0.7), and again reduced at alpha=1. Second, from a set of cosine similarity as 0% , 5%, 10%,...35%, 40%. We are suggesting that 10 % similarity threshold is better to get enough diversity, and better summary (as per Rouge Score). Third, Hub/Authority based ranking is the same as Eigenvalue based centrality. Fourth,

Subgraph based centrality measure is performing better to all, the reason of this is the higher score for small subgraph which recognizes a small subgraph and this cover various subgraph (can be considered as cluster-like structure). Fifth, We are not suggesting any particular WSD is better all the time.

# Chapter 5: Modeling Automatic Text Document Summarization as multi objective optimization

## 5.1 Introduction

Text document summary may be achieved from many features, for examples location, TF-IDF, length of summary, relatedness, redundancy based scores etc. These stand-alone features will not generate a good quality summary, so it can be hybridized to get better summary. When we take care of multiple features, then many solutions are possible, and it is time taking. Hence, we always try to optimize these functions. In this work, we are presenting multi objective function to get a better summary. Our optimization function maximizes diversity and minimizes redundancy. Here, constraints are defined in terms of summary length. Since objective function and constraints are linear, so we are using ILP (Integer Linear Programming) to find an optimized solution i.e. which sentences to extract for the summary.

## 5.2 Literature Work

In ***section*** 5.2.1 we are presenting related work that is categorized according to the research area, and in ***section*** 5.2.2 we are mentioning the MDR model (Baseline model) used for comparison that is proposed by (McDonald, 2007).

### 5.2.1 Related Work

(Alguliev, Aliguliyev, & Mehdiyev, 2011) have proposed a multiobjective based approach for multi-document summarization. Their model is based on reducing redundant information. They formalize sentence extraction based multi document summarization as an optimization problem. Formulation of the model depends on three aspects redundancy, content coverage, summary length, and shown in Equation-5.1. Where, $S_i$ is $i^{th}$ sentence, "O" is the mean vector of the collection $D = \{s_1, s_2, . . . , s_n\}$, sim() is cosine similarity, $x_{ij}$ is a binary variable which is one if a pair of sentences $s_i$ and $s_j$ are selected to be included in the summary, else zero. In Equation-5.2, L is required length summary, $l_i$ is the length of $i^{th}$ sentence, tolerance is represented $\varepsilon$ sign. In this model denominator evaluating the correlation between the sentences $s_i$ and $s_j$. The numerator provides the coverage of the main content of the document collection, while the denominator reduces the redundancy in the summary. Model is using Sentence to sentence, and sentence to document relations.

$$\max(F(x)) = \frac{\sum_{i=1}^{n-1}\sum_{j=i+1}^{n}[sim(S_i,O)+sim(S_j,O)]X_{ij}}{\sum_{i=1}^{n-1}\sum_{j=i+1}^{n} sim(S_i,S_j)X_{ij}} \tag{5.1}$$

$$subjecto \quad \text{L-} \in \leq \sum_{i=1}^{n-1}\sum_{j=i+1}^{n}(L_i,L_j)X_{ij} \leq \text{L+} \in \tag{5.2}$$

$$Xij \in \{0,1\} \quad \forall\ i, j$$

(Alguliev, Aliguliyev, Hajirahimova, & Mehdiyev, 2011) have proposed multi document summarization as an integer linear programming. They state this model presented by Equation-5.5, which is suitable for single and multi-document summarization. This system optimizes three property Redundancy, Relevance, and Length. Similarity function which is used here is cosine-based similarity. For better calculation author calculate $S_i$ as feature vector by removing stop words. In combined objective function $F_{cos}$, and $F_{NGT}$ stands for cosine similarity measure, and NGT (normalized google distance) based similarity measure used in the objective function. The model is represented by Equation-5.3, 5.4, 5.5 5.6 and 5.7. The Equation-5.3 is representing objective of maximization of similarity (may be drawn by cosine or NGT based) between summary "S" and document "D", Equation-5.4 for length constraints, Equation-5.5 representing combined function to maximize similarity between summary and documents but minimum among summary sentences, Equation-5.6 is about length constraints, and Equation-5.7 represents function F to optimize, where alpha can be decided as per user preference. F is maximized the combination of cosine and NGT based function. Here, $X_{ij}$ is a binary variable.

$$Max \quad \text{Sim}(D,S) \tag{5.3}$$

$$s.t. \quad \text{len(S)} \leq \text{L} \tag{5.4}$$

This optimization function Equation-5.3 w.r.t constraint in Equation-5.4, is elaborated by following Equations,

$$\max(f) = \sum_{i=1}^{n-1}\sum_{j=i+1}^{n}[sim(D,S_i)+sim(D,S_j)-sim(S_i,S_j)]X_{ij} \quad (5.5)$$

$$s.t.\sum_{i=1}^{n-1}\sum_{j=i+1}^{n}[len(S_i)+len(S_j)]X_{ij} \le L \quad (5.6)$$

$$Max\ F_\alpha = \alpha F_{\cos} + (1-\alpha)F_{NGT} \qquad \alpha \in [0,1] \quad (5.7)$$

In another work, (Alguliev, Aliguliyev, & Hajirahimova, 2012) have optimized multi document summarization issues as an aspect of coverage and redundancy. In this formulation, coverage is represented by Equation-5.8, and redundancy is represented by Equation-5.9. Equation-5.10, and 5.11 are representing a combined function. Let $x_i$ be binary variable, $x_i$ = 1 when $S_i$ is selected. Equation-5.12 is representing length summary length constraints, where L is summary length, $L_i$ is $S_i^{th}$ sentence's length. Weight w is decided based on the importance of coverage or redundancy.

$$F_{\text{cov}}(s_i) = sim(s_i, O) \quad (5.8)$$

$$F_{red}(s_i, s_j) = 1 - sim(s_i, s_j) \quad (5.9)$$

$$F(x) = w.F_{\text{cov}}(x) + (1-w)F_{red}(x) \quad (5.10)$$

$$= w \cdot \sum_{i=1}^{n-1} sim(s_i, O)x_i + (1-w)\sum_{i=1}^{n-1}\sum_{j=i+1}^{n}(1-sim(s_i,s_j))x_i x_j \quad (5.11)$$

$$s.t.\sum_{i=1}^{n} l_i x_i \le L \qquad \mathrm{x_i} \in \{0,1\}, \forall i \quad (5.12)$$

(Alguliev et al., 2013a) have proposed Constraint-driven document summarization and the constraints are defined in term of, diversity in summarization, and sufficient coverage. The model is formulated as a quadratic integer programming (QIP) problem, to solve the problem discrete PSO is used. Diversity constraint is defined by Equations 5.13, 5.14, 5.15, and 5.16 that covers the relation between a sentence and document, and content coverage is defined by Equation-5.17, 5.18, and 5.19 this shows sentence to sentence relations. Where, $S_i$ $i^{th}$ sentence, "O" is the mean vector of the collection D = {$s_1$, $s_2$, . . . , $s_n$}, Sim() is cosine similarity, $x_{ij}$ is a binary variable which is one if a pair of sentences $s_i$ and $s_j$ are selected to be included to in summary, else 0. L is the required length summary, $l_i$ is length of $i_{th}$ sentence, $\Theta_{diver}$ specifies high diversity in summary, parameter $\Theta_{cont}$ we can control the content coverage in summary.

$$\max(F(x)) = \sum_{i=1}^{n} \frac{sim(S_i, O)}{l_i} X_i \quad (5.13)$$

$$s.t. \quad \sum_{i=1}^{n} l_i x_i \leq L \quad (5.14)$$

$$sim(S_i, S_j) X_i X_j \leq \Theta_{diver} \quad (5.15)$$

$$Xi \in \{0,1\} \qquad \forall i \quad (5.16)$$

$$Max(F(x)) = \sum_{i=1}^{n-1} \sum_{j=i+1}^{n} (1 - sim(S_i, S_j)) X_i X_j \quad (5.17)$$

$$s.t. \sum_{i=1}^{n} l_i x_i \leq L \quad (5.18)$$

$$\frac{sim(Si, O)}{l_i} x_i \geq \Theta_{cont} \qquad \mathrm{xi} \in \{0,1\} \quad (5.19)$$

(Alguliev, Aliguliyev, & Isazade, 2012) proposed a modified p-median problem, for multi document summarization. This approach expresses sentence-to-sentence, summary-to-document, and summary-to-subtopics relationships. They cover four aspects of summarization relevance, content coverage, diversity, length, and jointly optimizing all aspects. Formally, their assumption is considering all vertices of a graph as potential medians. P-median is defined as a subset of vertices with p cardinality. $S_j$ , $i^{th}$ sentence, $d_{ij}$ distance between vertices "i" and "j," $X_{ij}$, $Y_i$ are binary variables, "O" is representing the center of collection documents, S is summary. In this formulation, the objective is to find out the binary assignment X = [$x_{ij}$] such that high relevancy, best content coverage, and the less redundancy/ high diversity w.r.t summary length is at most L. The P Median problem can be expressed by Equations 5.19, 5.20, 5.21,5.22.

$$\min \sum_{i=1}^{n} \sum_{j=1}^{n} d_{ij} X_{ij} \quad (5.19)$$

$$s.t. \sum_{j=1}^{n} X_{ij} = 1 \forall i = 1, 2, \ldots n \quad (5.20)$$

$$\sum_{j=1}^{n} Y_i = p \quad (5.21)$$

$$X_{ij} < Y_j \quad (5.22)$$

$$Y_j \in \{0,1\} \qquad \mathrm{Y_j} = 1 \; if \; \mathrm{S_j} \text{ is selected as median}$$

$$X_{ij} \in \{0,1\}$$

The model used for summarization is formulated by Equation-5.23, 5.24, 5.25, and 5.26.

$$\max(f(x,y)) = sim(S,O)\cdot\sum_{j=1}^{n} sim(S,Sj)Yj + \sum_{i=1}^{n}\sum_{j=1}^{n} sim(Si,Sj)Xij +$$
$$\sum_{i=1}^{n}\sum_{j=1}^{n}(1-sim(S_i,S_j))Y_iY_j \quad (5.23)$$
$$s.t.\sum_{j=1}^{n} X_{ij} = 1 \quad \forall i=1,...n \quad (5.24)$$
$$\sum_{j=1}^{n} Y_j l_j \le L \quad (5.25)$$
$$X_{ij} \le Y_j, \quad (5.26)$$
$$X_{ij}, Y_i \in \{0,1\}$$

### 5.2.2 Baseline (MDR Model)

We compare our results with (McDonald, 2007) work, Author used a Multi-objective optimization technique to optimize various criteria, and for that they used ILP. Objective criteria taken by them is given in following Equation-5.27. For simplicity, we represent the document collection simply as the set of all textual units from all the documents in the collection, i.e., D = {$t_1$, . . . , $t_n$} where ti ϵ D iff ∃ $t_i$ ϵ $D_j$ ϵ D. We let S ⊆ D be the set of textual units constituting a summary. Equation-5.28 representing length constraints, and Equation-5.29, 5.30 and 5.31make sure about the selection of binary variables $X_i$ band $Y_{ij}$. In implementation to determine relevance, we are using PageRank and subgraph based centrality.

$$\max(x,y)\left[\sum_{i=1}^{N} \mathrm{Re}\,levance(S_i)X_i - \sum_{i=1}^{N}\sum_{j=1}^{N} \mathrm{Re}\,dundancy(S_i,S_j)Y_{ij}\right] \quad 5.27$$

$s.t.$

$$\sum_{i=1}^{n} l_i x_i \le L_{\max} \quad (5.28)$$
$$Y_{ij} - X_i \le 0 \quad (5.29)$$
$$Y_{ij} - X_j \le 0 \quad (5.30)$$
$$Y_i + X_j - Y_{ij} \le 1 \quad (5.31)$$

In objective function Equation-5.27 function, relevance (), and redundancy () can describe as follows. Relevance: Summaries should contain informative textual units that are relevant to the user. Redundancy: Summaries should not contain multiple textual units that convey the same information. Length ($L_{max}$): Summaries are bounded in length. The similarity between two sentences is calculated by cosine similarity.

Where N total number of the sentence, importance (Si) decide the importance of i$^{th}$ sentence higher value given higher importance, $l_i$ represent the length of i$^{th}$ sentences, and $L_{max}$ required max length summary. The $X_i$ binary variable indicates whether or not the corresponding sentence $s_i$ is selected in summary. In the same way, the $Y_{i,j}$ binary variables indicating whether or not both $s_i$ and $s_j$ are included in the summary.

## 5.3 Background

In this ***section***, we are presenting some intermediate techniques followed in this work. Handling with Text is not easy because output depends on how text is handled, how preprocessing and post-processing done. Preprocessing is required before Lex-Network (Lexical-Network) creation. Some preprocessing task is to remove stop words, convert to lower, remove punctuation, remove the number, remove whitespaces, stemming & lemmatization, and WSD (Word Sense Disambiguation). Here, we are not presenting detailed WSD techniques, and centrality measures because that are explained in the previous Chapter. Introduction to Linear Programming is explained here.

### 5.3.1 Linear programming

Linear programming (LP) also called linear optimization is a method to achieve the best outcome (minimum cost, maximum profit) in a mathematical model in which requirements are represented by linear relationships. In linear programming contains three parts, Decision Variable, Optimization function to maximize or minimize, and constraints. While an Integer programming (IP) problem is a mathematical optimization or feasibility program in which some or all of the variables are restricted to be integers. In other words ILP, in which the given objective function and the constraints (other than the integer constraints) are linear. Integer programming is NP-hard. There is various method to solve ILP. Broadly, this can be categorized into three parts, (1) Exact algorithms that guarantee to find an optimal solution, but time complexity may be exponential examples are Cutting Planes, Branch and bound, and Dynamic Programming. (2) Heuristic Algorithms provide a suboptimal solution, but without a guarantee on its quality. Running time is not guaranteed to be polynomial, but provide a fast solution. (3) Approximation algorithms provide a solution in polynomial time a suboptimal solution together with a bound on the degree of sub-optimality. Let an example given by Equation-5.32 we want to maximize given Z.

$$\text{Maximize } Z = 50x + 18y \tag{5.32}$$

Subject to the constraints,

$$x + y = 80, \tag{5.33}$$

$$2x + y = 100, \tag{5.34}$$

$$\text{Where } x, y >= 0 \tag{5.35}$$

Using Graphical Method solving Equation 5.32 w.r.t 5.33, 5.34, 5.35 in Figure-5.1, Solution is shown at point (20, 60).

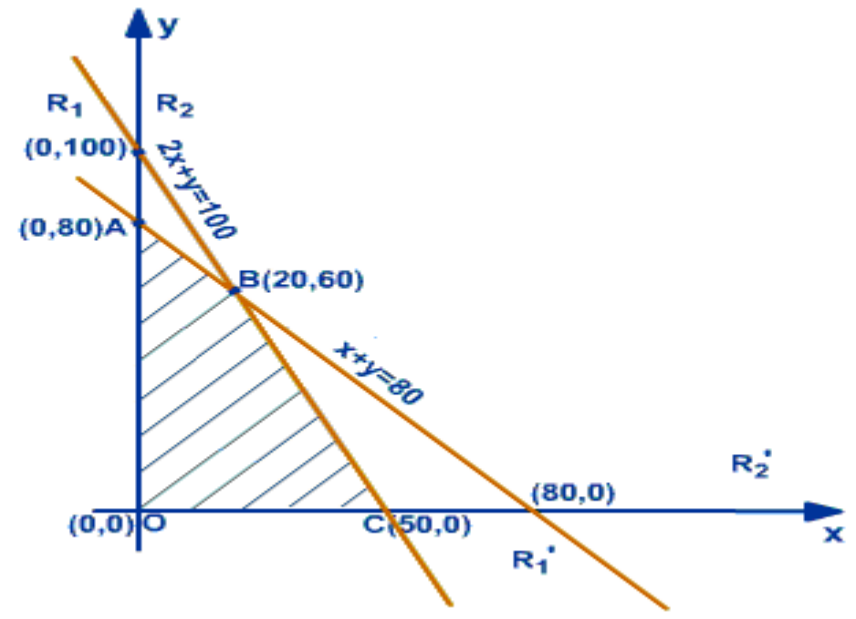

**Figure-5.1**: Graphical method of Solution of 1, w.r.t (2, 3, and 4)

If the above optimization function is represented by

$$Max\ (f) = \max\left[\sum_{i,j=1}^{n} 5 \times X_i + Y_j\right] \tag{5.36}$$

$$\textit{such that,}$$

$$X_i + Y_j = 80 \tag{5.37}$$

$$2X_i + Y_j = 100; \quad i > j \tag{5.38}$$

Then to solve this process should follow if n=3

$$f = 5X_1 + Y_2 + Y_3 + 5X_2 + Y_3 + 5X_3 \tag{5.39}$$

$$X_1 + Y_2 = 80, \tag{5.40}$$

$$X_2 + Y_3 = 80, \tag{5.41}$$

$$2X_1 + Y_2 = 100, \tag{5.42}$$

$$2X_2 + Y_3 = 100 \tag{5.43}$$

The Limitation of ILP is that when we want to solve the function given in Equation-5.36, with constraints in Equation-5.37, and 5.38, then we need to expand "f" per Equation-5.39, and constraints as 5.40, 5.41, 5.42, and 5.43.

## 5.4 Proposed Multi-objective optimization Model

In this section, we are presenting a basic outline of our model. This modeling is based on Chapter 4 i.e. based on "Lexical chain Network".

### 5.4.1 Outline of Proposed Model

In Figure 5.2 a diagram is represented to show a brief introduction. In this Figure, it is shown that how the output of the last chapter (lexical network) is used as input for this work. Lexical Network creation, and finding centrality score of each sentence is fetched from there. Here Centrality sore is used to decide the importance of sentence, edged present in LexicalNetwork is used to find relatedness. We want to maximize information/coverage and minimization of redundancy.

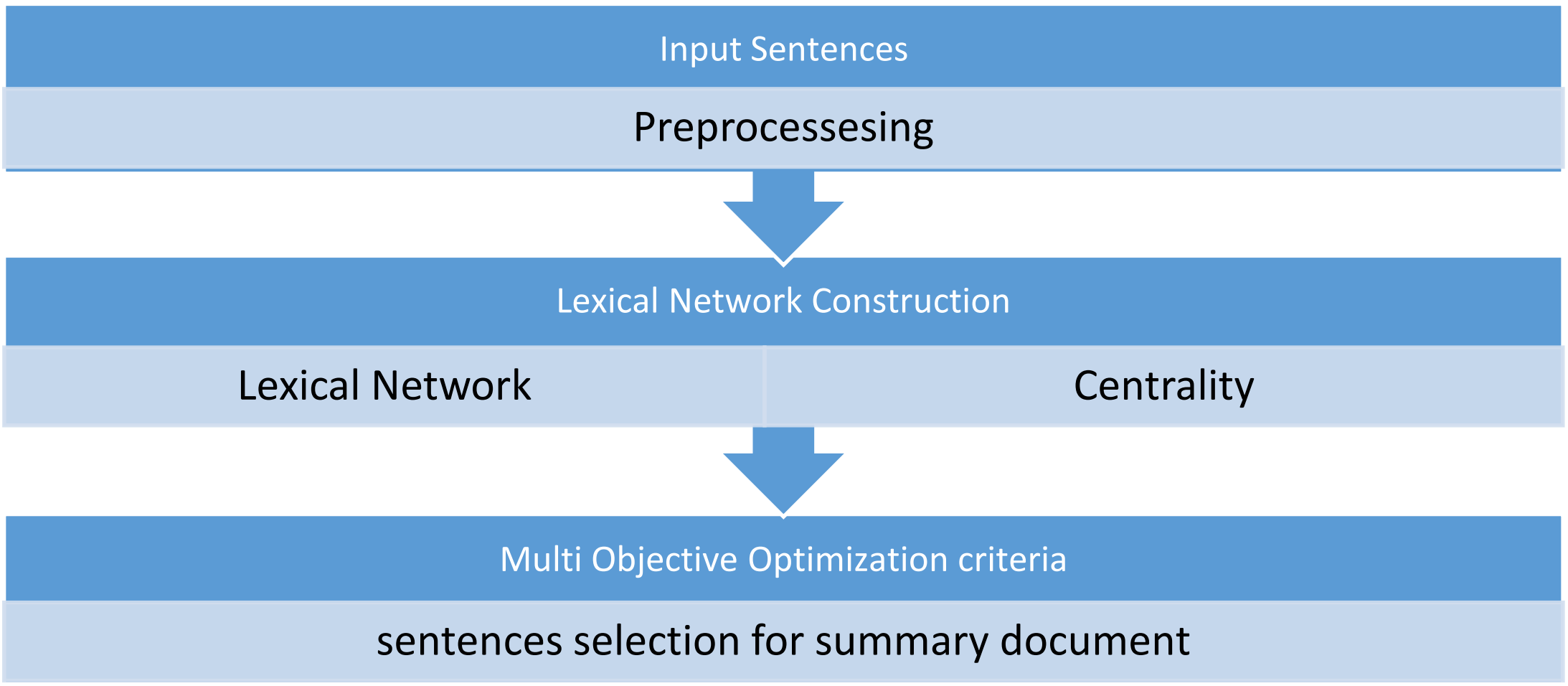

Figure 5.2: Outline of Summarizer System

Optimization function is given by Equation-5.44, and constraints are given by Equation 5.45, 5.46, and 5.47.

$$\max(x,y)\left[\sum_{i=1}^{N} Centrality(S_i)X_i - \sum_{i=1}^{N}\sum_{j=1}^{N} Lexical_Network(S_i,S_j)Y_{ij}\right] \quad 5.44$$

$$s.t.$$

$$Y_{ij} - X_i \leq 0 \quad (5.45)$$

$$Y_{ij} - X_j \leq 0 \quad (5.46)$$

$$Y_i + X_j - Y_{ij} \leq 1 \quad (5.47)$$

This Objective function contains two parts, part-1 before minus sign, guaranteed to maximize the highest informative sentences, and the second part after minus sign, is take care of redundant sentences. Between(Si), Subgraph(Si), Hub_Authority(Si), Closeness(Si), Bonpow(Si), EigenValCentrality (Si) are representing the importance of $i^{th}$ sentence decided by centrality measure, and Lexical_Network($S_i$, $S_j$) represents the weight of the edge between sentence $S_i$ and $S_j$ from lexicalNetwork. $X_i$, $X_j$, $Y_{ij}$ are binary variables to decide no two same sentences will be part of the summary.

### 5.4.2 Detail description of proposed model

This proposed work has been divided into three modules. The outline of this framework is shown in Figure-5.3. Model-1 is comprised sentences selection, tagging, and creation of another set of sentences with stop-word removed, i.e. that is later used for Lexical Network creation. In Model-2 we are creating Lexical Network, and computing centrality score using degree centrality and Betweenness Centrality. (Subgraph centrality performing better in the past?), the outcome of this step later used in summary generation (Module-3).

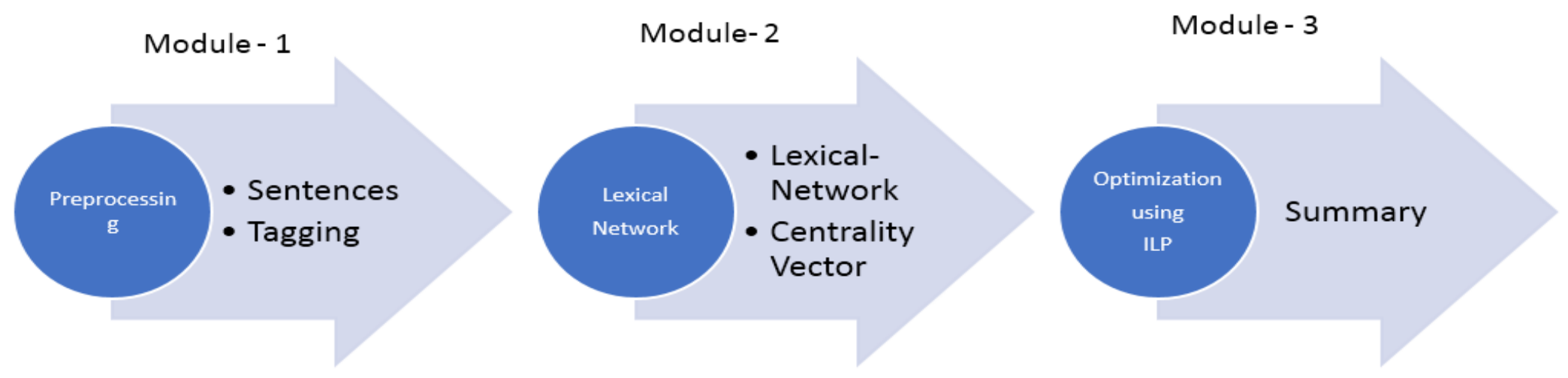

**Figure-5.3**: Three Module Process for Automatic Text Summarization

**Module 1:**

In a text document, there are many sentences which are not properly readable by the system. First, we should apply any regular expression for proper handling of sentences so that words/tokens like Mr., Dr., U.S., " ", '', ',' , '.' etc. take care properly in sentence extraction. We are doing some preprocessing, and then we are tagging these sentences using NLTK (Natural Language Tool Kit) parser. This is explained by an example in Figure-5.4. After sentence identification tagging is done, then only significant units i.e. verb and nouns, are considered for lexical network construction that are represented in ModifiedSentences, shown in Figure-5.4.

**Module 2:**

In this module, we are constructing a Lexical Network of N×N matrix. Where N is a number of sentences, i.e. N= |Sentences|. We are picking each sentence on by one and extracting only Nouns, and Verbs from this sentence because only these provides information, and this is added to the modified sentence.

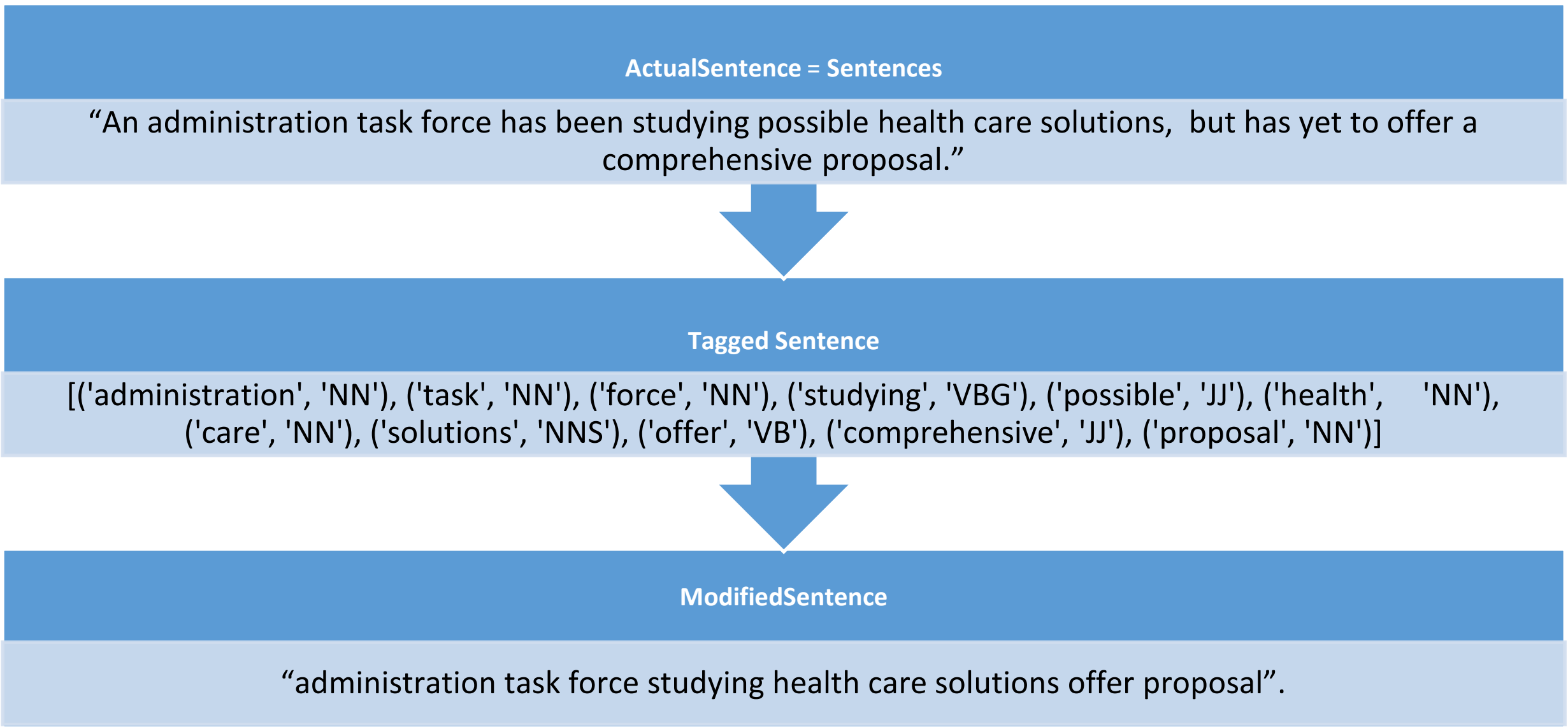

**Figure-5.4**: Showing Procedure 2 Stepwise, example from DUC 2002, HealthCare Data Set, File number WSJ910528-0127.txt

The actual Lexical Network is construed for ModifiedSentences here |ModifiedSentences|=N =|Sentences|. We are finding is any relation either Lexical, or Semantic is present between two

ModifiedSentence. To find the relation present between sentences we need to disambiguate the sense of the word this is done with the help of Lesk Algorithm. Here Lesk Algorithm takes two argument word to disambiguate sense and $i^{th}$ Actual sentence. Based on that sense this algorithm finds Hyper, Hypo, Synonyms, Meronym, and Antonym relation present in other remaining sentences. At a time only one relation will available, and according to us, every relation is treated of the same importance, so we will assign an edge between sentence, and each time increase by one if more relation is present between sentences.

After getting a Lexical Network represented by Figure-4.3, for given sentences in **Table** 4.1, we are applying Centrality-based measures for sentence scoring. Sentence with high centrality score is decided more important compared to low centrality score, and Output from this module (Lexical Network, and Centrality Score) given input in next Module 3.

**Module 3:**

In this module after getting Sentences, LexicalNetwork, and Centrality score, we are going to optimize our function given by Equation-5.44. This objective function divided into two parts, the first section of optimization function that is centrality($S_i$), is to maximizing coverage. Maximization of coverage can be linked by maximum informative sentences. The second section that is Lexical_Network($S_i$, $S_j$), in this we are trying to reduce redundancy between selected sentences for the summary. In Equation 5.44, N is the total number of sentences, $X_i$, $X_j$, and $Y_{ij}$ are binary variables. $X_i$ equals one if $i^{th}$ sentences selected and added in summary, $Y_{ij}$ equals one will donate that sentences $i^{th}$ and $j^{th}$ are present in summary, and i is not equal to j. These objective criteria are represented by Equation-5.45, 5.46 and 5.47 that will guarantee to add a new sentence with optimality condition.

## 5.5 Experiments and Results

In this section, we are presenting different experiments and results. In section-5.5.1 coverage or importance of sentences is decided by our proposed centrality-based measures, and in section-5.5.2 importance is decided by cosine-based measure.

### 5.5.1 Experiment-5.1

In this experiment we are maximizing the baseline objective function (A Section), and our proposed model (B section). In both, the objective function the relevance is decided by centrality measures. In the baseline, method redundancy is decided by cosine measure, and in the proposed model, the lexical network is considered for minimization of redundancy. In Tables we have presented Precision (P), Recall (R), and F-score (F), with 95 % confidence interval. For example, if confidence interval is 0.644 - 0.671, with 95 % confidence, then the significance of this can interpreted like, there is a 95 % chance that the actual value of unstandardized coefficient is between 0.644 – 0.671.

### A) Cosine based criteria to minimize redundancy

In this experiment, we are using a baseline model in which we are trying to maximize the centrality score and minimize cosine similarity, with given length constraints. The combined optimization function is represented by Equation-5.44. In Table-5.3 Precision, Recall and F score are presented with 95% confidence. From the paper (McDonald, 2007) it is not clearly mentioned how to decide the importance of the sentences, but in our implementation, we are using subgraph and PageRank based centrality measure.

**Table-5.1:** Precision, Recall and F-Score of model MDR Model with subgraph centrality

| **Rouge** | **R-Cosine** | **P-Cosine** | **F-Cosine** | **95% Confidence R (Cosine)** | **95% Confidence P (Cosine)** | **95% Confidence-F (Cosine)** |
|---|---|---|---|---|---|---|
| **ROUGE-1** | 0.45654 | 0.22449 | 0.29959 | 0.38279 - 0.53288 | 0.19111 - 0.25345 | 0.25479 - 0.34064 |
| **ROUGE-2** | 0.16998 | 0.08186 | 0.10995 | 0.11417 - 0.23229 | 0.05704 - 0.10670 | 0.07588 - 0.14441 |
| **ROUGE-L** | 0.43680 | 0.21401 | 0.28601 | 0.36017 - 0.52110 | 0.18018 - 0.24450 | 0.24042 - 0.33097 |
| **ROUGE-W-1.2** | 0.17884 | 0.14709 | 0.16051 | 0.14843 - 0.21462 | 0.12350 - 0.16854 | 0.13485 - 0.18711 |
| **ROUGE-S*** | 0.19225 | 0.04462 | 0.07150 | 0.12999 - 0.26687 | 0.03251 - 0.05672 | 0.05189 - 0.09298 |
| **ROUGE-SU*** | 0.20051 | 0.04750 | 0.07580 | 0.13817 - 0.27485 | 0.03508 - 0.05960 | 0.05569 - 0.09739 |

**Table-5.2:** Precision, Recall and F-Score of model MDR Model with PageRank centrality

| **Rouge** | **R-Cosine** | **P-Cosine** | **F-Cosine** | **95% Confidence R** | **95% Confidence P** | **95% Confidence-F** |
|---|---|---|---|---|---|---|

| | | | | (Cosine) | (Cosine) | (Cosine) |
|---|---|---|---|---|---|---|
| **ROUGE-1** | 0.22104 | 0.30454 | 0.25543 | 0.16282 - 0.28416 | 0.22001 - 0.38948 | 0.18719 - 0.32607 |
| **ROUGE-2** | 0.08044 | 0.11170 | 0.09326 | 0.04545 - 0.11881 | 0.06199 - 0.16738 | 0.05209 - 0.13798 |
| **ROUGE-L** | 0.20951 | 0.28905 | 0.24225 | 0.15674 - 0.26823 | 0.21039 - 0.37202 | 0.17987 - 0.30955 |
| **ROUGE-W-1.2** | 0.09532 | 0.21849 | 0.13245 | 0.07188 - 0.12232 | 0.16044 - 0.28424 | 0.09962 - 0.17062 |
| **ROUGE-S*** | 0.04577 | 0.08768 | 0.05964 | 0.02376 - 0.07195 | 0.04632 - 0.14177 | 0.03073 - 0.09537 |
| **ROUGE-SU*** | 0.05246 | 0.09916 | 0.06804 | 0.02919 - 0.07978 | 0.05557 - 0.15473 | 0.03764 - 0.10488 |

From **Table**-5.2 and **Table**-5.3 it is clear that we are getting improved results.

## B) Lexical based criteria to reduce redundancy

In this experiment, we are using our proposed model in which we are trying to maximize the relevance score, that is found out by various degree centrality measures and minimize lexical similarity getting from the lexical network, w.r.t. given constraints. The combined optimization function is represented by Equation- 5.44. In **Tables**-5.3-5.9 Precision, Recall and F score are presented with 95% confidence.

**Table-5.3:** Precision, Recall and F-Score of our proposed model when centrality based measure is subgraph centrality

| **Rouge** | **R-Lexical** | **P-Lexical** | **F-Lexical** | **95% Confidence – R (Lexical)** | **95% Confidence-P (Lexical)** | **95% Confidence-F (Lexical)** |
|---|---|---|---|---|---|---|
| **ROUGE-1** | 0.45826 | 0.22476 | 0.30037 | 0.38663 - 0.53288 | 0.19139 - 0.25388 | 0.25489 - 0.34127 |
| **ROUGE-2** | 0.16998 | 0.08151 | 0.10972 | 0.11417 - 0.23229 | 0.05681 - 0.10661 | 0.07559 - 0.14421 |
| **ROUGE-L** | 0.43852 | 0.21440 | 0.28686 | 0.36264 - 0.52110 | 0.18041 - 0.24497 | 0.24110 - 0.33147 |
| **ROUGE-W-1.2** | 0.17945 | 0.14719 | 0.16090 | 0.14879 - 0.21533 | 0.12370 - 0.16864 | 0.13524 - 0.18728 |
| **ROUGE-S*** | 0.19304 | 0.04465 | 0.07170 | 0.13077 - 0.26767 | 0.03251 - 0.05672 | 0.05201 - 0.09319 |
| **ROUGE-SU*** | 0.20134 | 0.04752 | 0.07600 | 0.13902 - 0.27568 | 0.03510 - 0.05962 | 0.05589 - 0.09739 |

**Table-5.4:** Precision, Recall and F-Score of our proposed model when centrality based measure is betweenness centrality

| **Rouge** | **R-Lexical** | **P-Lexical** | **F-Lexical** | **95% Confidence – R (Lexical)** | **95% Confidence-P (Lexical)** | **95% Confidence-F (Lexical)** |
|---|---|---|---|---|---|---|
| **ROUGE-1** | 0.23944 | 0.30226 | 0.26662 | 0.18437 - 0.30322 | 0.23709 - 0.37610 | 0.20799 - 0.33607 |
| **ROUGE-2** | 0.08341 | 0.10513 | 0.09277 | 0.04643 - 0.13044 | 0.05990 - 0.15902 | 0.05244 - 0.14311 |
| **ROUGE-L** | 0.23106 | 0.29180 | 0.25733 | 0.17781 - 0.29558 | 0.22884 - 0.36624 | 0.20027 - 0.32602 |
| **ROUGE-W-1.2** | 0.09877 | 0.20973 | 0.13397 | 0.07694 - 0.12683 | 0.16518 - 0.26262 | 0.10464 - 0.16956 |
| **ROUGE-S*** | 0.05677 | 0.08817 | 0.06862 | 0.02996 - 0.09160 | 0.04903 - 0.13617 | 0.03700 - 0.10895 |
| **ROUGE-SU*** | 0.06240 | 0.09661 | 0.07533 | 0.03465 - 0.09818 | 0.05633 - 0.14531 | 0.04263 - 0.11666 |

Table-5.5: Precision, Recall and F-Score of our proposed model when centrality based measure is PageRank

| **Rouge** | **R-Lexical** | **P-Lexical** | **F-Lexical** | **95% Confidence – R (Lexical)** | **95% Confidence-P (Lexical)** | **95% Confidence-F (Lexical)** |
|---|---|---|---|---|---|---|
| **ROUGE-1** | 0.23098 | 0.29711 | 0.25951 | 0.17791 - 0.29067 | 0.23108 - 0.36973 | 0.19847 - 0.32574 |
| **ROUGE-2** | 0.08004 | 0.10313 | 0.08996 | 0.04364 - 0.12230 | 0.05701 - 0.15511 | 0.04914 - 0.13691 |
| **ROUGE-L** | 0.22407 | 0.28841 | 0.25182 | 0.17056 - 0.28393 | 0.22242 - 0.36240 | 0.19332 - 0.31841 |
| **ROUGE-W-1.2** | 0.09637 | 0.20791 | 0.13148 | 0.07508 - 0.12228 | 0.16152 - 0.26095 | 0.10193 - 0.16558 |
| **ROUGE-S*** | 0.05229 | 0.08513 | 0.06451 | 0.02695 - 0.08346 | 0.04566 - 0.13235 | 0.03383 - 0.10228 |
| **ROUGE-SU*** | 0.05810 | 0.09406 | 0.07153 | 0.03186 - 0.09037 | 0.05337 - 0.14256 | 0.03989 - 0.11025 |

**Table-5.6:** Precision, Recall and F-Score of our proposed model when centrality-based measure is EvenCentrality

| **Rouge** | **R-Lexical** | **P-Lexical** | **F-Lexical** | **95% Confidence – R (Lexical)** | **95% Confidence-P (Lexical)** | **95% Confidence-F (Lexical)** |
|---|---|---|---|---|---|---|
| **ROUGE-1** | 0.22823 | 0.30671 | 0.26130 | 0.17142 - 0.29269 | 0.22882 - 0.39450 | 0.22882 - 0.39450 |
| **ROUGE-2** | 0.07951 | 0.10788 | 0.09139 | 0.04357 - 0.12292 | 0.05853 - 0.16565 | 0.04974 - 0.14082 |
| **ROUGE-L** | 0.21828 | 0.29355 | 0.24998 | 0.16386 - 0.28163 | 0.21826 - 0.37798 | 0.18762 - 0.32175 |
| **ROUGE-W-1.2** | 0.09614 | 0.21538 | 0.13273 | 0.07311 - 0.12362 | 0.16199 - 0.28030 | 0.10021 - 0.17043 |
| **ROUGE-S*** | 0.04905 | 0.08888 | 0.06295 | 0.02535 - 0.07928 | 0.04571 - 0.14014 | 0.03245 - 0.10120 |
| **ROUGE-SU*** | 0.05511 | 0.09880 | 0.07045 | 0.03044 - 0.08627 | 0.05413 - 0.15125 | 0.03892 - 0.11015 |

**Table-5.7:** Precision, Recall and F-Score of our proposed model when centrality-based measure is Authority_Hub

| **Rouge** | **R-Lexical** | **P-Lexical** | **F-Lexical** | **95% Confidence – R (Lexical)** | **95% Confidence-P (Lexical)** | **95% Confidence-F (Lexical)** |
|---|---|---|---|---|---|---|
| **ROUGE-1** | 0.22745 | 0.30774 | 0.26112 | 0.16780 - 0.29411 | 0.22745 - 0.39866 | 0.19207 - 0.33852 |
| **ROUGE-2** | 0.08109 | 0.11081 | 0.09348 | 0.04510 - 0.12339 | 0.06182 - 0.16832 | 0.05190 - 0.14212 |
| **ROUGE-L** | 0.21607 | 0.29262 | 0.24814 | 0.15965 - 0.28019 | 0.21485 - 0.38108 | 0.18242 - 0.32155 |
| **ROUGE-W-1.2** | 0.09550 | 0.21562 | 0.13219 | 0.07196 - 0.12357 | 0.16004 - 0.28170 | 0.09922 - 0.17104 |
| **ROUGE-S*** | 0.04893 | 0.08966 | 0.06301 | 0.02490 - 0.07820 | 0.04587 - 0.14134 | 0.03235 - 0.10045 |
| **ROUGE-SU*** | 0.05519 | 0.10006 | 0.07079 | 0.03017 - 0.08573 | 0.05498 - 0.15257 | 0.03900 - 0.10930 |

Table-5.8: Precision, Recall and F-Score of our proposed model when centrality based measure is closeness

| **Rouge** | **R-Lexical** | **P-Lexical** | **F-Lexical** | **95% Confidence – R (Lexical)** | **95% Confidence-P (Lexical)** | **95% Confidence-F (Lexical)** |
|---|---|---|---|---|---|---|
| **ROUGE-1** | 0.22691 | 0.30695 | 0.26045 | 0.16740 - 0.29301 | 0.22703 - 0.39645 | 0.19060 - 0.33643 |
| **ROUGE-2** | 0.08195 | 0.11200 | 0.09447 | 0.04548 - 0.12407 | 0.06286 - 0.16907 | 0.05298 - 0.14371 |
| **ROUGE-L** | 0.21528 | 0.29147 | 0.24718 | 0.15843 - 0.27862 | 0.21341 - 0.37995 | 0.18155 - 0.31971 |
| **ROUGE-W-1.2** | 0.09577 | 0.21555 | 0.13243 | 0.07216 - 0.12403 | 0.16122 - 0.28034 | 0.09961 - 0.17186 |
| **ROUGE-S*** | 0.04934 | 0.09005 | 0.06342 | 0.02434 - 0.07947 | 0.04605 - 0.14267 | 0.03161 - 0.10167 |
| **ROUGE-SU*** | 0.05565 | 0.10059 | 0.07128 | 0.02937 - 0.08691 | 0.05491 - 0.15468 | 0.03821 - 0.11096 |

**Table-5.9:** Precision, Recall and F-Score of our proposed model when centrality based measure is Bonpow

| **Rouge** | **R-Lexical** | **P-Lexical** | **F-Lexical** | **95% Confidence – R (Lexical)** | **95% Confidence-P (Lexical)** | **95% Confidence-F (Lexical)** |
|---|---|---|---|---|---|---|
| **ROUGE-1** | 0.22332 | 0.30409 | 0.25690 | 0.16789 - 0.28233 | 0.22849 -.38874 | 0.19323 - 0.32594 |
| **ROUGE-2** | 0.07972 | 0.10951 | 0.09205 | 0.04548 - 0.11791 | 0.06258- 0.16185 | 0.05267 - 0.13580 |
| **ROUGE-L** | 0.21032 | 0.28659 | 0.24201 | 0.15887 - 0.26709 | 0.21452 - 0.36720 | 0.18252 - 0.30751 |
| **ROUGE-W-1.2** | 0.09417 | 0.21327 | 0.13042 | 0.07269 - 0.11906 | 0.16255 - 0.27251 | 0.10032 - 0.16591 |
| **ROUGE-S*** | 0.04654 | 0.08614 | 0.06000 | 0.02405 - 0.07291 | 0.04638 - 0.13505 | 0.03151 - 0.09397 |
| **ROUGE-SU*** | 0.05308 | 0.09721 | 0.06818 | 0.02946 - 0.08045 | 0.05523 - 0.14749 | 0.03824 - 0.10349 |

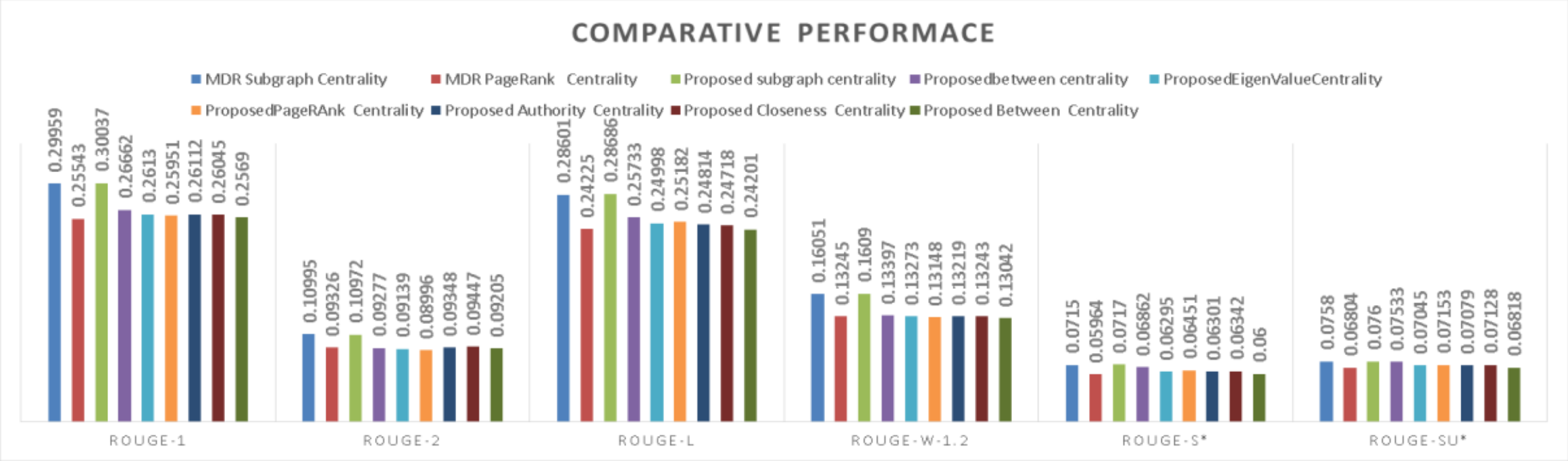

Figure 5.5: ROUGE-1 comparative performance of proposed and MDR (baseline) model

From Table 5.3 – 5.9 and Figure 5.5, we can say that in proposed model when subgraph-based centrality is used to measure relevance and lexical graph used for redundancy then we are getting best mode performance. The second best model is MDR based with subgraph-based centrality measure. When comparison w.r.t MDR based page rank model, all the proposed model's performance is improved.

### 5.5.2 Experiment-5.2

In this experiment, we are finding a correlation between different centrality-based measure. The significance of this experiment highlights the relation between centrality measures. Since the different centrality measure returns different summary sentences, and we are not doing summary to summary analysis i.e. what kind of sentences return by different measures. This experiment gives a glimpse of the summary to summary analysis.

#### A) Co-relation analysis between different centrality based proposed model

Our optimization function is described by Equation-5.44, In that after changing the centrality we are getting different results, and here we are finding the correlation between these methods. In statistics, correlation coefficients are used to measure relationship is between two given variables. It may be strong and poor. The value return is between +1 to -1. Where +1 indicates a positive correlation, -1 indicates negative correlation, and 0 indicates no correlation or better to say no relation exists between these two variables. In our study, we used Pearson's correlation, Spearman's correlation, Kendall's co-relation. Co-relation can be shown by the following examples, (available at goo.gl/sP7uAW). We are using Equation-5.48, calculating the average correlation r.

$$r=\frac{\sum_{i=1}^{n} r}{n} \qquad (5.48)$$

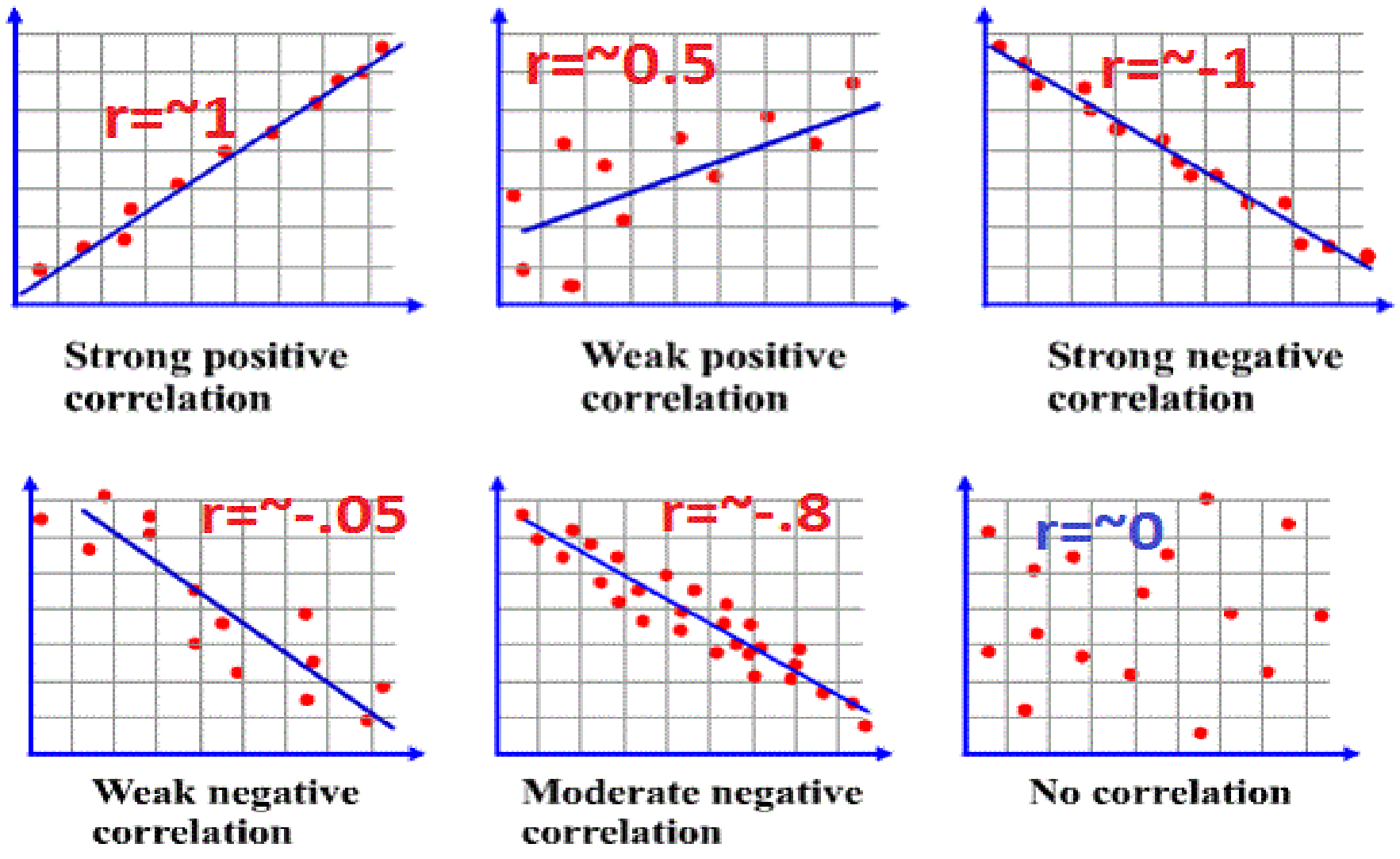

**Figure-5.6:** Different scatter plots showing different directions and strengths of correlation

*5.5.2.1 Pearson correlation:*

It is also known by PPMC (Pearson Product Moment Correlation), It shows the linear relationship between two sets of data. PPMC answers are it possible to draw a line graph to represent the data? A potential problem with PPMC is that it is not able to differentiate between dependent and independent variables. This can be expressed by Equation-5.49. where x and y denote two classes of observation with n cardinality.

$$r = \frac{n\sum xy - \sum x \sum y}{\sqrt{[n\sum x^2 - (\sum x)^2] \times [n\sum y^2 - (\sum y)^2]]}} \quad (5.49)$$

**Table-5.4: showing pairwise Pearson's correlation coefficient (symmetric)**

| | **Authority** | **Between** | **Bonpow** | **Closeness** | **Evencen** | **Hub** | **Pagerank** | **Subgraph** |
|---|---|---|---|---|---|---|---|---|
| **Authority** | 1 | -0.20635 | 0.038674 | -0.04362 | 0.999997 | 1 | 0.98223 | 0.837617 |
| **Between** | -0.20635 | 1 | -0.04618 | 0.670767 | -0.20649 | -0.20635 | -0.14568 | 0.075254 |
| **Bonpow** | 0.038674 | -0.04618 | 1 | -0.06675 | 0.038851 | 0.038674 | 0.011554 | 0.043898 |
| **Closeness** | -0.04362 | 0.670767 | -0.06675 | 1 | -0.0444 | -0.04362 | -0.02543 | 0.286333 |
| **Evencen** | 0.999997 | -0.20649 | 0.038851 | -0.0444 | 1 | 0.999997 | 0.982199 | 0.837563 |
| **Hub** | 1 | -0.20635 | 0.038674 | -0.04362 | 0.999997 | 1 | 0.98223 | 0.837617 |
| **Pagerank** | 0.98223 | -0.14568 | 0.011554 | -0.02543 | 0.982199 | 0.98223 | 1 | 0.840349 |
| **Subgraph** | 0.837617 | 0.075254 | 0.043898 | 0.286333 | 0.837563 | 0.837617 | 0.840349 | 1 |

### *5.5.2.2 Spearman coefficient*

As per (goo.gl/aVLjmP), "Spearman's correlation determines the strength and direction of the monotonic relationship between your two variables rather than the strength and direction of the linear relationship between your two variables, which is what Pearson's correlation determines." This can represent by Equation-5.50. $\bar{x}, and\ \bar{y}$ , denotes mean of x and y, and $x_i$ and $y_i$ is $i^{th}$ observation.

$$r = \frac{\sum_i (x_i - \bar{x})(y_i - \bar{y})}{\sqrt{\sum_i (x_i - \bar{x})^2 \sum_i (y_i - \bar{y})^2}} \qquad (5.50)$$

**Table-5.5:** showing pairwise Spearman's correlation coefficient (symmetric)

| | **Authority** | **Between** | **Bonpow** | **Closeness** | **Evencen** | **Hub** | **Pagerank** | **Subgraph** |
|---|---|---|---|---|---|---|---|---|
| **Authority** | 1 | -0.097 | 0.044488 | 0.036072 | 0.999818 | 1 | 0.97878 | 0.900631 |
| **Between** | -0.097 | 1 | -0.03831 | 0.845987 | -0.09713 | -0.097 | -0.05419 | 0.12077 |
| **Bonpow** | 0.044488 | -0.03831 | 1 | 0.022796 | 0.045115 | 0.044488 | 0.020087 | 0.035691 |
| **Closeness** | 0.036072 | 0.845987 | 0.022796 | 1 | 0.035726 | 0.036072 | 0.050246 | 0.257558 |
| **Evencen** | 0.999818 | -0.09713 | 0.045115 | 0.035726 | 1 | 0.999818 | 0.978504 | 0.900406 |
| **Hub** | 1 | -0.097 | 0.044488 | 0.036072 | 0.999818 | 1 | 0.97878 | 0.900631 |
| **Pagerank** | 0.97878 | -0.05419 | 0.020087 | 0.050246 | 0.978504 | 0.97878 | 1 | 0.914697 |
| **Subgraph** | 0.900631 | 0.12077 | 0.035691 | 0.257558 | 0.900406 | 0.900631 | 0.914697 | 1 |

### *5.5.2.3 Kendall*

The advantage of Kendall test is that this not only provides the relationship between variables, but also provides distribution-free test of independence. "Spearman's rank correlation is satisfactory for testing a null hypothesis of independence between two variables, but it is difficult to interpret when the null hypothesis is rejected.” The Kendall's rank correlation (Equation-5.51) improves upon this by reflecting the strength of the dependence between the variables being compared. Kendall’s rank correlation coefficient depends on concordant, discordant. Concordant ($n_c$) is defined by $x^i > y^i$, else ($x^i$, $y^i$) pair are discordant ($n^d$), and here n is the total number of observations, $n_c$ and $n_d$ are counting of concordant and discordant.

$$r = \frac{n_c - n_d}{n(n-1)/1} \tag{5.51}$$

**Table-5.6:** showing pairwise Kendall’s correlation coefficient (symmetric)

| | **Authority** | **Between** | **Bonpow** | **Closeness** | **Evencen** | **Hub** | **Pagerank** | **Subgraph** |
|---|---|---|---|---|---|---|---|---|
| **Authority** | 1 | -0.07639 | 0.031027 | 0.008746 | 0.998618 | 1 | 0.893236 | 0.751449 |
| **Between** | -0.07639 | 1 | -0.0252 | 0.681275 | -0.07667 | -0.07639 | -0.04549 | 0.090989 |
| **Bonpow** | 0.031027 | -0.0252 | 1 | 0.023437 | 0.032093 | 0.031027 | 0.012295 | 0.022971 |
| **Closeness** | 0.008746 | 0.681275 | 0.023437 | 1 | 0.007388 | 0.008746 | 0.02239 | 0.192801 |
| **Evencen** | 0.998618 | -0.07667 | 0.032093 | 0.007388 | 1 | 0.998618 | 0.891857 | 0.750076 |
| **Hub** | 1 | -0.07639 | 0.031027 | 0.008746 | 0.998618 | 1 | 0.893236 | 0.751449 |
| **Pagerank** | 0.893236 | -0.04549 | 0.012295 | 0.02239 | 0.891857 | 0.893236 | 1 | 0.769953 |
| **Subgraph** | 0.751449 | 0.090989 | 0.022971 | 0.192801 | 0.750076 | 0.751449 | 0.769953 | 1 |

Using **Table**-5.4, **Table**-5.5, and **Table**-5.6 we can interpret that there are eight different measures to find the importance of sentence, and rank it. Using **Table-5.**4 we find that the following pair are highly correlated as, (Authority, Evencen), (Authority, Hub), (Authority, PageRank), (Authority, Subgraph), (Between, Closeness), (Evencen, Hub), (Evencent, PageRank), (Evencen, Subgraph), (Hub, PageRank), (Hub, Subgraph), and (PageRank, Subgraph). Those pairs are highly related, their combination will not give better results in the case when we use its combination.

### 5.5.3 Experiment-5.3

In this experiment, we are hybridizing different centrality measure to maximize the relevance of the sentences, and minimization of redundancy has done by the lexical network. Since, we have eight different measures, so total $2^8$-9=247 hybridize combinations are possible. In limited time constraints, we cannot check all the combination possible. To decide a better combination we are considering experiment 5.2. There we will combine those measures which are minimum correlated because highly correlated will return the same results. To make it simple here we are combining two centrality measures. From experiment 5.1 we can conclude that the best performer is subgraph based centrality. As per our requirement, to get a better combination we can select (subgraph, bonpow) and (subgraph, between) based measures.

**Table-5.7:** Precision, Recall and F-Score of our proposed model when relevance is decided by Hybridizing subgraph and bonpow.

| Rouge | R-Lexical | P-Lexical | F-Lexical | 95% Confidence – R (Lexical) | 95% Confidence- P (Lexical) | 95% Confidence- F (Lexical) |
|---|---|---|---|---|---|---|
| **ROUGE-1** | .290 | .344 | .314 | .24576-.33910 | .29324-.39869 | .26766-.36659 |
| **ROUGE-2** | .084 | .100 | .091 | .05347-.11740 | 06520-.13880 | .05875-.12763 |
| **ROUGE-L** | .25488 | .303 | .276 | .209-30464 | .25280-.35750 | .22812-.32929 |
| **ROUGE-W-1.2** | .093 | .206 | .128 | .07549-.11295 | .16923-.24692 | .10430-.15471 |
| **ROUGE-S*** | .072 | .102 | .084 | .05466-.10371 | .07804-.14160 | .06427-.11919 |
| **ROUGE-SU*** | .078 | .109 | .0911 | .05466-.10371 | .07804-.14160 | .06427-.11919 |

**Table-5.8:** Precision, Recall and F-Score of our proposed model when relevance is decided by Hybridizing subgraph and Betweenness.

| Rouge | R-Lexical | P-Lexical | F-Lexical | 95% Confidence – R (Lexical) | 95% Confidence- P (Lexical) | 95% Confidence- F (Lexical) |
|---|---|---|---|---|---|---|
| **ROUGE-1** | .297 | .351 | .322 | .25346-.34417 | .301602-.40285 | .27535-.37146 |
| **ROUGE-2** | .087 | .102 | .094 | .05674-.12082 | .06728-.14227 | .06162-.13108 |
| **ROUGE-L** | .261 | .308 | .283 | .21880-.30836 | .25868-.36104 | .06162-.13108 |
| **ROUGE-W-1.2** | .096 | .210 | .131 | .08005-.11406 | .17564-.24742 | .10991-.15596 |
| **ROUGE-S*** | .077 | .106 | .089 | .05335-.10138 | .07548-.13998 | .06263-.11779 |
| **ROUGE-SU*** | .082 | .113 | .095 | .05823-.10698 | .08196-.14718 | .06816-.12402 |

In Table 5.7 and 5.8, we have presented, Precision, Recall and F-Score of hybridized feature subgraph+bonpow and subgraph+between. Figure 5.4 representing, comparative performance of hybridize centrality w.r.t. independent centrality measure. From, Table-5.7, 5.8, and Figure-5.7 we can conclude that hybridize relevance measure returning better ROUGE-1, S*, SU* and for ROUGE-L, ROUGE-2, ROUGE-W-1.2 highest performance derived by subgraph-based centrality.

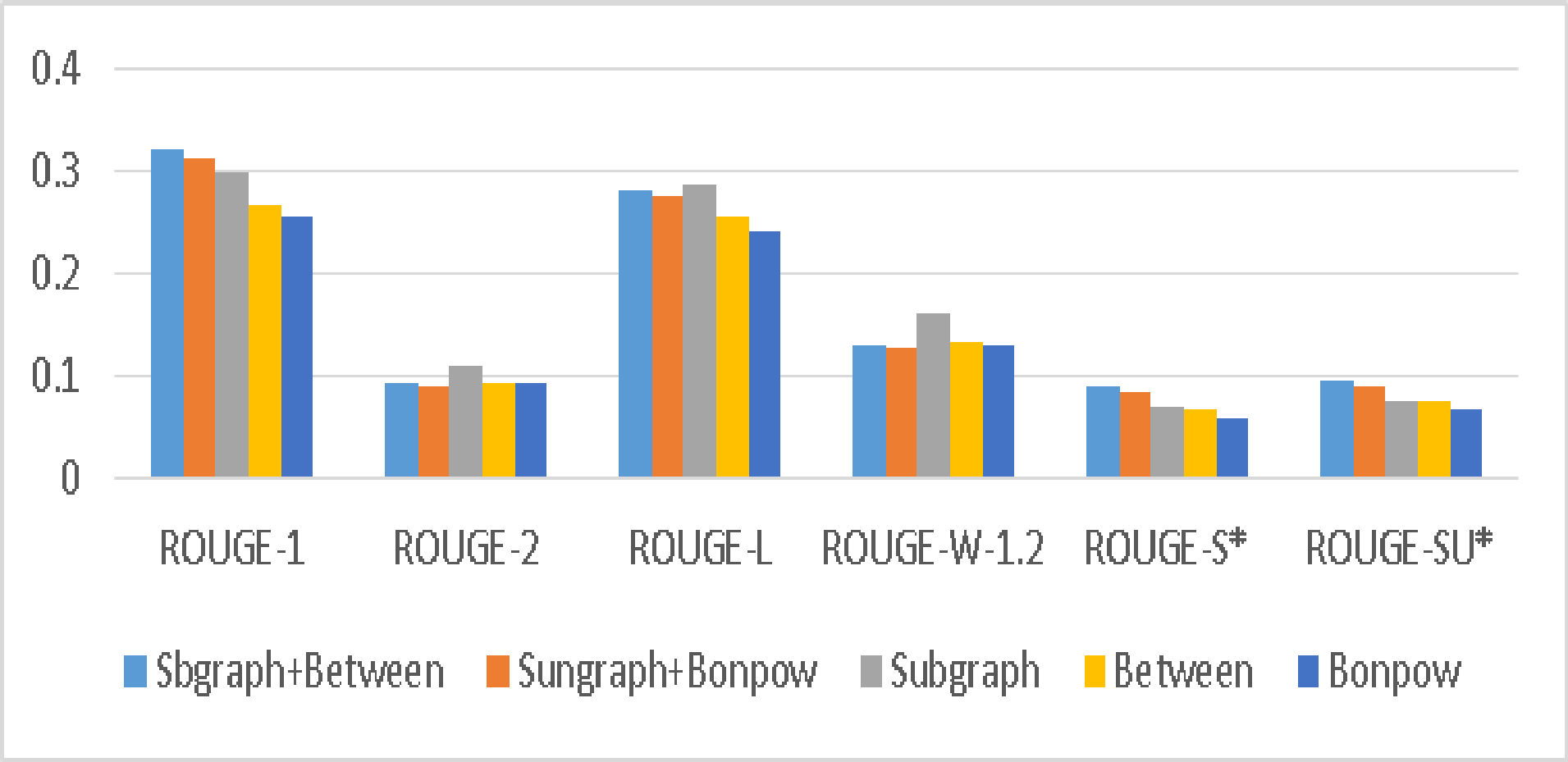

Figure 5. 7: Graph showing comparative performance of different relevance-based measure, where + sign denoting hybridization of different features.

## 5.6 Concluding Remark

In this Chapter, we have proposed optimization-based summarization that guarantees maximize coverage and minimum redundancy. We have tested our model by selecting centrality for coverage/relevance, and lexical network as a function of redundancy between sentences. We have compared our approach with cosine similarity-based redundancy measure between sentences. In both approaches (proposed and baseline) relevance measure has decided by centrality score.

We have performed three experiments, in the first experiments we have implemented the baseline model using subgraph and PageRank based centrality measure used for relevance measure, and cosine similarity matrix for redundancy. In this work, we have implemented all the centrality measure as relevance and lexical network for redundancy. Where we have found out that subgraph-based centrality is giving better results. Even, in the baseline model when subgraph centrality is used, it is giving respectable results. In the second experiment we have found out the correlation between the summary results when different centrality-based measure used for relevance and lexical network used for redundancy. Here we have suggested that the high correlation pair is not suitable for hybridization, and the low correlation pair is expected for better results. In the third experiment, we have hybridized different centrality measures, and improved results are shown by Tables and Figures. In the third experiment several feature combinations are possible, but to our limitation, we have tested only limited possibility.

# Chapter 6: Conclusion and Future work

This work our focus is on an important aspect of information retrieval's task automatic text document summarization. This work is divided into five chapters. The first chapter gives a brief introduction about text summarization, evaluating techniques and datasets on which this wide range of models developed and experimented.

In the second chapter "Hybrid Approach for Single Text Document Summarization using Statistical and Sentiment Features", we have presented a linear combination of different statistical measures and semantic measures. In our hybrid approach, we have taken statistical measures like sentence position, centroid, TF-IDF, and word level analysis based semantic approach that is sentiment analysis. The sentiment score of a sentence is computed as the sum of the sentiment score of every entity present in the sentence. Since, for any entity sentiment score have three polarities as Neutral, Negative and Positive. We are more interested in such kind of sentences those have a high semantic score either negative or positive. So, if entity sentiment is negative, then we are multiplying it by minus one to treat it as a positive score.

To generate several summaries of different length, we have used different approaches like MEAD, Microsoft, OPINOSIS and human based. In this chapter, we have done four experiments. In the first experiment, we have considered our summary (generate from a proposed algorithm) as a system summary and all others as a model summary. After evaluating this has shown that we are getting high precision almost every time, that denotes we covered most relevant results. In the second experiment, we have compared different system generated summary (MEAD, Microsoft, OPINOSIS, our algorithm) to the Model summary (human-generated). In this we find that our explained algorithm performed well for 24% generated summary for almost every time but, in 40% MEAD system generates a summary leading in some way but here also we are getting higher RECALL to compare to MEAD. The third experiment is showing the contribution of sentiment score in the selection of most informative sentences. We have shown that when we are adding sentiment score as a feature we are getting improved results to compare to without a sentiment score.

Initially, for all experiments have done by assigning equal importance for every feature. To score a sentence we took the sum of all the feature's score and pickup highest score sentence and added that into the summary. In the next step, we are selecting next sentence based on the next highest score and add it to the summary if, the similarity between summary and sentence is lower that threshold to maintain redundancy and coverage. We will repeat this iterative process until the desired length summary is achieved. In experiment four we have extended this approach. We have suggested, and tested a better combination of feature weights. Here parameter estimation is done by regression and random forest.

In third chapter “A new Latent Semantic Analysis and Entropy-based Approach for Automatic Text Document Summarization”, we have proposed two new approaches (three new models) for automatic text document summarization and a novel Entropy-based approach for summary evaluation. Both the approaches for summary generation is based on SVD based decomposition. In the first approach (proposed_model-1), we are using right singular matrix “$V^T$” for processing and selects a concept one by one (top to bottom till required). Previous approaches are focused on selecting only one sentence of the highest information content. In our approach, we are selecting two sentences w.r.t each concept such that sentence-1 is highest related to concept and sentence-2 least related to the concept. This approach is based on assumption that by doing this we are covering two different topics. As a result, it leads to more coverage and diversity. In future, we can increase the selection of number of sentences at a glass. The second approach is based on Entropy, which formulate into two different models (proposed_model-2 and proposed_model-3). In proposed_model-2, first we are selecting a higher informative concept and from that concept, we are selecting summary sentences. In proposed_model-3 repeatedly we are selecting highest informative sentences, i.e. a sentence which is related to all the concepts with a high score. The advantage of the Entropy-based model is that these are not length dominating models, giving a better ROUGE score, statistically closer to standard/gold summary.

During an experiment, we have found out that ROUGE score depends only on the count of matched words, an increasing the summary length, sometimes ROUGE score decreases, and on increasing redundancy ROUGE score also increases. We have pointed out that ROUGE score doesn’t measure redundancy i.e. count matched sentences. We have also realized the need of a new measure for summary evaluation that provide a tradeoff between redundancy & countmatch, and Entropy-based criteria are proposed. During testing of the new proposed measure on different summary generated by previous models, and our proposed models we have find that our entropy based summary is closer to standard summary. From the experiment results it is clear that our model works well for summary evaluation (especially for higher length summary), because as summary length increases redundancy also increases and, in this measure, we are measuring redundancy. Currently we are giving equal importance to all n-gram, but theoretically and practically we should give more weight to higher n-gram because of high redundancy of information (in case of repetition). In future we may assign different feature weights to get better results.

In fourth chapter "LexNetwork based summarization and a study of: Impact of WSD Techniques, and Similarity Threshold over LexNetwork", we have presented a Lexical Network concept for Automatic Text Document summarization. This approach is a little bit different from previously proposed lexical chain based techniques. In previous techniques author concentrate to create a number of lexical chains, that creates ambiguity which chain to prefer, even this problem efficiently can handle with chain scoring techniques. Still, lexical chains have problem that if one particular word *let "Donald Trump"* is coming in two or more sentences then which sentence to prefer. Another problem with this technique that they consider only nearby sentences for a chain construction i.e. window of two or three sentences was selected, so this is unable to handle long term relationship between sentences. Our Lex-Net Handle long term relationship between sentences. Nodes are scored and high score sentence given priority. So, both problems are handled in this way.

Our Lexical network is based on the number of Lexical and semantic relations. To decide the importance of sentences we have used centrality-based measure on Lexical Network. Since human language is highly ambiguous (here English) so we need to find a correct sense of a particular word in given context. The solution to this ambiguity problem done with Simplified Lesk, Cosine Lesk, and Adapted Lesk Algorithm. In this work, we have studied the impact of WSD techniques and cosine similarity threshold (Θ= 0%, 5%, 10%, ….35%,40%). Less value of Θ represents more divesity compare to high valued. More diversity (if Θ is less) is not good for summary because it will maintain diversity, but the less relatedness between sentences which is harming good summary property. For Comparison purpose we are used Semantrica-Lexalytics algorithm in which we find that System proposed by us is working better many times. From number of experiments we have find out that subgraph based centrality is performing best among all.

From this work we have reached on number of conclusions, (1) for alpha centrality when alpha is .1 to .5 (inclusion) the performance of summarizer system is arbitrary up and down, but after that

alpha= .6 to .9 (inclusion) for all centrality measures, for different value of threshold performance is continuously increasing (some time exception at alpha=0.7), and again reduced at alpha=1; (2) From a set of cosine similarity as 0% , 5%, 10%,...35%, 40%. We are suggesting that 10 % similarity threshold is better to get enough diversity, and better summary (as per Rouge Score); (3) Hub/Authority based ranking is same as Eigenvalue based centrality; (4) Subgraph based centrality measure is performing better to all, the reason of this is higher score for small subgraph which recognizes small subgraph and this cover various subgraph (can be considered as cluster-like structure); (5) We are not suggesting any particular WSD is better all time.

During LexNetwork creation, we have used lexical, semantic relations, and in this work we have assigned equal weight assigns for each relation presents between sentences. In literature, different priority assigned to all relations. In the future that may be considered for Network creation, and it may improve the ranking of sentences. During the experiment, we have find out some corelation between Eigen Value based centrality and Authority/ Hub based centrality measures, at present this is not objective of this work why it is, in future, we will try to answer this.

In fifth chapter "Modeling Automatic Text Document Summarization as multi objective optimization", we have proposed optimization-based summarization that guarantee maximize coverage and minimum redundancy. We have tested our model by selecting coverage as centrality-based score, and similarity function as relatedness between sentences. We have compared our proposed lexnetwork based approach with cosine similarity-based redundancy, and relevance has been measured by centrality based measures in both the approaches. In this work, we have performed three experiments, in the first experiments we have implemented the baseline model using subgraph and PageRank based centrality measure used for relevance measure, and cosine similarity matrix for redundancy. In this work, we have implemented all the centrality measure as relevance and lexical network for redundancy. Where we have found out that subgraph-based centrality is giving better results. Even, in the baseline model when subgraph centrality is used, it is giving respectable results. In the second experiment we have found out the correlation between the summary results when different centrality-based measure used for relevance and lexical network used for redundancy. Here we have suggested that the high correlation pair is not suitable for hybridization, and the low correlation pair is expected for better results. In the third experiment, we have hybridized different centrality measures, and improved results are shown by Tables and Figures. In future this work, can be extended to other combinations, and an aggregate centrality-based model may be proposed and tested along with optimal combination using soft computing techniques.

# References

Alguliev, R. M., Aliguliyev, R. M., & Hajirahimova, M. S. (2012). Expert Systems with Applications GenDocSum + MCLR : Generic document summarization based on maximum coverage and less redundancy. *Expert Systems With Applications*, *39*(16), 12460–12473. https://doi.org/10.1016/j.eswa.2012.04.067

Alguliev, R. M., Aliguliyev, R. M., Hajirahimova, M. S., & Mehdiyev, C. A. (2011). Expert Systems with Applications MCMR : Maximum coverage and minimum redundant text summarization model. *Expert Systems With Applications*, *38*(12), 14514–14522. https://doi.org/10.1016/j.eswa.2011.05.033

Alguliev, R. M., Aliguliyev, R. M., & Isazade, N. R. (2012). DESAMC+ DocSum: Differential evolution with self-adaptive mutation and crossover parameters for multi-document summarization. *Knowledge-Based Systems*, *36*, 21–38.

Alguliev, R. M., Aliguliyev, R. M., & Isazade, N. R. (2013a). Expert Systems with Applications CDDS : Constraint-driven document summarization models. *Expert Systems With Applications*, *40*(2), 458–465. https://doi.org/10.1016/j.eswa.2012.07.049

Alguliev, R. M., Aliguliyev, R. M., & Isazade, N. R. (2013b). Formulation of document summarization as a 0–1 nonlinear programming problem. *Computers & Industrial Engineering*, *64*(1), 94–102.

Alguliev, R. M., Aliguliyev, R. M., & Mehdiyev, C. A. (2011). Sentence selection for generic document summarization using an adaptive differential evolution algorithm. *Swarm and Evolutionary Computation*, *1*(4), 213–222. https://doi.org/10.1016/j.swevo.2011.06.006

Balaji, J., Geetha, T. V, & Parthasarathi, R. (2016). Abstractive summarization: A hybrid approach for the compression of semantic graphs. *International Journal on Semantic Web and Information Systems (IJSWIS)*, *12*(2), 76–99.

Banerjee, S., & Pedersen, T. (2002). An adapted Lesk algorithm for word sense disambiguation using WordNet. In *International conference on intelligent text processing and computational linguistics* (pp. 136–145). Springer.

Barzilay, R., & Elhadad, M. (1997). Using Lexical Chains for Text Summarization.

Baxendale, P. B. (1958). Machine-made index for technical literature—an experiment. *IBM Journal of Research and Development*, *2*(4), 354–361.

Beliga, S., Meštrović, A., & Martinčić-Ipšić, S. (2016). Selectivity-based keyword extraction method. *International Journal on Semantic Web and Information Systems (IJSWIS)*, *12*(3), 1–26.

Bonacich, P. (1987). Power and centrality: A family of measures. *American Journal of Sociology*, *92*(5), 1170–1182.

Boulesteix, A., Janitza, S., Kruppa, J., & König, I. R. (2012). Overview of random forest methodology and practical guidance with emphasis on computational biology and bioinformatics. *Wiley Interdisciplinary Reviews: Data Mining and Knowledge Discovery*, *2*(6), 493–507.

Breiman, L. (2001). Random forests. *Machine Learning*, *45*(1), 5–32.

Chen, Y.-N., Huang, Y., Yeh, C.-F., & Lee, L.-S. (2011). Spoken lecture summarization by random walk over a graph constructed with automatically extracted key terms. In *Twelfth Annual Conference of the International Speech Communication Association*.

Chen, Y., Liu, B., & Wang, X. (2007). Automatic text summarization based on textual cohesion. *Journal of Electronics (China)*, *24*(3), 338–346.

Chen, Y., Wang, X., & Guan, Y. (2005). Automatic text summarization based on lexical chains. In *International Conference on Natural Computation* (pp. 947–951). Springer.

Chiru, C.-G., Rebedea, T., & Ciotec, S. (2014). Comparison between LSA-LDA-Lexical Chains. In *WEBIST (2)* (pp. 255–262).

Creation, A., & Abstracts, L. (1958). The Automatic Creation of Literature Abstracts *, (April), 159–165.
Deerwester, S., Dumais, S. T., Furnas, G. W., Landauer, T. K., & Harshman, R. (1990). Indexing by latent semantic analysis. *Journal of the American Society for Information Science*, *41*(6), 391–407.
Dolbear, C., Hobson, P., Vallet, D., Fernández, M., Cantadorz, I., & Castellsz, P. (2008). Personalised multimedia summaries. In *Semantic Multimedia and Ontologies* (pp. 165–183). Springer.
Doran, W., Stokes, N., Carthy, J., & Dunnion, J. (2004). Comparing lexical chain-based summarisation approaches using an extrinsic evaluation. *GWC 2004*, *112*.
Edmundson, H. P. (n.d.). New Methods in Automatic Extracting, *16*(2).
Elhadad, M. (2012). No Title. Retrieved from https://cs-serv.cs.bgu.ac.il/cs_service/subsys.html
Ercan, G., & Cicekli, I. (2007). Using lexical chains for keyword extraction. *Information Processing & Management*, *43*(6), 1705–1714.
Erekhinskaya, T. N., & Moldovan, D. I. (2013). Lexical Chains on WordNet and Extensions. In *FLAIRS Conference*.
Estrada, E., & Rodriguez-Velazquez, J. A. (2005). Subgraph centrality in complex networks. *Physical Review E*, *71*(5), 56103.
Freeman, L. C. (1977). A set of measures of centrality based on betweenness. *Sociometry*, 35–41.
Freeman, L. C. (1978). Centrality in social networks conceptual clarification. *Social Networks*, *1*(3), 215–239.
Ganapathiraju, K., Carbonell, J., & Yang, Y. (2002). Relevance of Cluster size in MMR based Summarizer: A Report 11-742: Self-paced lab in Information Retrieval.
Ganesan, K., Zhai, C., & Han, J. (2010). Opinosis : A Graph-Based Approach to Abstractive Summarization of Highly Redundant Opinions, (August), 340–348.
Gleich, D. F. (2015). PageRank beyond the Web. *SIAM Review*, *57*(3), 321–363.
Goldstein, J., Mittal, V., Carbonell, J., & Callan, J. (2000). Creating and evaluating multi-document sentence extract summaries. In *Proceedings of the ninth international conference on Information and knowledge management* (pp. 165–172). ACM.
Gong, Y., & Liu, X. (2001). Creating Generic Text Summaries, 903–907.
Gonzàlez, E., & Fort, M. F. (2009). A New Lexical Chain Algorithm Used for Automatic Summarization. In *CCIA* (pp. 329–338).
Gurevych, I., & Nahnsen, T. (2005). Adapting lexical chaining to summarize conversational dialogues. In *Proceedings of the Recent Advances in Natural Language Processing Conference* (pp. 287–300).
Hahn, U., & Mani, I. (2000). The challenges of automatic summarization. *Computer*, *33*(11), 29–36.
Halliday, M. A. K., & Hasan, R. (1976). Cohesion in. *English, Longman, London*.
Hariharan, S. (2010). Multi document summarization by combinational approach. *International Journal of Computational Cognition*, *8*(4), 68–74.
Hoskinson, A. (2005). Creating the ultimate research assistant. *Computer*, *38*(11), 97–99.
Hovy, E., & Lin, C. (1999). Automated Text Summarization in SUMMARIST.
Jagadeesh, J., Pingali, P., & Varma, V. (2005). Sentence extraction based single document summarization. *International Institute of Information Technology, Hyderabad, India*, *5*.
Karanikolas, N. N., & Galiotou, E. (2012). A workbench for extractive summarizing methods. https://doi.org/10.1109/PCi.2012.67
Katz, L. (1953). A new status index derived from sociometric analysis. *Psychometrika*, *18*(1), 39–43.
Katz, S. M. (1996). Distribution of content words and phrases in text and language modelling. *Natural Language Engineering*, *2*(1), 15–59.
Kim, J.-H., Kim, J.-H., & Hwang, D. (2000). Korean text summarization using an aggregate similarity. In *Proceedings of the fifth international workshop on on Information retrieval with Asian languages* (pp. 111–118). ACM.
Kleinberg, J. M. (1999). Authoritative sources in a hyperlinked environment. *Journal of the ACM (JACM)*,

*46*(5), 604–632.
Koschützki, D., Lehmann, K. A., Peeters, L., Richter, S., Tenfelde-Podehl, D., & Zlotowski, O. (2005). Centrality indices. In *Network analysis* (pp. 16–61). Springer.
Kulkarni, A. R., & Apte, S. S. (2014). An Automatic Text Summarization using lexical cohesion and correlation of sentences. *International Journal of Research in Engineering and Technology*, *3*(06).
Kupiec, J., Pedersen, J., & Chen, F. (1995). A trainable document summarizer. In *Proceedings of the 18th annual international ACM SIGIR conference on Research and development in information retrieval* (pp. 68–73). ACM.
Lesk, M. (1986). Automatic sense disambiguation using machine readable dictionaries: how to tell a pine cone from an ice cream cone. In *Proceedings of the 5th annual international conference on Systems documentation* (pp. 24–26). ACM.
Lin, C.-Y. (2004). Rouge: A package for automatic evaluation of summaries. *Text Summarization Branches Out*.
Lin, C., & Rey, M. (2004). R OUGE : A Package for Automatic Evaluation of Summaries, (1).
Luhn, H. P. (1958). The automatic creation of literature abstracts. *IBM Journal of Research and Development*, *2*(2), 159–165.
Luo, W., Zhuang, F., He, Q., & Shi, Z. (2010). Effectively Leveraging Entropy and Relevance for Summarization. In *Asia Information Retrieval Symposium* (pp. 241–250). Springer.
Mani, I., & Maybury, M. T. (1999). Advances in Automatic Text Summarization Reviewed by Mark Sanderson University of Sheffield, *26*(2), 280–281.
Mani, I., & Maybury, M. T. (2001). Automatic summarization.
McDonald, R. (2007). A study of global inference algorithms in multi-document summarization. In *European Conference on Information Retrieval* (pp. 557–564). Springer.
McKeown, K., Barzilay, R., Chen, J., Elson, D., Evans, D., Klavans, J., … Sigelman, S. (2003). Columbia's newsblaster: new features and future directions. *Companion Volume of the Proceedings of HLT-NAACL 2003-Demonstrations*.
Medelyan, O. (2007). Computing lexical chains with graph clustering. In *Proceedings of the 45th Annual Meeting of the ACL: Student Research Workshop* (pp. 85–90). Association for Computational Linguistics.
Morris, J., & Hirst, G. (1991). Lexical cohesion computed by thesaural relations as an indicator of the structure of text. *Computational Linguistics*, *17*(1), 21–48.
Murray, G., Renals, S., & Carletta, J. (2005). Extractive summarization of meeting recordings.
Newman, M. E. J. (2008). The mathematics of networks. *The New Palgrave Encyclopedia of Economics*, *2*(2008), 1–12.
normalisation statistics. (2018). wikipedia.
Ou, S., Khoo, C. S. G., & Goh, D. H.-L. (2009). Automatic text summarization in digital libraries. In *Handbook of Research on Digital Libraries: Design, Development, and Impact* (pp. 159–172). IGI Global.
Ouyang, Y., Li, W., Lu, Q., & Zhang, R. (2010). A study on position information in document summarization. In *Proceedings of the 23rd international conference on computational linguistics: Posters* (pp. 919–927). Association for Computational Linguistics.
Ozsoy, M. G., Alpaslan, F. N., & Cicekli, I. (2011). Journal of Information Science, (June). https://doi.org/10.1177/0165551511408848
Ozsoy, M. G., Cicekli, I., & Alpaslan, F. N. (2010). Text summarization of turkish texts using latent semantic analysis. In *Proceedings of the 23rd international conference on computational linguistics* (pp. 869–876). Association for Computational Linguistics.
PadmaLahari, E., Kumar, D. V. N. S., & Prasad, S. (2014). Automatic text summarization with statistical and linguistic features using successive thresholds. In *Advanced Communication Control and*

*Computing Technologies (ICACCCT), 2014 International Conference on* (pp. 1519–1524). IEEE.
Page, L., Brin, S., Motwani, R., & Winograd, T. (1999). *The PageRank citation ranking: Bringing order to the web.* Stanford InfoLab.
Plaza, L., Stevenson, M., & Díaz, A. (2012). Resolving ambiguity in biomedical text to improve summarization. *Information Processing & Management*, *48*(4), 755–766.
Pourvali, M., & Abadeh, M. S. (2012). Automated text summarization base on lexicales chain and graph using of wordnet and wikipedia knowledge base. *ArXiv Preprint ArXiv:1203.3586*.
Precision and recall. (n.d.).
Radev, D. R., Blair-Goldensohn, S., & Zhang, Z. (2001). Experiments in single and multi-document summarization using MEAD. *Ann Arbor*, *1001*(48109).
Radev, D. R., Hovy, E., & McKeown, K. (2002). Introduction to the special issue on summarization. *Computational Linguistics*, *28*(4), 399–408.
Radev, D. R., Jing, H., Sty, M., & Tam, D. (2004). Centroid-based summarization of multiple documents, *40*, 919–938. https://doi.org/10.1016/j.ipm.2003.10.006
Rambow, O., Shrestha, L., Chen, J., & Lauridsen, C. (2004). Summarizing email threads. In *Proceedings of HLT-NAACL 2004: Short Papers* (pp. 105–108). Association for Computational Linguistics.
Rautray, R., Balabantaray, R. C., & Bhardwaj, A. (2015). Document summarization using sentence features. *International Journal of Information Retrieval Research (IJIRR)*, *5*(1), 36–47.
Roul, R. K., Sahoo, J. K., & Goel, R. (2017). Deep Learning in the Domain of Multi-Document Text Summarization. In *International Conference on Pattern Recognition and Machine Intelligence* (pp. 575–581). Springer.
Sakai, T., & Sparck-Jones, K. (2001). Generic summaries for indexing in information retrieval. In *Proceedings of the 24th annual international ACM SIGIR conference on Research and development in information retrieval* (pp. 190–198). ACM.
Sankarasubramaniam, Y., Ramanathan, K., & Ghosh, S. (2014). Text summarization using Wikipedia. *Information Processing and Management*, *50*(3), 443–461. https://doi.org/10.1016/j.ipm.2014.02.001
Sarkar, K. (2010). Syntactic trimming of extracted sentences for improving extractive multi-document summarization. *Journal of Computing*, *2*(7), 177–184.
Shannon, C. E. (1948). A Mathematical Theory of Communication, Bell System Technical Journal, vol. 27, 379-423 & 623-656, July & October.
Sharan, A., Siddiqi, S., & Singh, J. (2015). Keyword Extraction from Hindi Documents Using Statistical Approach. In *Intelligent Computing, Communication and Devices* (pp. 507–513). Springer.
Shimada, K., Tadano, R., & Endo, T. (2011). Multi-aspects review summarization with objective information, *27*(Pacling), 140–149. https://doi.org/10.1016/j.sbspro.2011.10.592
Silber, H. G., & McCoy, K. F. (2000). An efficient text summarizer using lexical chains. In *Proceedings of the first international conference on Natural language generation-Volume 14* (pp. 268–271). Association for Computational Linguistics.
Sood, A. (2013). Towards summarization of written text conversations. *International Institute of Information Technology, India*.
Steinberger, J., & Ježek, K. (2004). Text summarization and singular value decomposition. In *International Conference on Advances in Information Systems* (pp. 245–254). Springer.
Steinberger, J., Poesio, M., Kabadjov, M. A., & Jez, K. (2007). Two uses of anaphora resolution in summarization, *43*, 1663–1680. https://doi.org/10.1016/j.ipm.2007.01.010
Stokes, N. (2004). Applications of lexical cohesion analysis in the topic detection and tracking domain. University College Dublin Department of Computer Science.
Takale, S. A., Kulkarni, P. J., & Shah, S. K. (2016). An Intelligent Web Search Using Multi-Document Summarization. *International Journal of Information Retrieval Research (IJIRR)*, *6*(2), 41–65.

Tan, L. (2013). Examining crosslingual word sense disambiguation. *Nanyang Technological University, Nanyang Avenue*.

Tofighy, S. M., Raj, R. G., & Javad, H. H. S. (2013). AHP techniques for Persian text summarization. *Malaysian Journal of Computer Science*, *26*(1), 1–8.

Tombros, A., & Sanderson, M. (1998). Advantages of query biased summaries in information retrieval. In *Proceedings of the 21st annual international ACM SIGIR conference on Research and development in information retrieval* (pp. 2–10). ACM.

Torres-Moreno, J.-M. (2014). *Automatic text summarization*. John Wiley & Sons.

Vechtomova, O., Karamuftuoglu, M., & Robertson, S. E. (2006). On document relevance and lexical cohesion between query terms. *Information Processing & Management*, *42*(5), 1230–1247.

W. N. Venables, D. M. S., & Team,  and the R. C. (2018). *An Introduction to R*.

Wan, X. (2008). Using only cross-document relationships for both generic and topic-focused multi-document summarizations. *Information Retrieval*, *11*(1), 25–49.

White, R. W., Jose, J. M., & Ruthven, I. (2003). A task-oriented study on the influencing effects of query-biased summarisation in web searching. *Information Processing & Management*, *39*(5), 707–733.

Xiong, S., & Ji, D. (2016). Query-focused multi-document summarization using hypergraph-based ranking. *Information Processing & Management*, *52*(4), 670–681.

Yadav, C. S., & Sharan, A. (2015). Hybrid approach for single text document summarization using statistical and sentiment features. *International Journal of Information Retrieval Research (IJIRR)*, *5*(4), 46–70.

Yadav, C. S., Sharan, A., & Joshi, M. L. (2014). Semantic graph based approach for text mining. In *Proceedings of the 2014 International Conference on Issues and Challenges in Intelligent Computing Techniques, ICICT 2014*. https://doi.org/10.1109/ICICICT.2014.6781348

Yadav, C. S., Sharan, A., Kumar, R., & Biswas, P. (2016). *A new approach for single text document summarization*. *Advances in Intelligent Systems and Computing* (Vol. 380). https://doi.org/10.1007/978-81-322-2523-2_39

Yeh, J.-Y., Ke, H.-R., Yang, W.-P., & Meng, I.-H. (2005). Text summarization using a trainable summarizer and latent semantic analysis. *Information Processing & Management*, *41*(1), 75–95.

Yeh, J. (2005). Text summarization using a trainable summarizer and latent semantic analysis q, *41*, 75–95. https://doi.org/10.1016/j.ipm.2004.04.003

Zhang, R., Li, W., Gao, D., & Ouyang, Y. (2013). Automatic twitter topic summarization with speech acts. *IEEE Transactions on Audio, Speech, and Language Processing*, *21*(3), 649–658.